\documentclass[apj]{emulateapj}
\usepackage{natbib}
\usepackage{graphics}

\shorttitle{M31 INTEGRATED LIGHT ABUNDANCES}
\shortauthors{COLUCCI ET AL.}

\begin{document}

\defcitealias{2008ApJ...684..326M}{MB08}
\newcommand{\msol}{M_\odot}
\newcommand{\etal}{et al.\ }
\newcommand{\kms}{km~s$^{-1}$}
\newcommand{\rkms}{km~s$^{-1}$ \enskip}
\newcommand{\rAA}{{\AA \enskip}}
\newcommand{\ew}{W_\lambda}
\newcommand{\II}{\small{II} \normalsize}
\newcommand{\I}{\small{I} \normalsize}
\submitted{Received 2009 March 23, Accepted 2009 August 26}

\title{M31 Globular Cluster Abundances from High-Resolution, 
Integrated-Light Spectroscopy\footnotemark[1]}

\footnotetext[1]{The data presented herein were obtained at the W.M. Keck Observatory, which is operated as a scientific partnership among the California Institute of Technology, the University of California and the National Aeronautics and Space Administration. The Observatory was made possible by the generous financial support of the W.M. Keck Foundation.}

\author{Janet E. Colucci}
\affil{Astronomy Department, 500 Church Street,University of Michigan, Ann Arbor, MI 48109--1090;\\ and  UCO/Lick Observatory, University of California, Santa Cruz, CA 95064; jcolucci@ucolick.org}
%\email{jcolucci@ucolick.org}

\author{Rebecca A. Bernstein}
\affil{Department of Astronomy and Astrophysics, 1156  High Street, UCO/Lick Observatory,\\
 University of California, Santa Cruz, CA 95064; rab@ucolick.org}
%\email{rab@ucolick.org}

\author{Scott Cameron}
\affil{Astronomy Department, 500 Church Street, University of Michigan, Ann Arbor, MI 48109--1090; sacamero@umich.edu}
%\email{sacamero@umich.edu}

\author{Andrew McWilliam}
\affil{The Observatories of the Carnegie Institute of Washington, 813 Santa Barbara Street, Pasadena, CA 91101-1292; andy@ociw.edu}
%\email{andy@ociw.edu}

\and

\author{Judith G. Cohen}
\affil{Palomar Observatory, Mail Stop 105-24, California Institute of Technology, Pasadena, CA 91125; jlc@astro.caltech.edu}
%\email{jlc@astro.caltech.edu}

\begin{abstract}
We report the first detailed chemical abundances for 5 globular clusters (GCs) in M31 from high-resolution ($R\sim$ 25,000) spectroscopy of their integrated light. These GCs are the first in a larger set of clusters observed as part of an ongoing project to study the formation history of M31 and its globular cluster population. The data presented here were obtained with the HIRES echelle spectrograph on the Keck I telescope, and are analyzed using a  new integrated light spectra analysis method that we have developed. In these clusters, we  measure abundances for Mg, Al, Si, Ca, Sc, Ti, V, Cr, Mn, Fe, Co, Ni, Y, and Ba,   ages $\geq$10 Gyrs, and  a range in [Fe/H] of $-0.9$ to $-2.2$.  As is typical of Milky Way GCs, we find these M31 GCs to be enhanced in the $\alpha$-elements Ca, Si, and Ti relative to Fe.    We also find [Mg/Fe] to be low relative to other [$\alpha$/Fe], and  [Al/Fe] to be enhanced in the integrated light abundances. 
  These  results imply that abundances of Mg, Al (and likely O, Na) recovered from integrated light do display the inter- and intra-cluster abundance variations seen in individual Milky Way GC stars, and that special care should be taken in the future in interpreting low or high resolution integrated light abundances of globular clusters that are based on Mg-dominated absorption features.  Fe-peak and the neutron-capture elements Ba and Y also follow Milky Way abundance trends.
  We also present high-precision velocity dispersion measurements for all 5 M31 GCs, as well as independent constraints on the reddening toward the clusters from our analysis.

\end{abstract}


\section{Introduction}
\setcounter{footnote}{1}

\label{sec:introduction}

Stellar atmospheres largely retain the chemical composition of the gas
from which they formed, and thus contain a record of the gas chemistry
of a galaxy throughout its star formation history.  The abundances
relative to Fe of some key elements, $\alpha$-elements in particular,
can be used to identify the timescales and rates of star formation
over the lifetime of a galaxy. $\alpha-$elements (C, O, Ne, Mg, Si, S,
Ar, Ca, and Ti) are produced primarily in type II supernovae (SNII)
\citep[e.g.][]{1995ApJS..101..181W} from massive stars and so build
up quickly during active star formation epochs, while Fe-peak elements
are produced in all supernovae.  [$\alpha$/Fe]  abundance patterns  are therefore particularly useful and  have been central to developing our current picture of the
assembly and star formation history of  the
Milky Way  and spiral
galaxies in general  \citep[see][and references therein]{2005AJ....130.2140P}.

Bright, young stars record the current gas phase abundance in a
galaxy; to probe the earliest formation times, one must target older,
lower mass, fainter stars.  It is only recently that individual red
giant branch (RGB) stars in our nearest neighbor galaxies in the Local
Group have been within reach of the high resolution spectroscopy
needed for detailed chemical abundance analysis.  These stars in Local
Group dwarf galaxies show a much greater range of abundance ratios at
all metallicities compared to stars in the Milky Way halo
\citep{2004AJ....128.1177V,2001ApJ...548..592S,2003AJ....125..684S,2005AJ....129.1428G,2009arXiv0904.4505T},
suggesting that they have had a much more complicated star formation
history than the halo's progenitor(s).  Detailed
abundances beyond the Milky Way and its nearest remaining neighbors
are needed to establish the broader patterns of star formation
in galaxies of all masses.

\begin{deluxetable*}{rrrrrrrrrr}
\centering
\tabletypesize{\small} 
\tablecolumns{10}
\tablewidth{0pc}
\tablecaption{M31 Clusters \label{tab:clusterdata}}
\tablehead{
\colhead{Cluster} & \colhead{RA}   & \colhead{Dec} & \colhead{V}    & \colhead{E(B$-$V)\tablenotemark{a}} & \colhead{$M_{Vtot}$\tablenotemark{c}}&  \colhead{R$_{gc}$\tablenotemark{d}}&  \colhead{[Fe/H]\tablenotemark{e}} & \colhead{[$\alpha$/Fe]\tablenotemark{f}} & \colhead{HBR\tablenotemark{g}} \\ \colhead{}&\colhead{(2000)}&\colhead{(2000)}&\colhead{}&\colhead{}&\colhead{}&\colhead{(kpc)}&\colhead{}&\colhead{} &\colhead{}     }
\startdata

G108-B045 & 00 41 43.26 & 41 34 21.8 & 15.83  & 0.10&$-$8.95	&4.87  & $-$0.94$\pm$0.27 & 0.22 $\pm$ 0.19 & 0.14 \\
G219-B358 & 00 43 18.01 & 39 49 13.5 & 15.12  & 0.06&$-$9.53	&19.78 &$-$1.83$\pm$0.22 &0.00 $\pm$ 0.27 & 0.78 \\
G315-B381 & 00 46 06.47 & 41 21 00.2 & 15.76  & 0.17\rlap{$b$}&$-$9.24	&8.8& $-$1.22$\pm$0.43 &\nodata &\nodata \\
G322-B386 & 00 46 26.94 & 42 01 52.9 & 15.64  & 0.13&$-$9.24	& 14.02& $-$1.21$\pm$0.38& 0.25 $\pm$ 0.22 & 0.41\\
G351-B405 & 00 49 39.81 & 41 35 29.4 & 15.20  & 0.08&$-$9.52	& 18.2& $-$1.80$\pm$0.31&\nodata& 0.71\\

\enddata

\tablerefs{a. \cite{2005AJ....129.2670R},  b.  \cite{2008MNRAS.385.1973F}, c. Reddening corrected and using a distance modulus for M31 of ($m-M$)$=24.47$ \citep{1998AJ....115.1916H}, d. \cite{2002AJ....123.2490P} e. Low resolution spectroscopic metallicities of \cite{1991ApJ...370..495H}, f. Low resolution [$\alpha$/Fe] of \cite{2005A&A...434..909P}, g. Horizontal branch morphology
ratios (HBR), defined as the "Mironov Index" = $B/B+R$,  where $B$ and $R$ correspond to the number of horizontal
branch stars bluer or redder than $V-I=0.5$ in the observed CMD of  \cite{2005AJ....129.2670R}   Other data: \cite{2000AJ....119..727B}}
 
\end{deluxetable*}

Unfortunately, at $\sim$780 kpc  from the Milky Way \citep{1998AJ....115.1916H}, even  M31 is distant enough
that older RGB stars are too faint (V$\sim$23) to obtain the required high
signal-to-noise ratio (SNR$\sim 60$) and high spectral resolution
($\lambda/\Delta\lambda \sim 20,000$) to measure detailed abundances
in individual stars.  Fortunately, globular clusters (GCs) can also be
targeted. Unlike single stars, high resolution spectra can be obtained
of unresolved GCs out to $\sim$ 4 Mpc with current telescopes. GCs are
bright enough ($-10 < M_{V} < -6$ mag) and have low enough velocity
dispersions ($2-20$ \kms) that even weak lines ($\sim$15 m\AA) can be
detected in spectra of their integrated light. Detailed abundances
have never been obtained from unresolved GCs because techniques have
not existed to analyze them.  We have developed a new method for
analyzing high resolution integrated light (IL) spectra of single age,
chemically homogeneous stellar populations to obtain detailed element
abundances as described in
\cite{2002IAUS..207..739B,2005astro.ph..7042B} and
\cite{2008ApJ...684..326M}, hereafter
``\citetalias{2008ApJ...684..326M}.''

Our method has been developed and tested on a ``training set'' of
Milky Way and LMC GCs with well determined properties from studies of
individual stars.  IL spectra of the training set GCs were obtained by
scanning a 32$\times$32 arcsec$^2$  region of
the cluster cores in the Milky Way clusters, and a $12\times12$ arcsec$^2$ region in the LMC.  Note that slit
scanning is only necessary for the resolved GCs we target in our
training set, not for unresolved, extragalactic GCs.  The
training set GCs were chosen to cover the range of metallicity,
horizontal branch morphology, mass, velocity dispersion, and age
available in the Milky Way and LMC systems.  Using this training set,
we have compared abundances determined with our IL method and
abundances from the literature determined for individual RGB stars.
Based on the Milky Way training set, we estimate empirically that our
[Fe/H] abundances are accurate to within $<$ 0.1 dex, and all other
element ratios ([X/Fe]) accurate to within $<$ 0.1 dex (see \citetalias{2008ApJ...684..326M} and S. Cameron et al. 2009, in preparation).  We
also derive approximate ages ($>$10 Gyr) for our entire Milky Way
training set.  
Using the larger age range of the LMC clusters, which includes clusters
as young as $\sim$10 Myr, we have found that our method can clearly
distinguish clusters over a large range in age.  Our accuracies at the youngest ages are  described in
a forthcoming paper (J. Colucci et. al 2009, in preparation).

In addition to the fact that GCs are among the oldest stellar
populations in galaxies, there is ample evidence that GCs are a good
tool for tracing both the early formation of galaxies themselves and
star formation throughout their histories.  The fact that the number
of GCs in a galaxy scales with the total luminosity of the galaxy
\citep[e.g.][]{1991ARA&A..29..543H} suggests that the GCs trace total
star formation.  Additionally, young massive clusters are seen forming
in regions with high star formation rates, such as mergers of gas rich
galaxies \citep[e.g.][]{1998AJ....116.2206S,1995AJ....109..960W},
indicating that GCs form in major episodes of star formation,
throughout the lifetimes of galaxies.  While the old age of GCs alone
suggest that they record the earliest star formation episodes in
galaxies, there is additionally strong evidence that both blue
(metal-poor) and red (metal-rich) sub-populations of GCs in normal
galaxies correlate with the host galaxy's luminosity and overall
metallicity\citep[e.g.][]{2006ARA&A..44..193B}. This suggests that
both sub-populations are a record of the formation history of the
galaxy, with blue clusters  possibly tracing the
earliest star formation in dark matter halos and red clusters tracing the later formation after the gas is 
more  enriched.  Finally, GCs are also relatively easy spectroscopic
targets to analyze because, to first order, they are simple stellar
populations (SSPs), and can be approximated with a single age and
single metallicity.  For all of these reasons, detailed abundance
analysis of GC systems is a powerful tool for understanding the
formation of galaxies beyond the Milky Way.

M31 is the closest large galaxy to the Milky Way and an ideal
target for galaxy formation studies using GCs.  
Like the Milky Way, M31 has a large system of GCs, most of which are
old ($\geq10$ Gyrs), and has a bimodal metallicity distribution with
peaks at [Fe/H]$\sim-1.4$ and [Fe/H]$\sim-0.5$ and a mean of
[Fe/H]$\sim-1.2$ \citep{2000AJ....119..727B, 2002AJ....123.2490P}.
\cite{2007AJ....133.2764B} find that M31 and Milky Way GCs are
structurally similar, with similar mass-to-light (M/L) ratios, and
fall on a single GC fundamental plane.  However, there are notable
differences that have been observed between the two galaxies.  To
begin with, M31's GC system is more than a factor of 2 larger than the
Milky Way's \citep{2004A&A...416..917G}, has massive but diffuse GCs
\citep{2005MNRAS.360.1007H,2006ApJ...653L.105M}, and metal-poor and
compact GCs at very large projected galactocentric radii
\citep{2006MNRAS.371.1983M,2007ApJ...655L..85M}.  There is evidence
for young and/or intermediate age GCs in M31 unlike in the Milky Way
\citep{2004AJ....128.1623B,2005A&A...434..909P}, and a significant
population of GCs of all metallicities kinematically associated with
the thin disk \citep{2004ApJ...603...87M}.  There is also evidence for
some chemical differences in M31 GCs; compared to the Milky Way, M31
GCs show enhanced CN molecular lines
\citep[e.g.][]{1984ApJ...287..586B}, and the first estimates of
[$\alpha$/Fe] ratios in M31 GCs have indicated it may on average be
$\sim$0.1$-$0.2 dex lower than in the Milky Way
\citep{2005A&A...434..909P,2005AJ....129.1412B}.  More detailed studies are required to understand what these similiarities and differences imply for the formation history of M31.

Much of what is known about M31 comes from low resolution spectroscopic methods like the ``Lick'' system \citep[e.g.][]{1985ApJS...57..711F}, which 
 have been used to target spatially unresolved GCs
\citep[e.g.][]{1991ApJ...370..495H} to obtain the constraints to date on [Fe/H] and
[$\alpha$/Fe] ratios 
\citep[e.g.][]{2005A&A...434..909P,2005AJ....129.1412B}.  These line
index systems were originally developed for studying the integrated
light of galaxies, which have velocity dispersions of 100$-$300 \kms
\citep{1976ApJ...204..668F}.  With the low  velocity dispersions of GCs, individual spectral lines can be resolved with only
slightly greater line blending than one finds in individual RGB stars.  The accuracy of the line index systems depends on calibrations that are sensitive to abundance ratios and overall metallicity.  These limitations make detailed abundance analysis from individual lines, as in our method, an important next step.

In this paper we present the analysis of 5 GCs in M31 using high
resolution spectroscopy of their integrated light.  These represent
the first set of clusters which we have observed as part of an ongoing
project to study the GC population in M31 with the goal of
constraining the stellar populations and formation history of M31.
From these clusters, we derive detailed stellar abundances of old
populations in M31 for the first time. We obtain results for 14
elements: Mg, Al, Si, Ca, Sc, Ti, V, Cr, Mn, Fe, Co, Ni, Y, and Ba.
We also present ages, velocity dispersions, and reddening constraints
for this sample.  In \textsection~\ref{sec:observations}, we describe
our observations and data reduction.  In \textsection~\ref{sec:vdisp}
we present analysis of the velocity dispersions and implications for
our equivalent width abundance analysis.  In
\textsection~\ref{sec:analysis} we describe our abundance analysis
method and present abundance results in
\textsection~\ref{sec:abundances}.  In \textsection
~\ref{sec:discussion} we discuss the M31 GC chemical abundances in the
contexts of galaxy formation, low resolution spectra line index
abundances, and overall consistency with broadband photometric colors
and existing Hubble Space Telescope (HST) CMDs.

\begin{deluxetable}{rrrrrrrrrr}
\tablecolumns{10}
\tablewidth{0pc}
\tablecaption{Observation Log and Estimated SNR \label{tab:observations}}
\tablehead{
\colhead{Cluster} &\colhead{Exposure (s)}   &\multicolumn{3}{c}{SNR (pixel$^{-1}$)} \\&&\colhead{4400 \AA} &\colhead{6050 \AA} & \colhead{7550 \AA} }
\startdata

G108-B045 &16,200&  40   & 60 & 80 \\
G219-B358 &10,500 &  40 & 60 & 70\\
G315-B381 &14,300&  30 & 50 & 60\\
G322-B386 &12,600 & 30 & 50 & 60\\
G351-B405 &10,800& 40 & 60 & 80 \\

\enddata
\end{deluxetable}

\section{Observations and Reductions}
\label{sec:observations}

In selecting GC targets in M31, we have focused initially on clusters
which are spectroscopically confirmed, and are estimated from
low resolution indexes to have abundances in the range -1.8 $<$ [Fe/H]
$<$ -0.5 dex.  At the time of selection, this was the range in which
we were most confident of the accuracy of our analysis based on our
previous work with the Milky Way training set clusters described above (\citetalias{2008ApJ...684..326M} and S. Cameron et al. 2009).  The
targets were further selected to be relatively well studied, reasonably isolated and well
outside of M31's disk to reduce confusion and the chance of confusion from interloping sources.  We
also avoided the brightest, most massive GCs  while
still targeting GCs bright enough to get sufficient SNR in a few hours of observations, as they will have the
highest velocity dispersions and thus broader, less pronounced spectral lines,.  With this in mind, we have selected these GCs from the
Barmby catalog \citep{2000AJ....119..727B}. This initial set of GCs has magnitudes between
15 and 16.  
Magnitudes and spatial information are listed for all of the GCs  
in Table~\ref{tab:clusterdata}, 
along with low resolution abundance estimates from the  
literature and horizontal branch morphologies from HST imaging when available.

\begin{figure}
\centering
\includegraphics[scale=0.5]{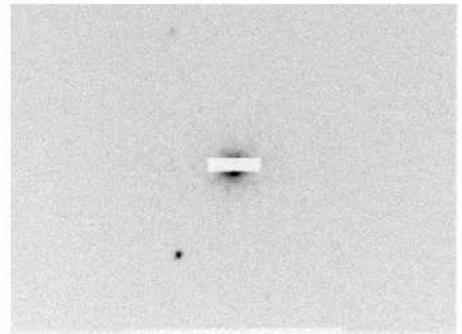}
\caption{A single frame taken with the Keck I guide camera is  
shown here to illustrate the relative size of a GC in M31 (half-light  
radius $\sim$1  arcsec) relative to the $1.7\times7$ arcsec$^{2}$ slit.  The guider  
images a reflection of the sky off the polished slit plate, so that  
the slit itself is clearly visible where no image of the sky is  
reflected.}
\label{fig:image} 
\end{figure}

\begin{figure*}
\centering
\includegraphics[scale=0.7]{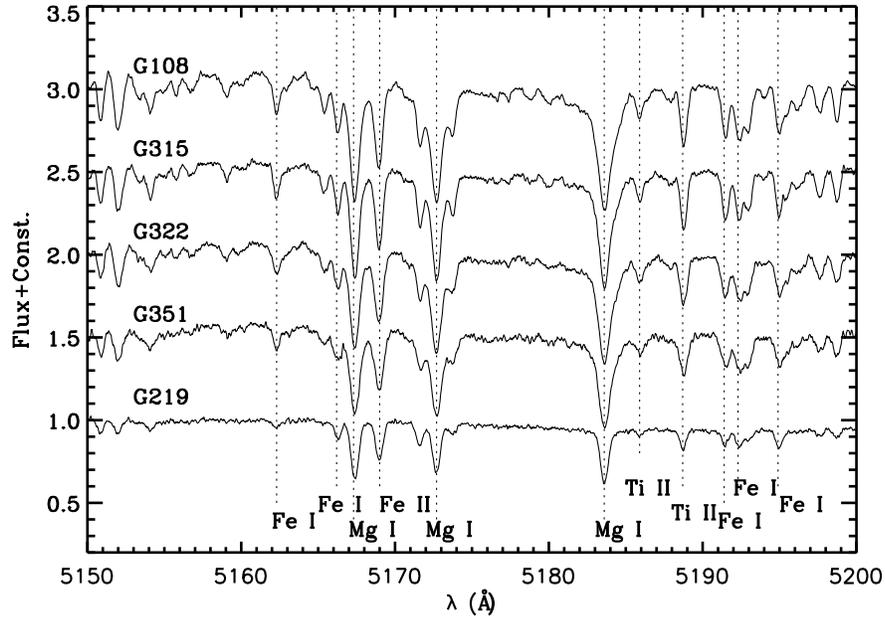}
\caption{Example M31 GC spectra in Lick Mgb region shown decreasing in [Fe/H] from our analysis (top to bottom).  Note that this region is dominated by saturated lines and is shown for illustration as a region familiar in low resolution spectra analyses.}
\label{fig:spectra} 
\end{figure*}

For this work we obtained high resolution IL spectra of the M31
globular clusters using the  HIRES spectrograph
\citep{1994SPIE.2198..362V} on the Keck I telescope over the dates
2006 September 10$-$14.  We used the D3 decker, which provides a slit size of $1".7\times7.0"$ and
spectral resolution of $R=24,000$.  The GCs in this sample
have half light radii (r$_{h}$) between $\sim$0.6"$-$1.1"
\citep{2007AJ....133.2764B}.  We calculate from surface brightness
profiles in \cite{2002AJ....123.1937B,2007AJ....133.2764B} that
$\sim$70$-$90$\%$ of the light fell in the $1".7\times7.0"$
slit for each cluster. An illustration of the observing setup is shown in Figure~\ref{fig:image}. 
 The wavelength coverage of the HIRES spectra is approximately
3800$-$8300 \AA.  Exposure times were between 3$-$4 hours for each GC
and are listed in Table~\ref{tab:observations} along with SNR
estimates at three wavelengths, one in each HIRES CCD.  Data was
reduced with the MAKEE software package available from
T. Barlow\footnote{MAKEE was developed by T. A.  Barlow specifically for reduction of Keck HIRES data.  It is freely available on the Web at http://spider.ipac.caltech.edu/staff/tab/makee/index.html .}.
In analysis of more recent observing runs, we have compared 
data analyzed with MAKEE and with the HiRes Redux pipeline produced
by J. X. Prochaska, which has many routines and strategies in common
with the MIKE Redux pipeline.  While HiRes Redux produces
lower noise spectra overall, and often traces weak orders more
accurately, the MAKEE results are accurate and sufficient
for our analysis, particularly since we explicitly avoid 
regions with sky emission or absorption features entirely.  Example spectra for the five M31 GCs are shown in Figure~\ref{fig:spectra}, where it can be seen that many individual Fe I, Fe II, Mg I and Ti II lines are easily identified.

\begin{deluxetable*}{rrrrrrrrrrrr}
\centering
\tablecolumns{10}
\tablewidth{0pc}
\tablecaption{Velocity Dispersion Measurements \label{tab:m31vdisp}}
\tablehead{
\colhead{Cluster} & \colhead{$\sigma_{obs}$}   & \colhead{RMS} & \colhead{$N_{ord}$} &\colhead{Error}  & \colhead{$\sigma_{best}$} &\colhead{RMS} &\colhead{$N_{best}$}&\colhead{Error} &  \colhead{$\sigma_{lit}$\tablenotemark{a}}  \\ \colhead{} & \colhead{(km s$^{-1}$)} & \colhead{} & \colhead{} & \colhead{} & \colhead{(km s$^{-1}$)} &&&&  \colhead{(km s$^{-1}$)}}
\startdata

G108-B045 & 10.24 & 0.44 & 26 & 0.09 & 10.14&0.37& 9&0.12& 9.82 $\pm$ 0.18  \\
G219-B358 & 10.99 & 1.00 & 13 &0.28& 10.36 &0.84&5&0.38& 8.11 $\pm$ 0.36 \\
G315-B381 & 9.87 & 0.43 & 28 &0.08& 9.65  &0.41&10&0.13& $<$ 10\tablenotemark{b} \\
G322-B386 & 11.42 & 0.58 & 19&0.13 & 11.35 &0.37&6&0.15& 11.49 $\pm$ 0.24\\
G351-B405 & 12.29 & 1.06 & 16 &0.27& 12.01 &0.41&7&0.15& 8.57 $\pm$ 0.45\\
\enddata

\tablerefs{(a) \cite{1997ApJ...474L..19D}, (b) \cite{1989ddse.work..161P}}
\tablecomments{All measured velocity dispersions are given without application of aperture corrections (see \textsection~\ref{sec:vdisp}). $\sigma_{obs}$ is the mean measured from all usable orders. $\sigma_{best}$ is the mean measured for higher SNR orders between 4800$-$5800 \AA. }

\end{deluxetable*}

\section{Velocity Dispersions}
\label{sec:vdisp}

One of the strengths of our ILS analysis is the amount of information
and the number of constraints available in the high resolution
spectra.  In addition to checks related to the abundance measurements
themselves (see \textsection~\ref{sec:best cmd}), we also have
overall photometric colors, magnitudes, and internal kinematics
(velocity dispersions).  Mass
estimates from velocity dispersions can also help to constrain the
basic initial mass function, contributing to the overall
consistency of our understanding of the stellar population.
Measurements of the velocity dispersions also tell us the spectral line resolutions we can
expect for our abundance analysis.  We compare these results to
measurements in the literature when available.

We measure velocity dispersions from the high resolution IL spectra of
the M31 GCs following the method described in
\cite{1979AJ.....84.1511T}, which is implemented in the
IRAF\footnote{IRAF is distributed by the National Optical Astronomy
  Observatories, which are operated by the Association of Universities
  for Research in Astronomy, Inc., under cooperative agreement with
  the National Science Foundation.} task $fxcor$. The GC spectrum is
cross correlated on an order by order basis with a suitable template
star.  The full width at half-maximum of the cross correlation peaks
(FWHM$_{cp}$) measured by $fxcor$ is then converted to a line-of-sight
velocity dispersion ( $\sigma_{obs}$) using an empirical relation
between the two. This relation is established by cross correlating the original
template star spectrum with artificially broadened versions of itself 
that are made by convolving it with Gaussian profiles corresponding to
$\sigma_{obs}$ of 2$-$25 \kms.  RGB stars are suitable template stars
for old stellar populations; in this analysis, we use a spectrum of
HR 8831 (type G8 III) taken during the 2006 September run as the
template.

In Table~\ref{tab:m31vdisp} we report $\sigma_{obs}$ and the
associated 1-$\sigma$ errors measured for orders between 4000$-$6800
\AA. Only orders with high SNR and weak atmospheric absorption are
used in the cross correlation.  We also avoid orders that include the
saturated Balmer lines.  Recently, \cite{2009arXiv0906.0397S} have
found a weak trend of $\sigma_{obs}$ with wavelength in a fraction of
the GCs they observed.  We do not find any correlations between
$\sigma_{obs}$ and wavelength for G108, G315, or G322.  However a small
correlation exists for G219 and G351.  Like
\cite{2009arXiv0906.0397S}, we find $\sigma_{obs}$ decreasing from
blue to red orders by $\sim$1 \kms.  It is unclear if this is
due to a color/metallicity mismatch of the GCs and the template star
(these two GCs are the more metal-poor ones in our sample) or
other systematic effect \citep[see][for a larger
discussion]{2009arXiv0906.0397S}.  Because of this issue, in
Table~\ref{tab:m31vdisp} we have reported $\sigma_{best}$, which is
the dispersion obtained for a subset of the highest SNR orders between
4800$-$5800 \rAA that do not show correlations with velocity
dispersion and also give the smallest RMS errors.  The number of
orders used in each measurement are recorded in columns 4 and 7 of
Table ~\ref{tab:m31vdisp}.

Four of the GCs observed here have $\sigma_{obs}$ measurements in the
literature from previous observations by \cite{1997ApJ...474L..19D}.
The spectra used by \cite{1997ApJ...474L..19D} were taken with a
slightly narrower slit ($1".15\times7.0"$), which will lead to a
velocity dispersion roughly 0.5 \rkms larger, which is comparable to
the measurement errors.  We find that our measurements for G108 and
G322 agree with those in \cite{1997ApJ...474L..19D} within the quoted
errors, but our observed velocity dispersions for G219 and G351 are
$2-4$ \rkms larger.  It is possible that the difference between our
results and those of \cite{1997ApJ...474L..19D} are due to a
difference in $v$sin$i$ of the template star used in the cross
correlation.  The template used by \cite{1997ApJ...474L..19D} has
$v$sin$i$= 7 \kms.  We measure a mean FWHM of 8 \rkms from line widths
in our template star.  Note that this is a combination of the
intrinsic stellar line width (including $v$sin$i$) and the
instrumental resolution, which is FWHM = 7.7 \kms based on an analysis
of arc lines taken through the appropriate slit.  This suggests that
the rotational velocity of the star is quite small (less than or equal
to roughly 1 \rkms when added in quadrature).  The lower $v$sin$i$ of
our template could therefore cause the higher inferred $\sigma_{obs}$
found here.  Both our measurements and those of
\cite{1997ApJ...474L..19D} require an additional correction of
approximately $+14\%$ to convert $\sigma_{obs}$ to a projected central
$\sigma_{v}$ as described in \cite{1997ApJ...474L..19D}.  This is the
geometrical correction appropriate to obtain a systemic line of sight
velocity dispersion from a velocity dispersion measured within a
radius of $2-3\times r_h$.

For completeness, we note that \cite{2007AJ....133.2764B} have predicted
aperture velocity dispersions for these same GCs by modeling surface
brightness profiles.  They have compared their predictions to
observations of velocity dispersions in the literature to derive an
empirical correction between the two.  We note that our velocity
dispersion measurements for G219 and G351 are more consistent with the
trend derived by \cite{2007AJ....133.2764B} than the measurements by
\cite{1997ApJ...474L..19D}.  A velocity dispersion of $12.1\pm1.3$
\rkms has also been measured by  \cite{2009arXiv0906.0397S},  which 
is more consistent with the value we measure. 

Finally, we report the first velocity dispersion measurement from high
SNR, high spectral resolution, IL spectra for G315-B381. Our
measurement of $9.65 \pm 0.13$ \rkms is consistent with the upper
limit of 10 \rkms found by \cite{1989ddse.work..161P}.

\begin{figure}
\centering
\includegraphics[angle=90,scale=0.3]{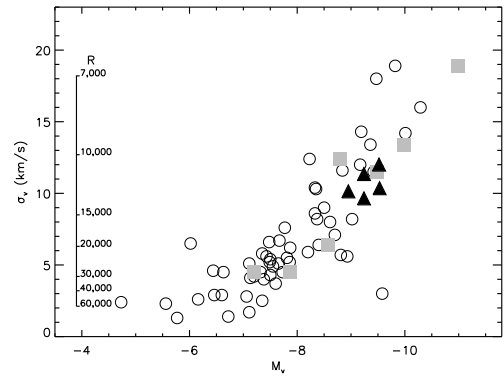}
\caption{M$_{V}$-$\sigma$ relation for Milky Way GCs (open circles), with data from
  \cite{1993ASPC...50..357P} and  \cite{1996AJ....112.1487H}. Training set Milky Way GCs are
  highlighted in gray squares and the five M31 GCs are shown as filled
  triangles. Training set and M31 GCs are plotted with reddening
  corrected V magnitudes and velocity dispersions from  \cite{1993ASPC...50..357P}.
 The line width parameter $R$ calculated for a range of velocity dispersions is shown on the inset axis.
 }
\label{fig:sigma} 
\end{figure}

Our velocity dispersions, when combined with GC r$_{h}$ measurements
in the literature, allow us to make order of magnitude estimates of
the cluster M/L ratios.  r$_{h}$ for all of the GCs in this sample,
with the exception of G315, can be found in
\cite{2007AJ....133.2764B}. A simple calculation, following
\cite{1987degc.book.....S}, assuming the virial theorem and an
isotropic velocity distribution for the GCs results in dynamical
M/L$\sim$1.5$-$2.5 for these four GCs.  Here we have assumed $M = 2.5
v^{2} r/ G$, where $r$, the half-mass radius, is 1.3$\times$r$_{h}$,
and $v$, the three-dimensional velocity dispersion is
$\sqrt{3}\times\sigma$.  We can also compare the dynamical M/L to
those predicted with population synthesis models.  In
\cite{2009ApJ...690..427P}, the Teramo group provides M/L ratios based
on their isochrones and an appropriate IMF from
\cite{2001MNRAS.322..231K}.  As we use the Teramo isochrones for our
abundance analysis (see \textsection ~\ref{sec:synthesis}), we can use
isochrones which are self-consistent with our later analysis and
well-matched to each cluster.  Using our own abundance and age
constraints, the most appropriate isochrones (ages and metallicities)
for our clusters have a M/L$\sim$1.9$-$2.9 based on the Teramo group's
population synthesis work.  It is typical to obtain systematically
different M/L from dynamical and population synthesis techniques; this
difference is probably due to the inclusion of low mass stars in the
population synthesis estimate, which are actually ejected in GCs due
to dynamical evolution.  We find a ratio between dynamical and
population synthesis M/L ratios of $0.84 \pm 0.27$, which is
consistent with the value of $0.73 \pm 0.25$ found by
\cite{2007AJ....133.2764B} for a larger sample of M31 GCs.
\cite{2005ApJS..161..304M} find a similar ratio of $0.82 \pm 0.07$ for
a sample of Milky Way and old LMC GCs.  These M/L values further
identify these four M31 GCs as consistent with the familiar Milky Way
GC population.

The relationship described above between light and velocity
dispersion is shown in Figure~\ref{fig:sigma}.  
The $\sigma_{obs}$ for
Milky Way GCs from \cite{1993ASPC...50..357P}, along with absolute
magnitudes ($M_{V}$) listed 
by \cite{1996AJ....112.1487H} are plotted as open circles. 
 For
comparison, we also plot the reddening-corrected magnitudes and velocity dispersions from \cite{1993ASPC...50..357P}  for our training set GCs as gray squares and those measured for the M31 GC sample
as black triangles.  From Figure~\ref{fig:sigma} we see that
the $\sigma_{obs}$ measured here are consistent with what we would
expect given the absolute magnitudes of this set of GCs.

The velocity dispersion that we measure above quantifies another
important point about the clusters, which is the limiting resolution
that we can obtain for individual spectral lines due to velocity
broadening.   The velocity dispersions
we measure for this set of M31 GCs of 9$-$12 \rkms give a line width parameter ($R=\lambda$/FWHM) $R=
$14,000$-$10,000, from which it is clear that we are fully sampling the line
profiles with the $R=24,000$ spectral resolution provided by the slit
configuration used here (the HIRES D3 decker). 
 However, with
dispersions this large we expect to have difficulty measuring
equivalent widths for elements with weak lines ($\lesssim$30 m\AA), as we discuss in \textsection~\ref{sec:abundances}.

\begin{deluxetable*}{rrrrrrrrr}
\tablecolumns{9}
\tablewidth{0pc}
\tablecaption{Line Parameters and Integrated-Light Equivalent Widths for M31 GCs \label{tab:linetable_stub}}
\tablehead{\colhead{Species} & \colhead{$\lambda$}   & \colhead{E.P.} & \colhead{log gf}   & \colhead{EW(m\AA)} & \colhead{EW(m\AA)} & \colhead{EW(m\AA)} & \colhead{EW(m\AA)} & \colhead{EW(m\AA)}\\ 
\colhead{} & \colhead{(\AA)} & \colhead{(eV)} &\colhead{} & \colhead{G108}&\colhead{G322}&\colhead{G315}&\colhead{G351}&\colhead{G219}}

\small
\startdata

 Mg  \I &4571.102 &  0.000 &$-$5.691 &  \nodata &120.0 &  \nodata & 86.7 & 34.5\\
 Mg  \I &4703.003 &  4.346 &$-$0.666 &  \nodata &139.2 &  \nodata & 99.8 &  \nodata\\
 Mg  \I &5528.418 &  4.346 &$-$0.341 &  \nodata &  \nodata &  \nodata &147.3 & 58.2\\
 Mg  \I &5528.418 &  4.346 &$-$0.341 &  \nodata &  \nodata &  \nodata &143.6 &  \nodata\\
\\
 Al  \I &3944.016 &  0.000 &$-$0.638 &  \nodata &  \nodata &  \nodata &  \nodata &134.9\\
 Al  \I &3944.016 &  0.000 &$-$0.638 &  \nodata &  \nodata &  \nodata &  \nodata &140.1\\
\\
 Si  \I &7405.790 &  5.610 &$-$0.660 & 58.4 &  \nodata &  \nodata &  \nodata &  \nodata\\
 Si  \I &7415.958 &  5.610 &$-$0.730 & 66.4 &  \nodata & 50.1 &  \nodata &  \nodata\\
 Si  \I &7423.509 &  5.620 &$-$0.580 & 84.3 & 66.5 & 82.4 &  \nodata &  \nodata\\

\enddata
\tablecomments{Table 4 is  presented in its entirety at the end of the text. ~ Lines listed twice correspond to those measured in adjacent orders with overlapping wavelength coverage.}
\normalsize
\end{deluxetable*}

\section{Abundance Analysis}
\label{sec:analysis}

Our new method for obtaining detailed abundances of from their
integrated light was developed and tested using a training set of
Milky Way and LMC globular clusters.  The basic method was described
in detail in \cite{2002IAUS..207..739B}, \cite{2005astro.ph..7042B},
\citetalias{2008ApJ...684..326M}, and the full training set is
presented in S. Cameron et al. (2009) and J. Colucci et
al. (2009). The method is briefly summarized below.

\subsection{Equivalent Widths and Line Lists}
\label{sec:EWs}

We measure absorption line equivalent widths (EWs) for
individual lines in the IL spectra using the semi-automated program GETJOB
\citep{1995AJ....109.2736M}, with which we fit  low order polynomials to continuum regions and single Gaussian profiles to individual lines and double or triple Gaussians to line blends when necessary. Line lists and oscillator strengths were
taken from \cite{1994ApJS...91..749M}, \cite{1995AJ....109.2757M},
\cite{1998AJ....115.1640M},  \citetalias{2008ApJ...684..326M} and \cite{2006ApJ...640..801J}. We measure fewer lines than in standard
individual RGB star analyses, as lines in IL spectra are broader and
weaker than in individual RGB stars due to the velocity dispersions of the GCs and the presence of continuum flux from warm stars.
 The lines
and EWs included in our final analysis are listed in Table
~\ref{tab:linetable_stub}. As expected, we find fewer clean lines for the GCs
with higher velocity dispersions.

\subsection{CMDs and EW Synthesis}
\label{sec:synthesis}

In order to synthesize IL EWs to compare to our observed IL EWs, we
next need to model the population using theoretical single age, single
metallicity isochrones. During analyses of our training set GCs, we
performed extensive testing of a variety of isochrones from the
Padova\footnote{Padova isochrones downloadable at
  http://pleiadi.pd.astro.it/} \citep{2000A&AS..141..371G} and
Teramo\footnote{Teramo isochrones downloadable at
  http://albione.oa-teramo.inaf.it/ }
\citep{2004ApJ...612..168P,2006ApJ...642..797P,2007AJ....133..468C}
groups \citepalias[see][]{2008ApJ...684..326M}.  Because we require a
large, self-consistent parameter space of both scaled-solar and
$\alpha-$enhanced isochrones, we have chosen to use the isochrones
from the Teramo group for our abundance analyses.  The isochrones
available cover abundances from Z=0.0001$-$0.04 for both scaled solar
and $\alpha-$enhanced ratios. We choose to use the recommended
canonical evolutionary tracks including an extended asymptotic giant
branch (AGB) and $\alpha-$enhanced low$-$temperature opacities
calculated according to \cite{ 2005ApJ...623..585F}.  We also choose
isochrones with mass$-$loss parameter of $\eta$=0.2 because comparison
with our training set GCs (particularly those of intermediate
metallicity) show that they more accurately match the CMD of GCs;
isochrones with $\eta$=0.4 over-predict the fraction of extreme blue
horizontal branch (HB) stars at intermediate metallicities.  Given
that we find blue HB stars are less critical to accurate abundance
analysis than red HB stars (see \textsection~\ref{sec:hbr}), it is
more important to our analysis that the isochrones accurately
reproduce the red horizontal branch than that they populate the blue
region of the horizontal branch when it may be present.

We apply an IMF according to the multiple$-$part power$-$law form
described in \cite{2002Sci...295...82K}, which changes index at
0.5$\msol$ and 0.08$\msol$.  Because we only observed the core
($\sim$0.1-0.2$\times$r$_{h}$) regions in our training set GCs, we
removed stars less massive than $\sim$0.7$\msol$ from the IMF to match
the present day core mass functions \citepalias{2008ApJ...684..326M}
that have experienced dynamical mass segregation and evaporation of
low mass stars \citep[e.g.][]{2003MNRAS.340..227B}.  In this set of
M31 GCs we have observed regions corresponding to $\sim$1.5-3r$_{h}$,
well beyond the core region where significant mass segregation is
expected.  Although we do expect present day GCs to be stripped of
stars less massive than $\sim$0.3$\msol$ due to dynamical evaporation,
the Teramo isochrones stop at 0.5$\msol$.  We do not believe that neglecting stars in the 0.3-0.5$\msol$ range is a problem for our
analysis, as stars less massive than 0.5$\msol$ contribute only
$\sim$1-2$\%$ to the total flux of the population and $<$1$\%$ in
absorption features.  By combining the model isochrones with
cluster-specific IMFs, we can create synthetic CMDs for the range of
possible ages and metallicities for which we have isochrones. Each
synthetic CMD is divided into $\sim$25 boxes of stars with similar
properties, with every box containing $\sim$4\% of the total $V$-band
flux.  The properties of a flux-weighted ``average'' star in each box
are used for the atmospheric parameters needed in synthesizing IL
flux-weighted EWs.

Flux-weighted synthesized EWs of lines are calculated using our
routine ILABUNDS \citepalias[see][]{2008ApJ...684..326M} which produces an integrated light EW
composed of the $\sim$25 representative stars in each CMD using
spectral synthesis routines from MOOG \citep{1973ApJ...184..839S} and
model stellar atmospheres from Kurucz
\citep[e.g.][]{2004astro.ph..5087C}\footnote{The models are available
  from R. L. Kurucz's Website at
  http://kurucz.harvard.edu/grids.html}. The synthesized EWs of
each of the $\sim$25 representative stars are averaged together, weighted by
their respective contribution to the total flux of the cluster. The
assumed abundance in the line synthesis is adjusted iteratively until
the synthetic flux-weighted EW matches the observed IL EW to
1\%. Initial abundance calculations are performed using scaled$-$solar
Teramo isochrones and Kurucz ODFNEW stellar atmospheres, and then
recalculated with $\alpha$-enhanced Teramo isochrones and AODFNEW
atmospheres when abundance results imply enriched $\alpha-$element ratios
are present. All five M31 GCs analyzed here were determined to be
$\alpha$-enhanced, and thus the abundances we report use
$\alpha$-enhanced isochrones and AODFNEW atmospheres in all cases.

All abundances were calculated under the assumption of local
thermodynamic equilibrium (LTE).  In the case of aluminum we also
discuss the non-LTE correction suggested by \cite{1997A&A...325.1088B}
in \textsection~\ref{sec:al}.

Our IL method as implemented here employs a fixed microturbulence law,
as described in \citetalias{2008ApJ...684..326M}.  Since we do not
adjust microturbulence values for individual stars in the synthetic
CMDs we must be careful of line saturation. For this reason we only
report abundances from lines with EW strengths less than $\sim$150
m\rAA for all elements except Fe II.  Our analysis indicates that
abundances from Fe II lines with EWs over $\sim$100 m\rAA start to
deviate significantly from the linear portion of the curve of growth.
To remain in the linear regime we avoid Fe II lines with EWs $>$100
m\rAA wherever possible.  In some GCs, like G351-B405, the only clean
Fe II lines we measure have EWs $>$100 m\AA.  Using these lines in
those cases will result in Fe II abundances that may be slightly
high.  In this work we will refer to lines we do not analyze because
of large EWs as ``saturated'' ($>$150 m\rAA\ in general, or $>$100
m\rAA for Fe II).  We note that, in principle, abundance upper limits
could be obtained for elements for which all lines are ``saturated.''
However, due to the flux-weighting of EWs from stars of different
types, this analysis requires special care and is not investigated
further here.

We have also calculated hyperfine splitting (hfs) abundances for Al,
Sc, V, Mn, Co, and Ba.  For lines with EWs $>$20$-$30 m\AA,
desaturation by hfs can significantly reduce the derived
abundances. We use the hfs line lists given in 
\citetalias{2008ApJ...684..326M}, \cite{2006ApJ...640..801J}, and references therein.  Typical hfs
abundances corrections here for Al, Sc, V, Mn, Co and Ba were $-$0.1,
$-$0.01, $-$0.1, $-$0.4, $-$0.2, and $-$0.15, respectively.

\begin{figure}
\centering
\includegraphics[angle=90,scale=0.3]{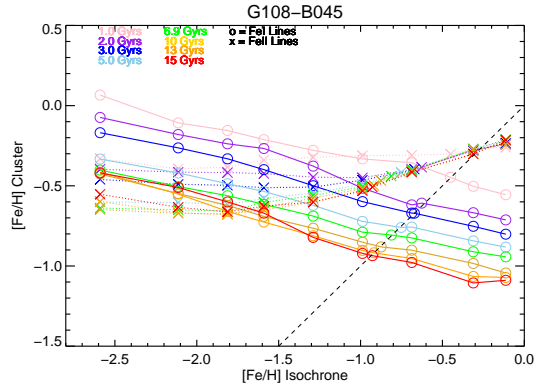}
\caption{Fe I (circles) and Fe II (crosses) abundance solutions.  The input [Fe/H] value of the isochrone (plotted on the x-axis) equals the output
  [Fe/H] value  of  our solution where circles lie on the dashed line.}
\label{fig:Fe plots} 
\end{figure}

\begin{figure}
\centering
\includegraphics[angle=90,scale=0.3]{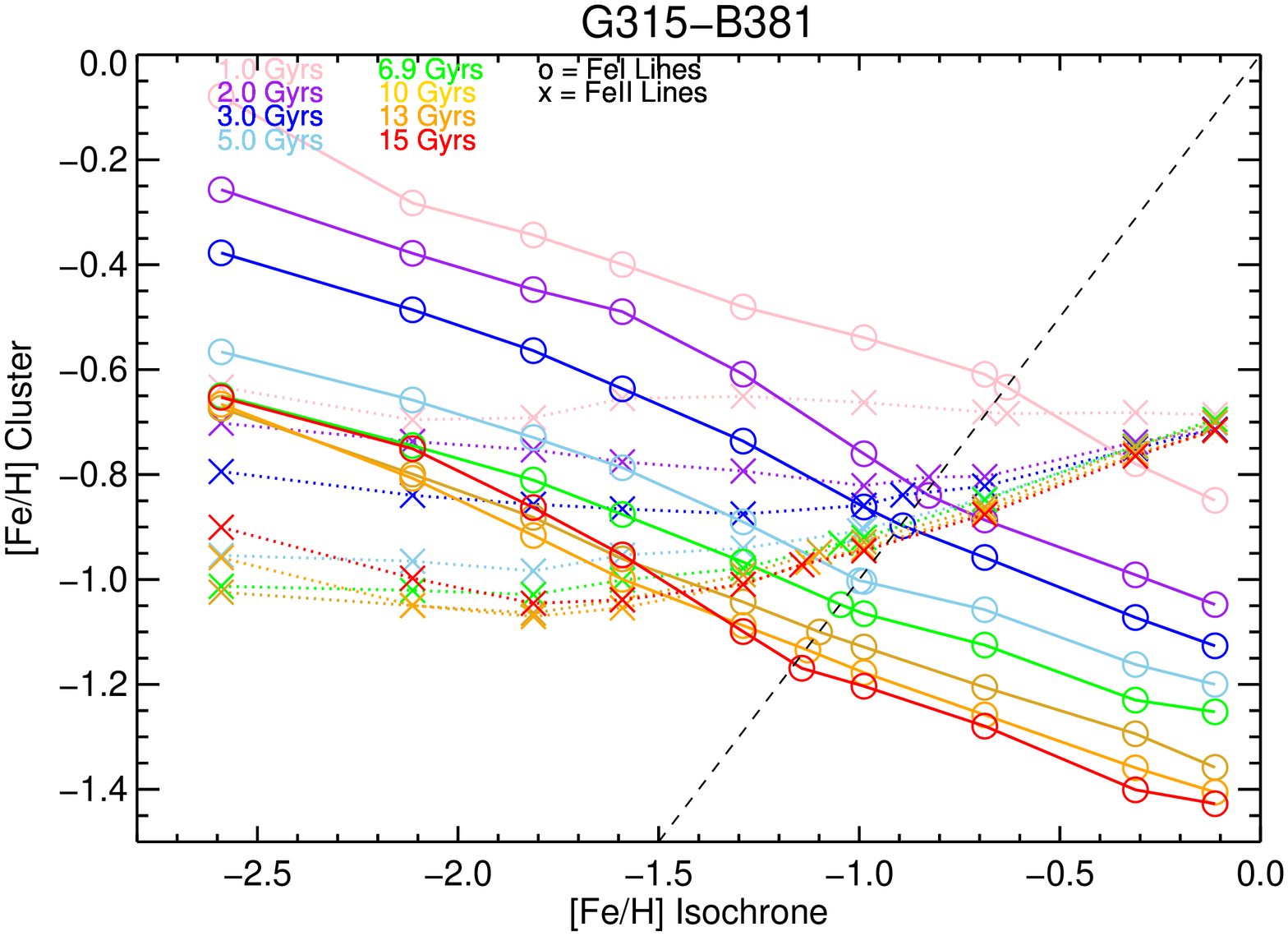}
\caption{Same as Figure~\ref{fig:Fe plots} for G315.}
\label{fig:Fe2 plots} 
\end{figure}

\begin{figure}
\centering
\includegraphics[angle=90,scale=0.3]{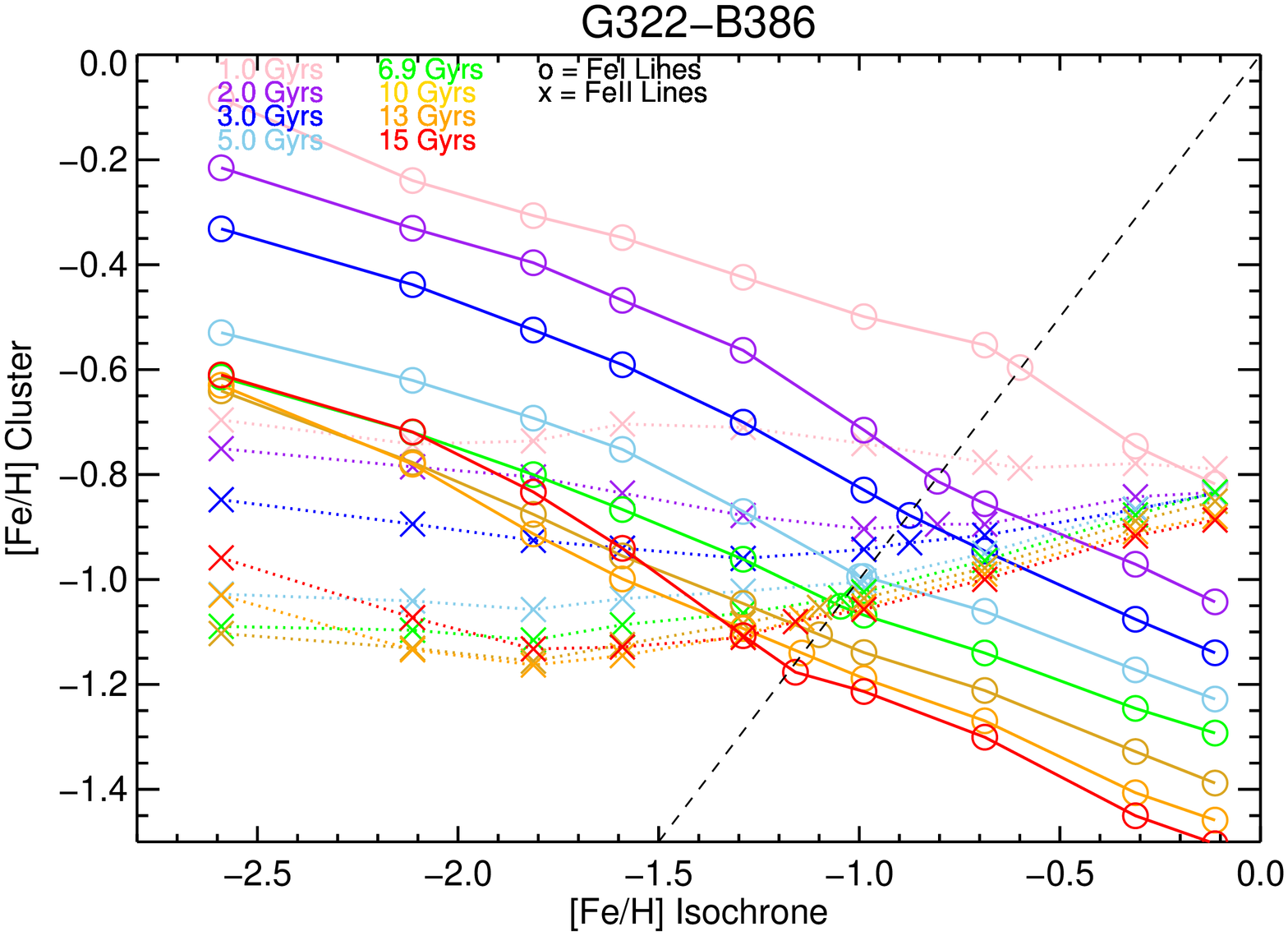}
\caption{Same as Figure~\ref{fig:Fe plots} for G322.}
\label{fig:Fe3 plots} 
\end{figure}

\begin{figure}
\centering
\includegraphics[angle=90,scale=0.3]{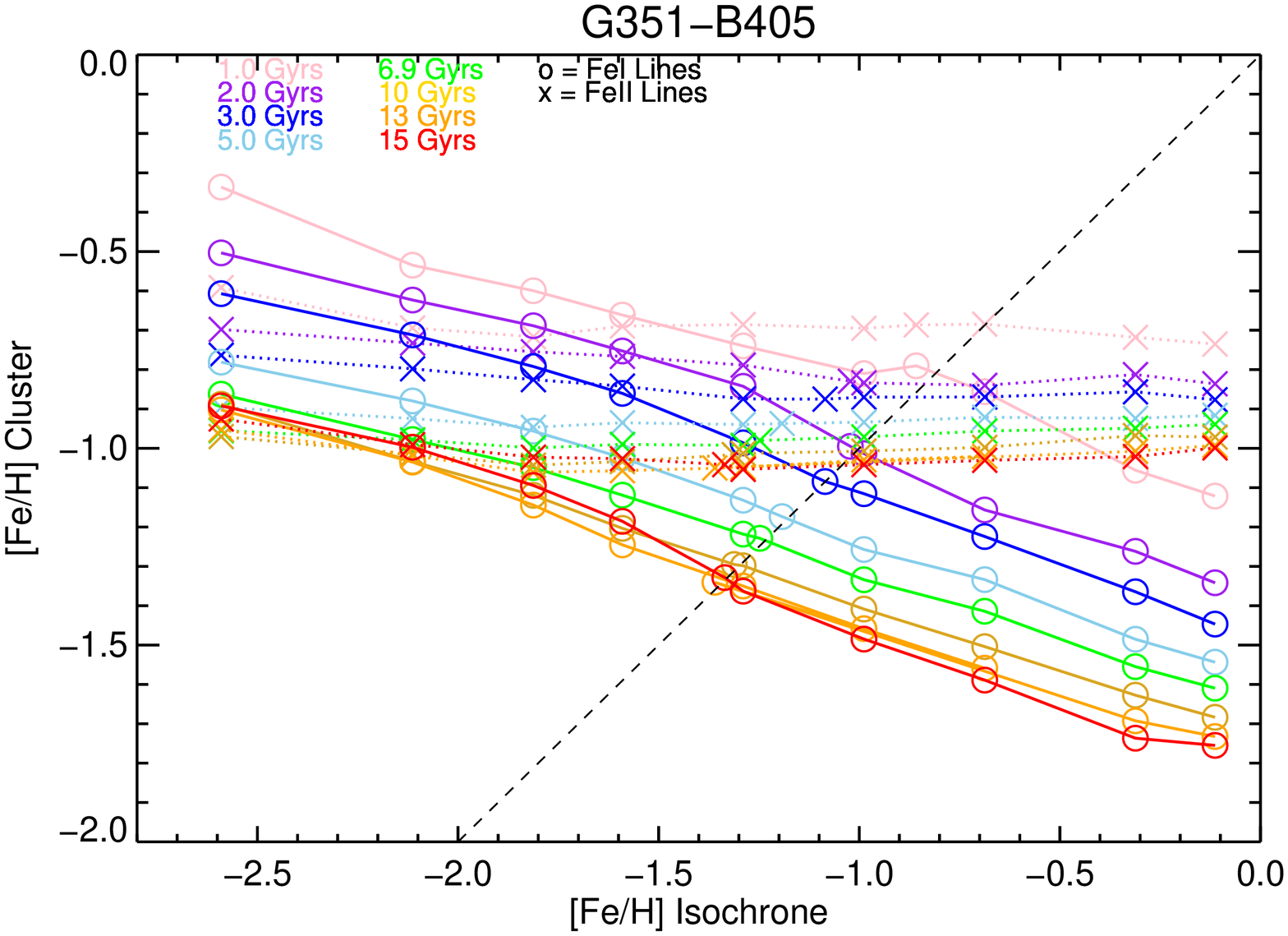}
\caption{Same as Figure~\ref{fig:Fe plots} for G351.}
\label{fig:Fe4 plots} 
\end{figure}

\begin{figure}
\centering
\includegraphics[angle=90,scale=0.3]{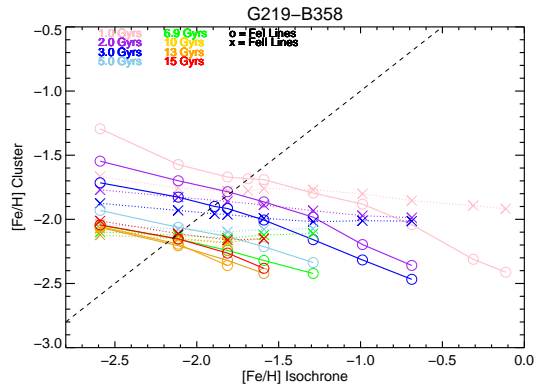}
\caption{Same as Figure~\ref{fig:Fe plots} for  G219.}
\label{fig:Fe5 plots} 
\end{figure}

\subsection{Finding the Best-Fitting CMD}
\label{sec:best cmd}

In the course of our work with the Milky Way and LMC training set GCs,
we have explored a variety of strategies for identifying the
best-fitting CMD --- the CMD which provides abundances that are most
consistent with those obtained from the spectral analysis of
individual RGB stars in those clusters.  Our first efforts to identify
the best-fitting CMD focused on obtaining a consistent [Fe/H] solution
from the Fe I and Fe II lines.  This strategy was discussed with
regard to the analysis of 47 Tuc in \cite{2002IAUS..207..739B, 2005astro.ph..7042B},
and \citetalias{2008ApJ...684..326M}.
However analysis of the full training set showed that the Fe I and Fe
II lines typically give a self-consistent solution at [Fe/H] values
that are frequently more metal-rich than those obtained from analysis
of individual stars.  There are several possible explanations for a
difference in [Fe/H] from Fe I and Fe II lines, although a detailed
study of this problem is beyond the scope of this paper.  We simply
note here that a difference between Fe I and Fe II abundances in
individual stars due to non-LTE overionization was noted by
\citep{2003PASP..115..143K}, and that inaccuracies in Fe II oscillator
strengths can lead to large uncertainties in abundances \citep[see
recent discussion in][]{2009arXiv0901.4451M}.  These results have led
us to focus on a different strategy.  Using the training set
spectra, we have found that the best-fitting CMD can be consistently
identified by taking advantage of the fact that the metallicity
dependence of RGB morphology is reasonably well understood
\citep{2005ARA&A..43..387G}.  After extensive testing, we have found
that we obtain consistent, accurate abundances by requiring that the
abundance used in calculating the isochrones themselves be consistent
with the abundance recovered by our analysis for the Fe I lines.
This is consistent with the fact that we find the RGB to have the
dominant influence on the strength of the Fe I spectral lines, more
so than on the Fe II lines, as discussed in
\citetalias{2008ApJ...684..326M}.  To clarify our analysis methods, we
describe below the procedure we follow for each GC.

For each CMD we calculate a mean [Fe/H] abundance from all available
Fe I and Fe II lines.  In this data set we measure 30$-$80 Fe I lines
and 2$-$10 Fe II lines per GC.  
Fe abundance results for all CMDs for each GC are plotted in Figures
~\ref{fig:Fe plots} through \ref{fig:Fe5 plots}.  Circles and crosses correspond to Fe I and Fe II
mean solutions.  The horizontal axis shows the [Fe/H] of the $\alpha-$enhanced Teramo
isochrones.  CMDs of the same age are connected by colored lines. Note that $\alpha-$enrichment affects the [M/H] value of the
isochrone; for clarity, the value plotted on the x-axis is the true
[Fe/H] value rather than the overall metallicity [M/H]
\citep[see][]{2006ApJ...642..797P}.

\begin{figure*}
\centering
\includegraphics[angle=90,scale=0.20]{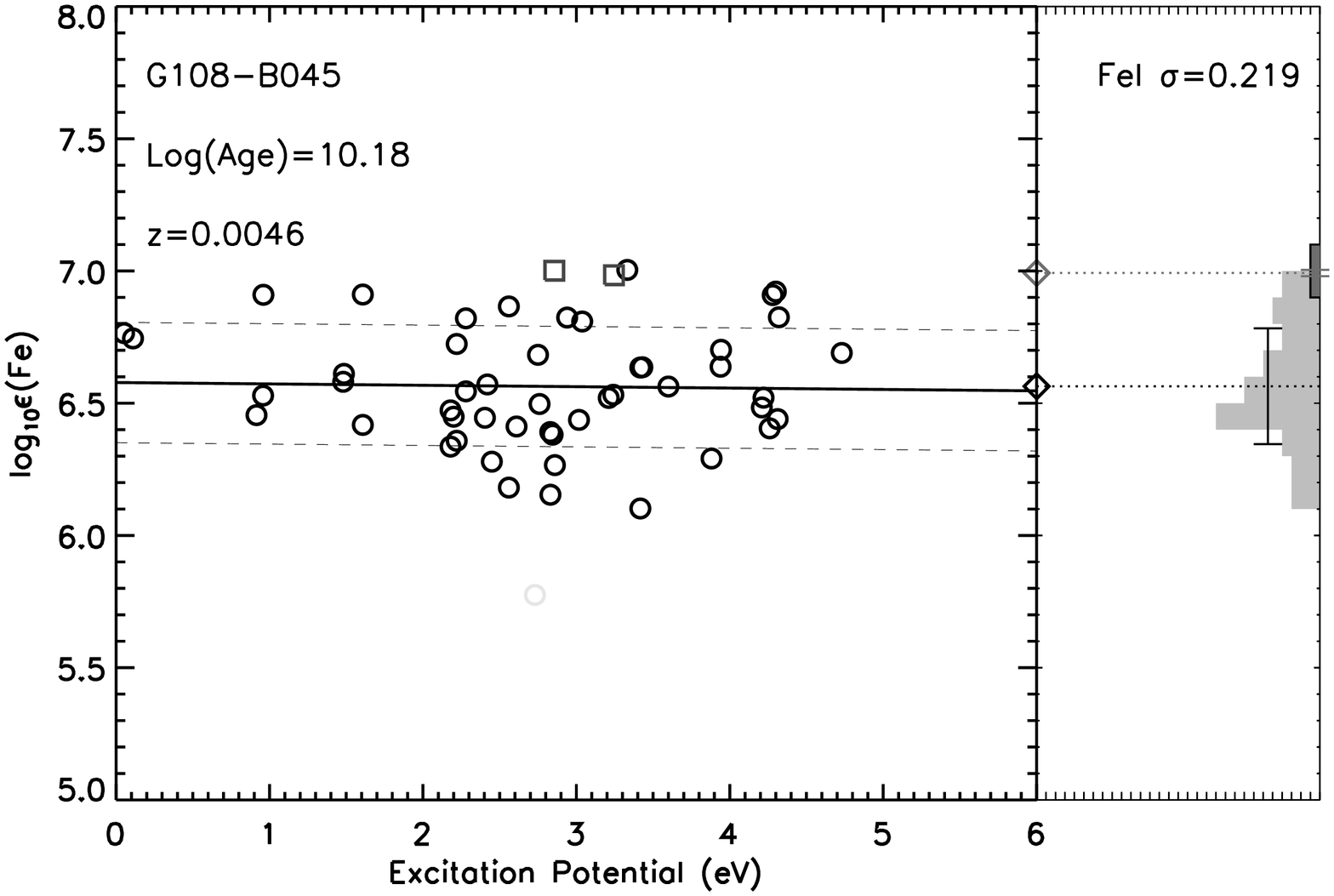}
\includegraphics[angle=90,scale=0.20]{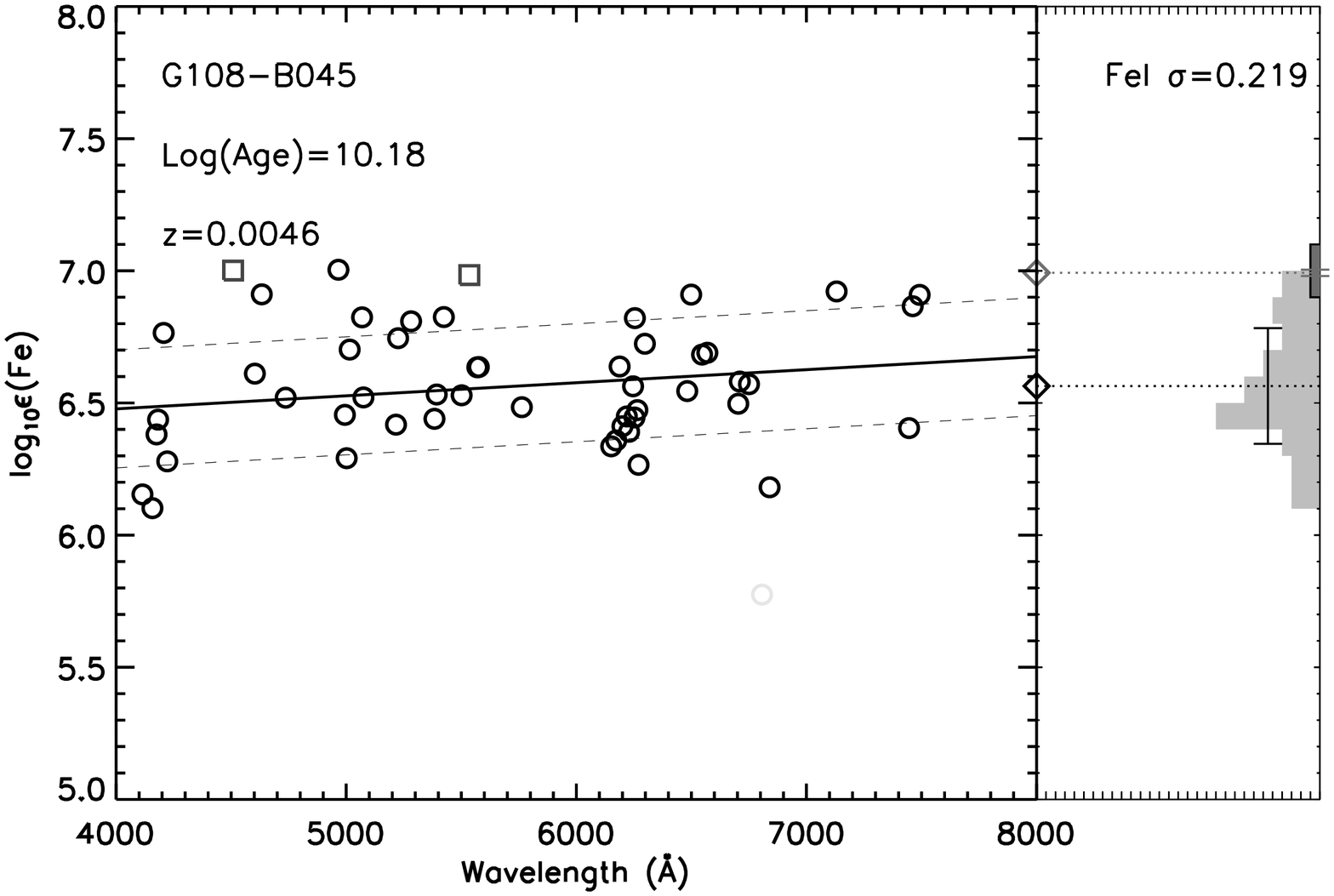}
\includegraphics[angle=90,scale=0.20]{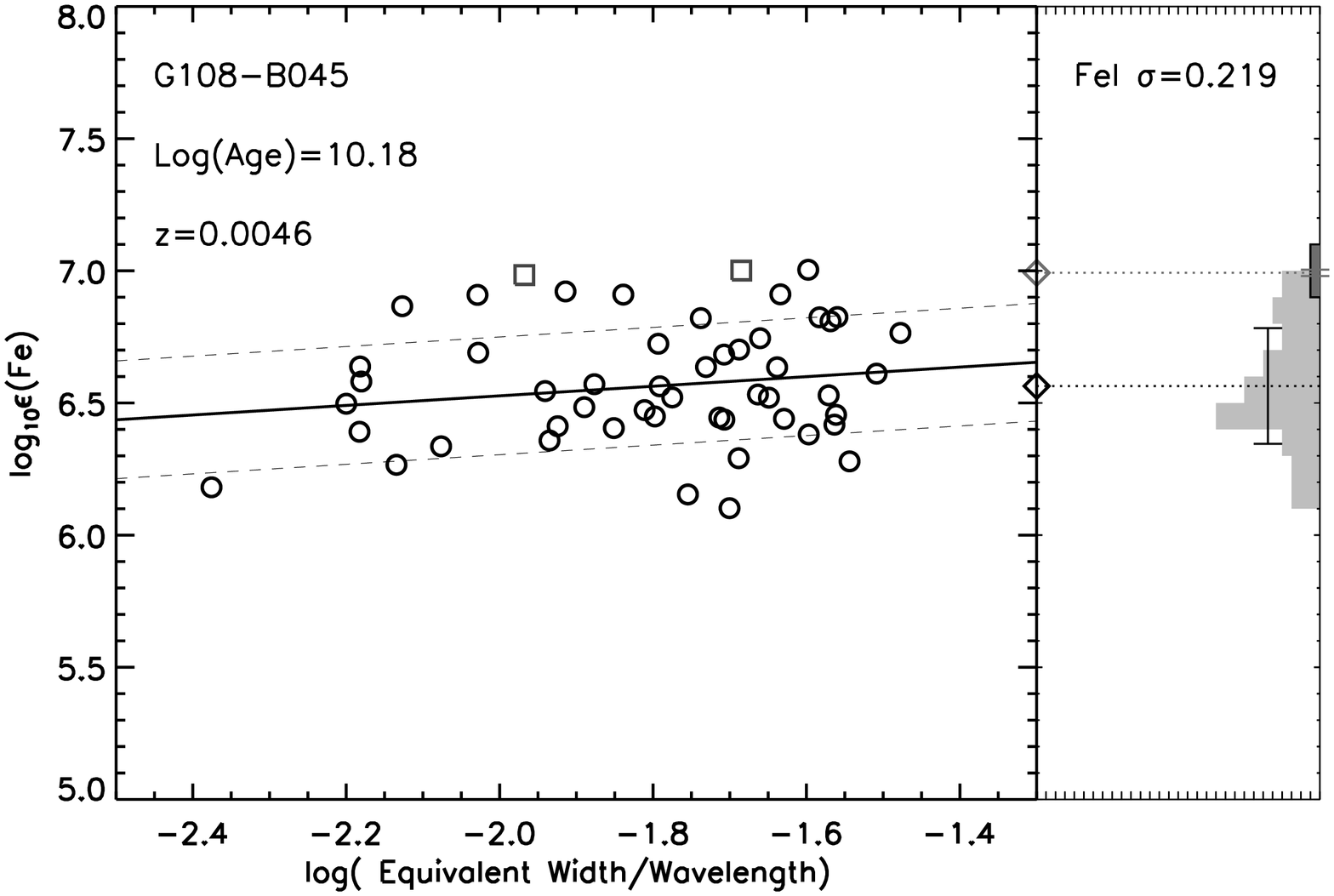}
\includegraphics[angle=90,scale=0.20]{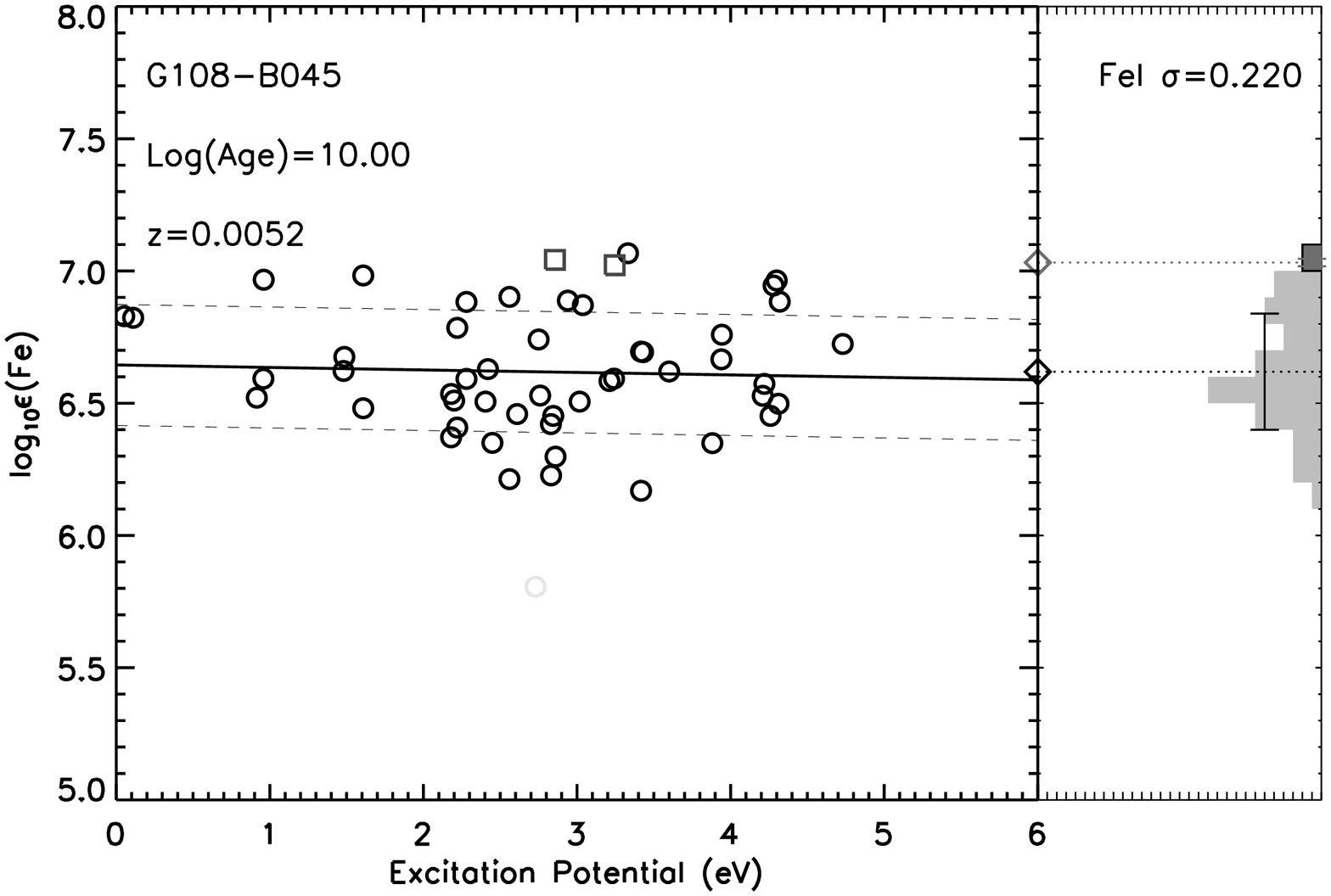}
\includegraphics[angle=90,scale=0.20]{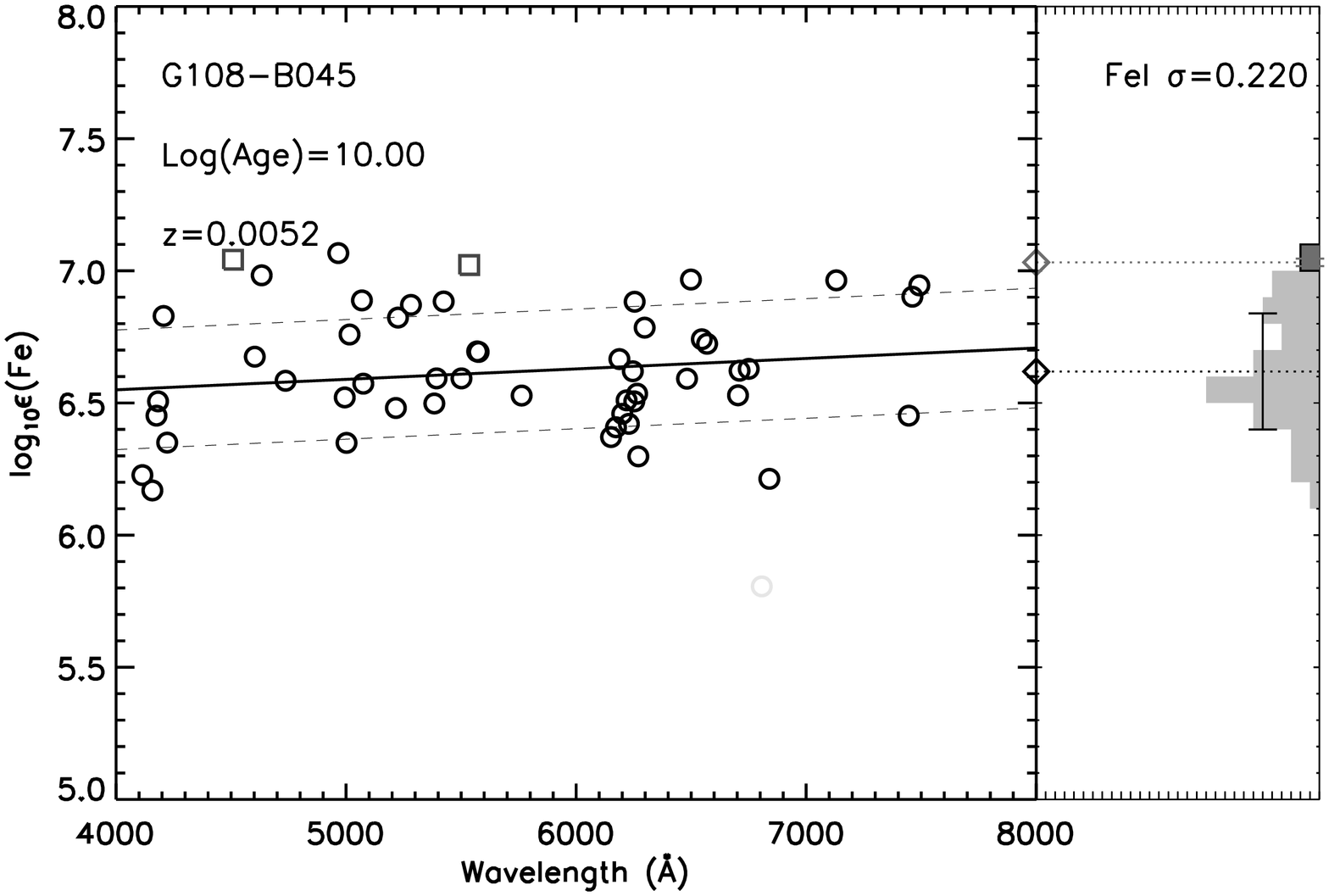}
\includegraphics[angle=90,scale=0.20]{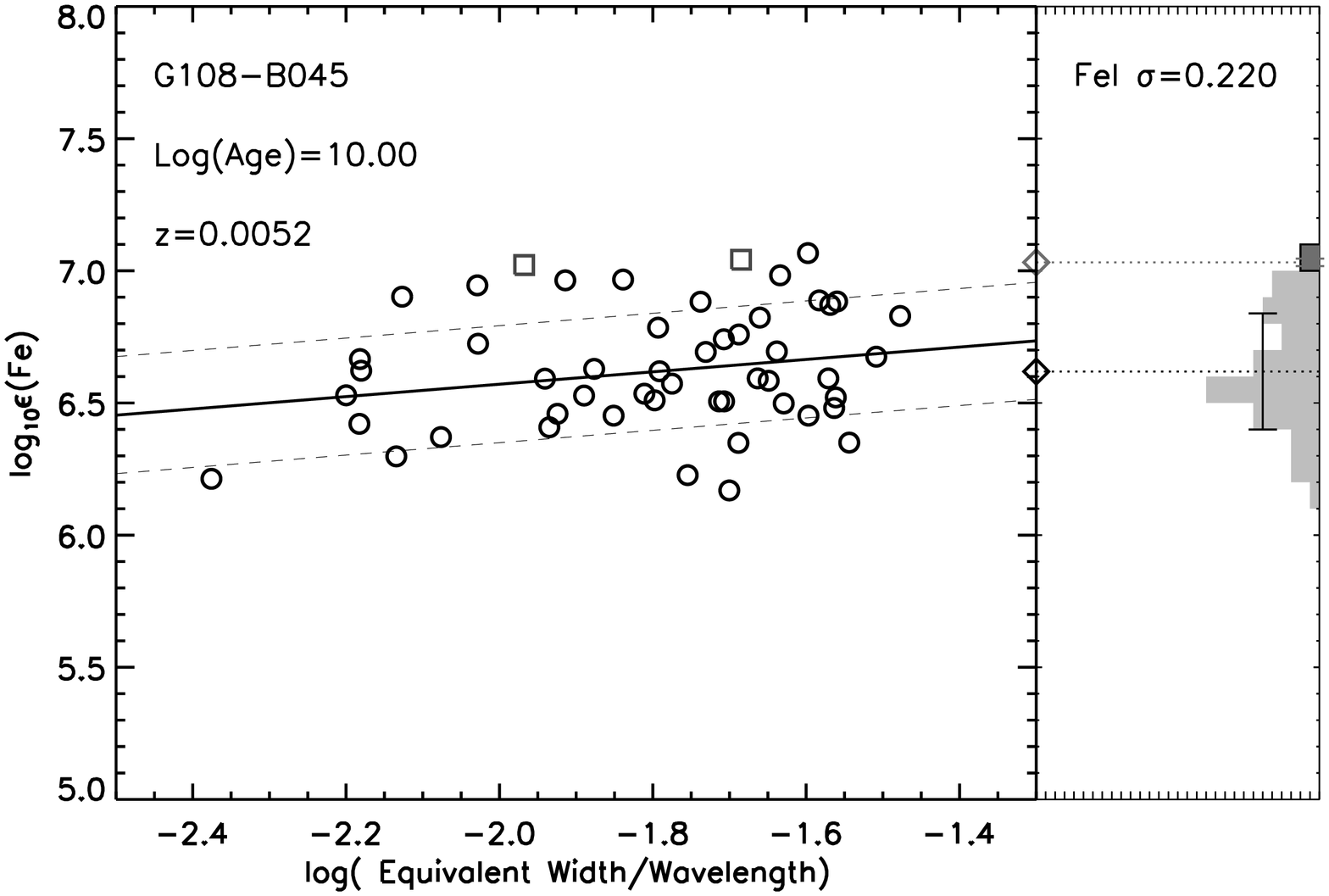}
\includegraphics[angle=90,scale=0.20]{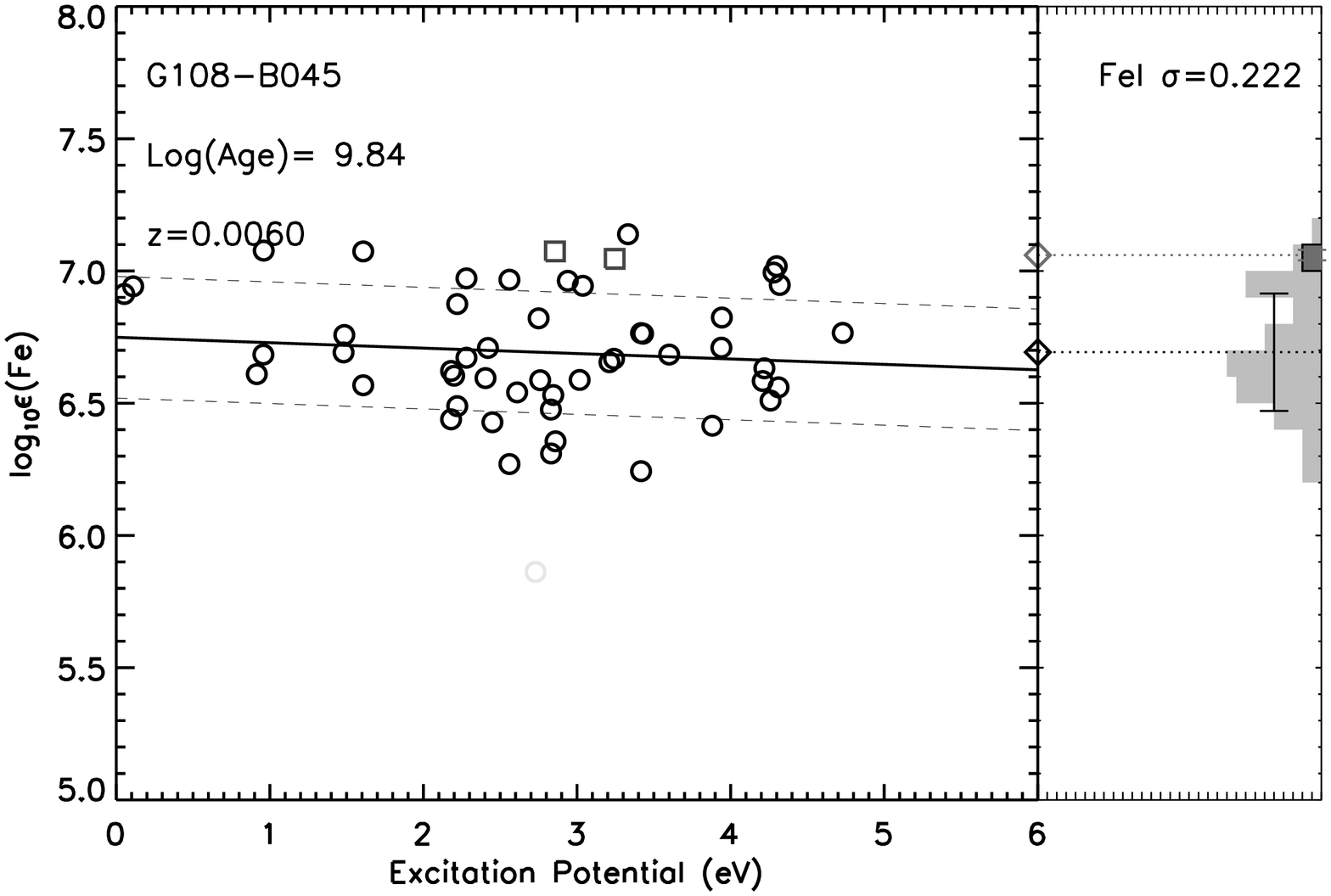}
\includegraphics[angle=90,scale=0.20]{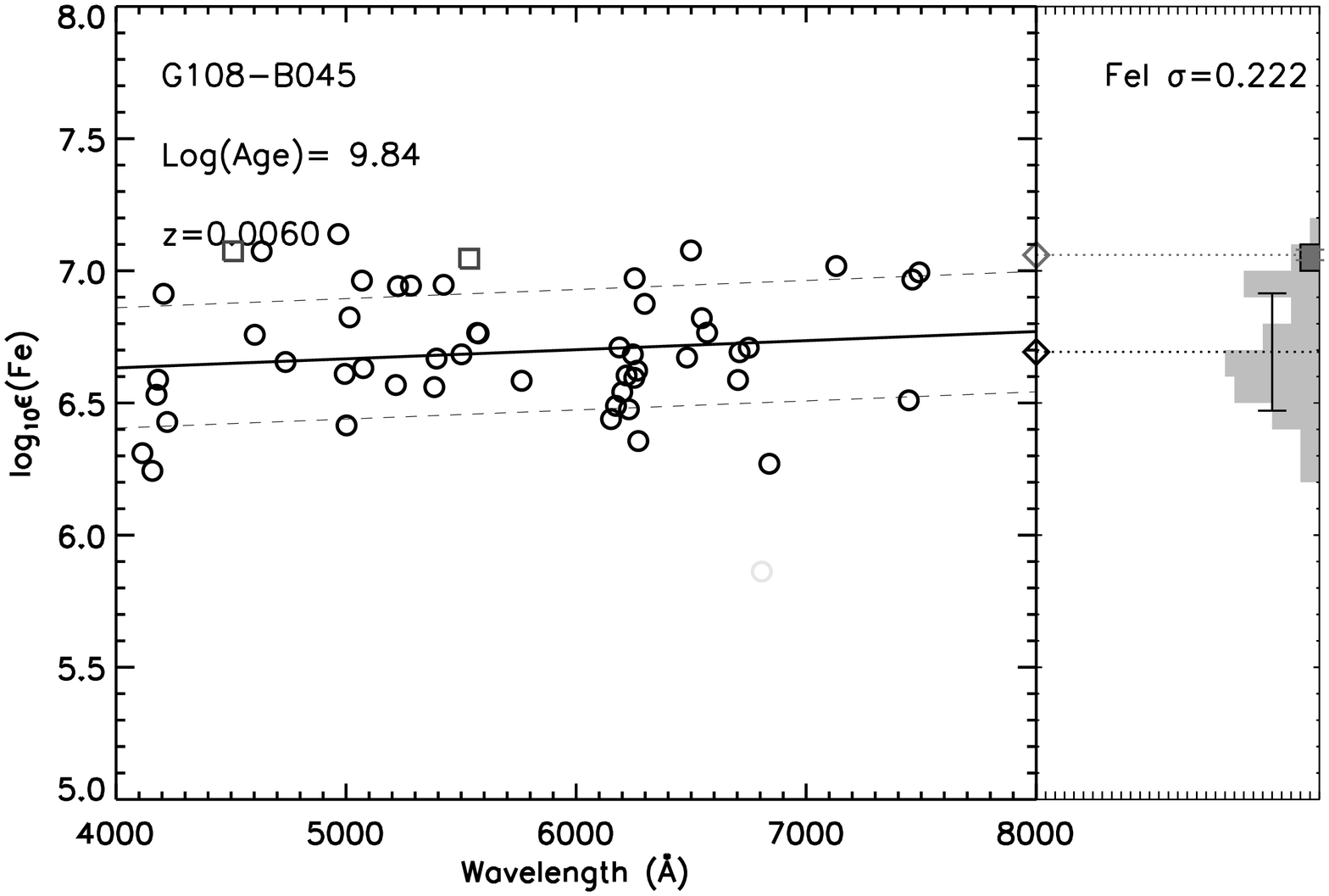}
\includegraphics[angle=90,scale=0.20]{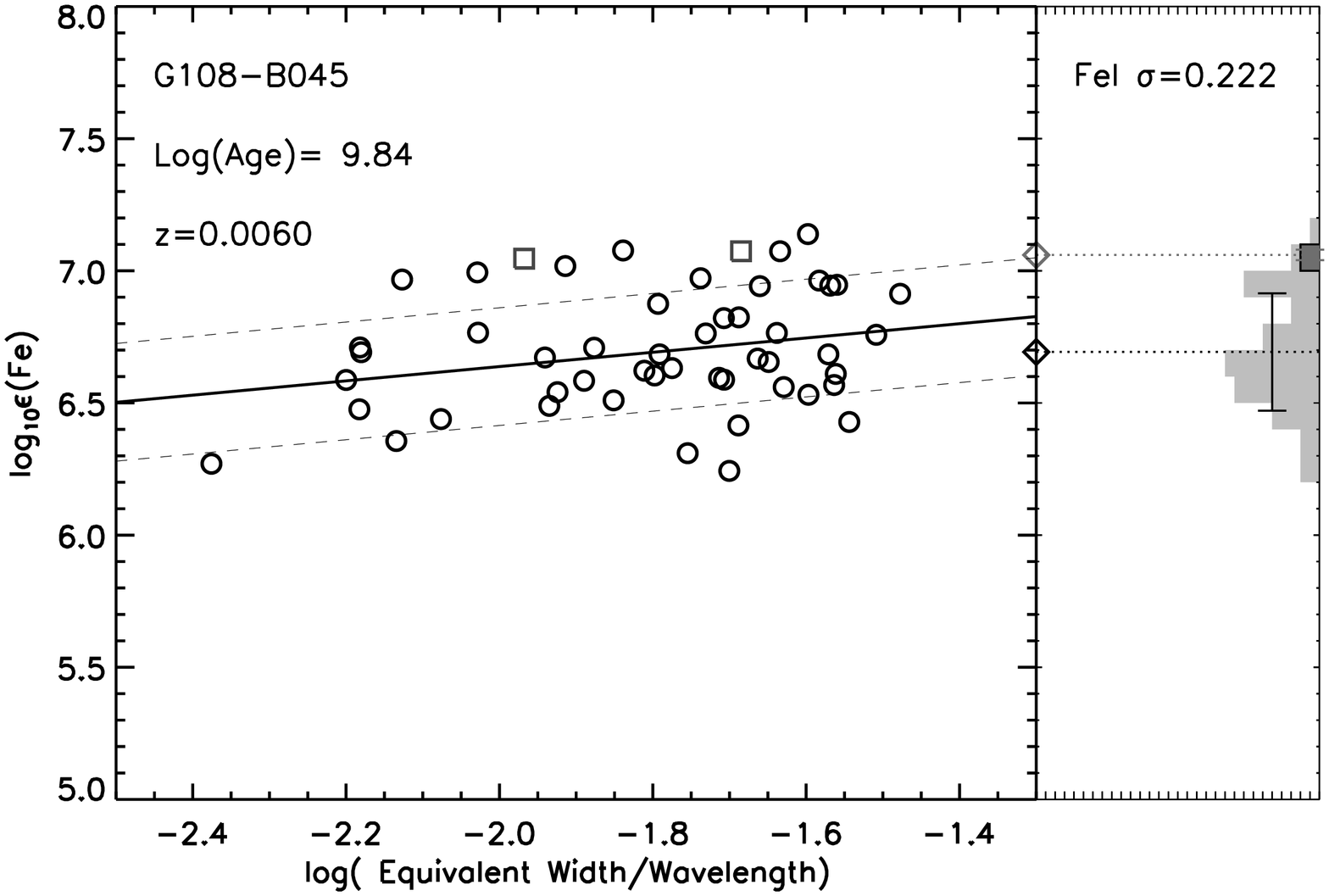}
\includegraphics[angle=90,scale=0.20]{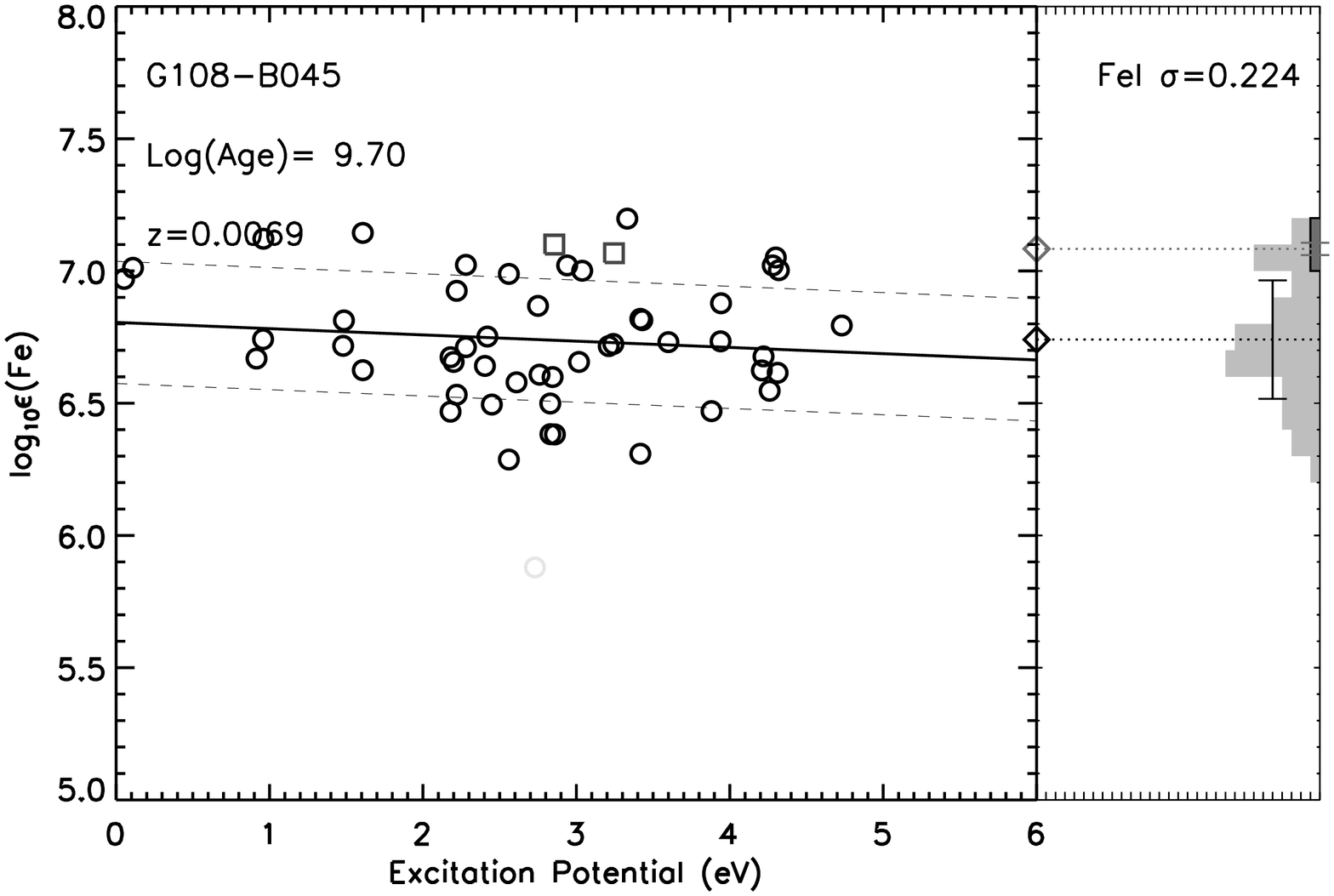}
\includegraphics[angle=90,scale=0.20]{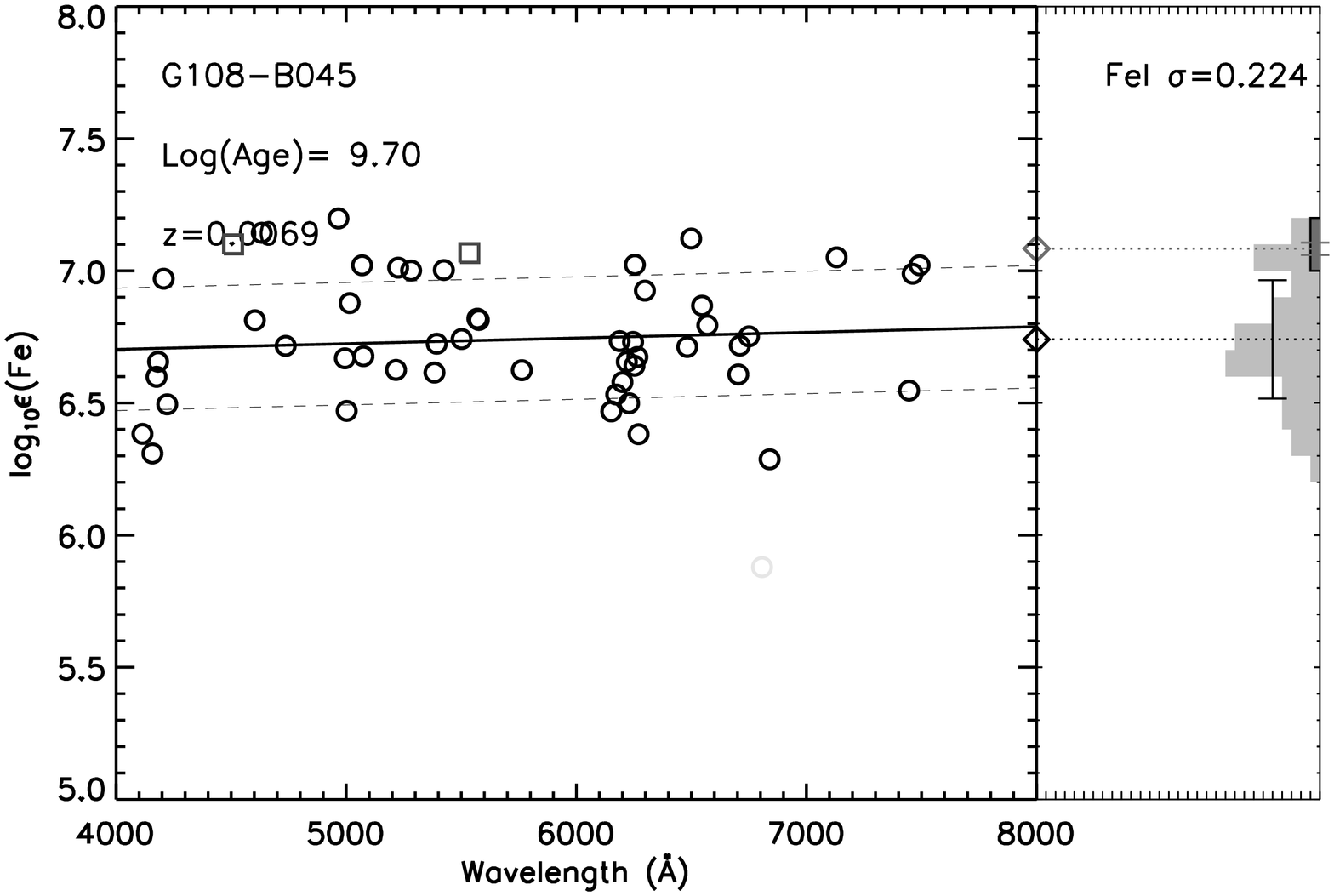}
\includegraphics[angle=90,scale=0.20]{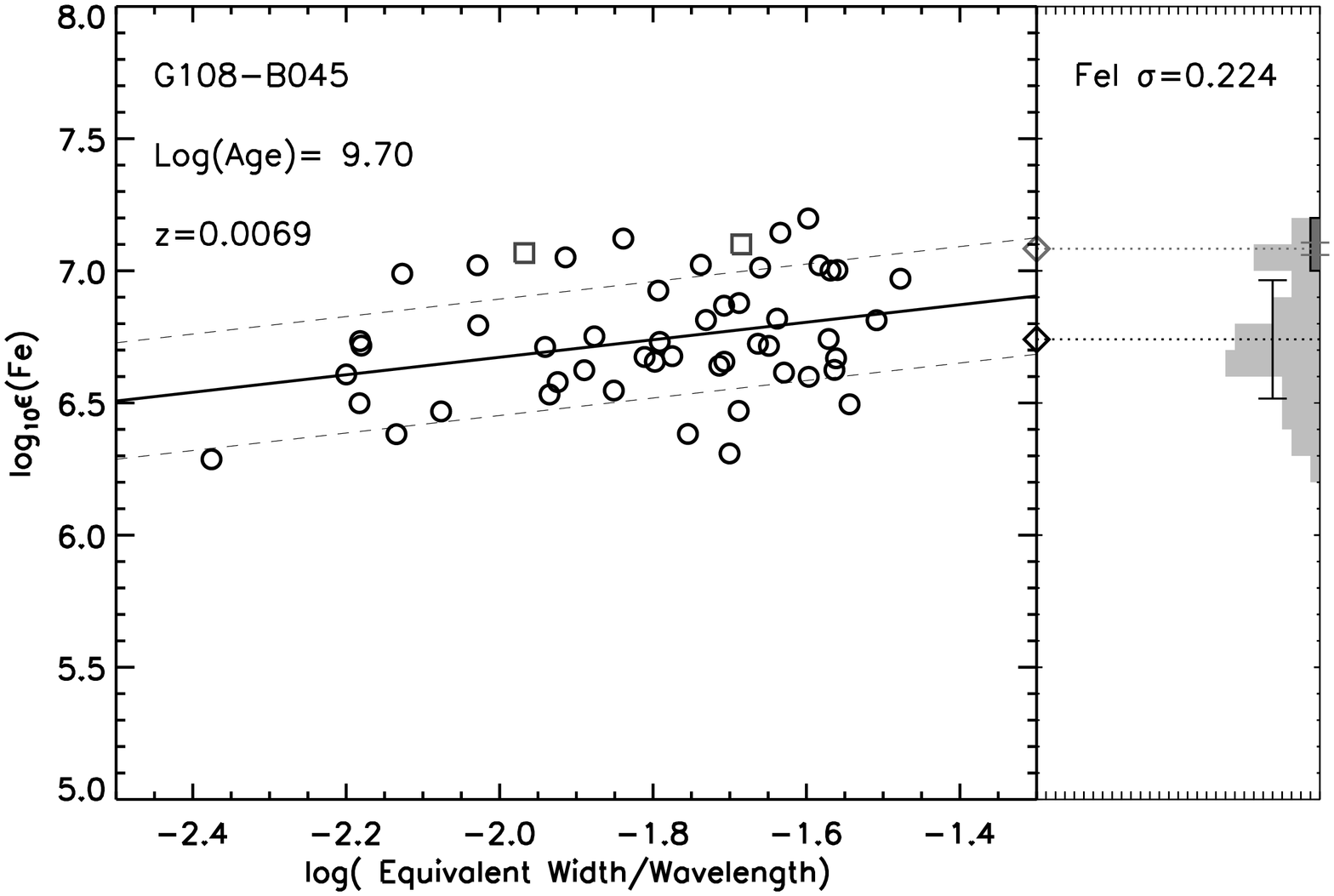}
\includegraphics[angle=90,scale=0.20]{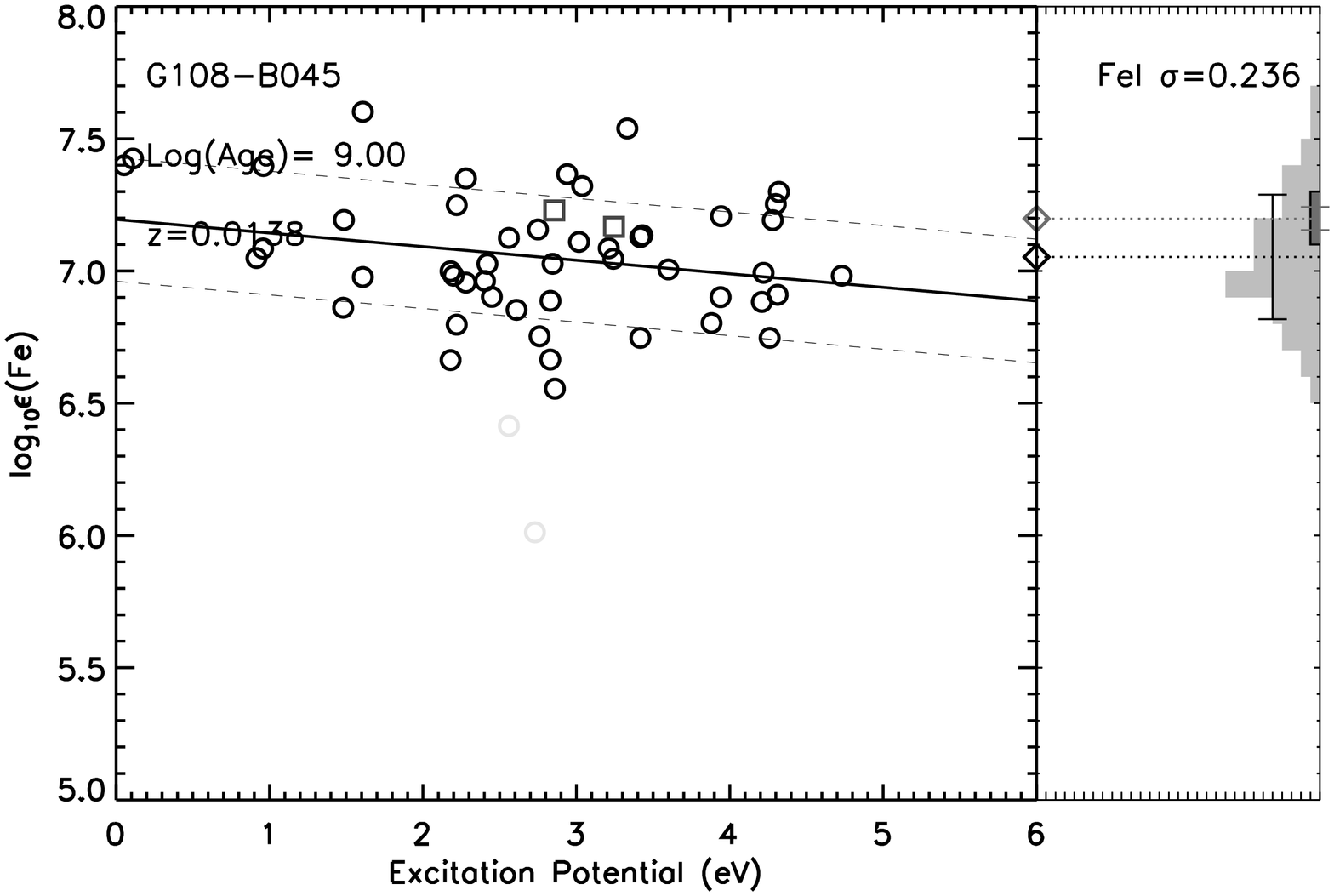}
\includegraphics[angle=90,scale=0.20]{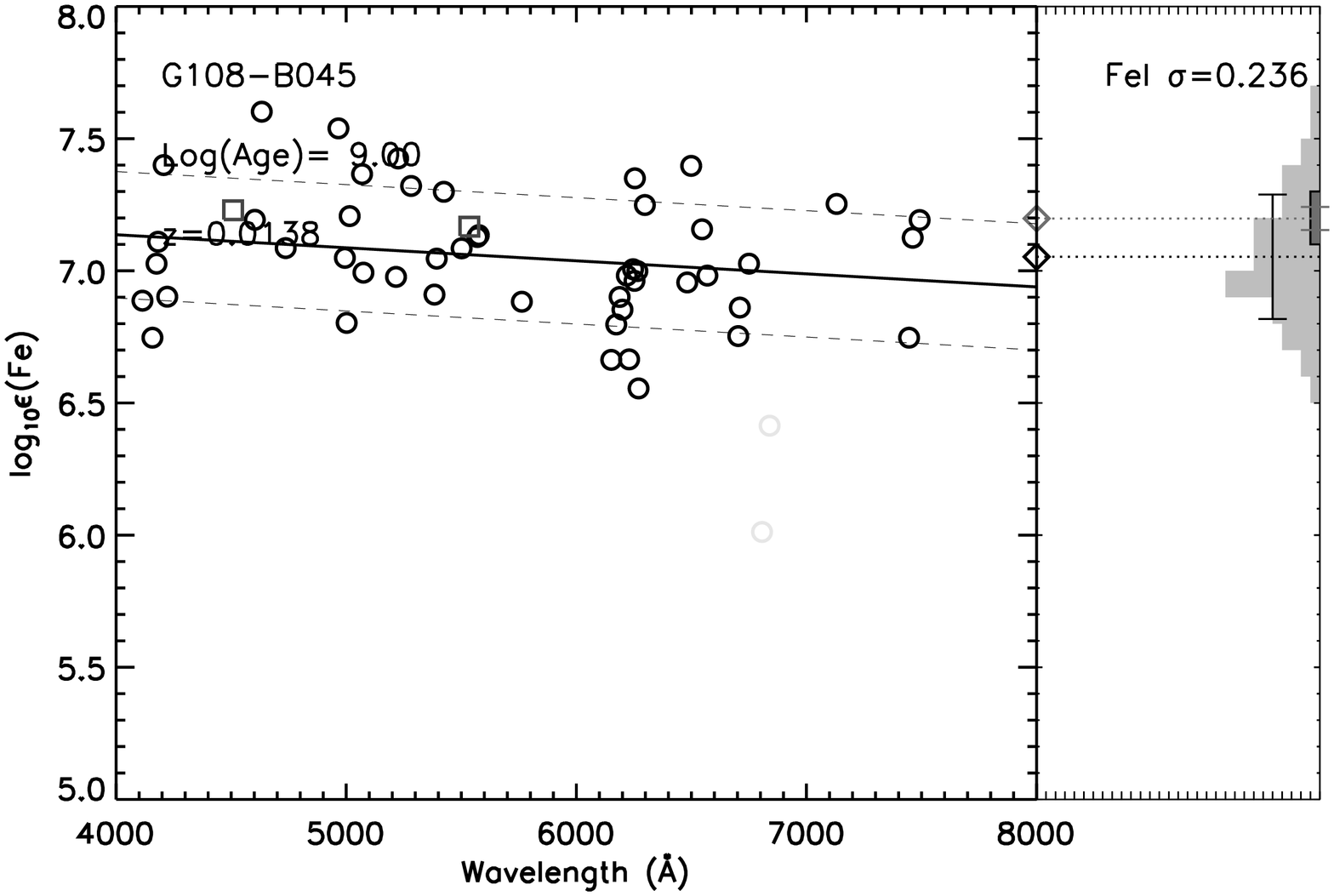}
\includegraphics[angle=90,scale=0.20]{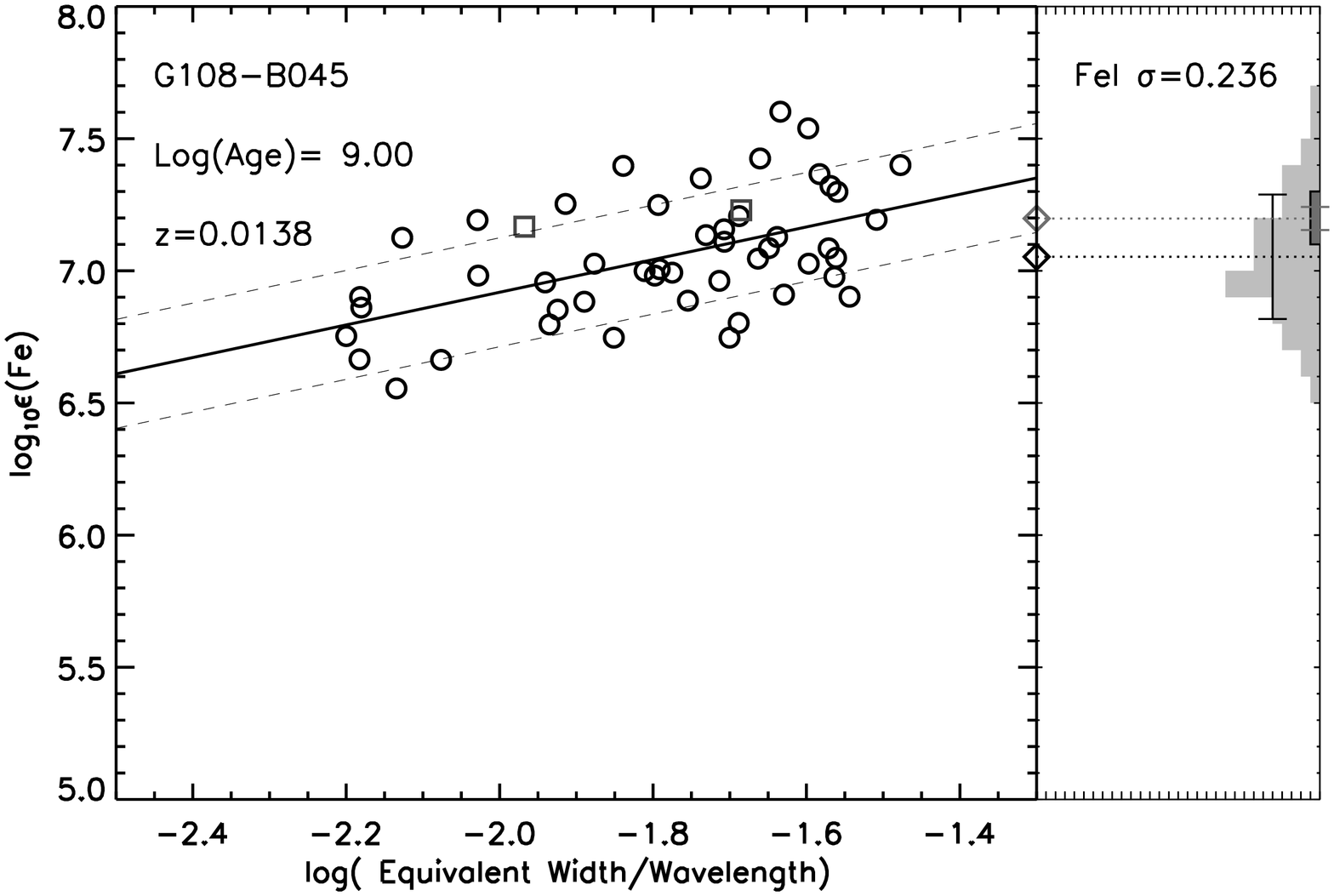}

\caption{
Diagnostic plots for G108-B045, for ages of 15, 10, 7, 5, and 1 Gyr (top to bottom). Oldest solutions have
  smallest Fe I standard deviation and smallest dependence on EP,
  wavelength, or  or reduced equivalent width ( log(EW/wavelength)) for this cluster. Fe I and Fe II lines are marked by dark circles and light squares, respectively. Gray points mark
  lines rejected by a sigma clipping routine when calculating the mean
  abundances. The solid line shows the linear fit to the Fe I lines and
  dashed lines show the 1 $\sigma$ deviation of points around the fit.
  Dark and light diamonds mark the final average Fe I and Fe II
  abundances. }
\label{fig:g108 diagnostics} 
\end{figure*}

\begin{figure*}
\centering

\includegraphics[angle=90,scale=0.20]{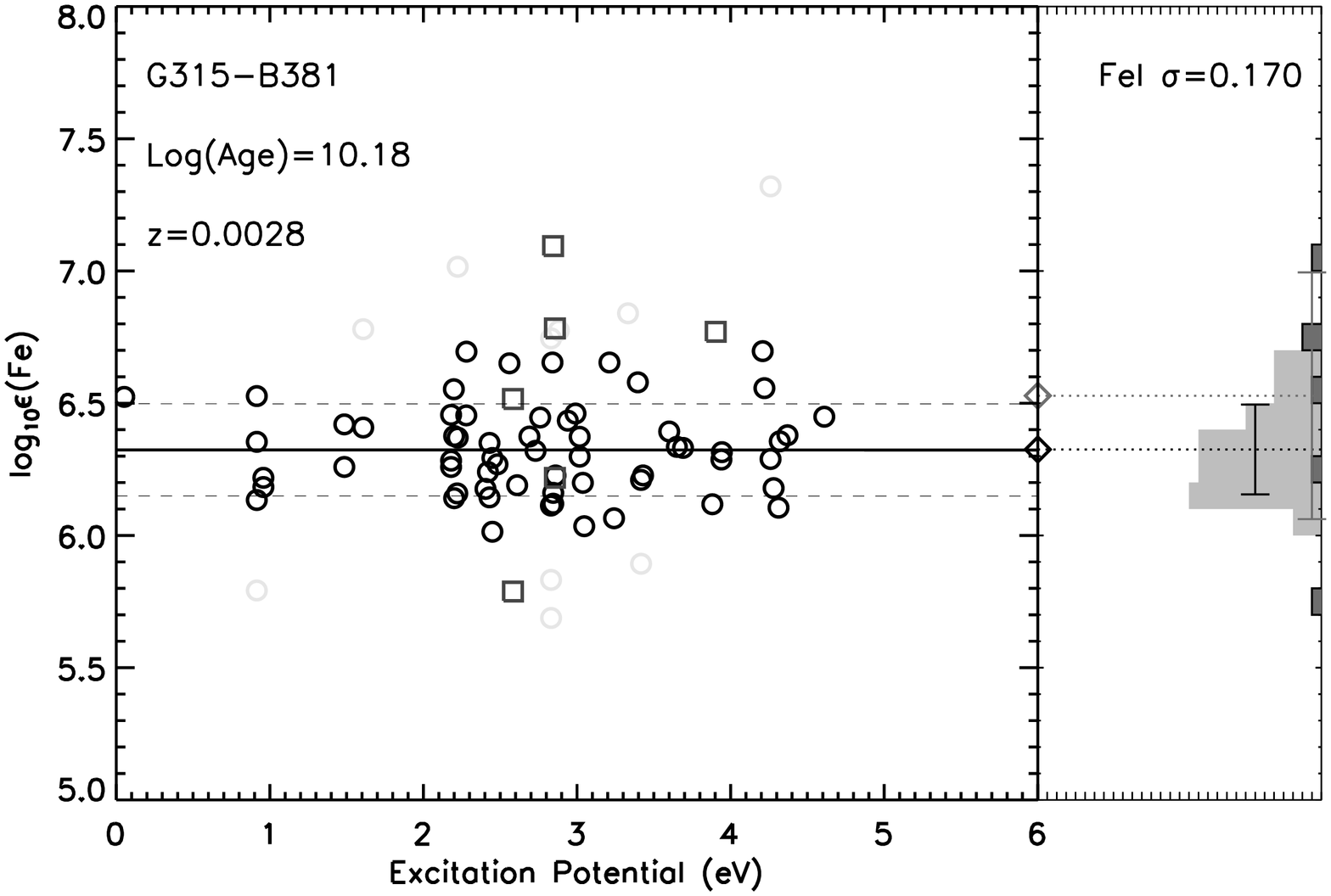}
\includegraphics[angle=90,scale=0.20]{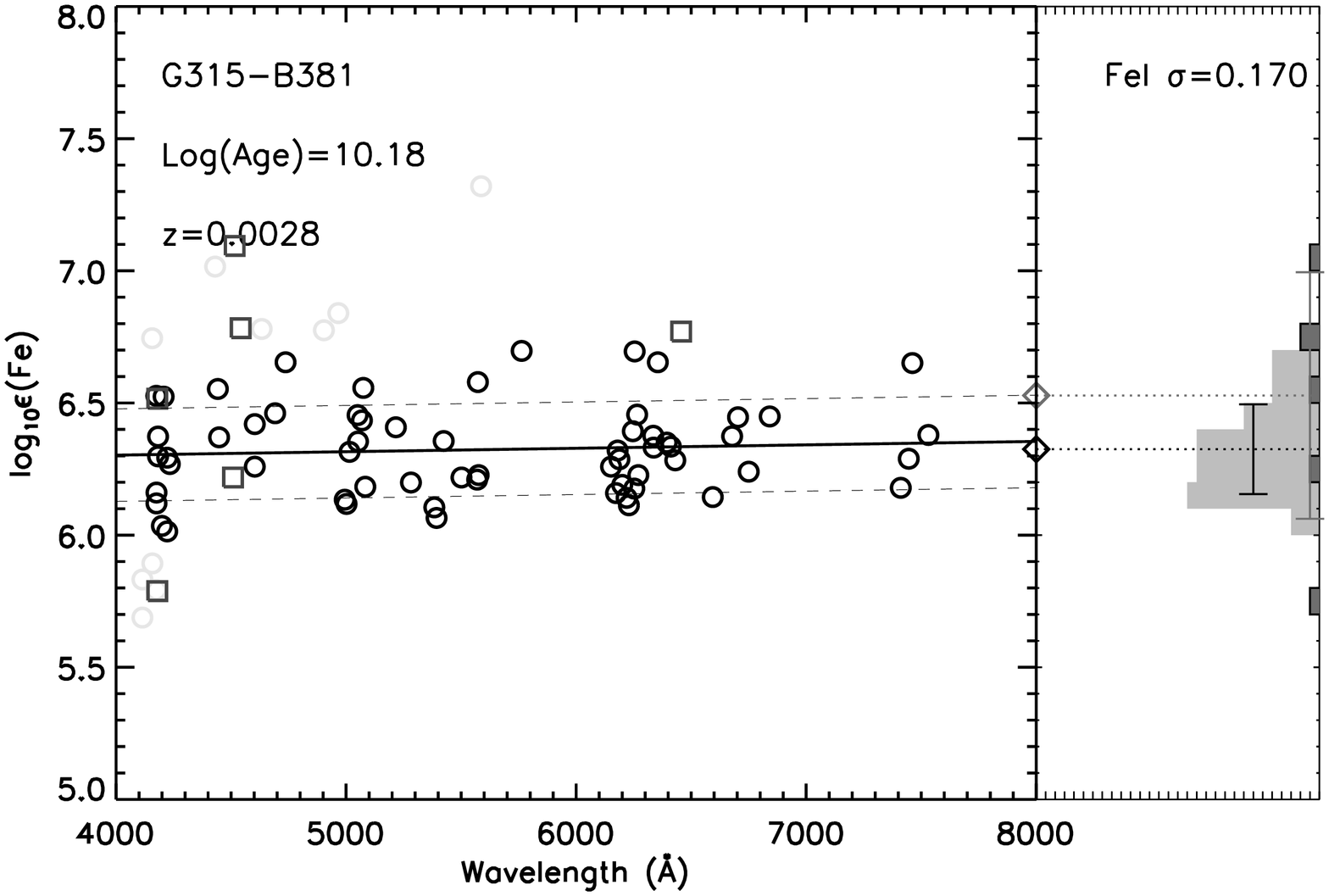}
\includegraphics[angle=90,scale=0.20]{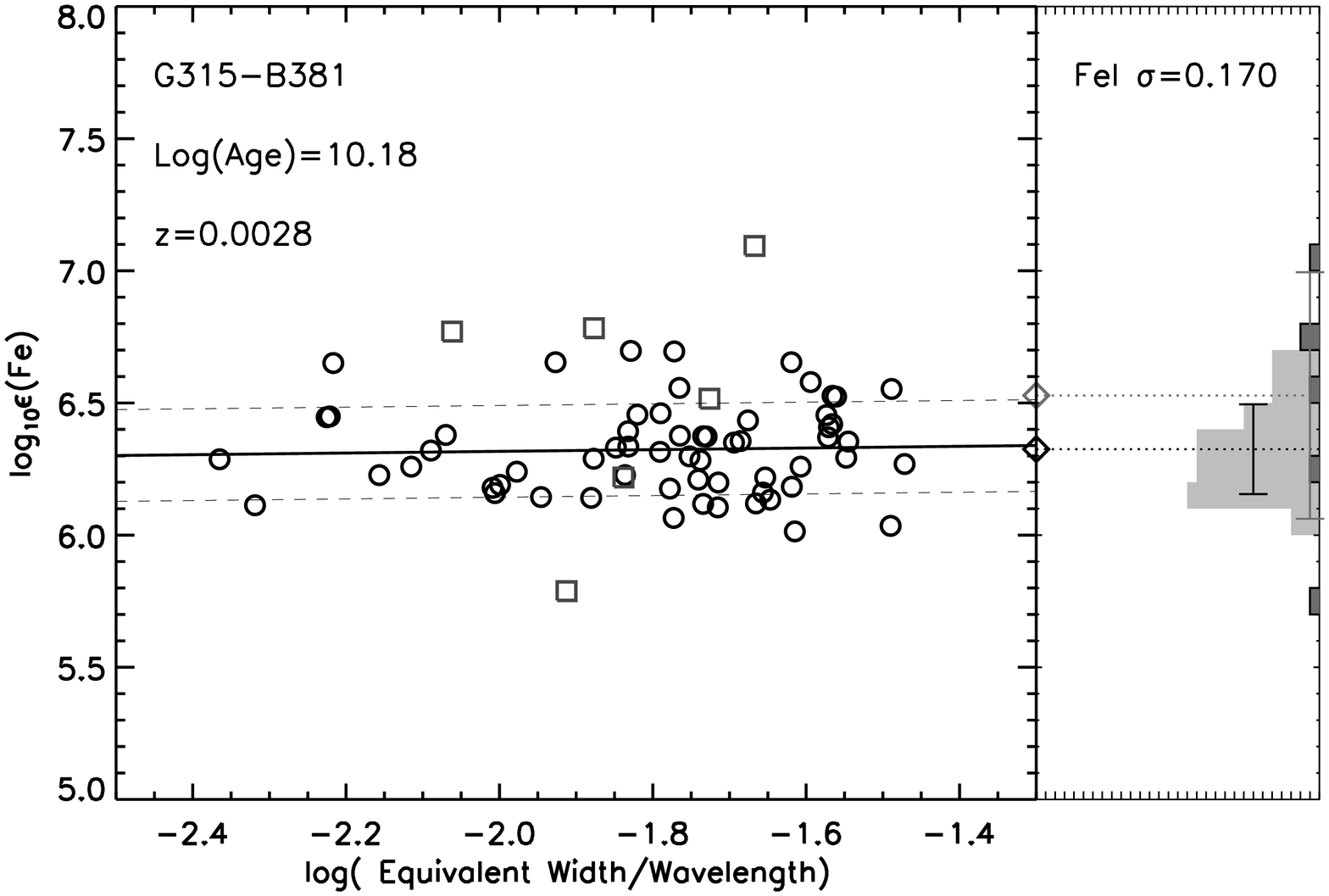}
\includegraphics[angle=90,scale=0.20]{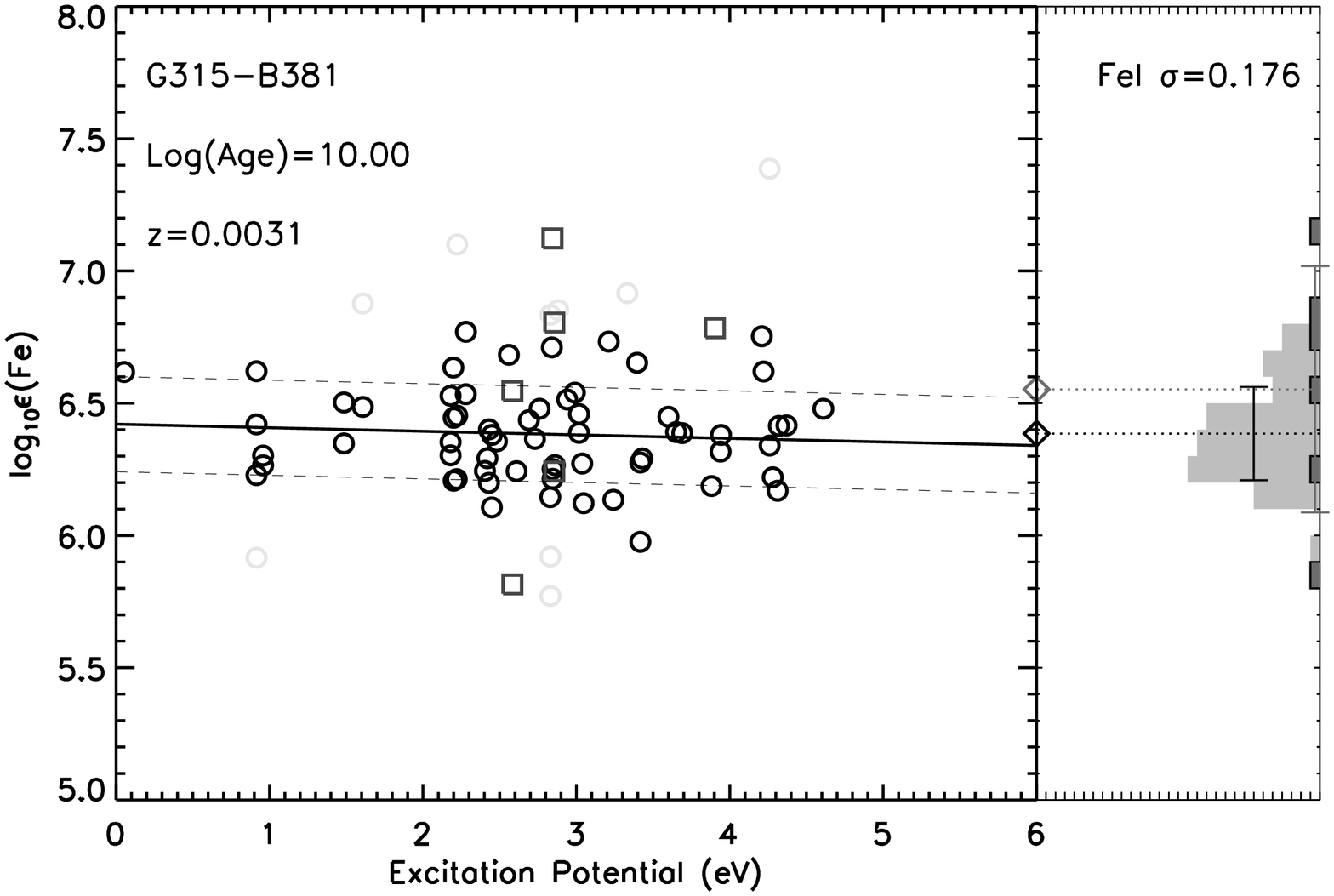}
\includegraphics[angle=90,scale=0.20]{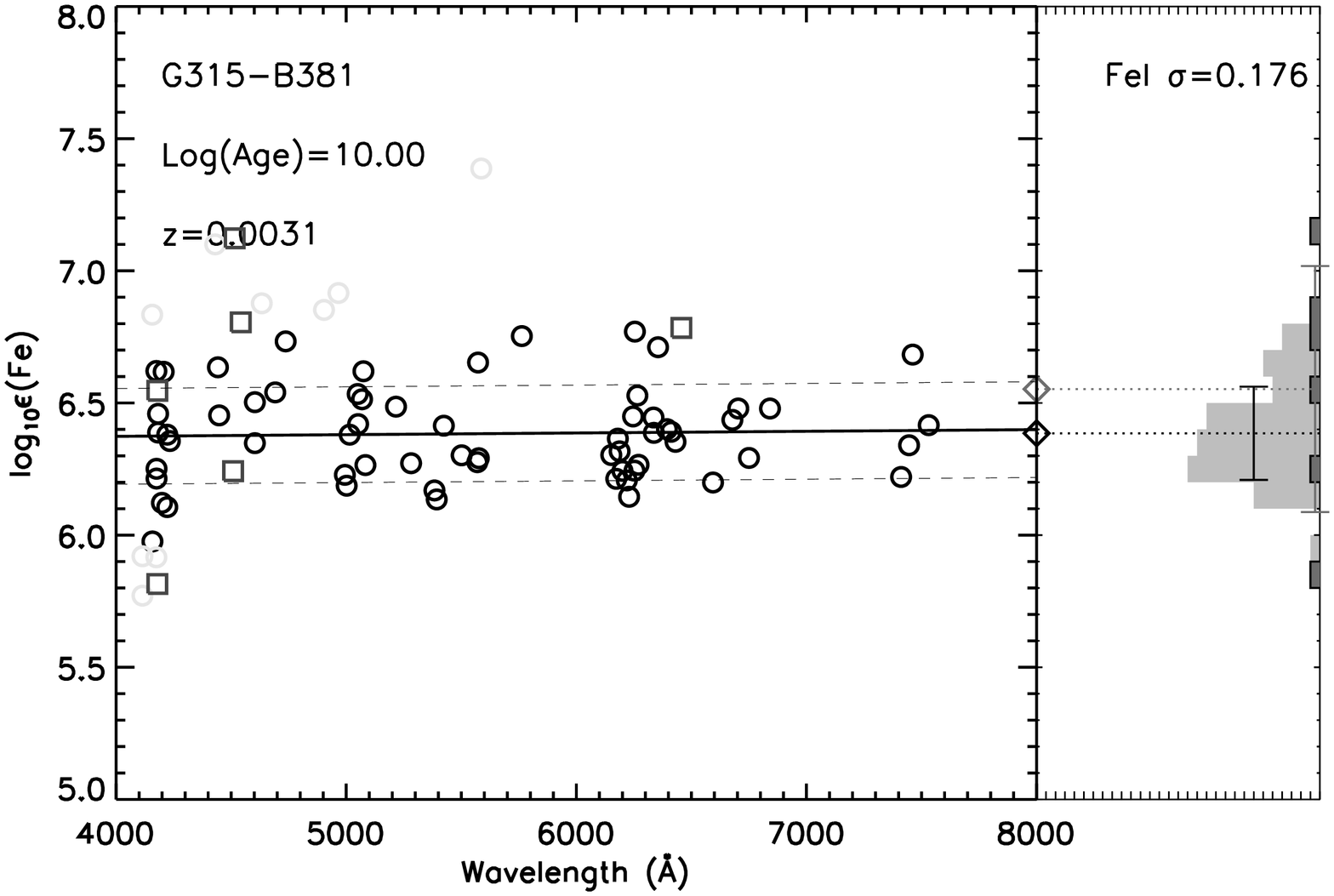}
\includegraphics[angle=90,scale=0.20]{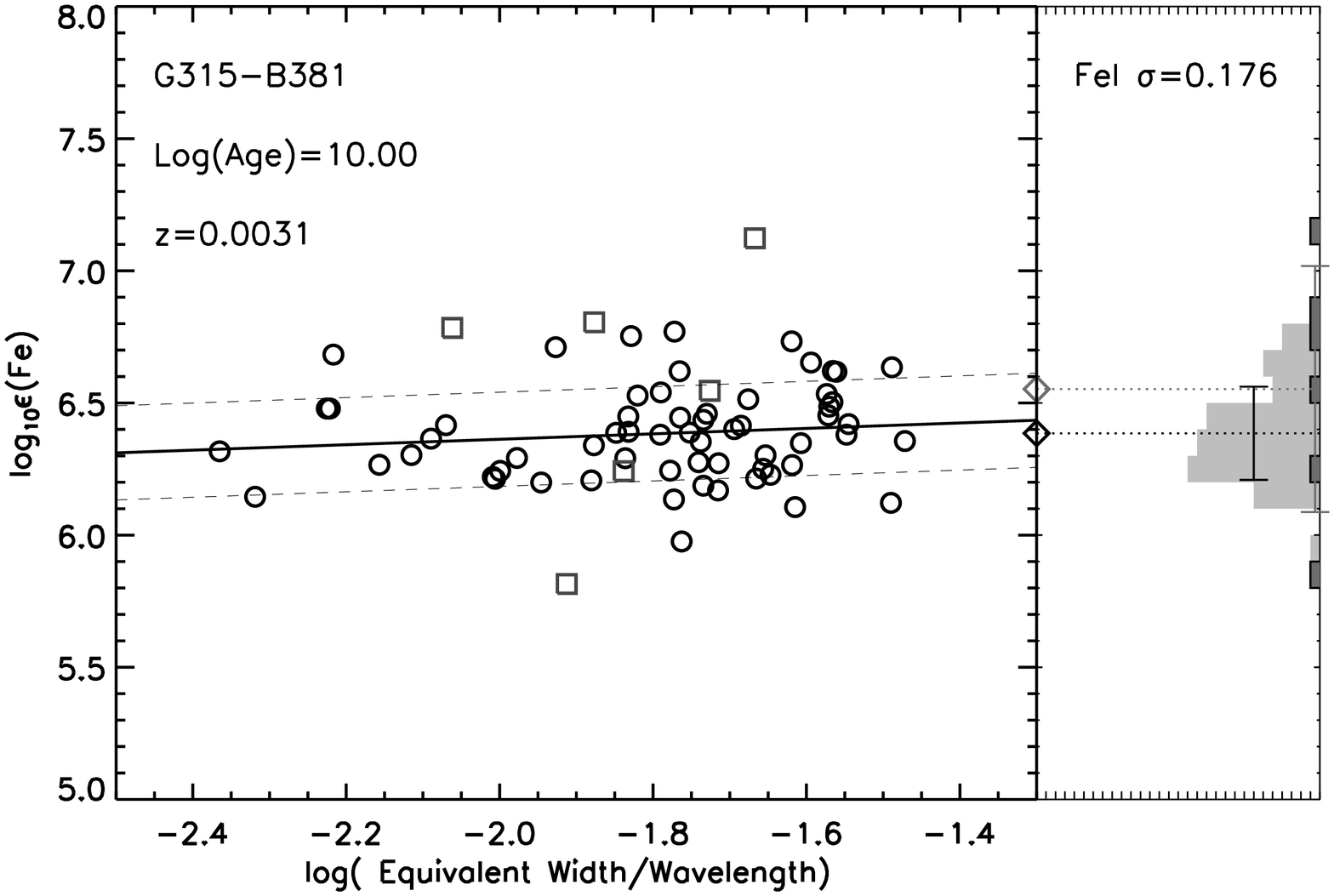}
\includegraphics[angle=90,scale=0.20]{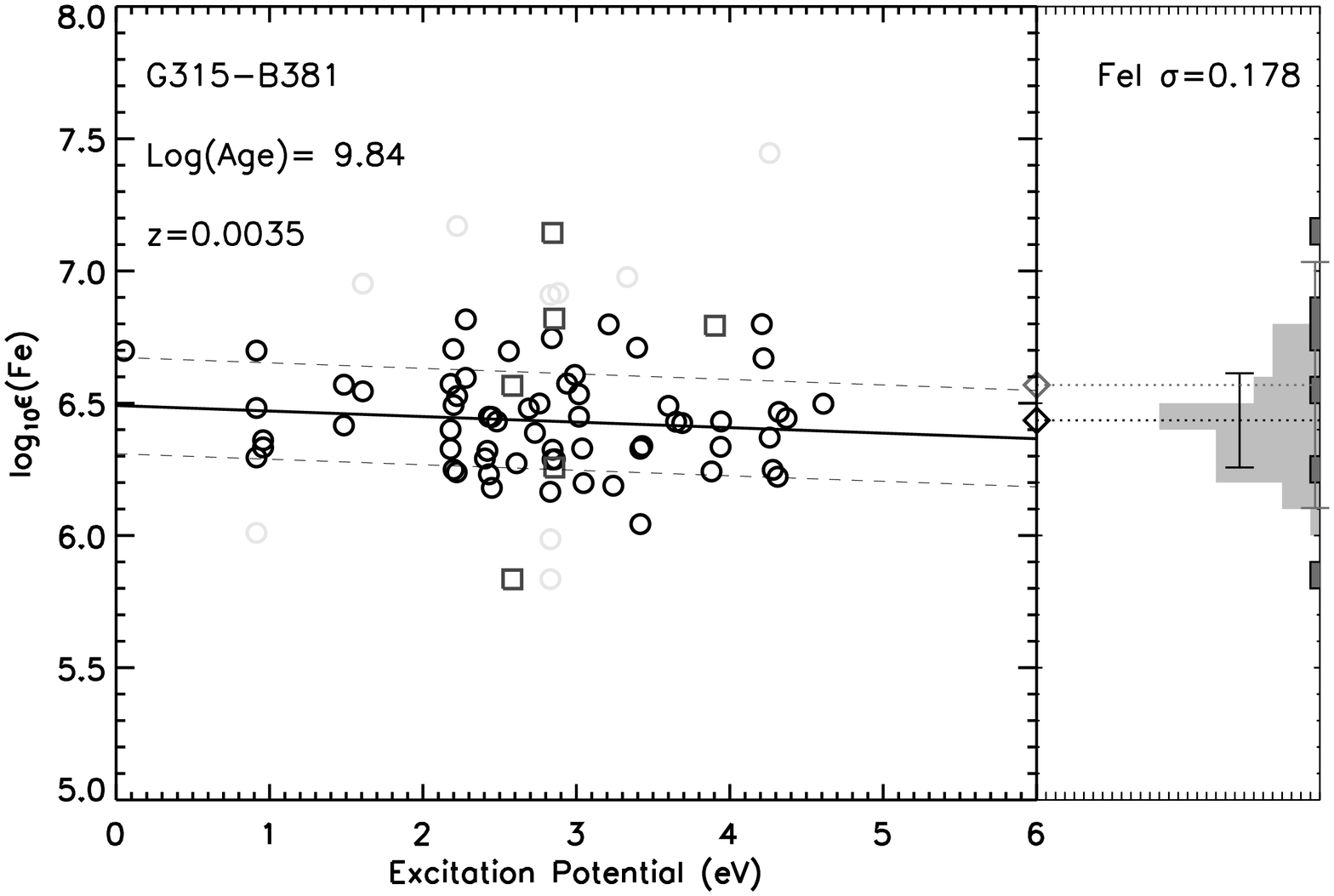}
\includegraphics[angle=90,scale=0.20]{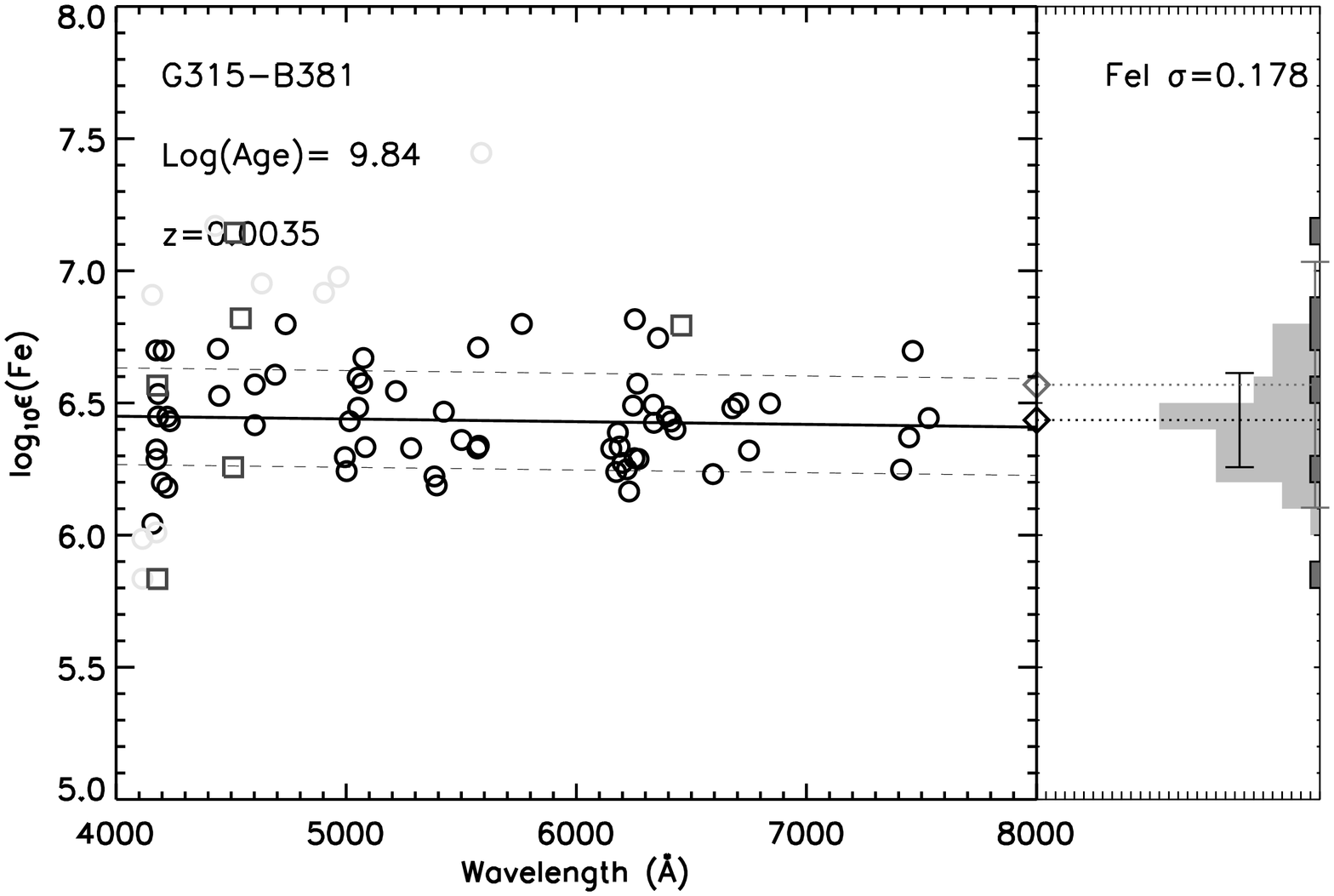}
\includegraphics[angle=90,scale=0.20]{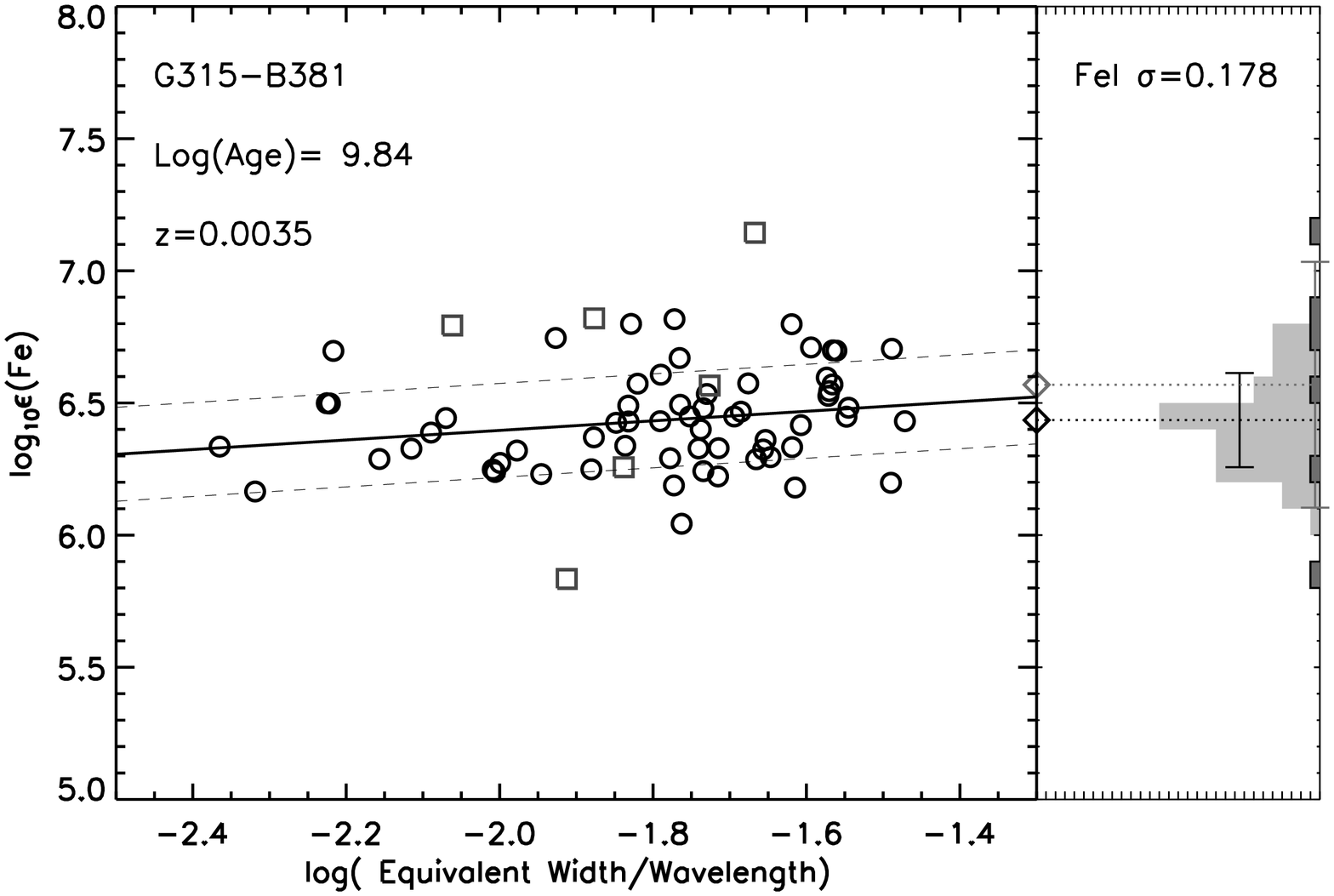}
\includegraphics[angle=90,scale=0.20]{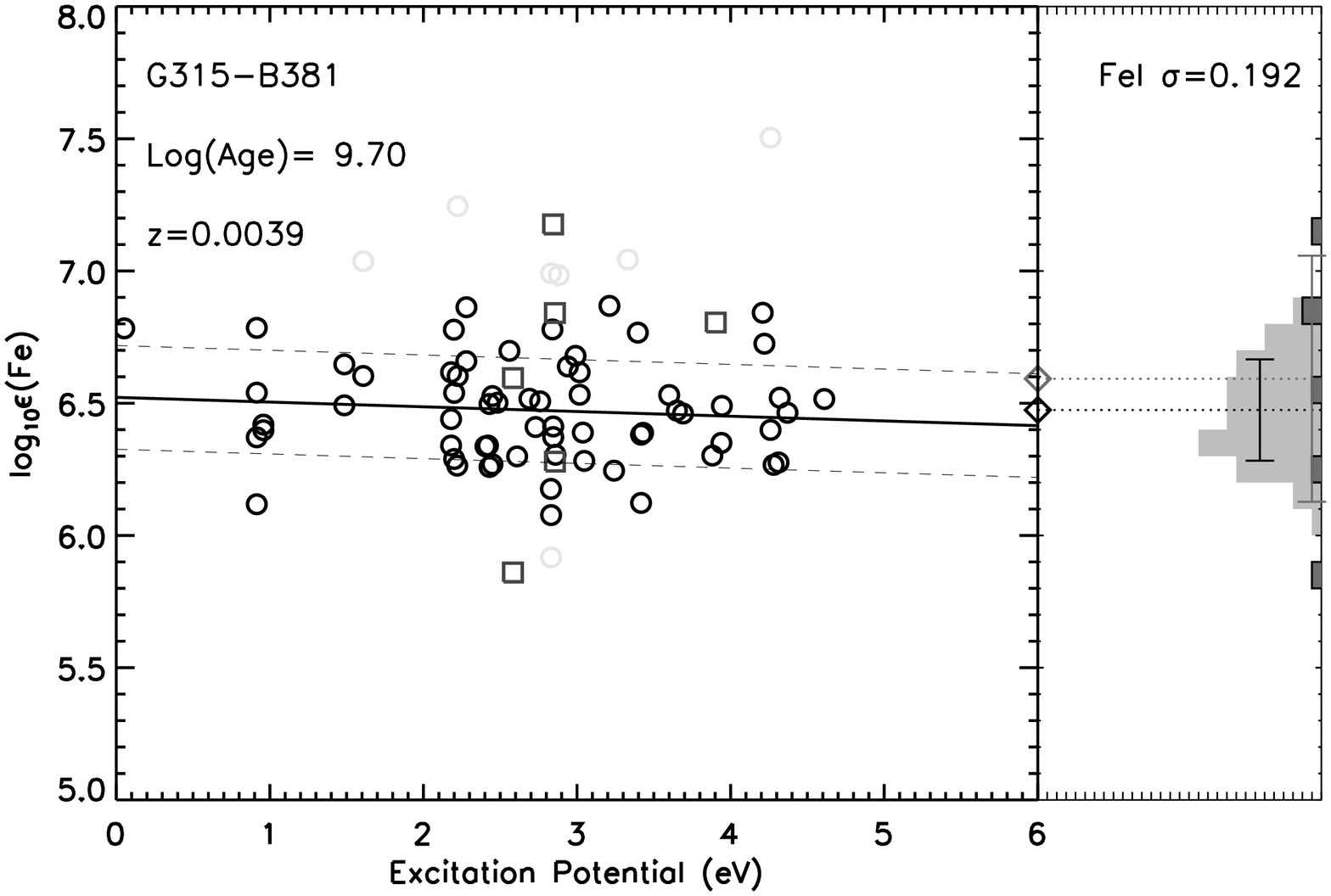}
\includegraphics[angle=90,scale=0.20]{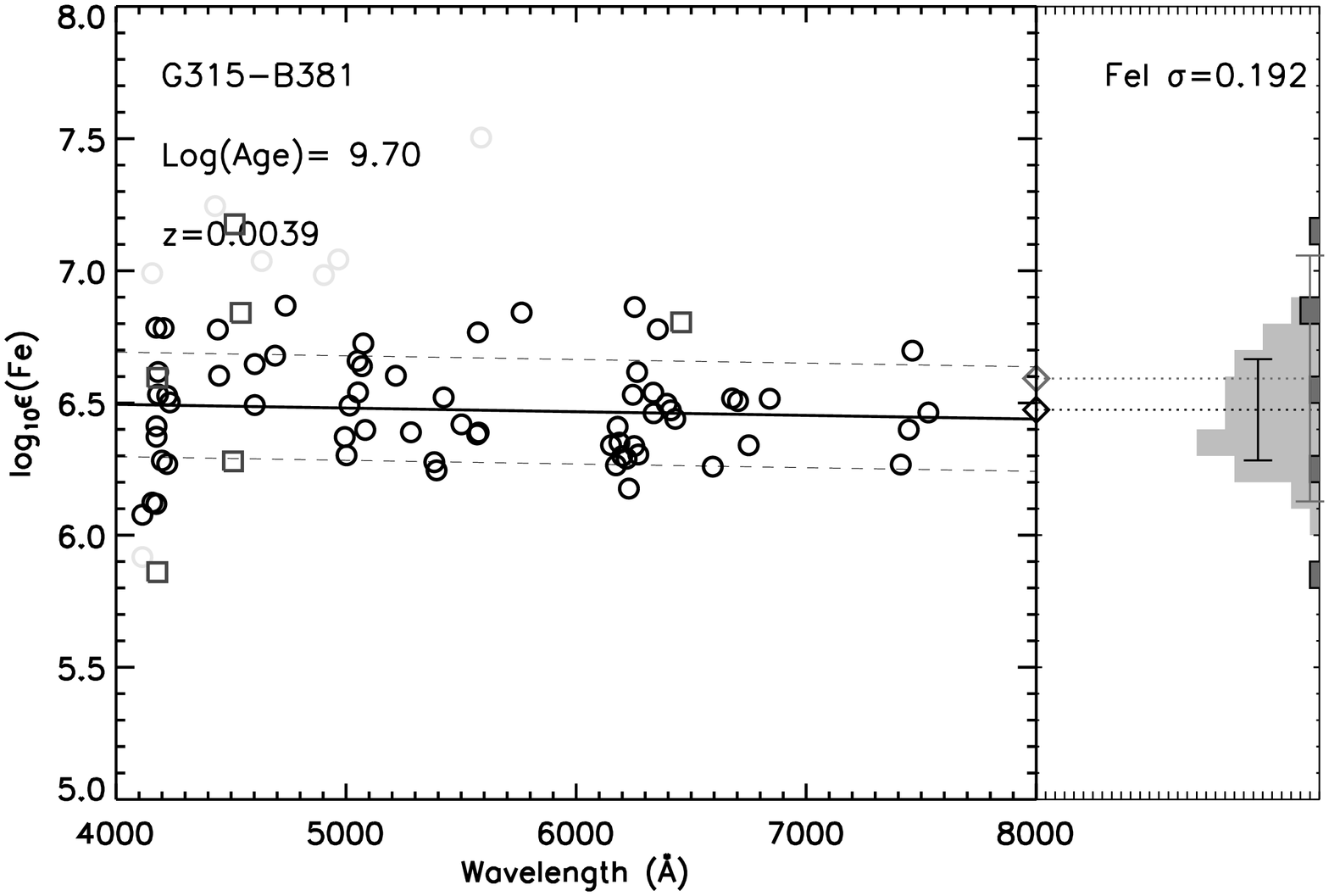}
\includegraphics[angle=90,scale=0.20]{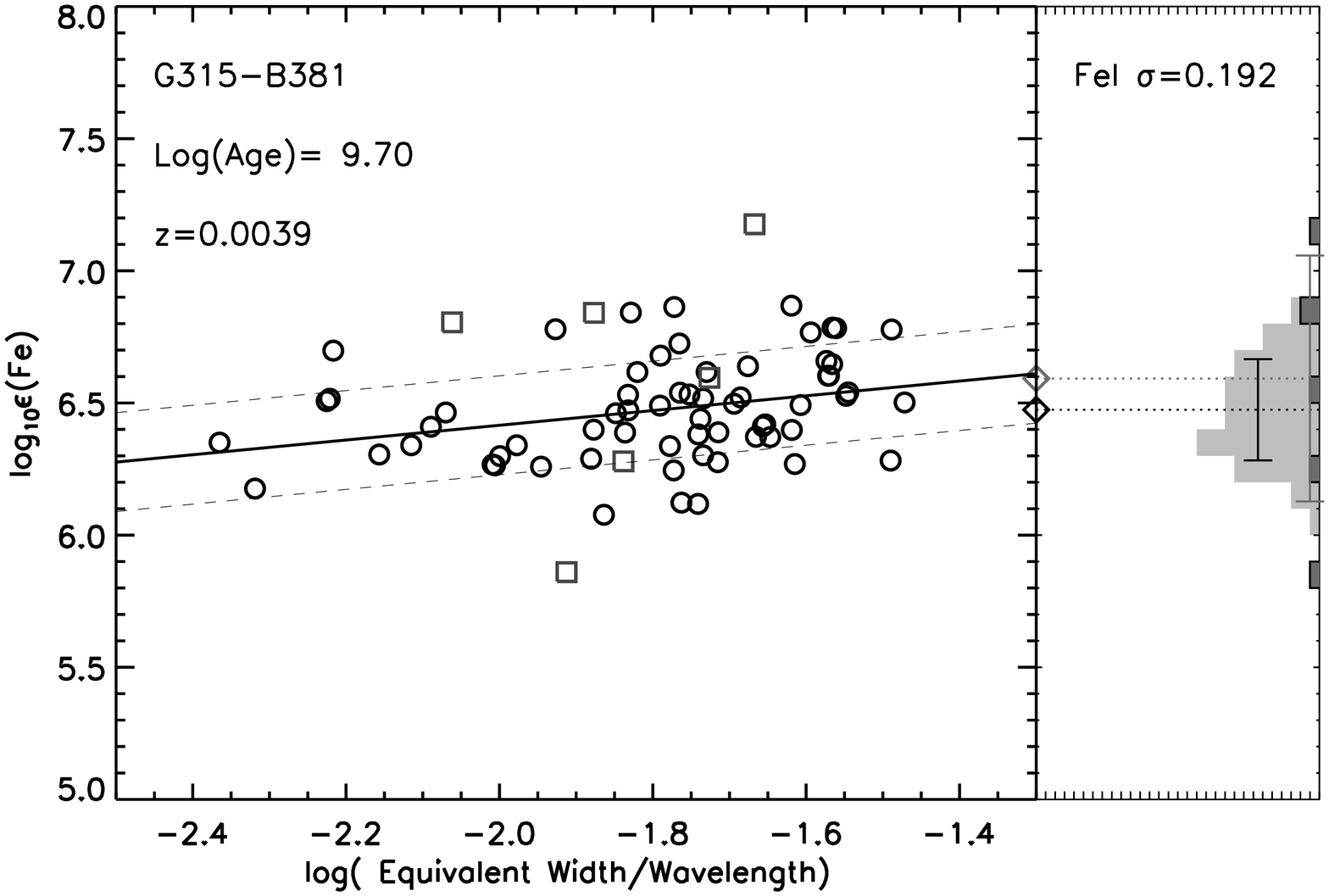}
\includegraphics[angle=90,scale=0.20]{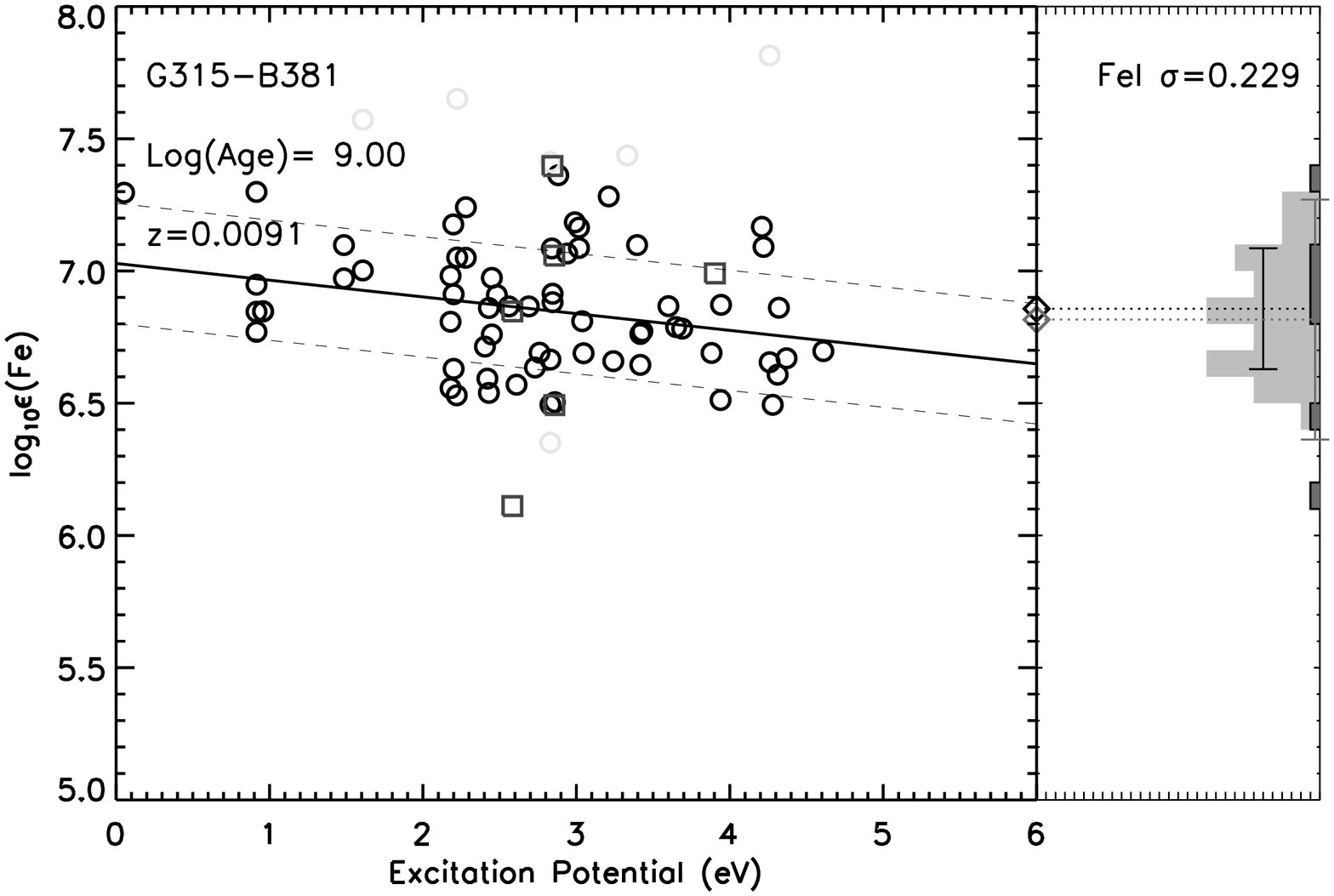}
\includegraphics[angle=90,scale=0.20]{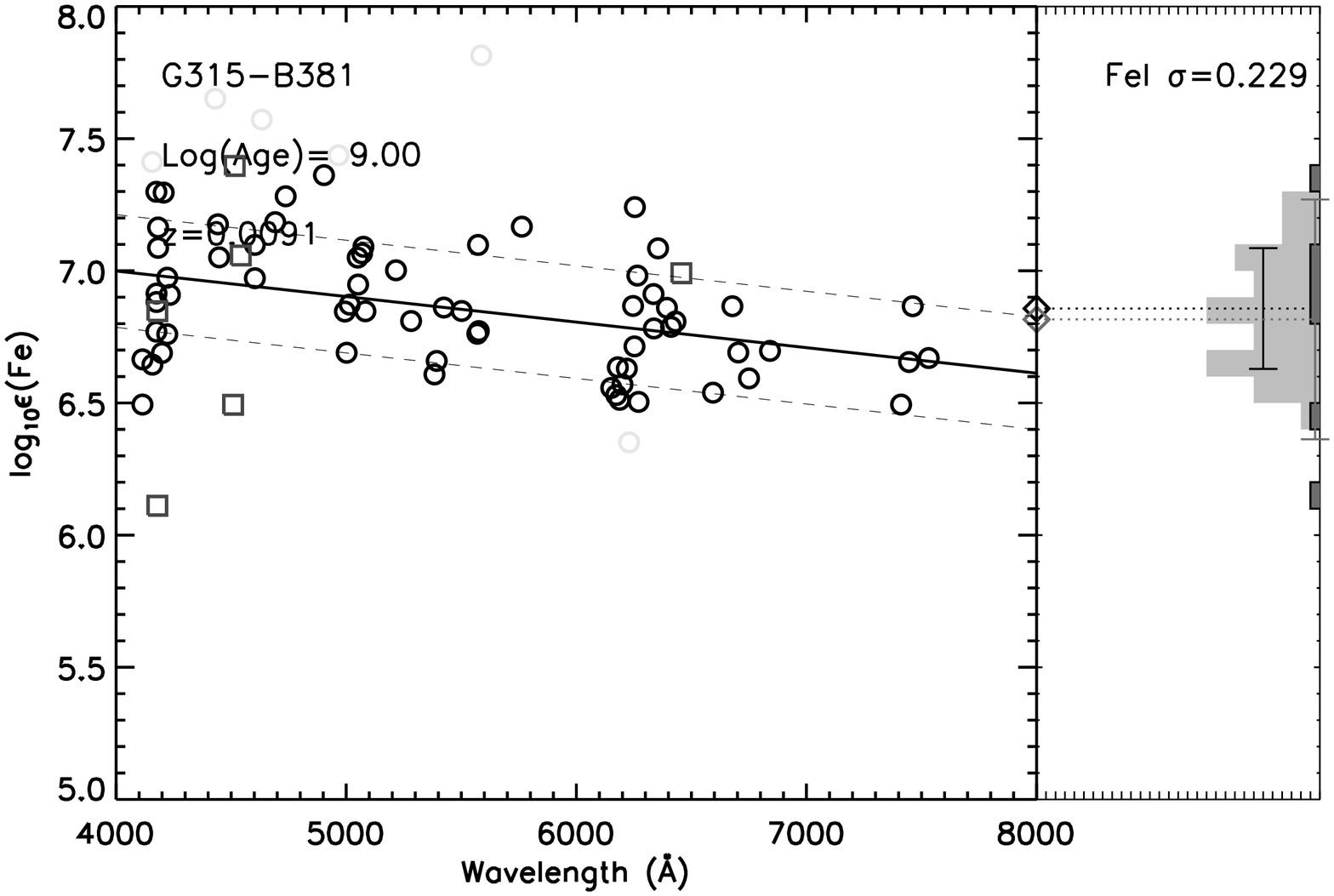}
\includegraphics[angle=90,scale=0.20]{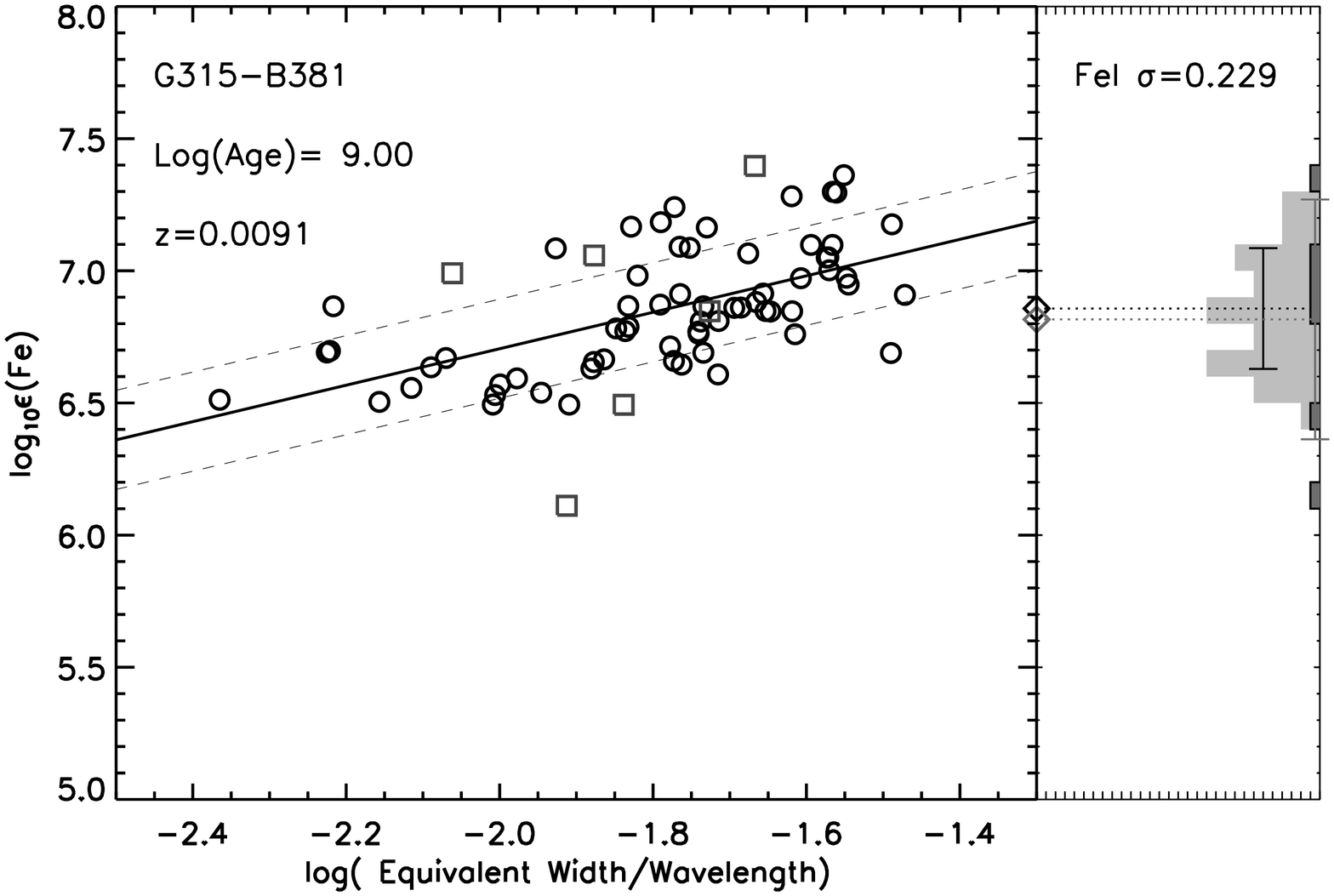}

\caption{Diagnostics for G315-B381.  The smallest Fe I standard deviation and smallest dependence on EP, wavelength, and observed equivalent width at ages of 10-15 Gyr. Symbols are the same as in Figure~\ref{fig:g108 diagnostics}.}
\label{fig:g315 diagnostics} 
\end{figure*}

\begin{figure*}
\centering
\includegraphics[angle=90,scale=0.20]{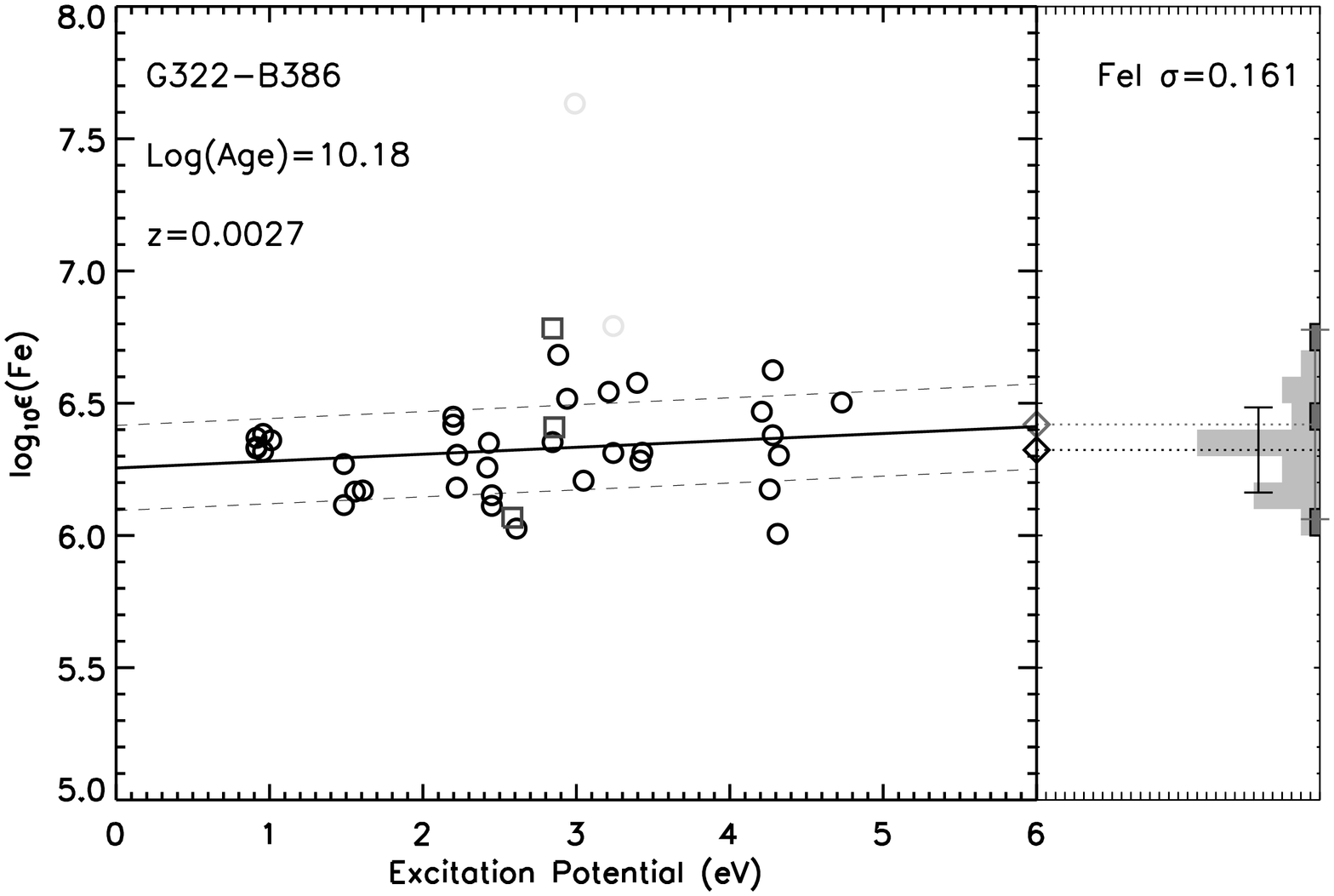}
\includegraphics[angle=90,scale=0.20]{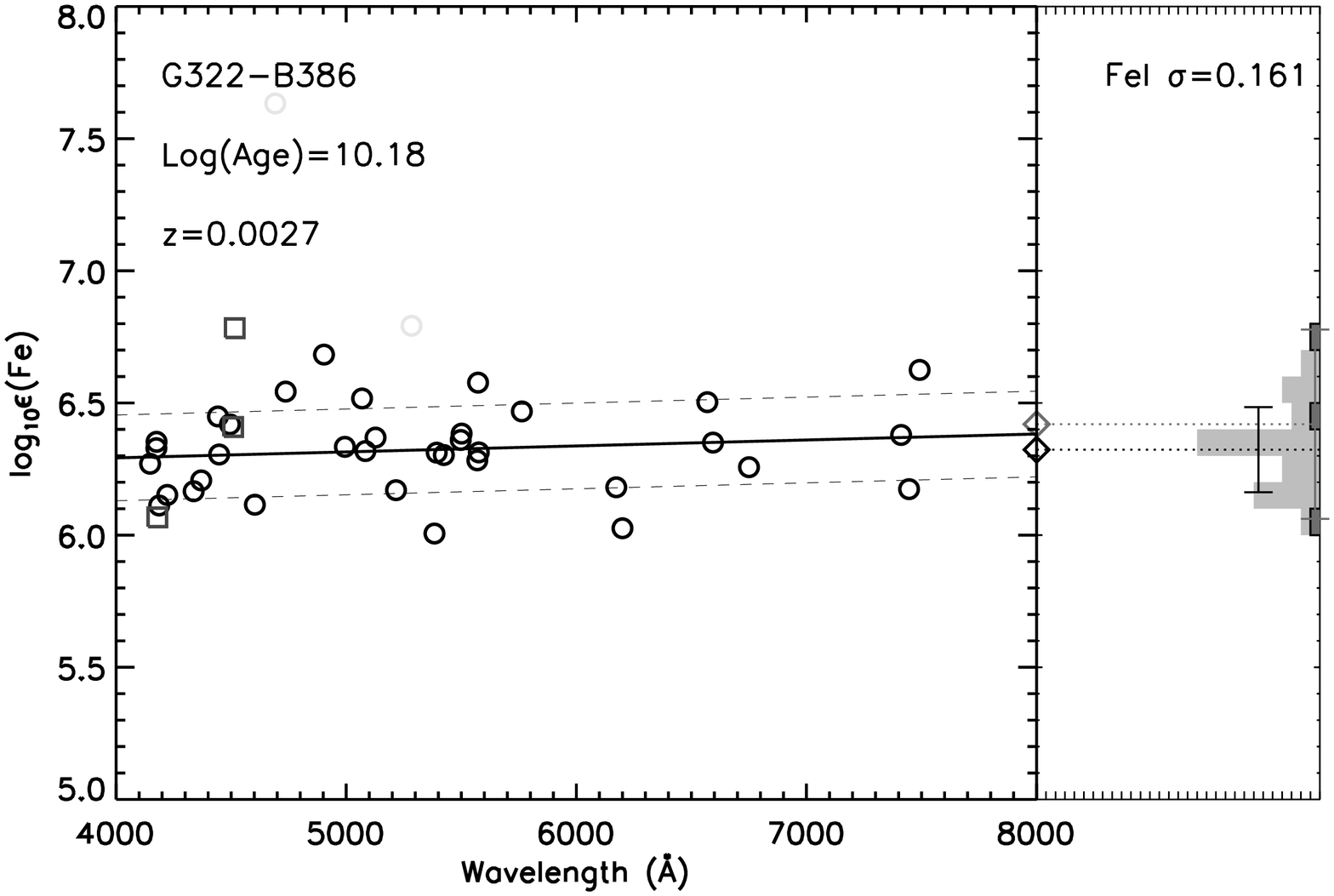}
\includegraphics[angle=90,scale=0.20]{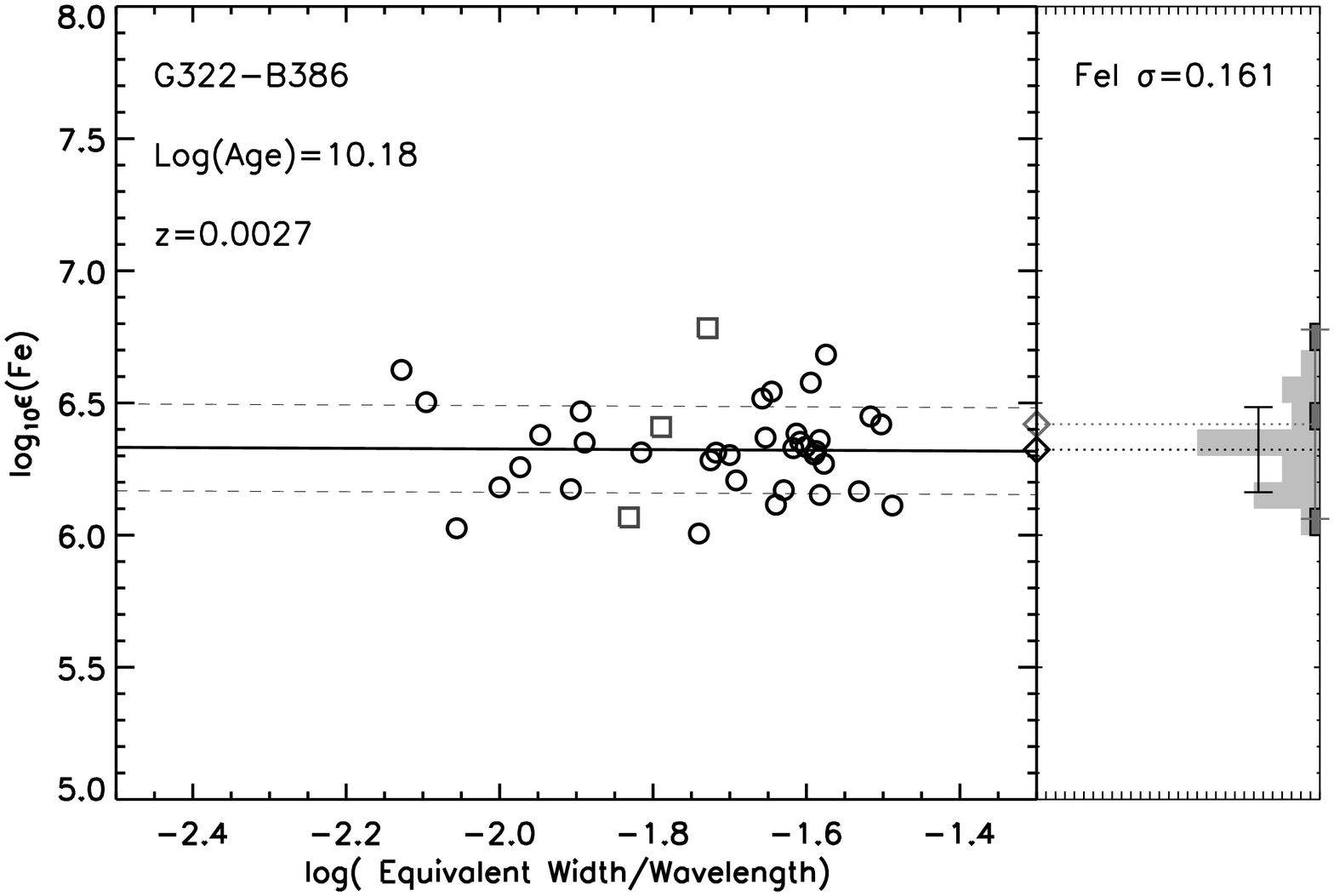}
\includegraphics[angle=90,scale=0.20]{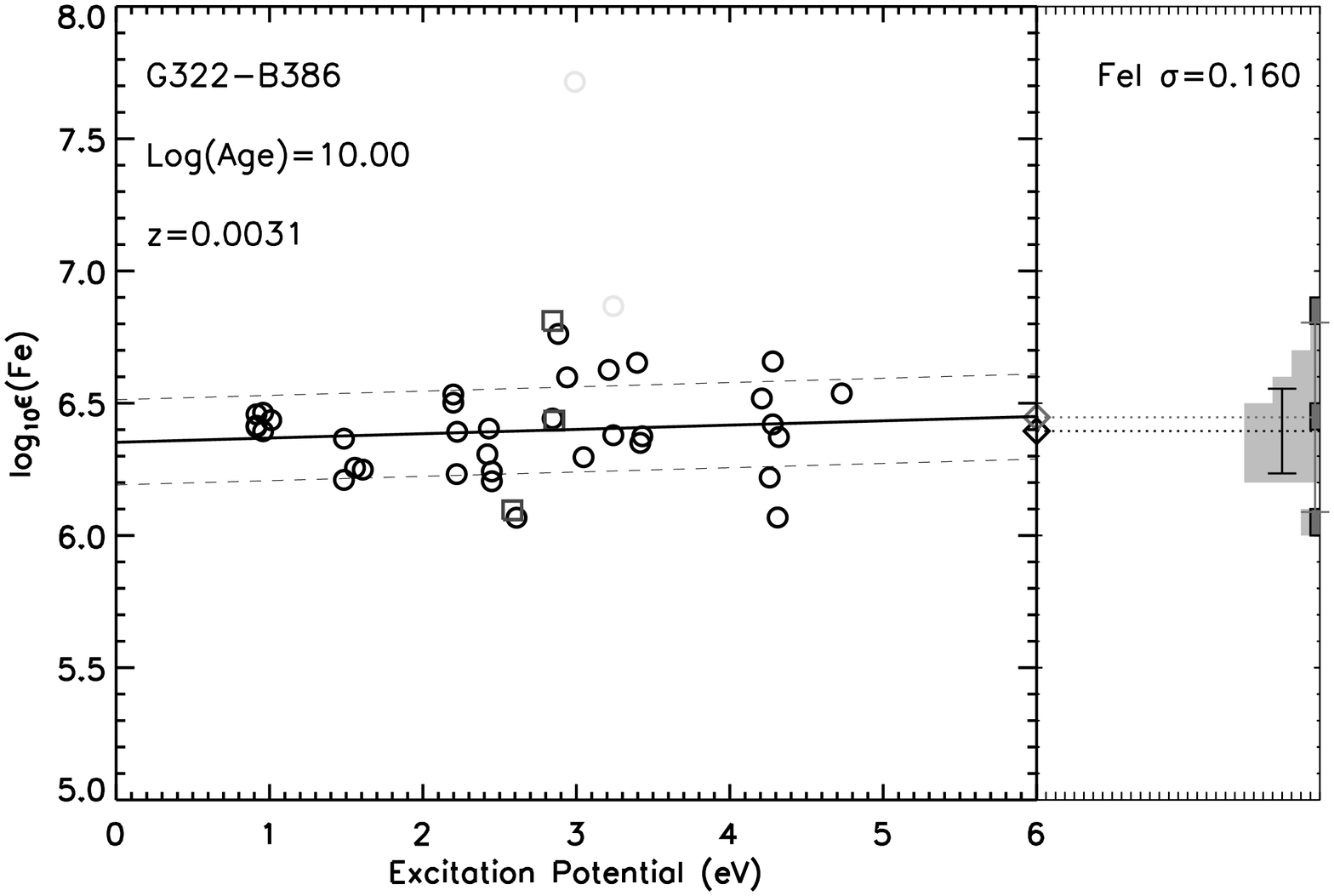}
\includegraphics[angle=90,scale=0.20]{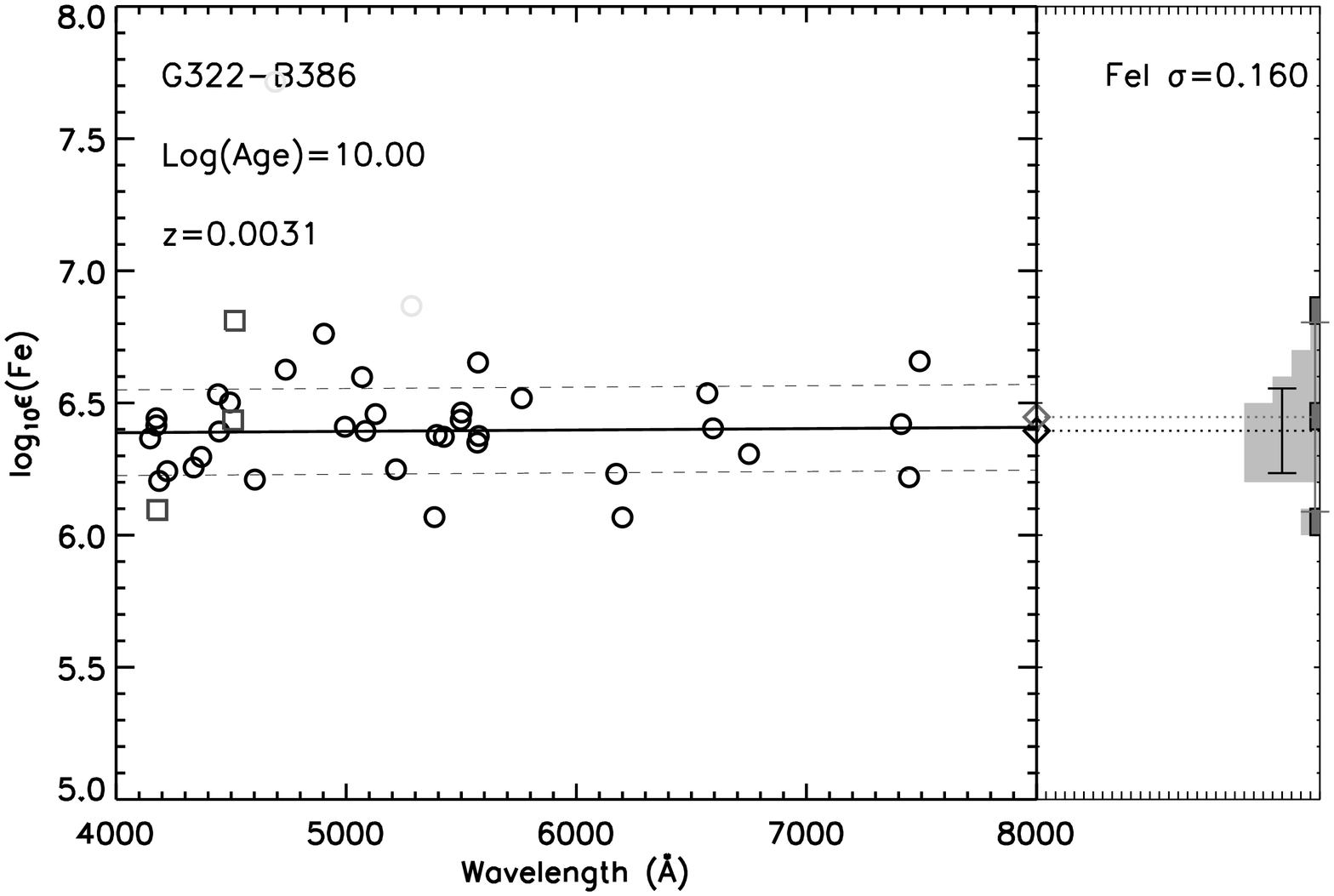}
\includegraphics[angle=90,scale=0.20]{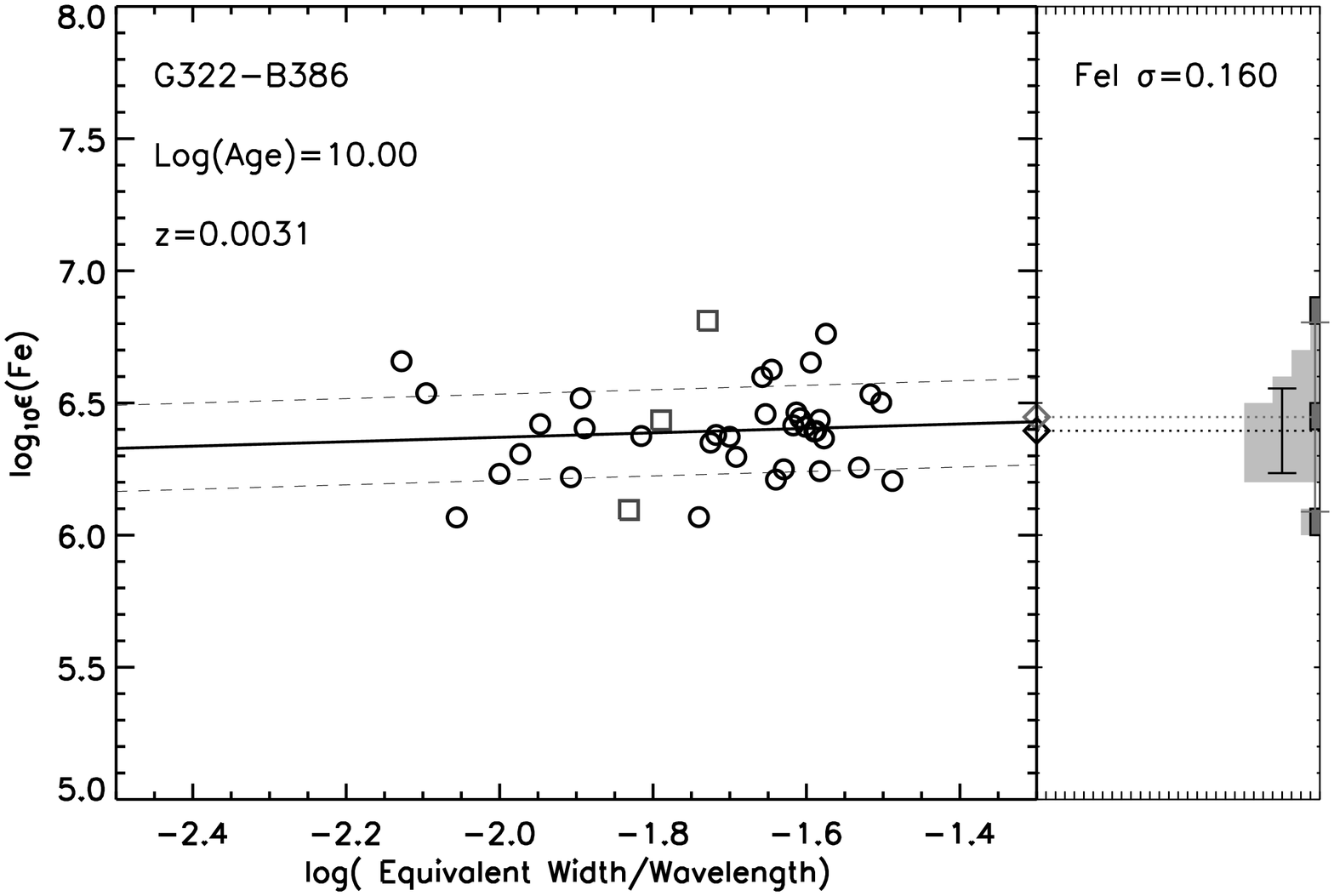}
\includegraphics[angle=90,scale=0.20]{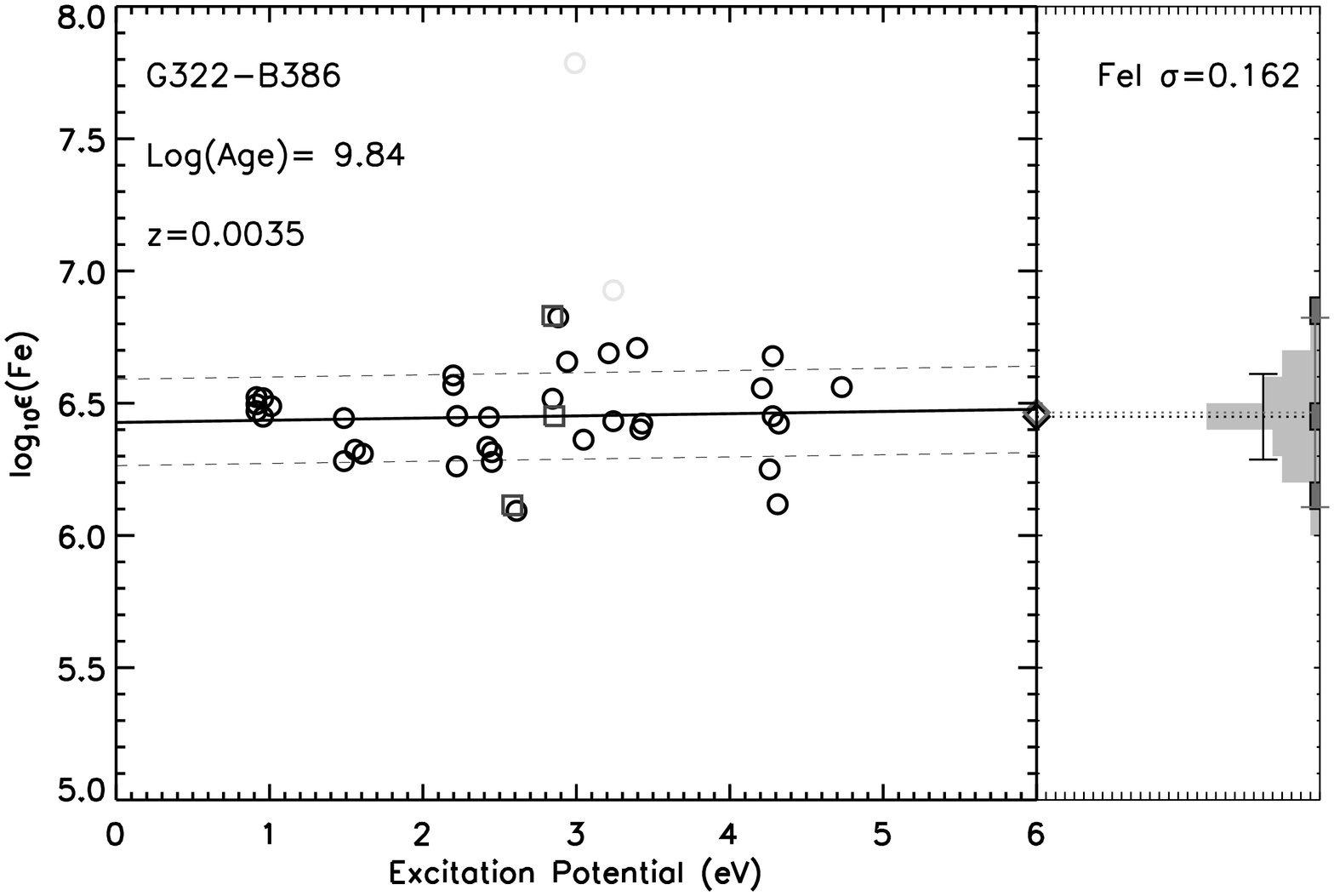}
\includegraphics[angle=90,scale=0.20]{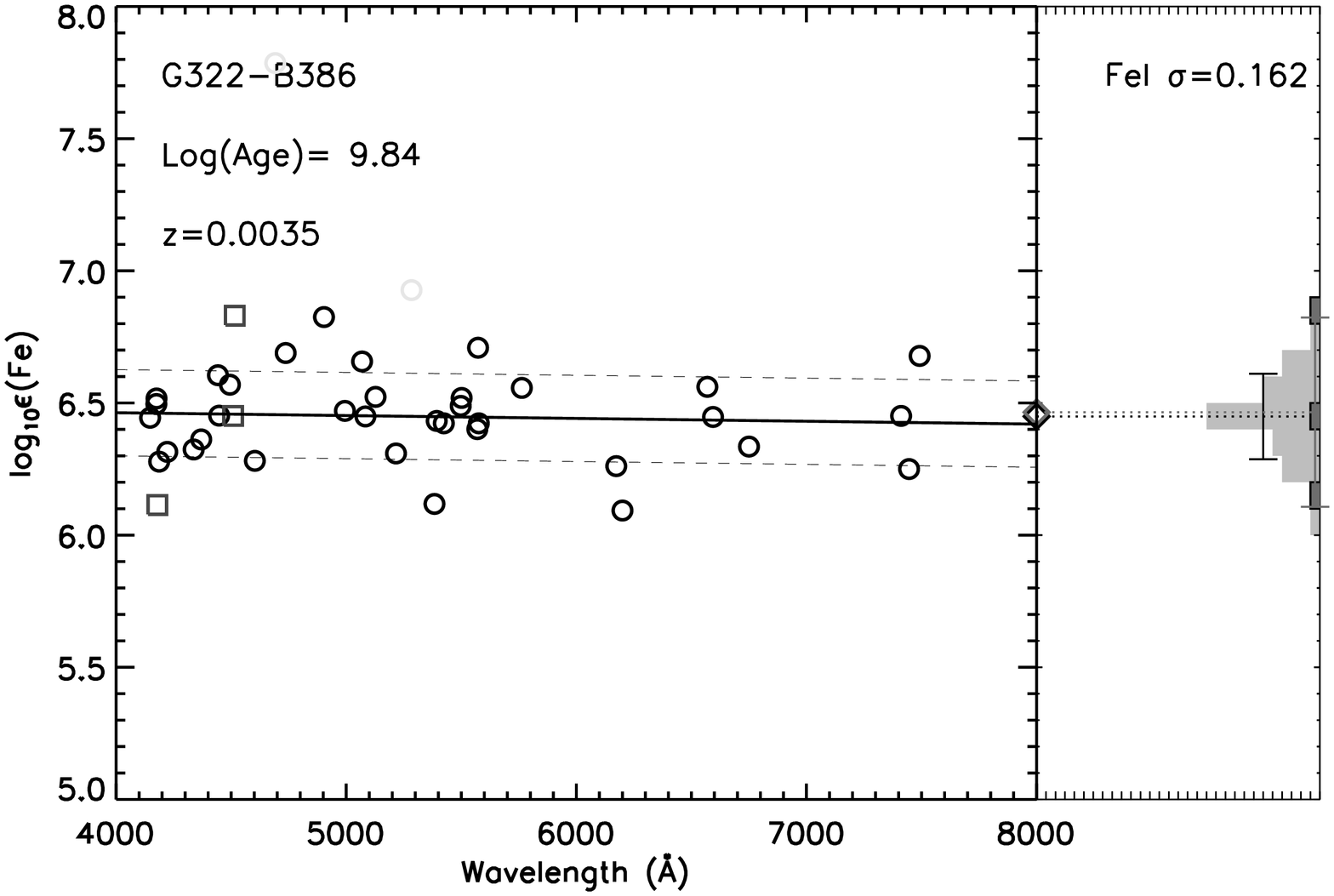}
\includegraphics[angle=90,scale=0.20]{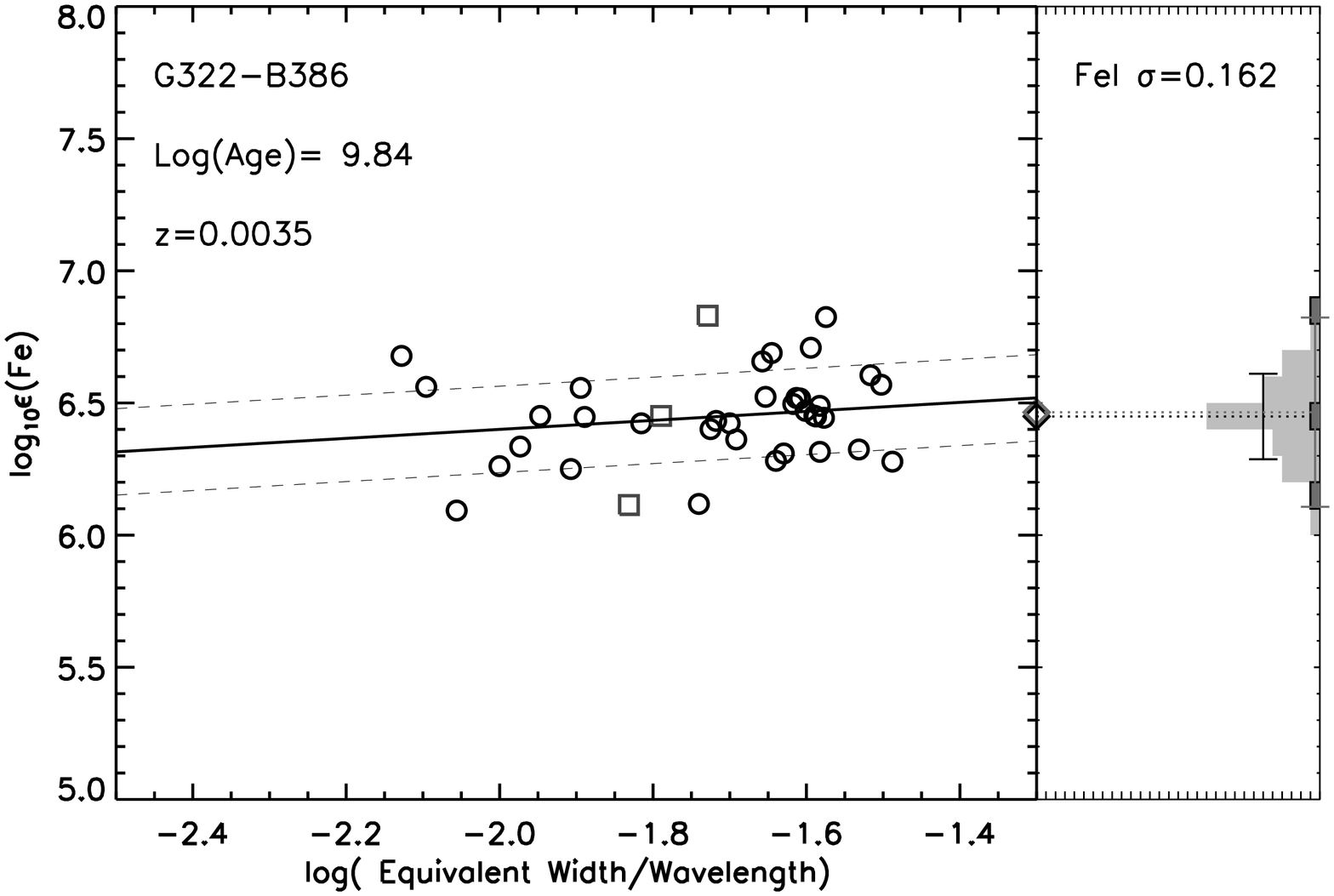}
\includegraphics[angle=90,scale=0.20]{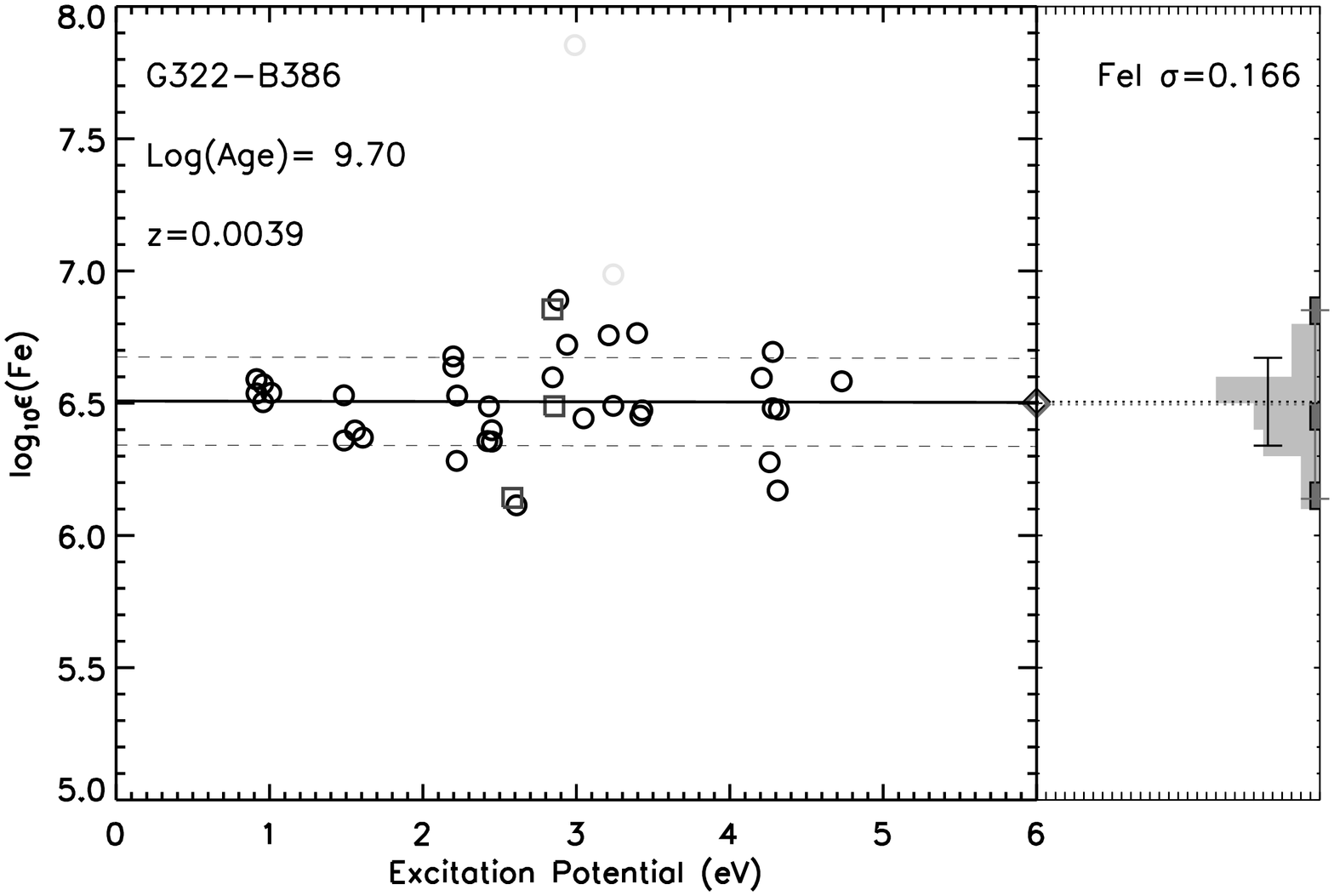}
\includegraphics[angle=90,scale=0.20]{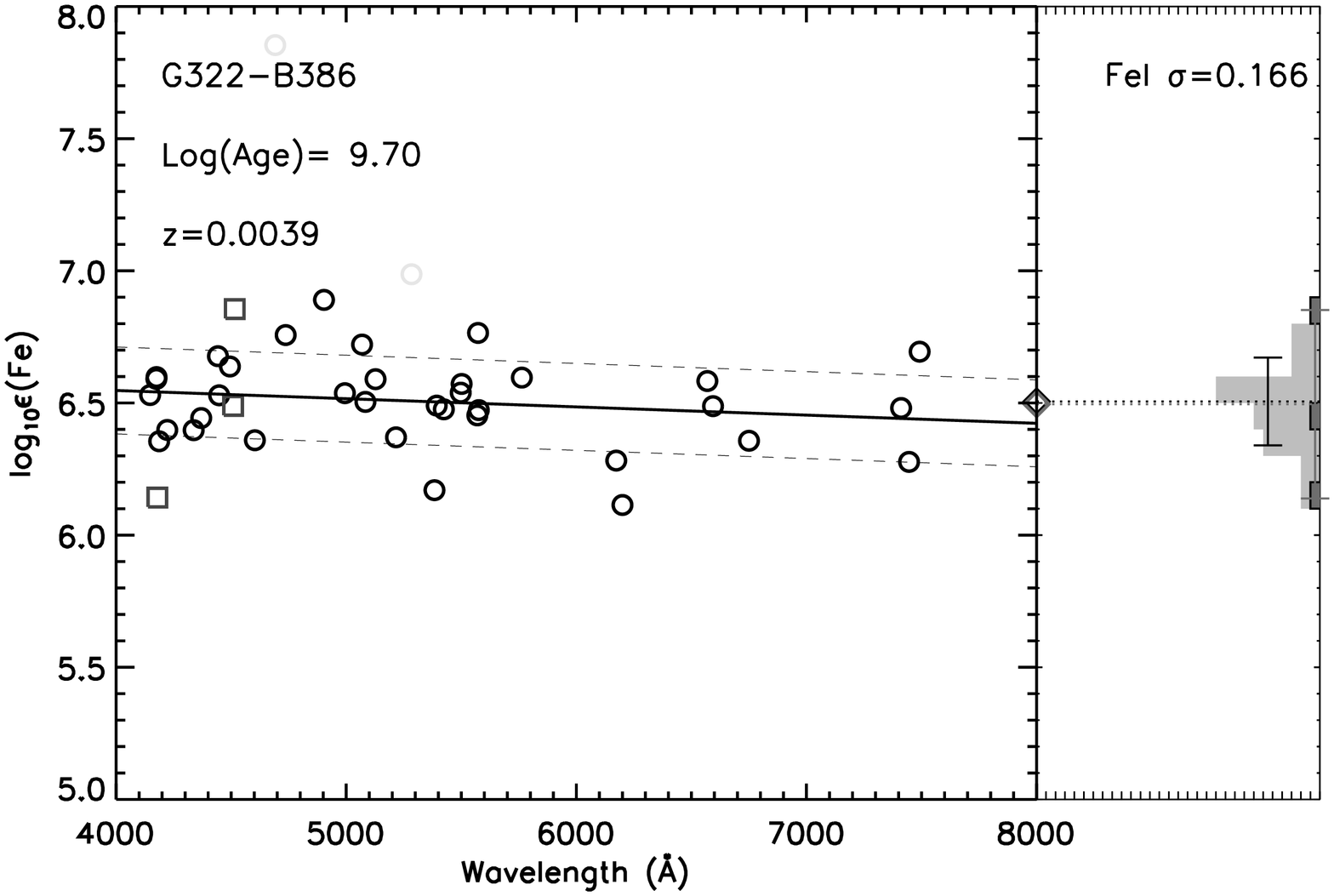}
\includegraphics[angle=90,scale=0.20]{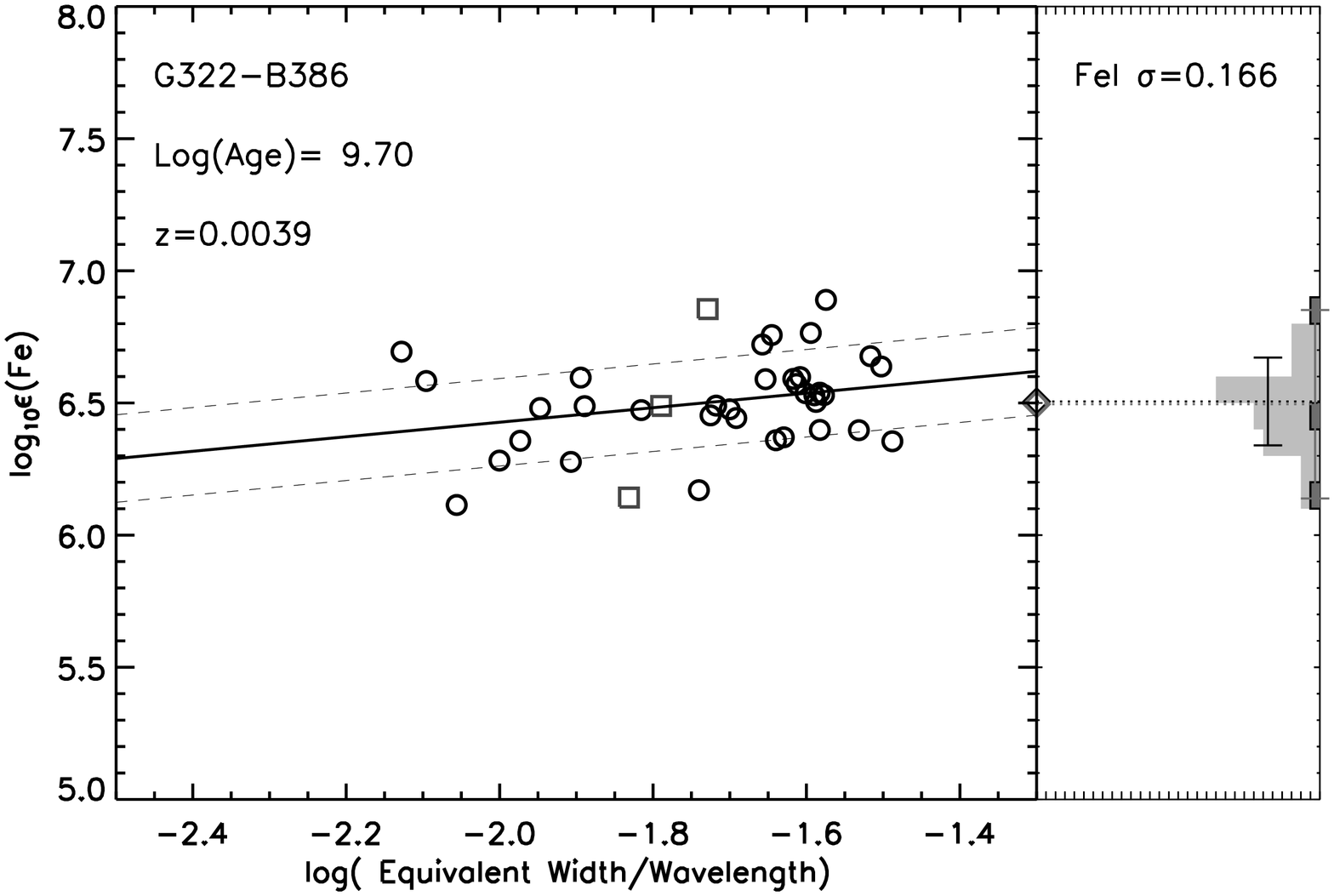}
\includegraphics[angle=90,scale=0.20]{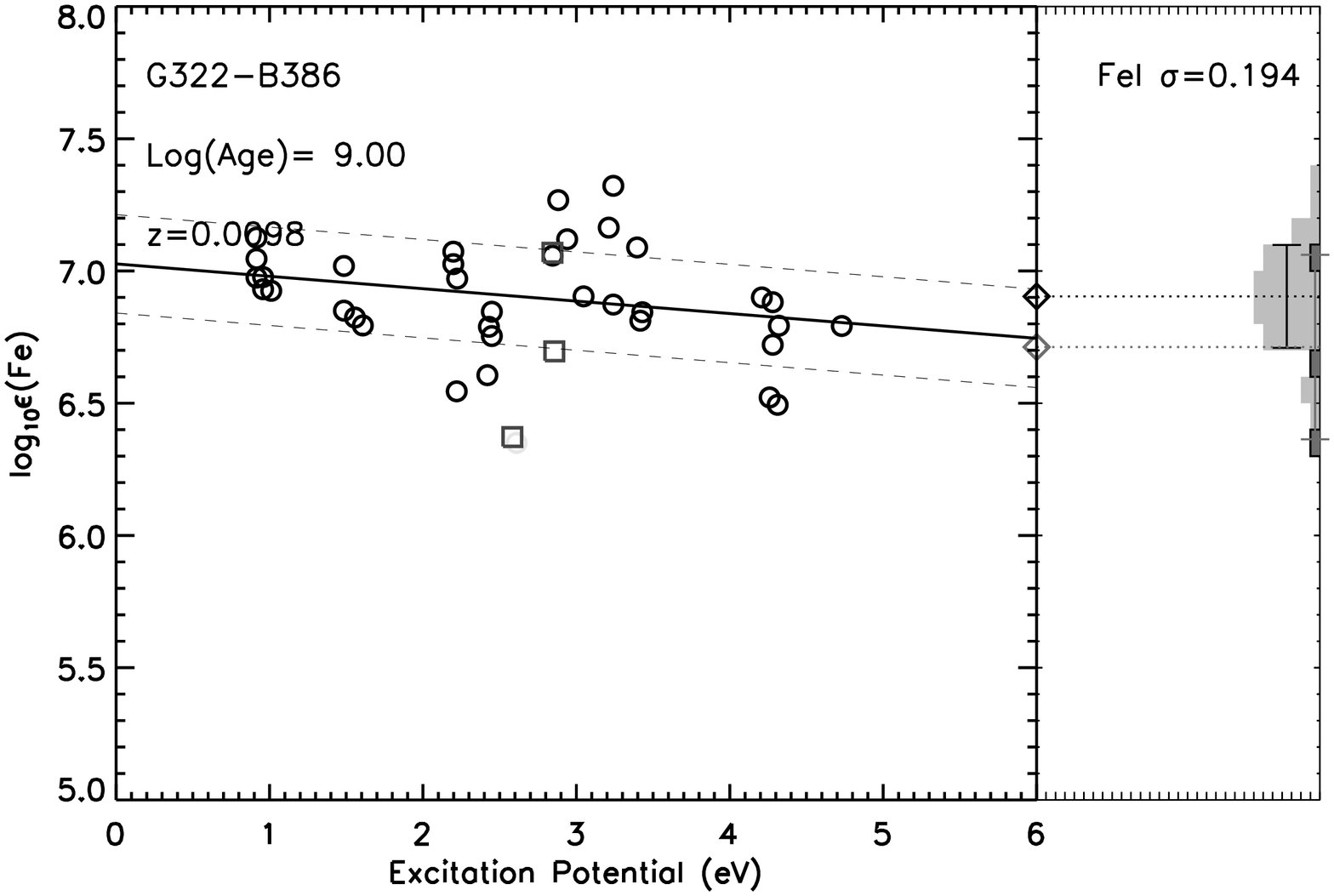}
\includegraphics[angle=90,scale=0.20]{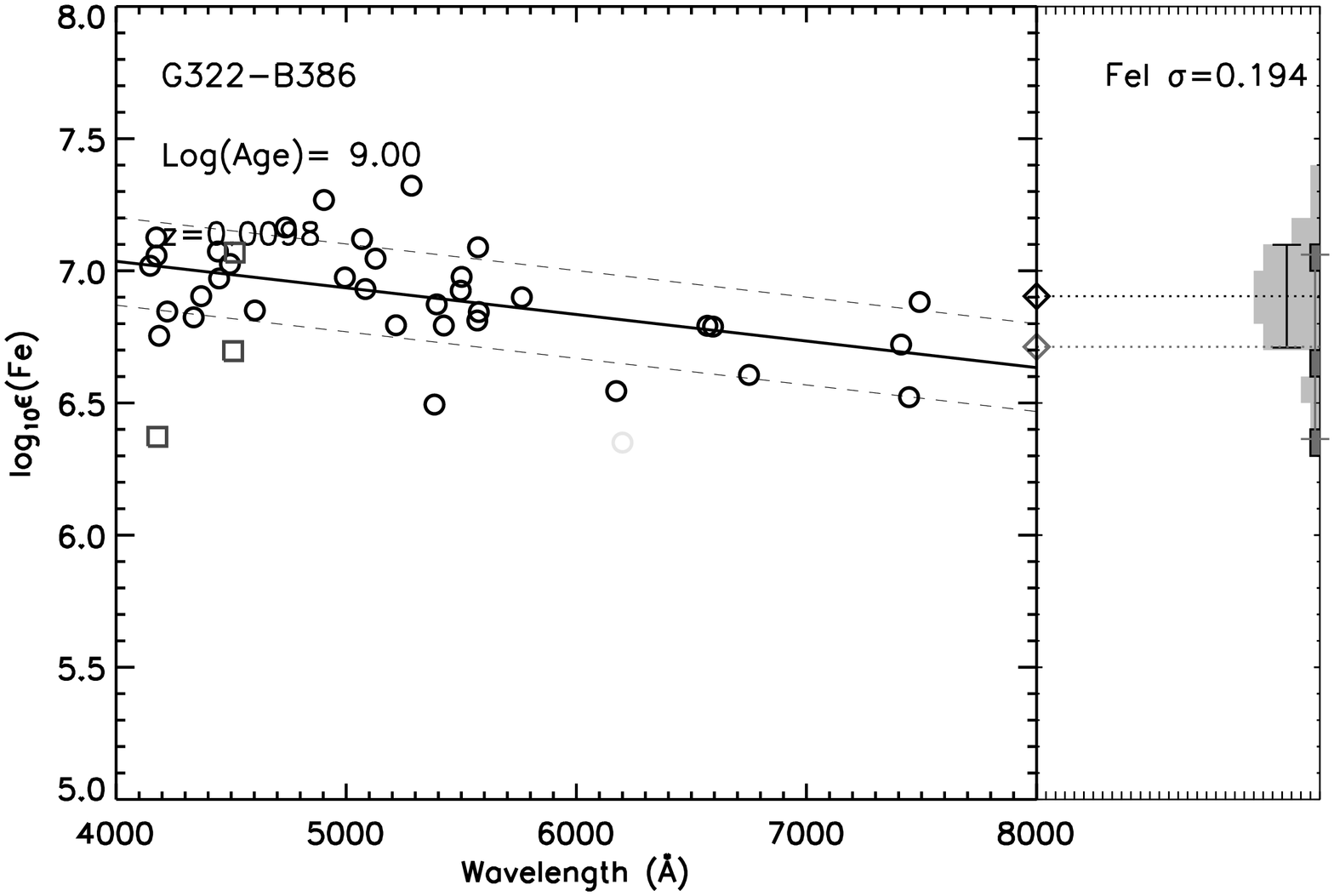}
\includegraphics[angle=90,scale=0.20]{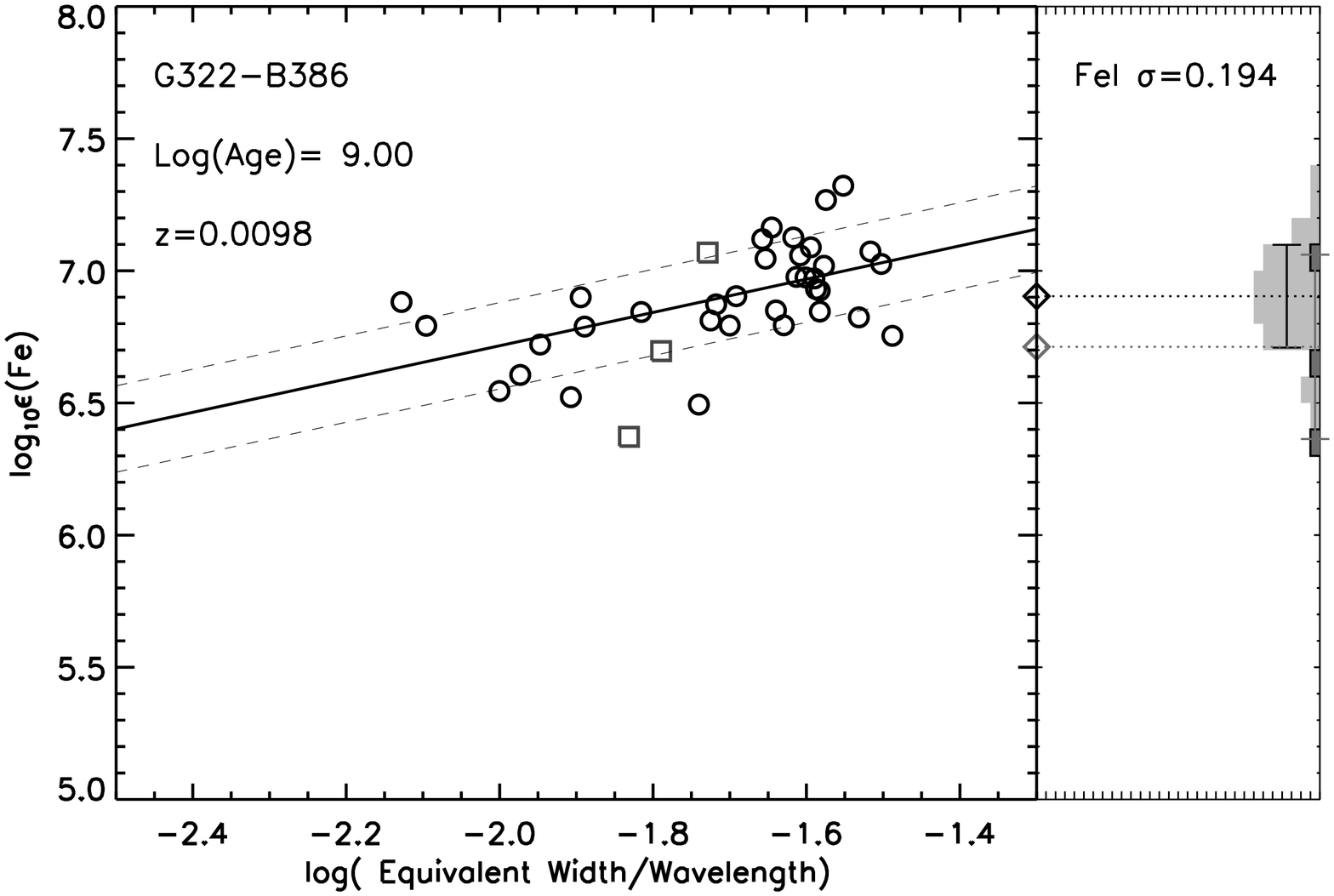}
\caption{Diagnostics for G322-B386. The smallest Fe I standard deviation and smallest dependence on EP, wavelength, and observed equivalent width at ages of 7-13 Gyr. Symbols are the same as in Figure~\ref{fig:g108 diagnostics}.}

\label{fig:g322 diagnostics} 
\end{figure*}

\begin{figure*}
\centering

\includegraphics[angle=90,scale=0.20]{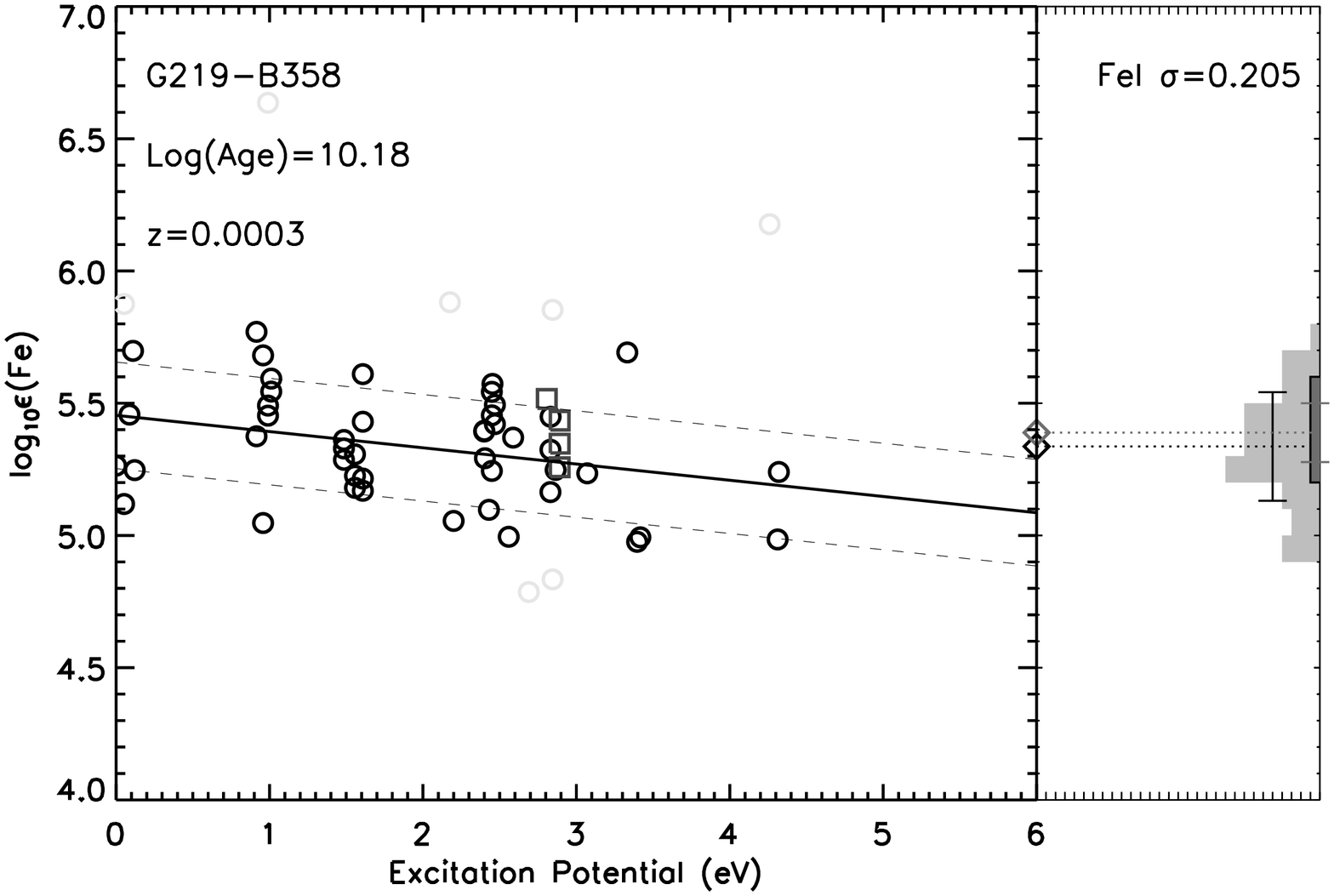}
\includegraphics[angle=90,scale=0.20]{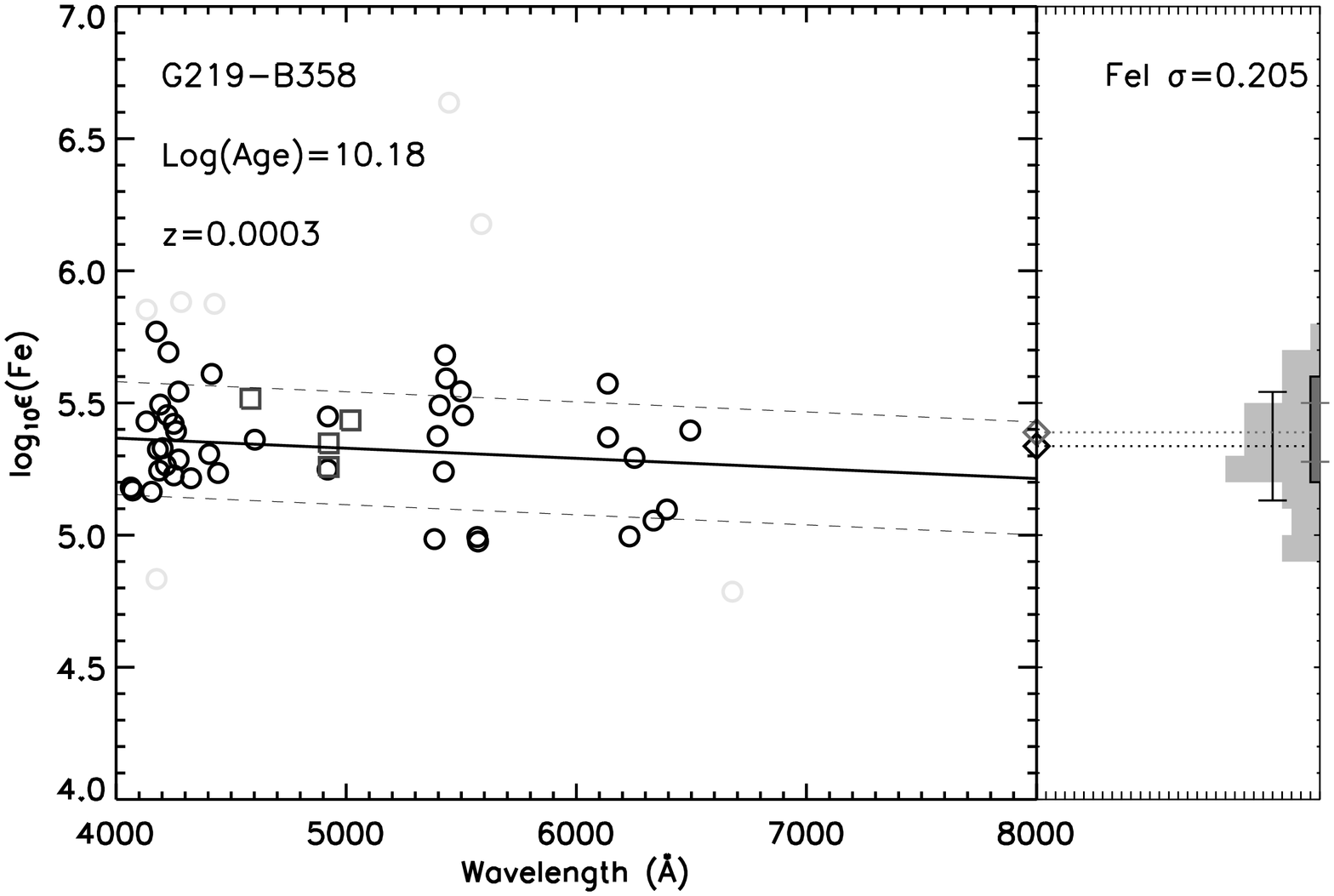}
\includegraphics[angle=90,scale=0.20]{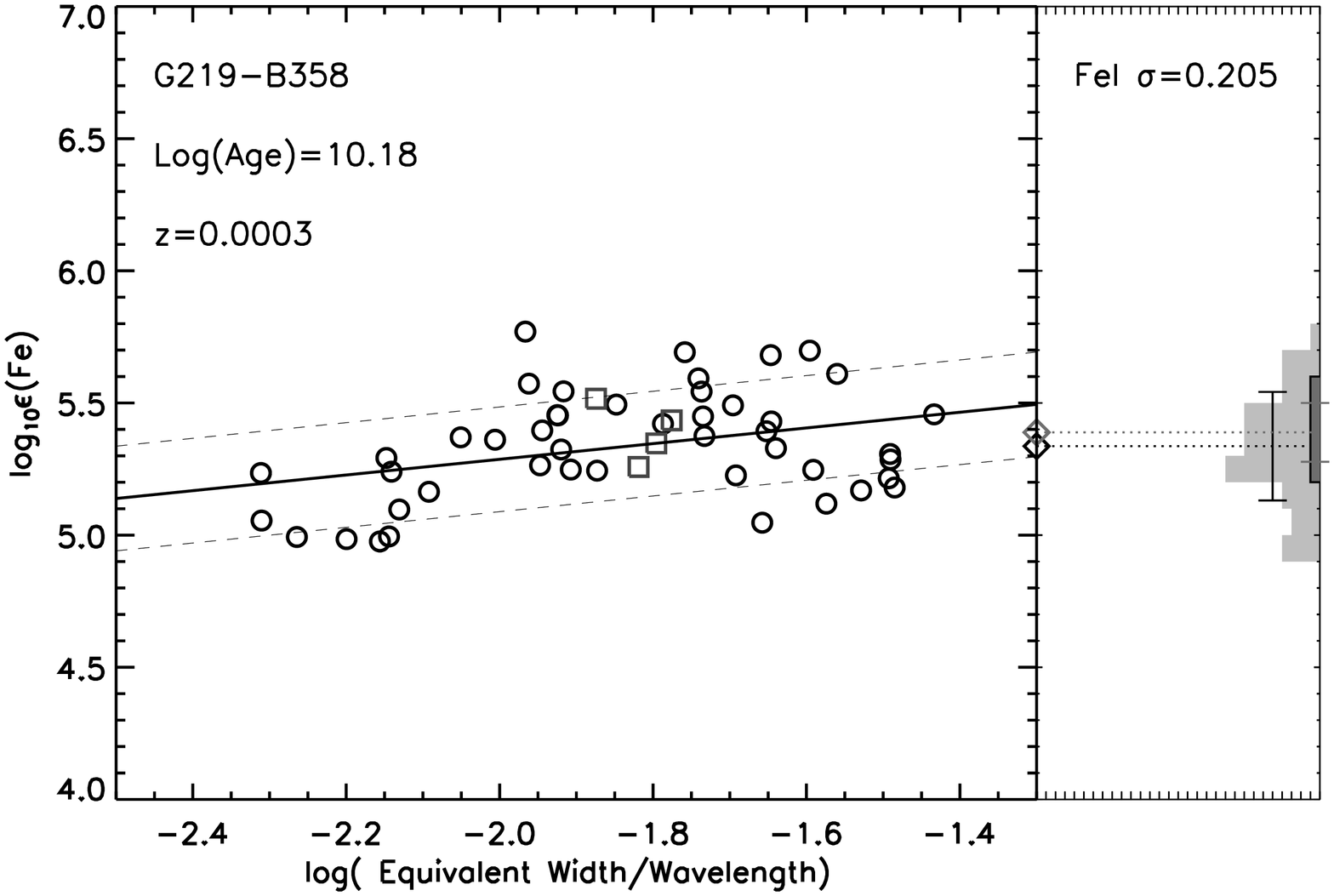}
\includegraphics[angle=90,scale=0.20]{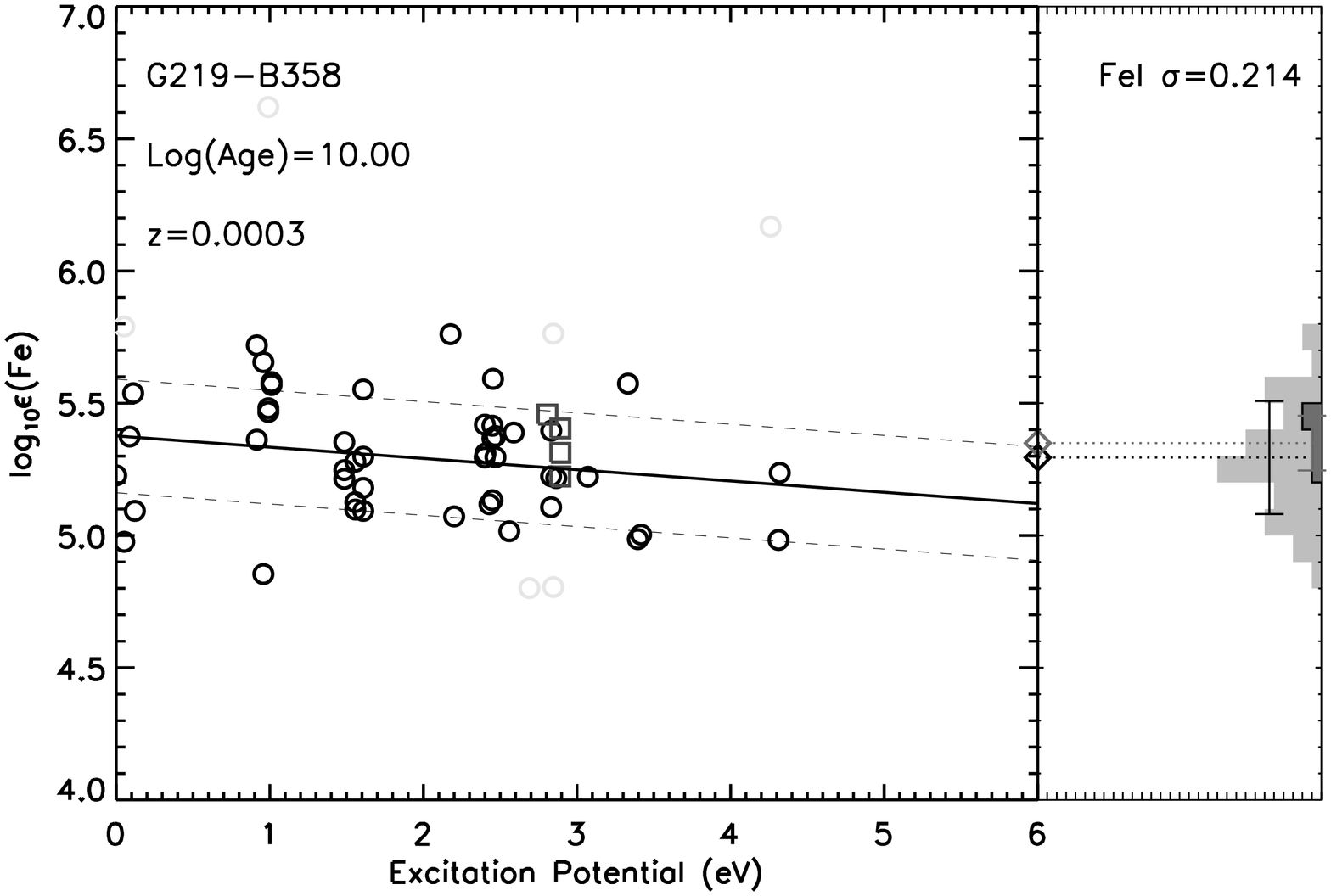}
\includegraphics[angle=90,scale=0.20]{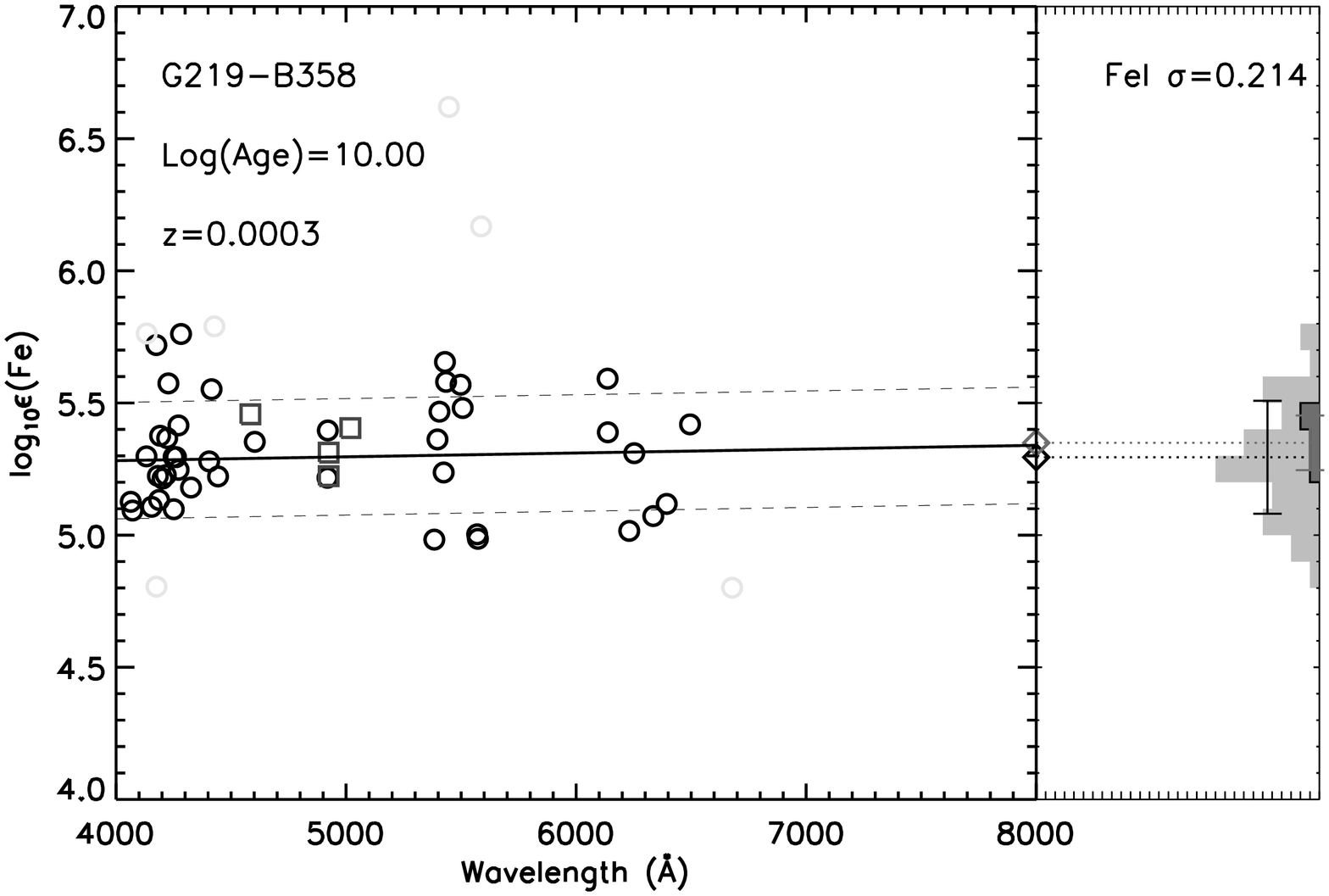}
\includegraphics[angle=90,scale=0.20]{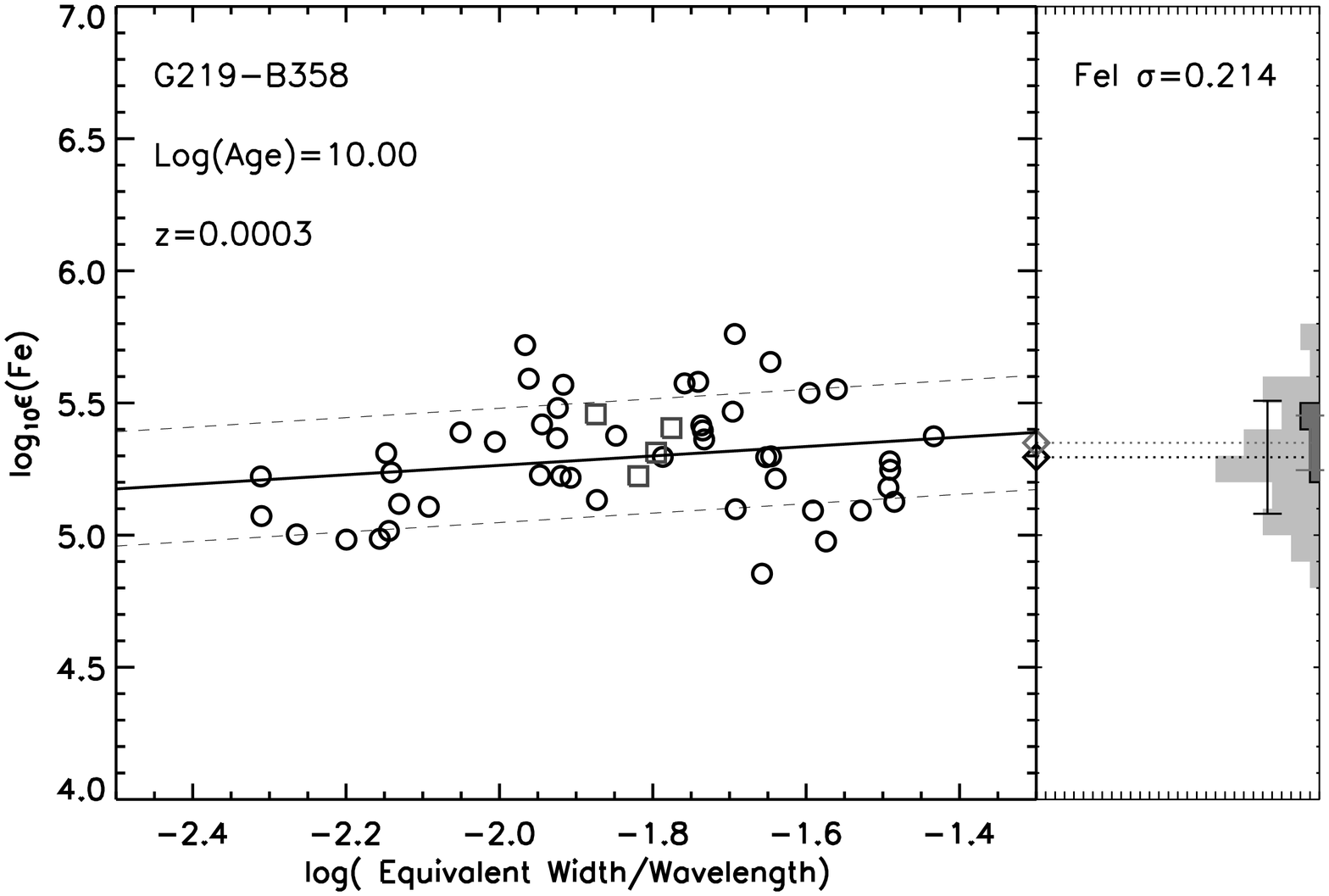}
\includegraphics[angle=90,scale=0.20]{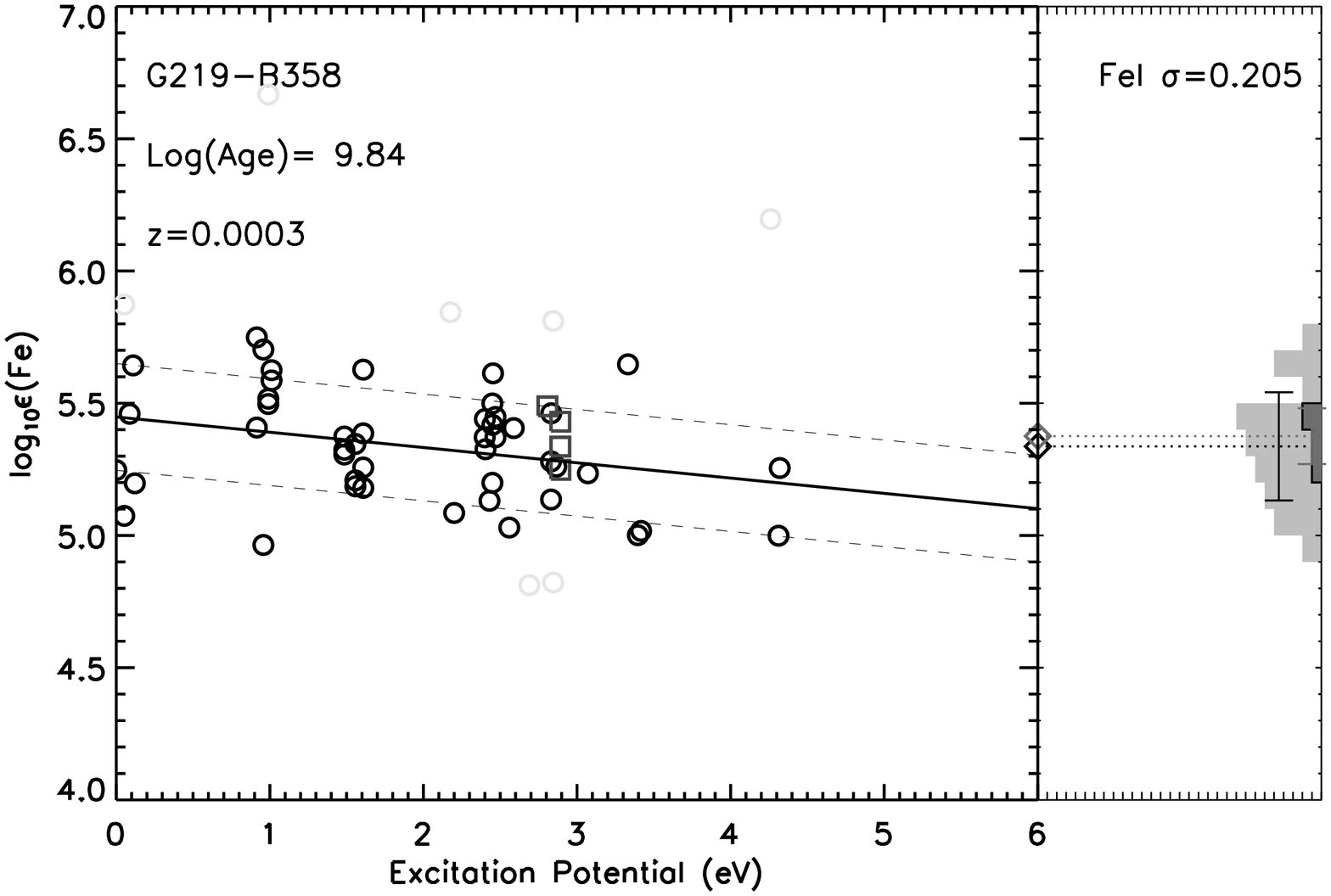}
\includegraphics[angle=90,scale=0.20]{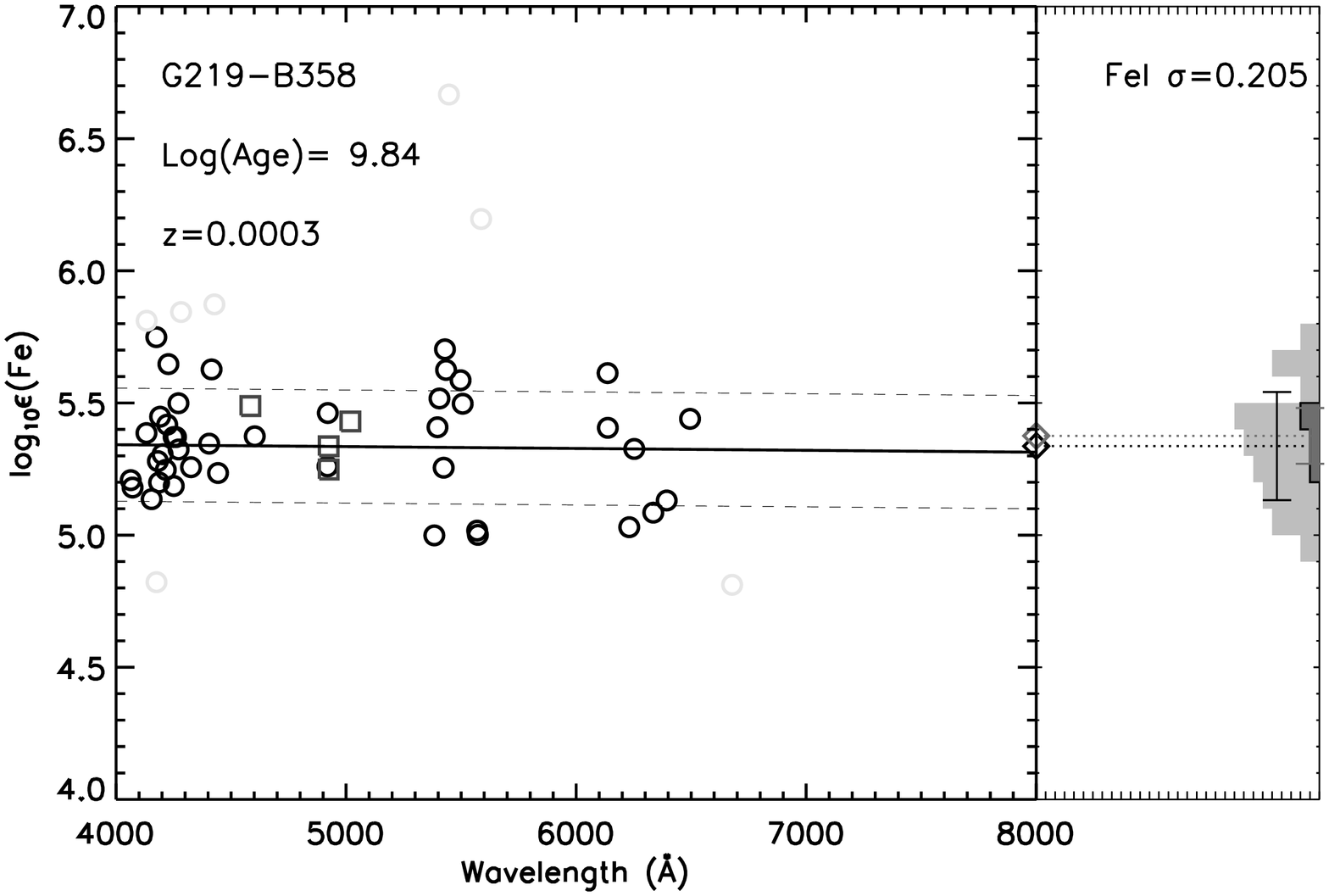}
\includegraphics[angle=90,scale=0.20]{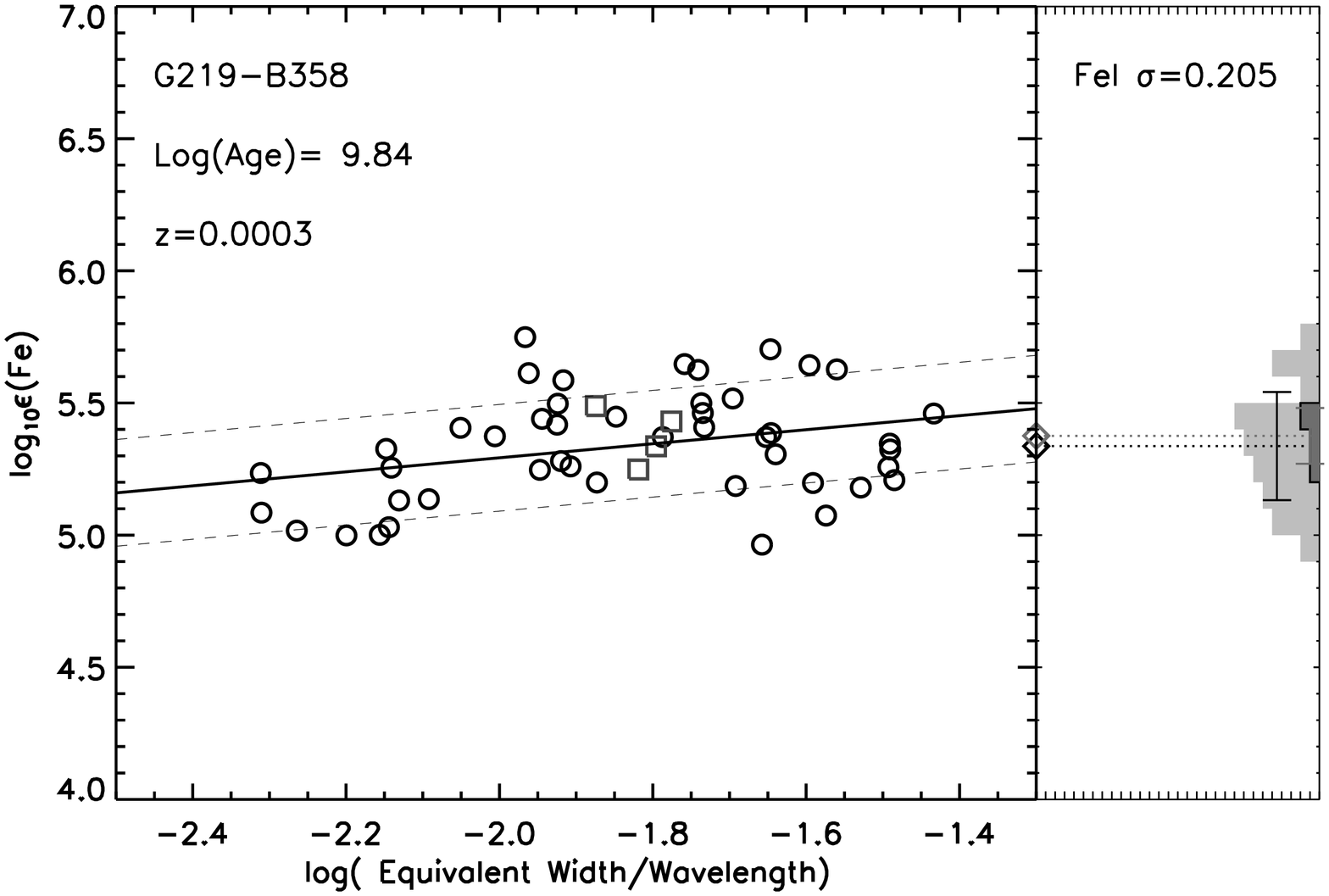}
\includegraphics[angle=90,scale=0.20]{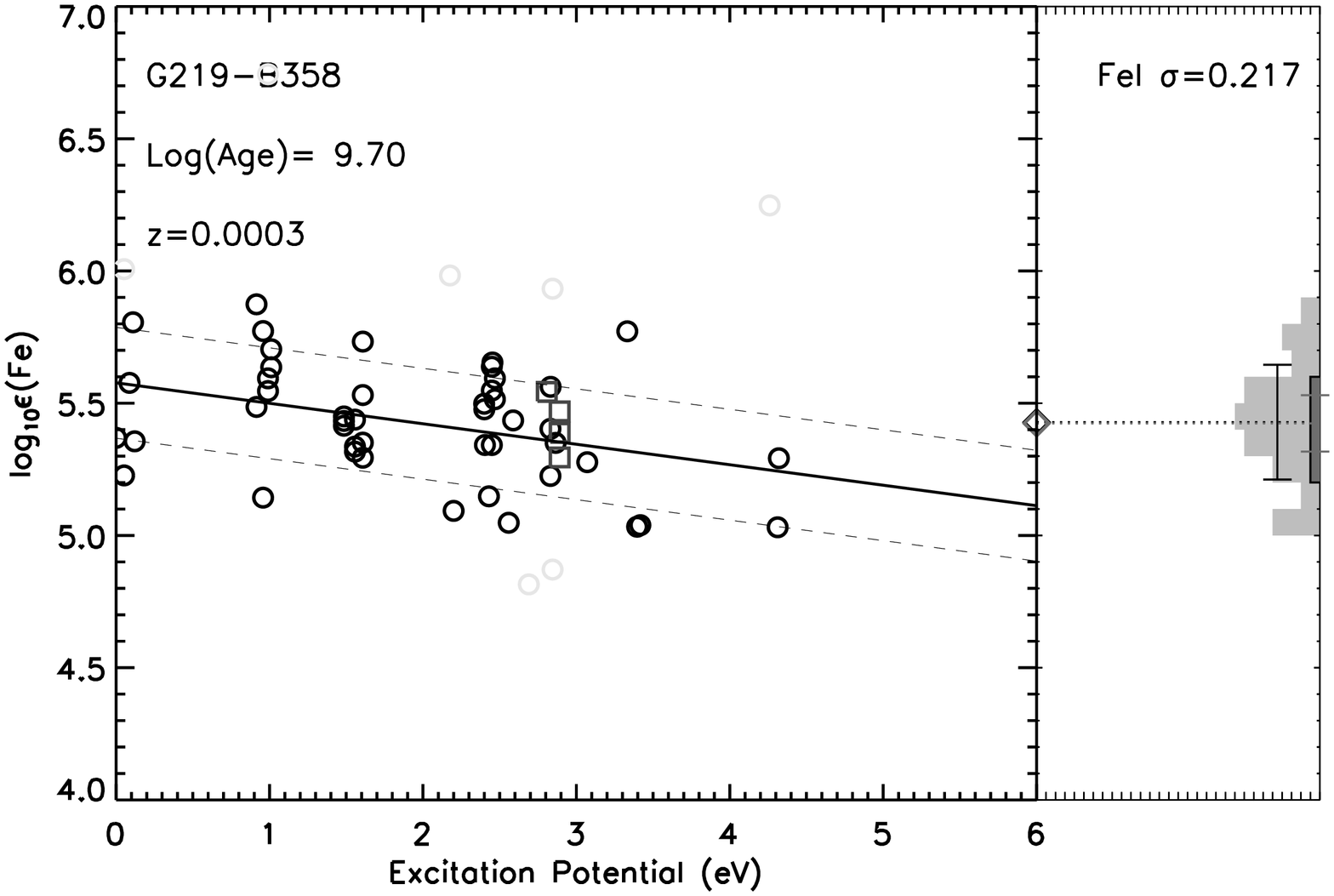}
\includegraphics[angle=90,scale=0.20]{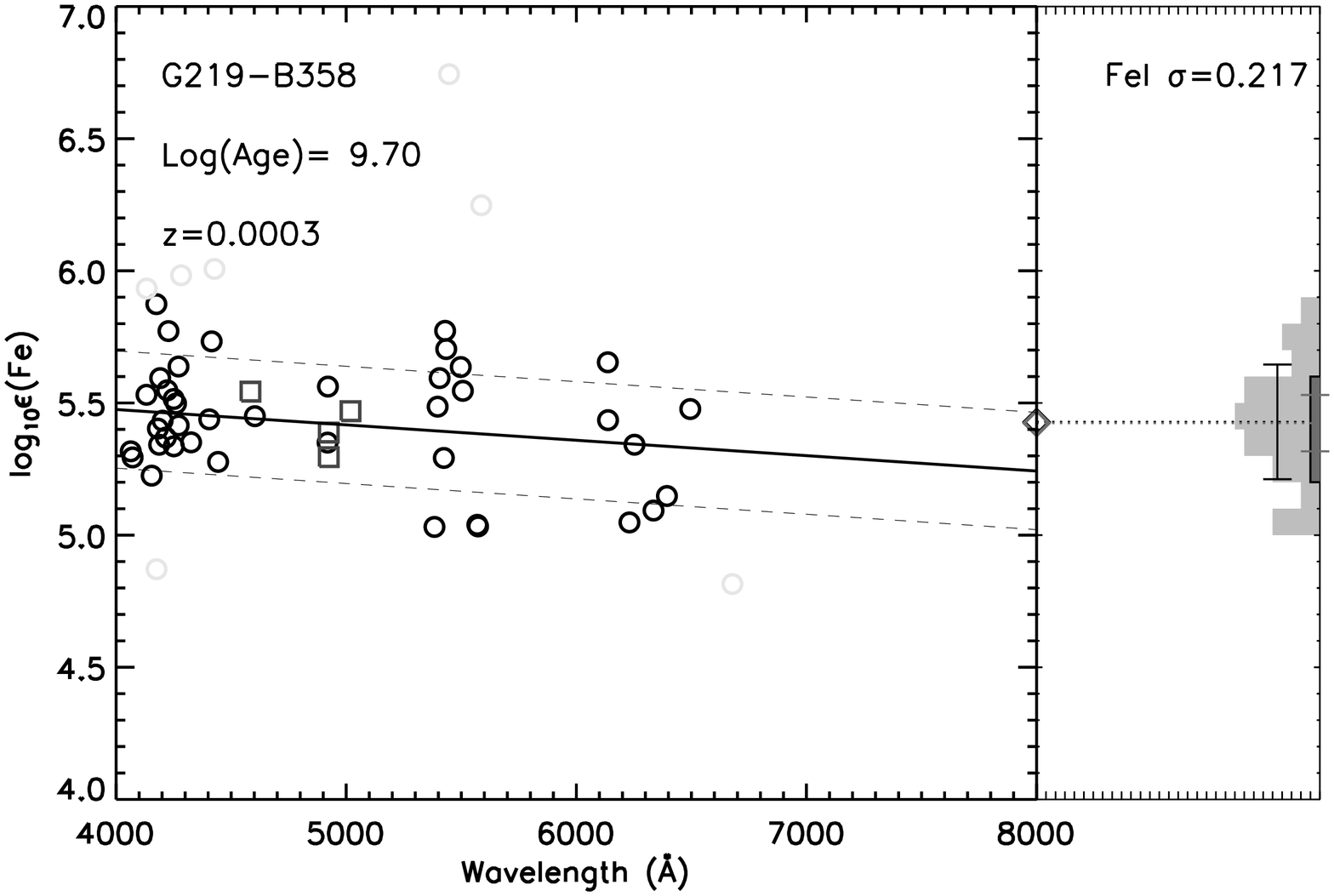}
\includegraphics[angle=90,scale=0.20]{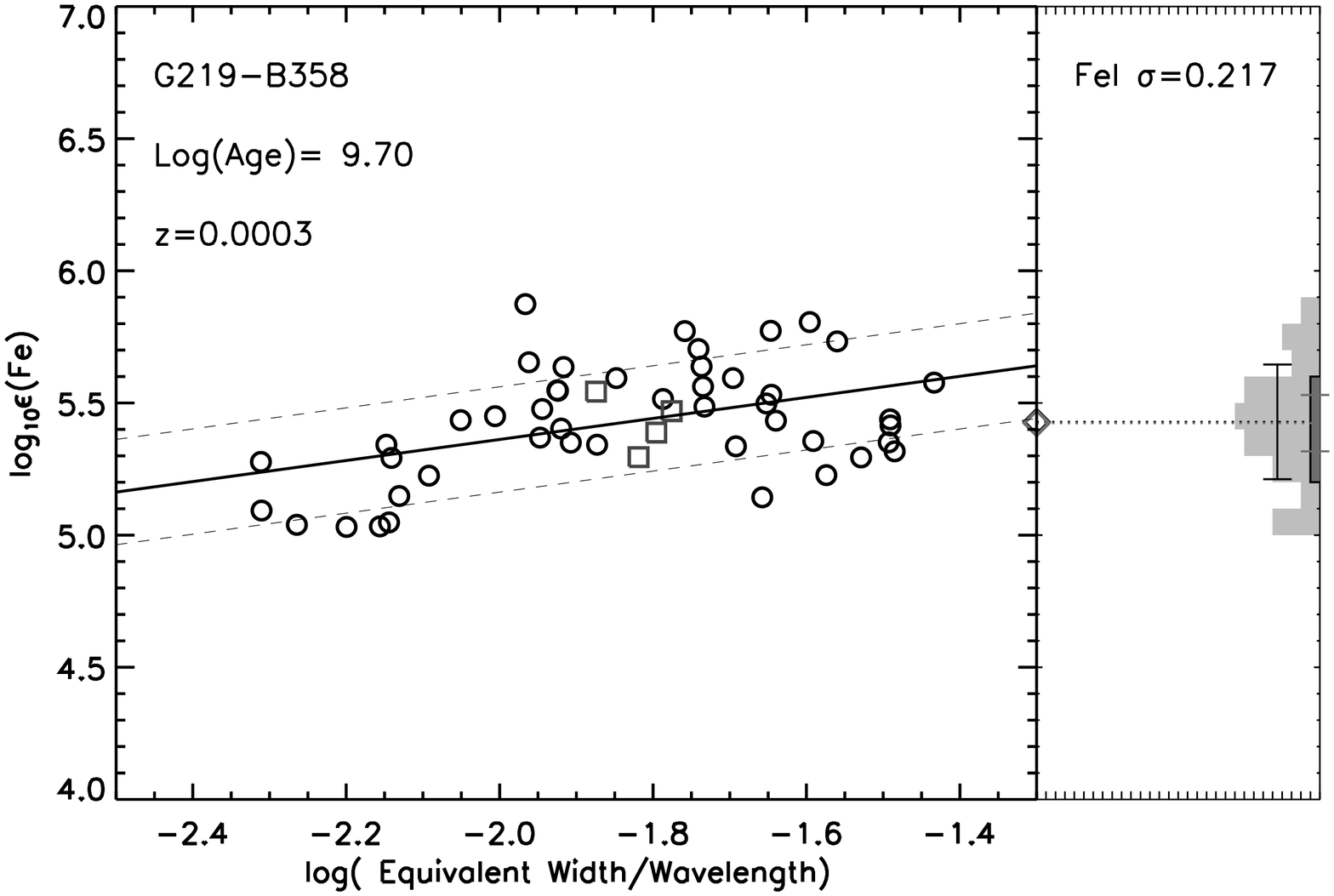}
\includegraphics[angle=90,scale=0.20]{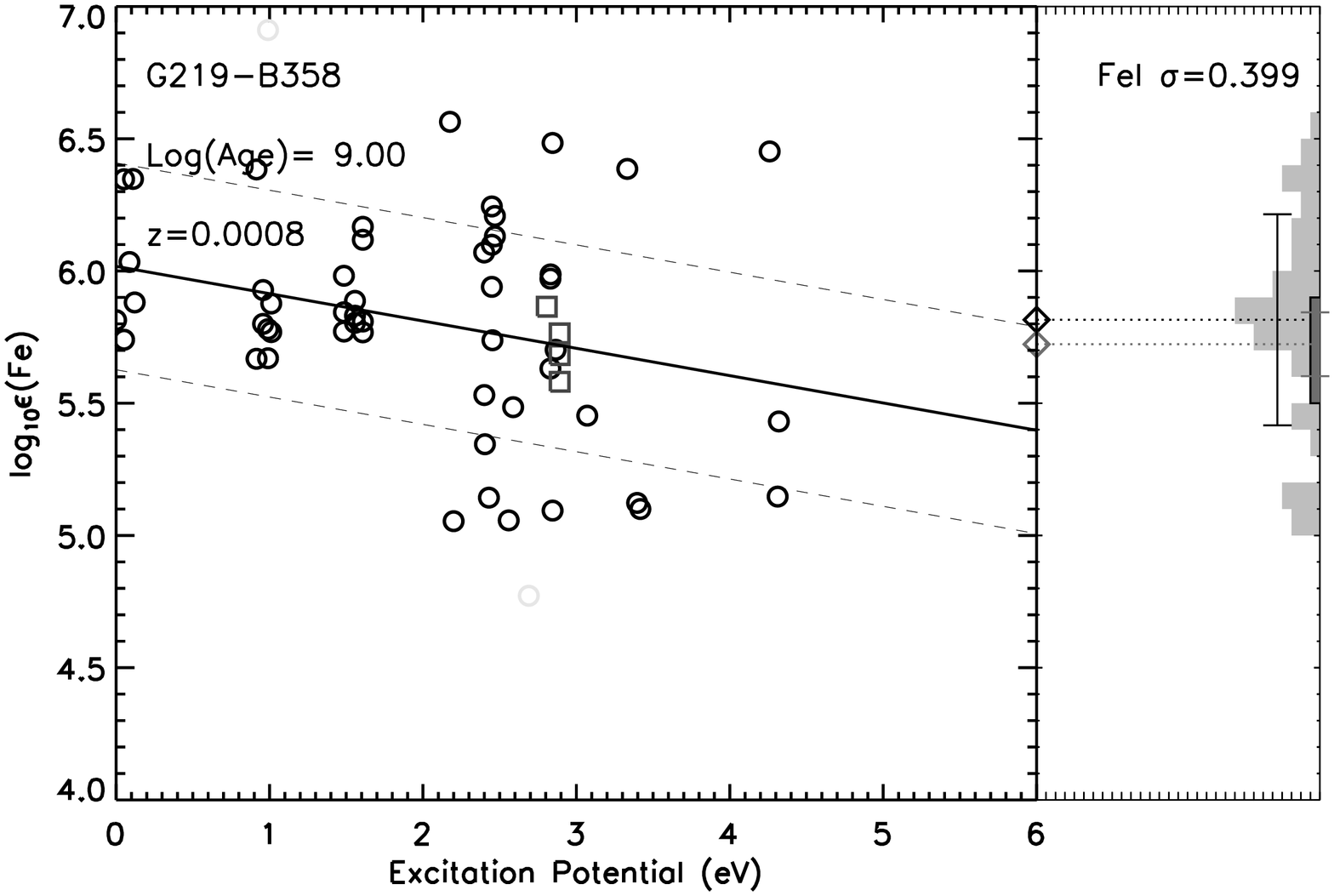}
\includegraphics[angle=90,scale=0.20]{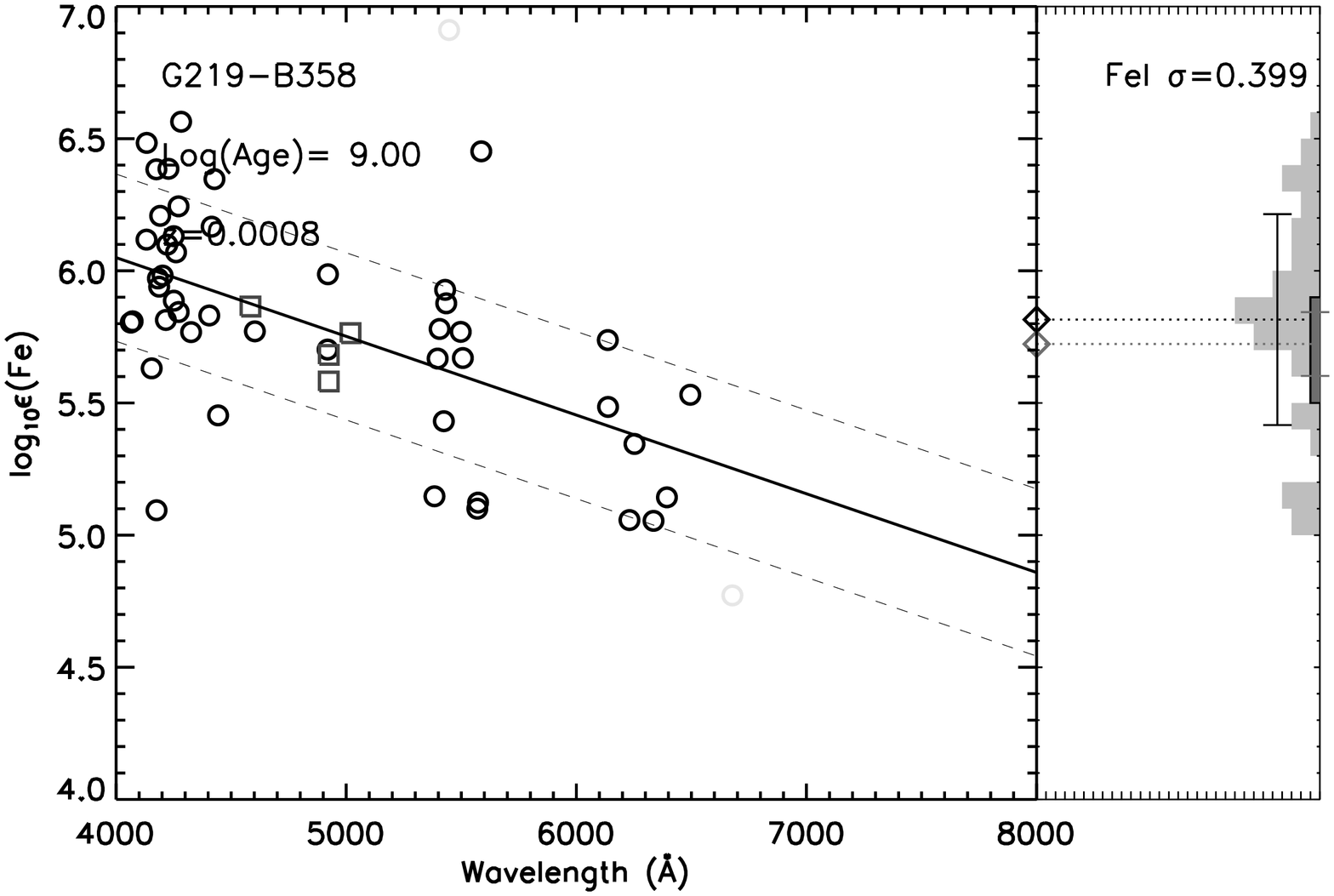}
\includegraphics[angle=90,scale=0.20]{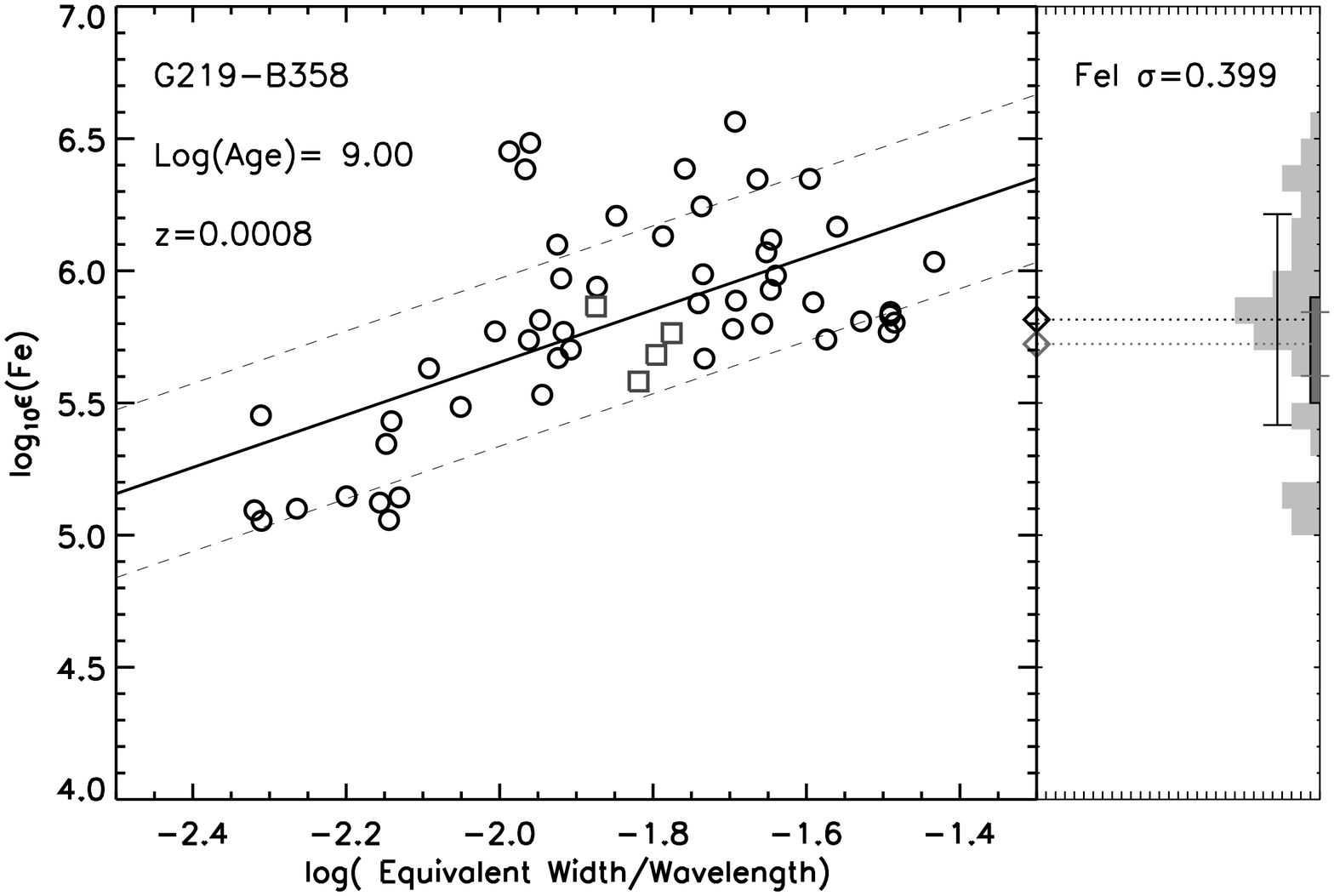}
\caption{Diagnostics for G219-B358.  The smallest Fe I standard deviation and smallest dependence on EP, wavelength, and observed equivalent width at ages of 7-13 Gyr. Symbols are the same as in Figure~\ref{fig:g108 diagnostics}.}

\label{fig:g219 diagnostics} 
\end{figure*}

\begin{figure*}
\centering
\includegraphics[angle=90,scale=0.20]{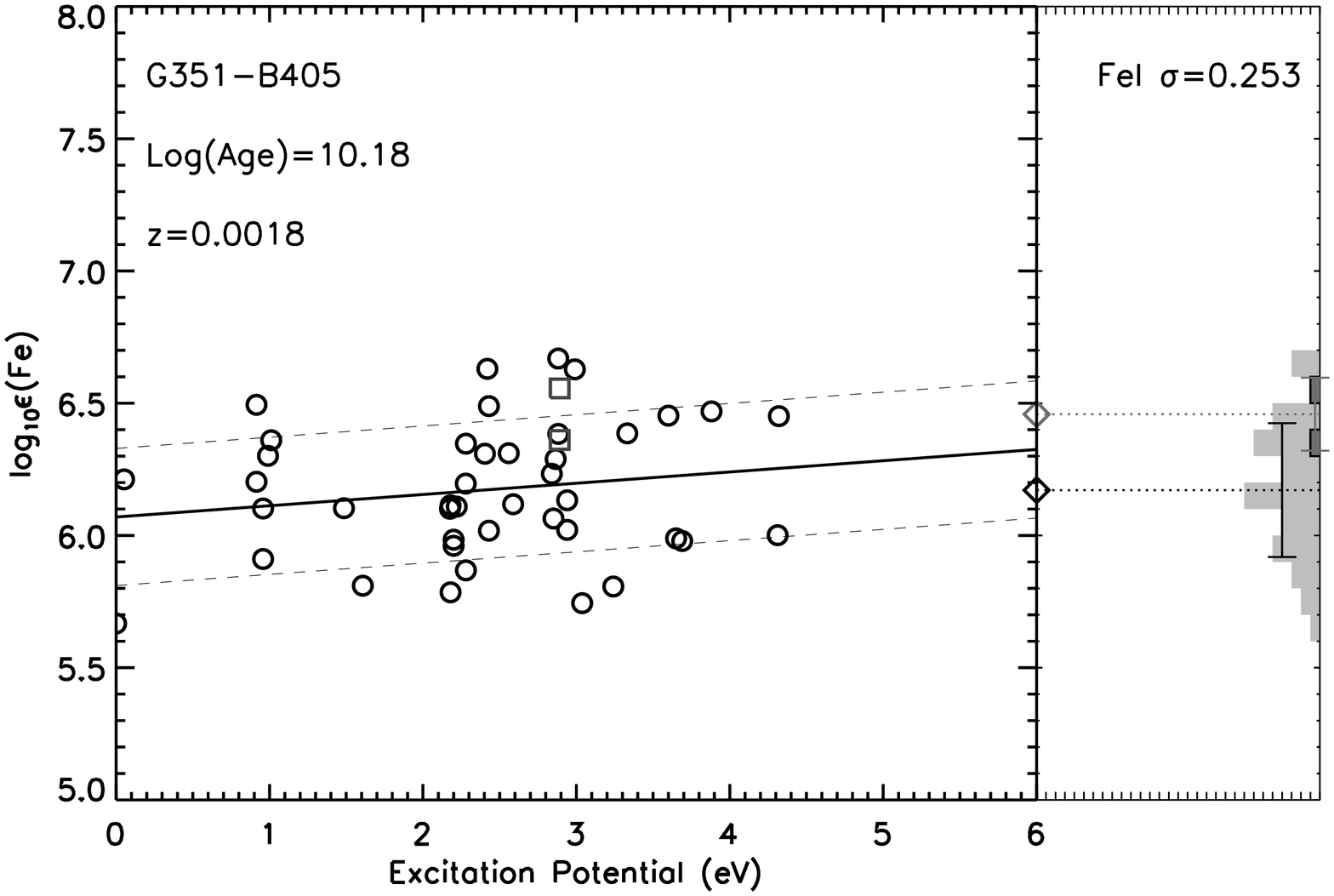}
\includegraphics[angle=90,scale=0.20]{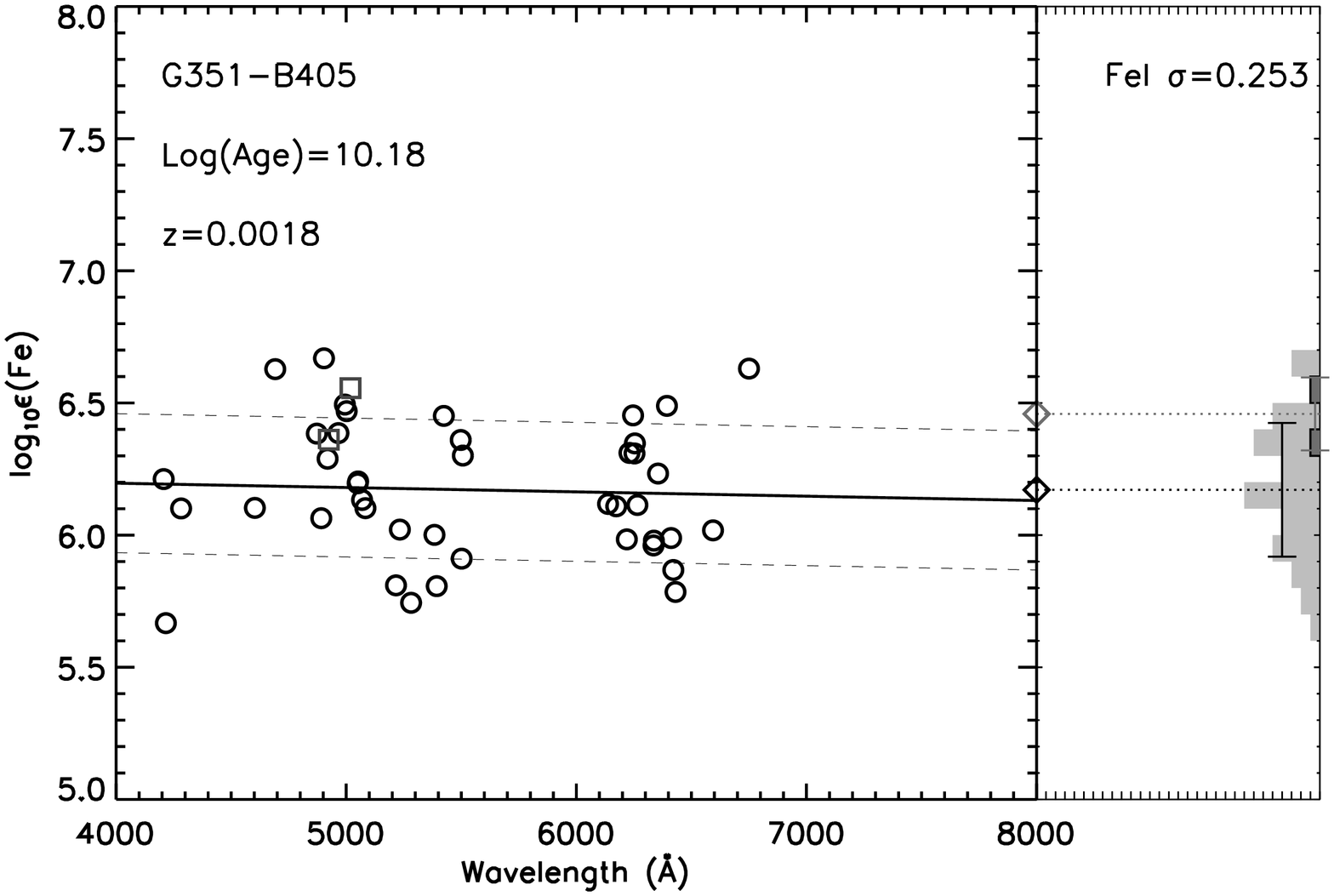}
\includegraphics[angle=90,scale=0.20]{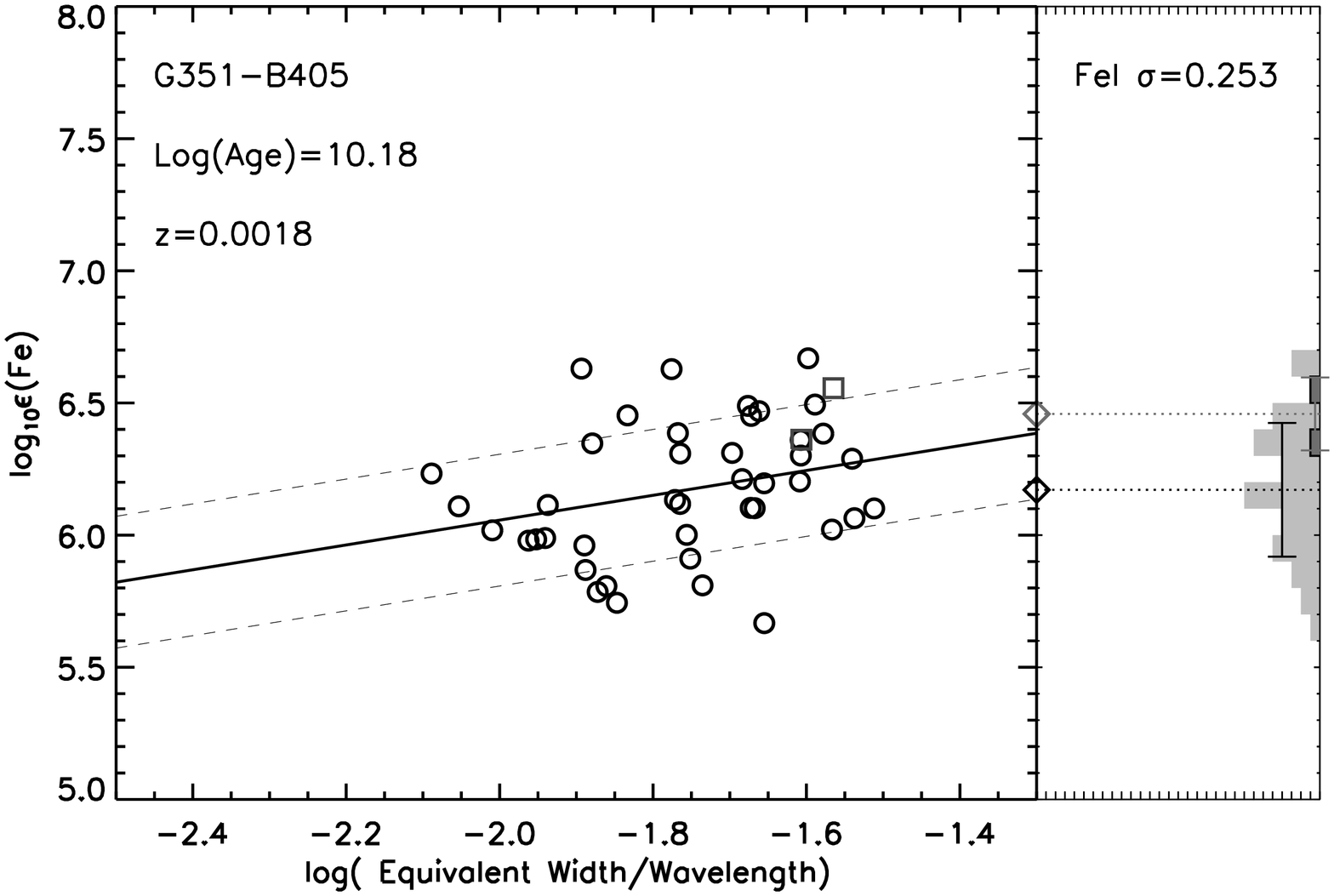}
\includegraphics[angle=90,scale=0.20]{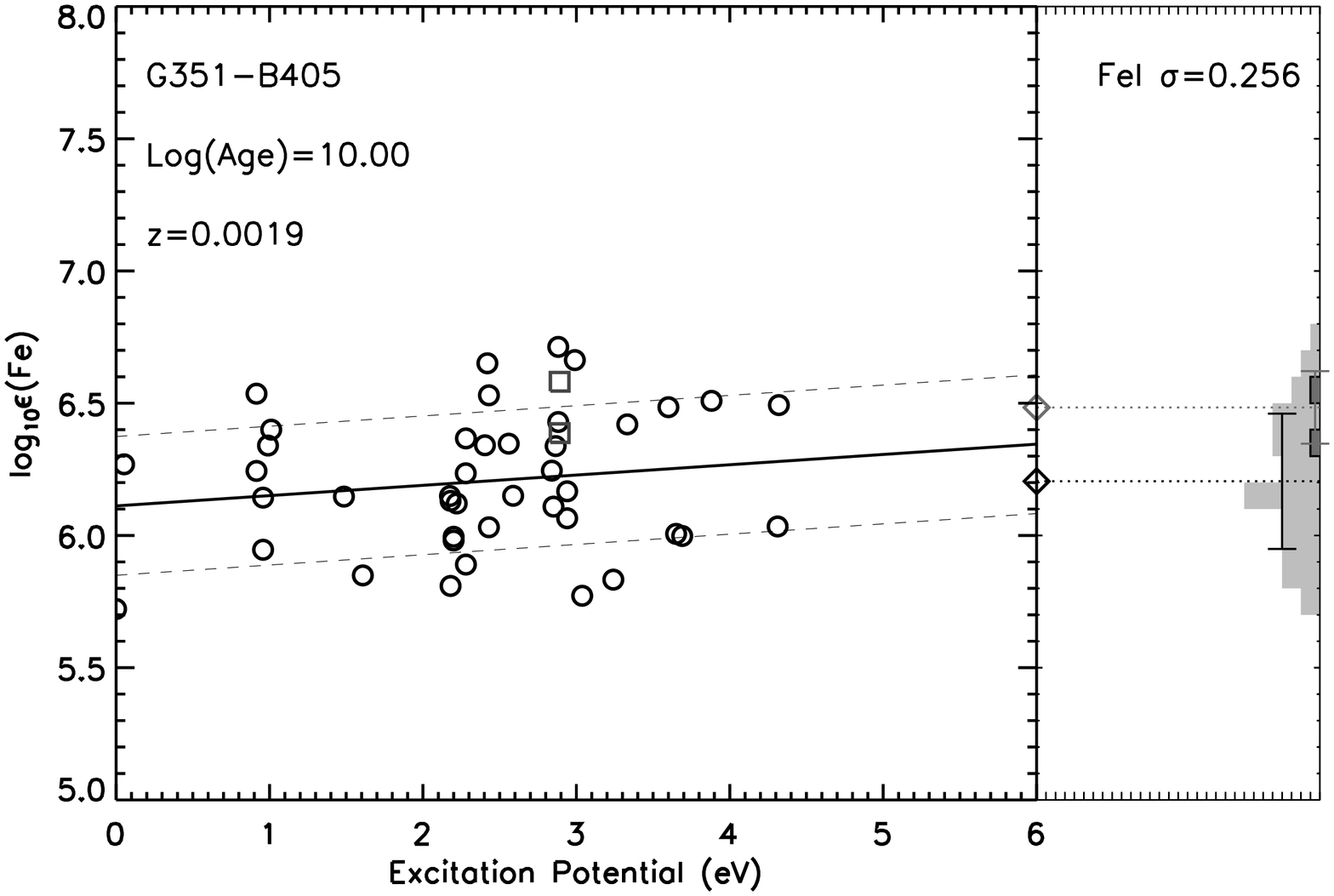}
\includegraphics[angle=90,scale=0.20]{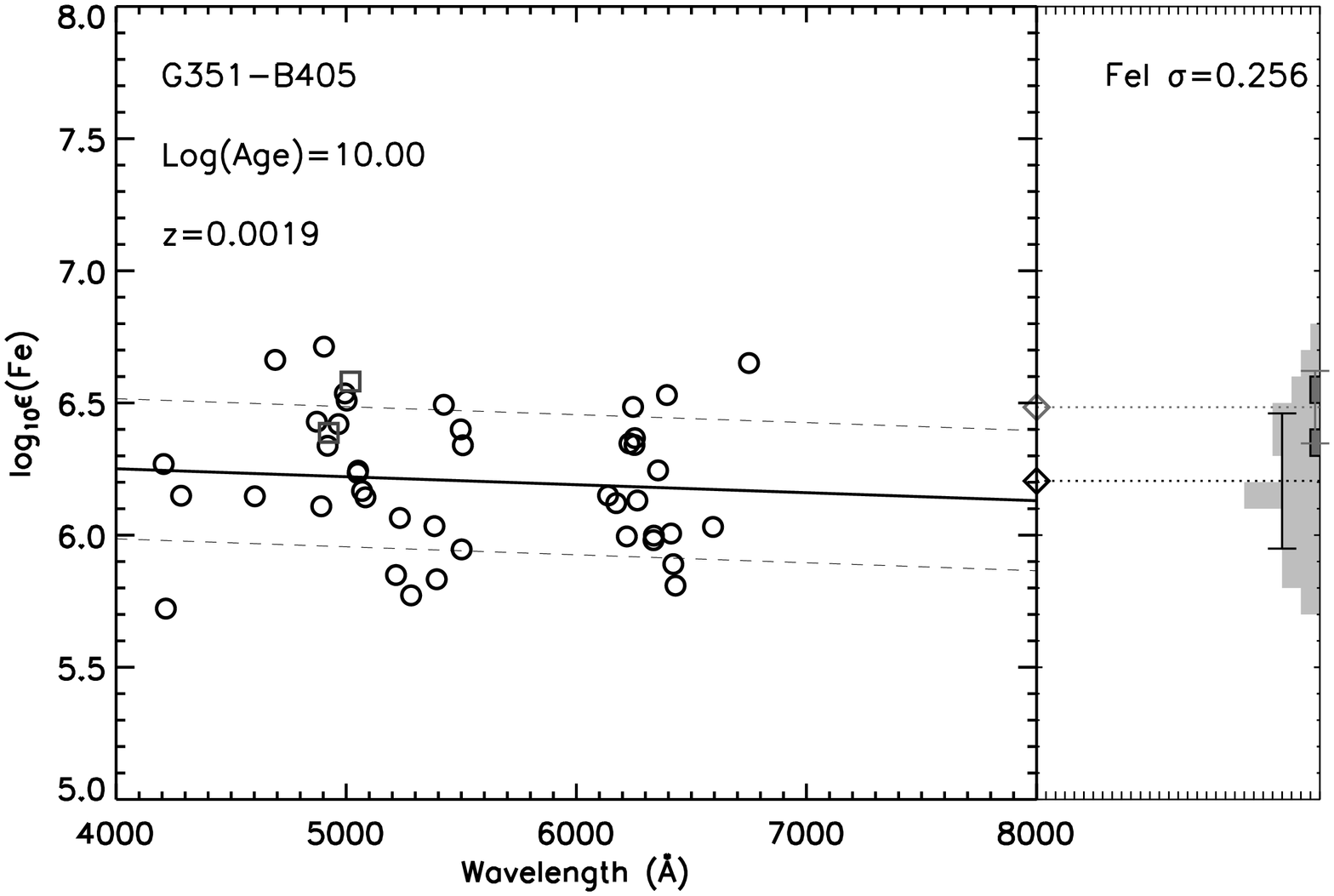}
\includegraphics[angle=90,scale=0.20]{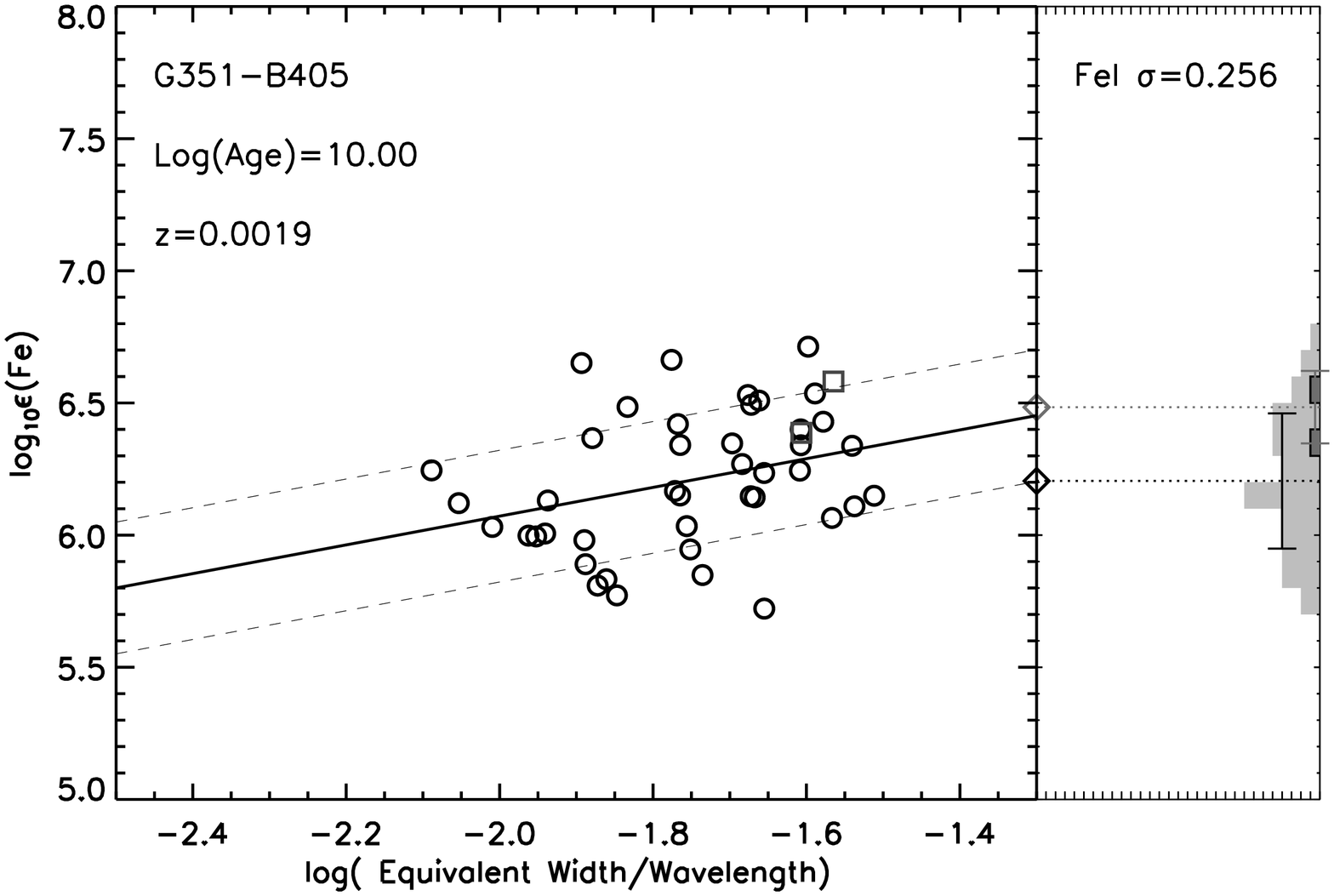}
\includegraphics[angle=90,scale=0.20]{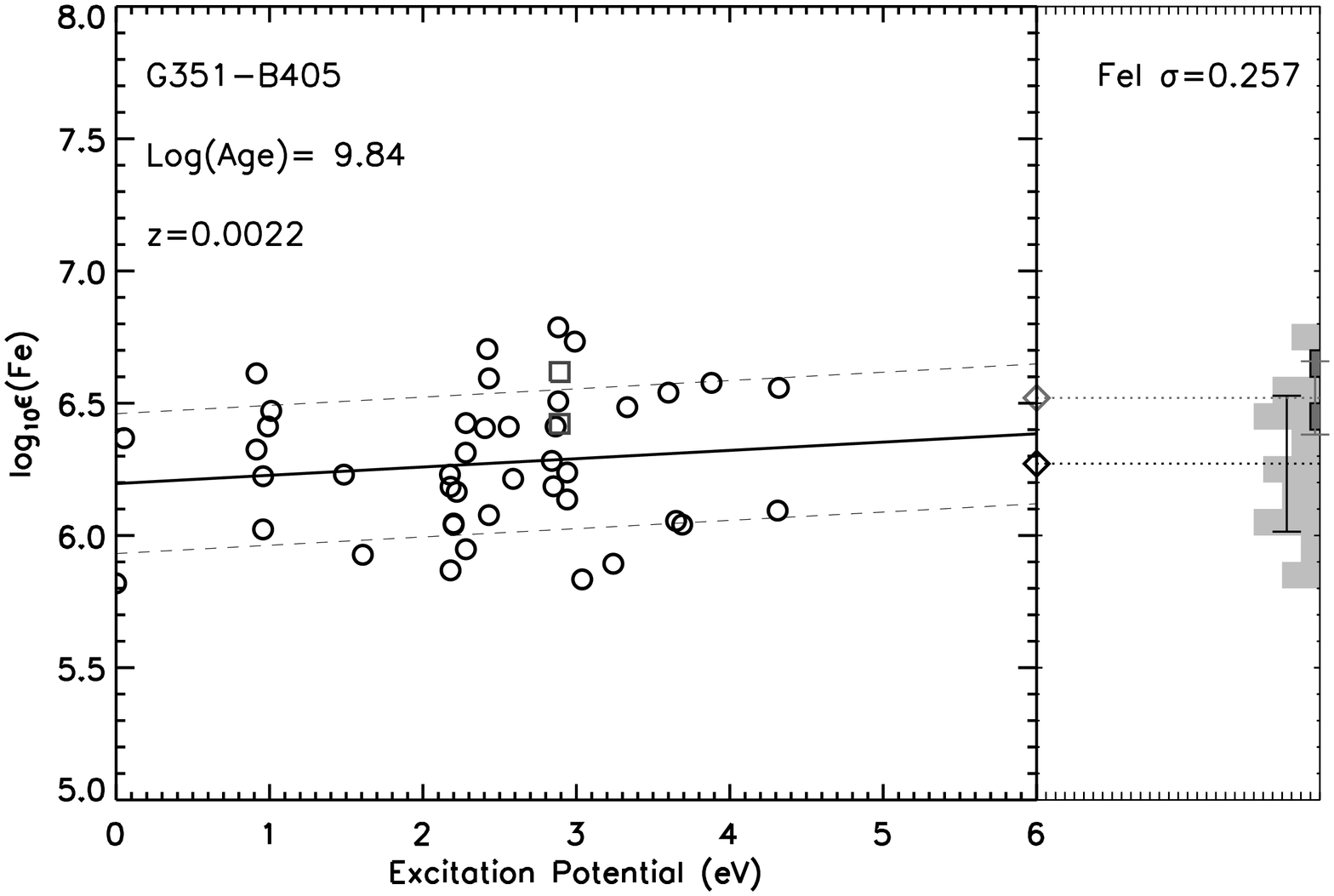}
\includegraphics[angle=90,scale=0.20]{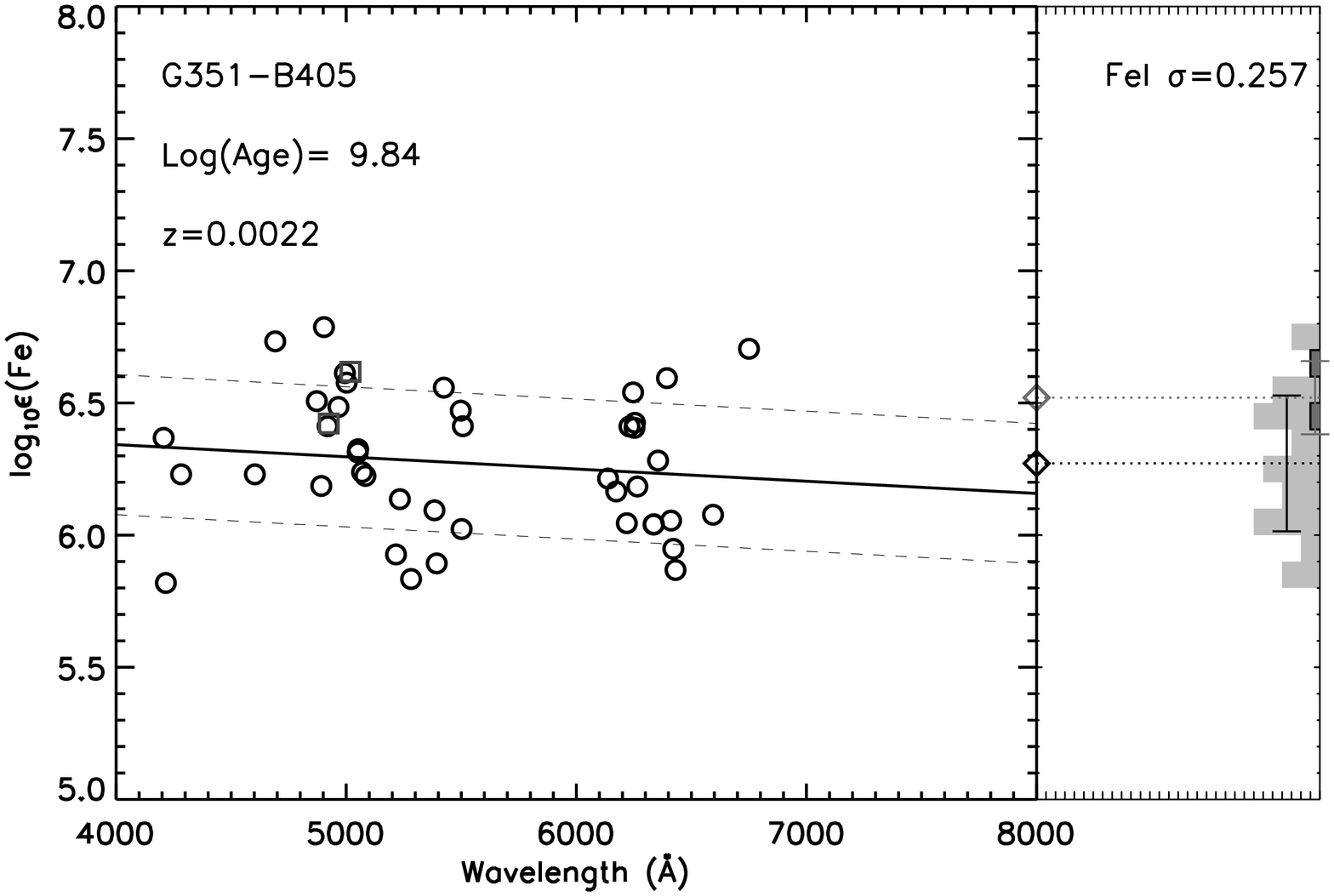}
\includegraphics[angle=90,scale=0.20]{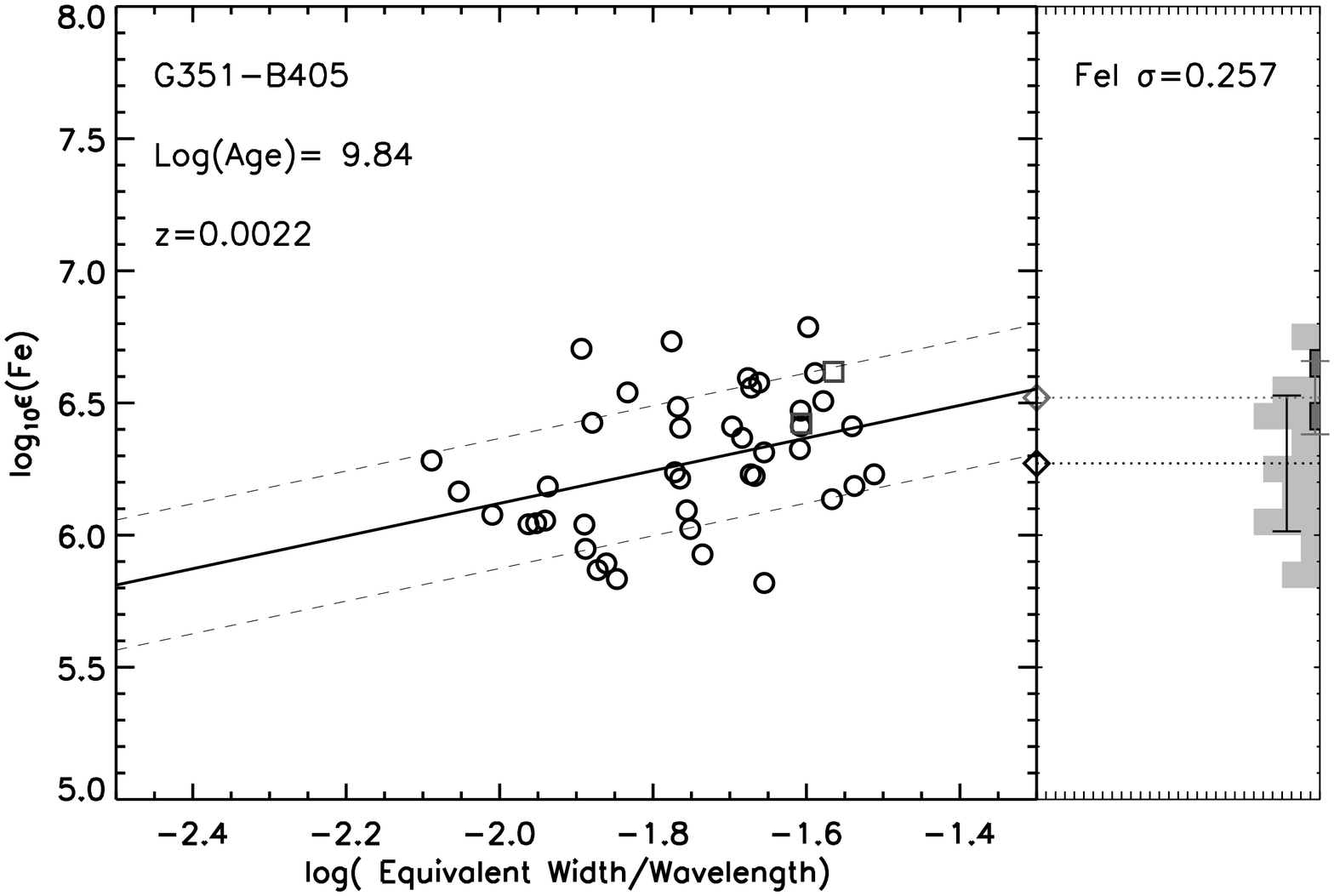}
\includegraphics[angle=90,scale=0.20]{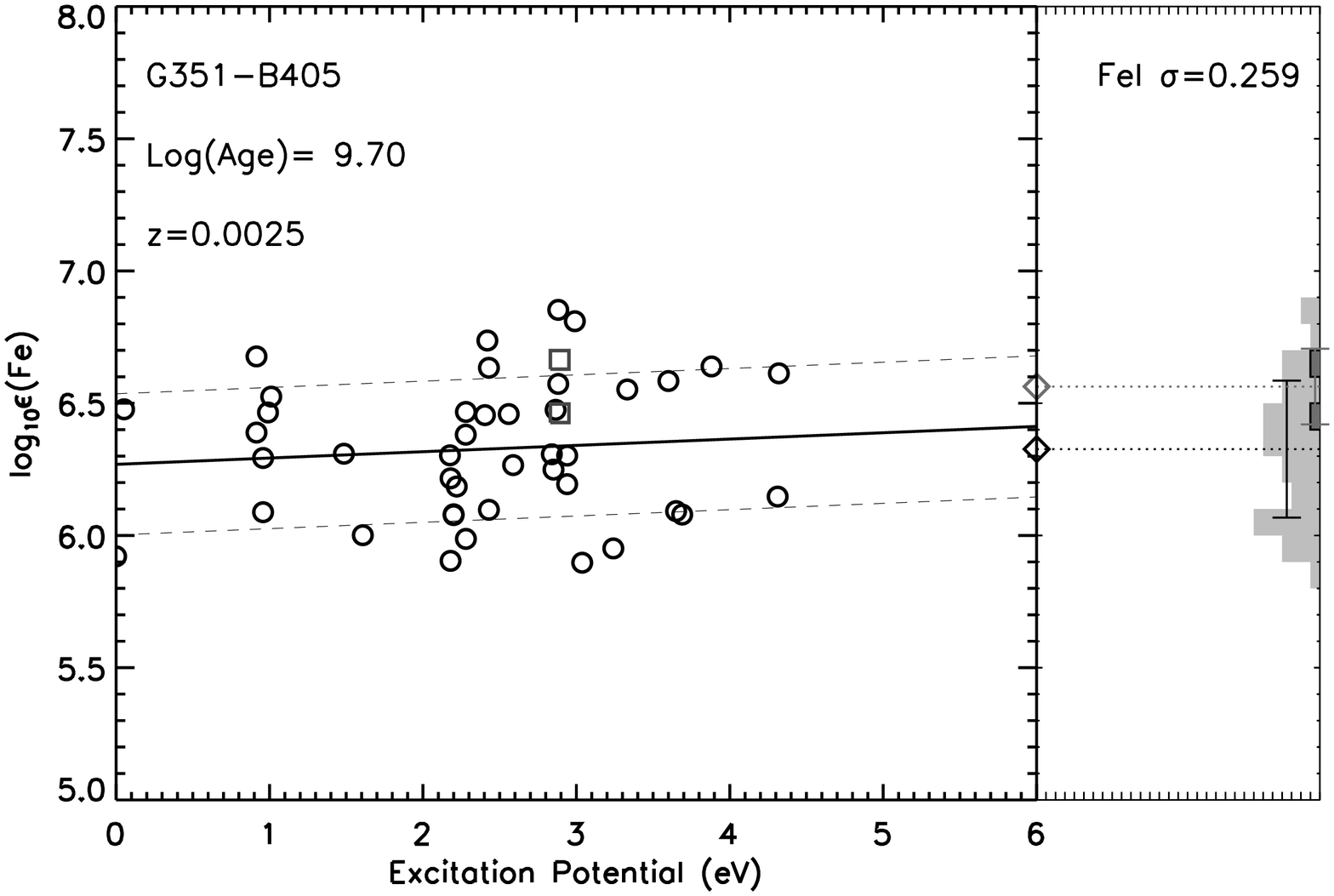}
\includegraphics[angle=90,scale=0.20]{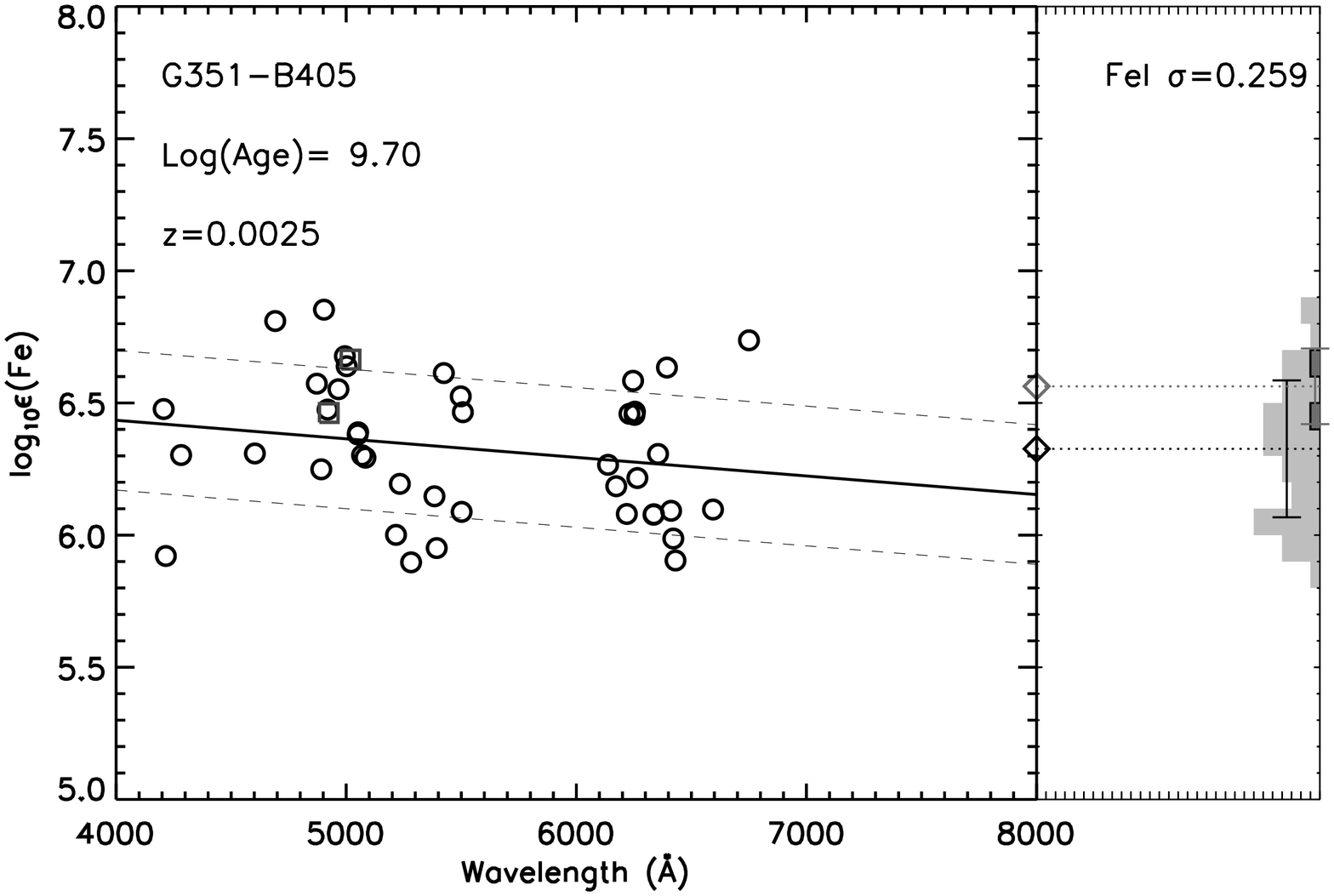}
\includegraphics[angle=90,scale=0.20]{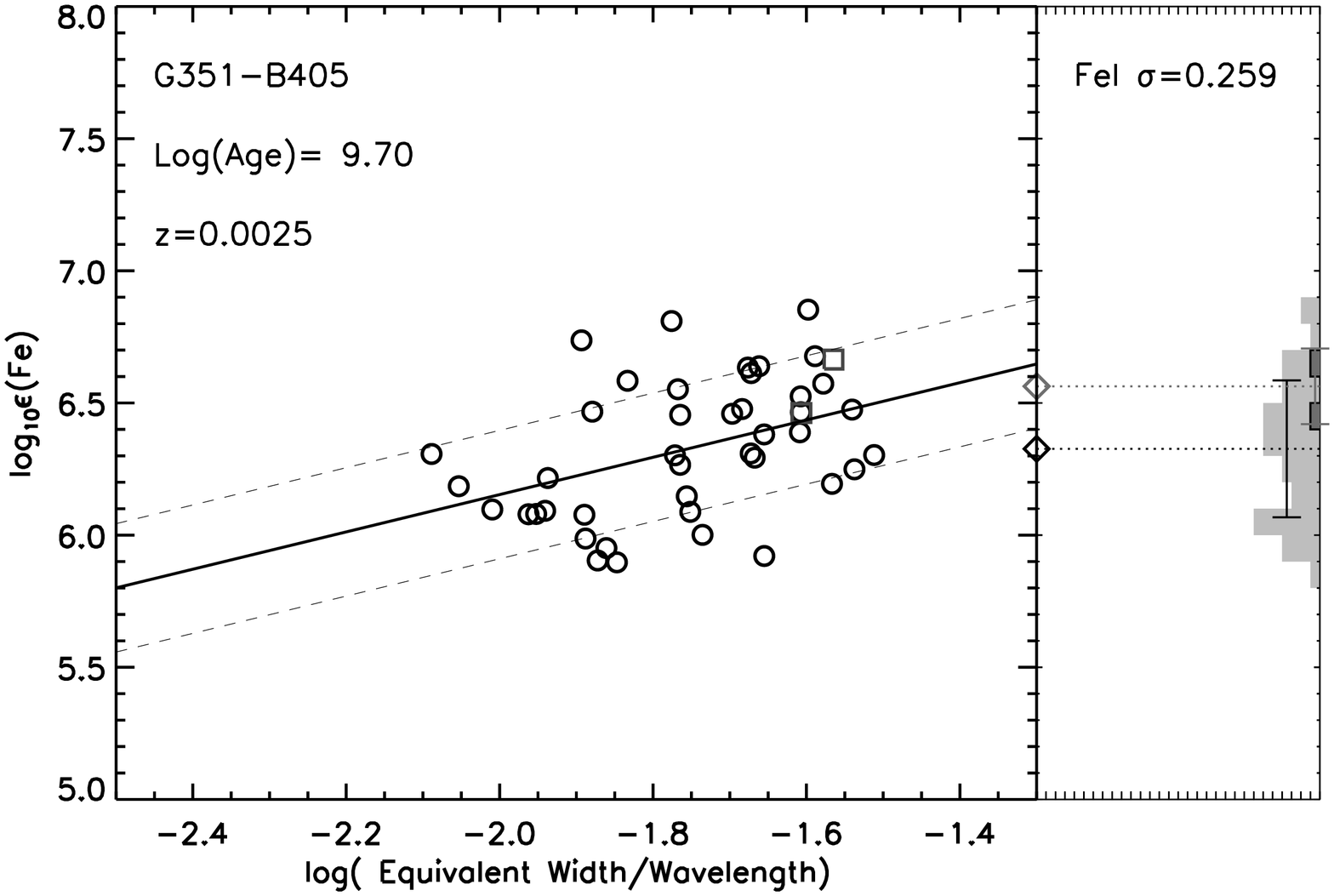}
\includegraphics[angle=90,scale=0.20]{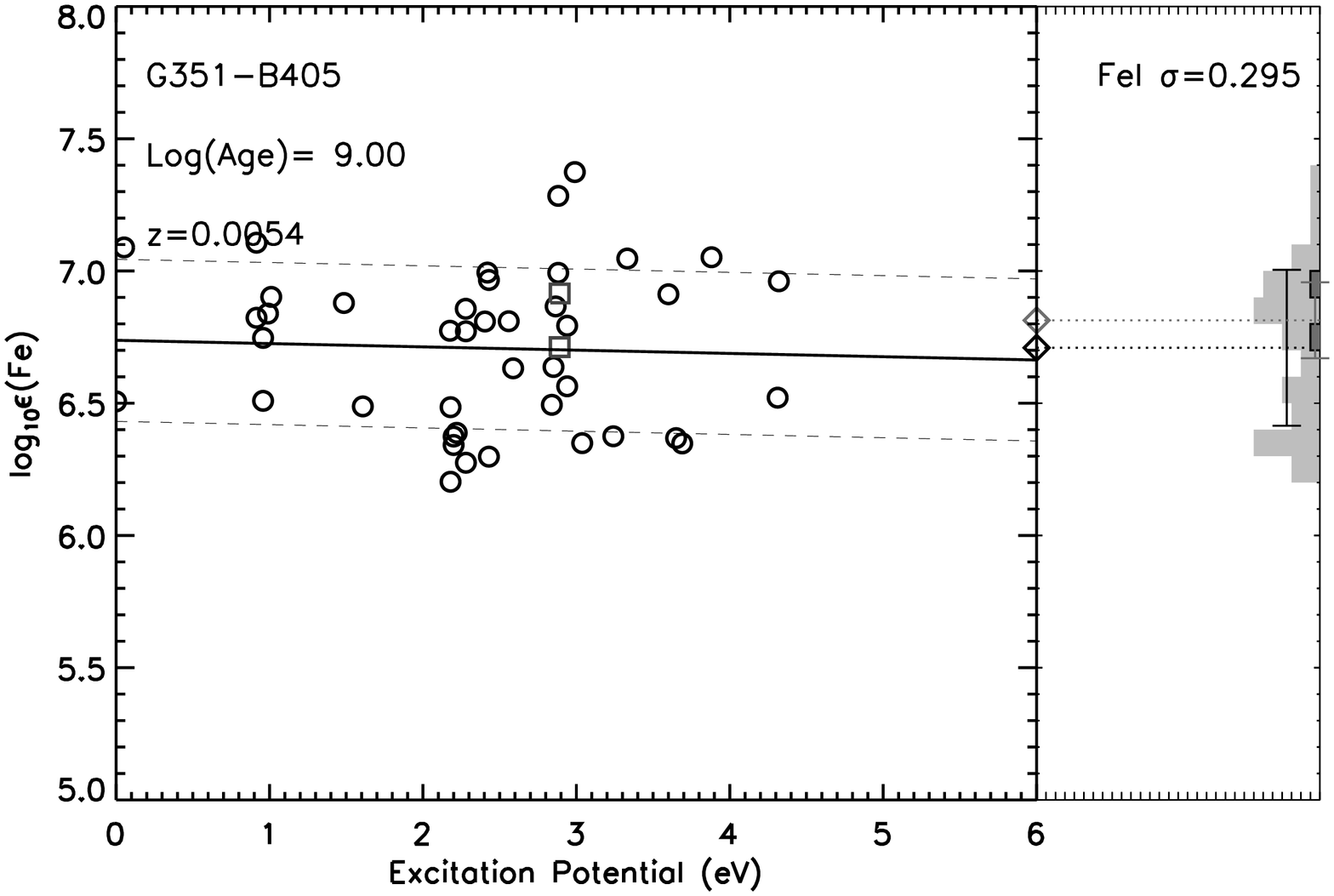}
\includegraphics[angle=90,scale=0.20]{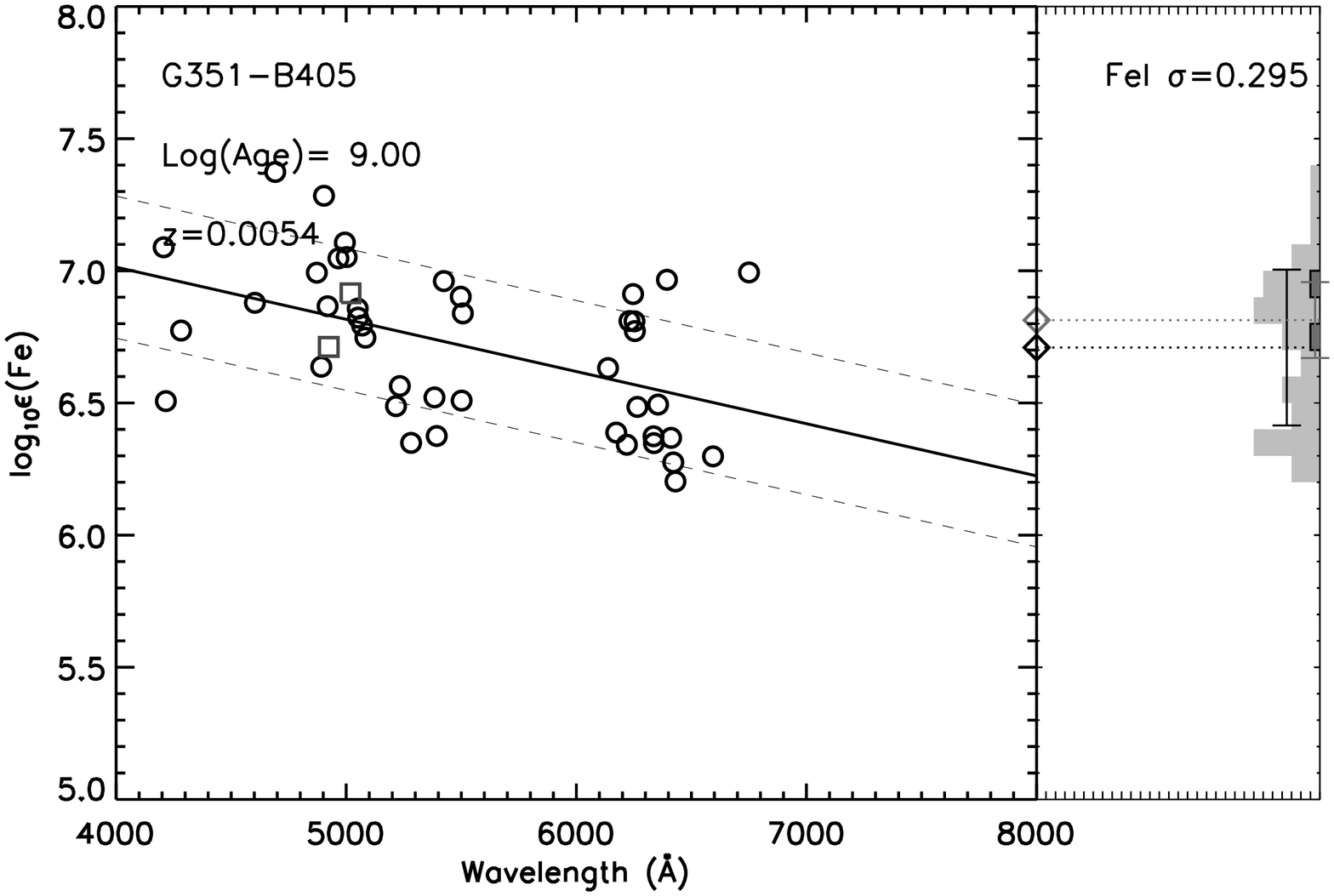}
\includegraphics[angle=90,scale=0.20]{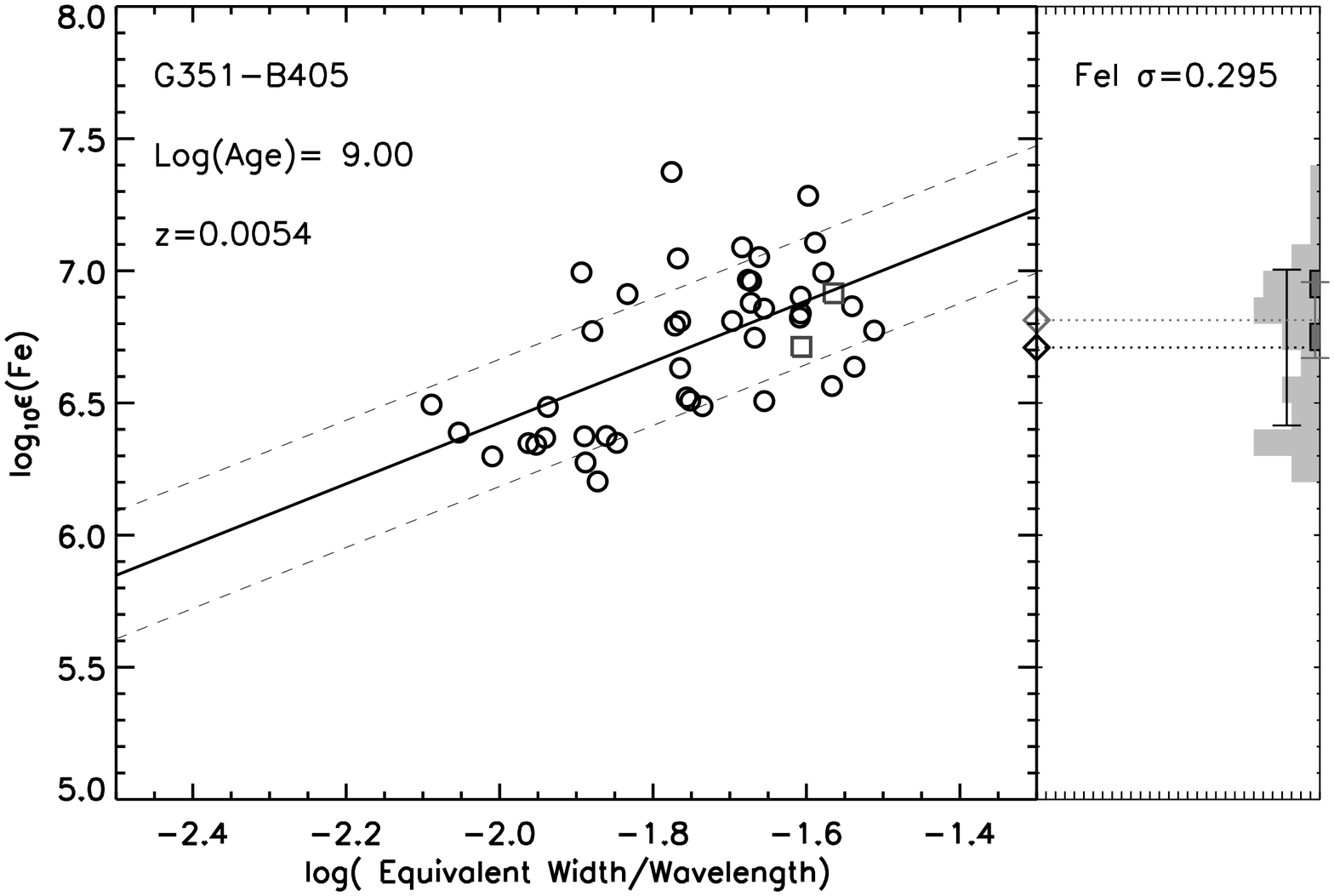}
\caption{Diagnostics for G351-B405.  The smallest Fe I standard deviation and smallest dependence on EP, wavelength, and observed equivalent width at ages of 10-15 Gyr. Symbols are the same as in Figure~\ref{fig:g108 diagnostics}.}

\label{fig:g351 diagnostics} 
\end{figure*}

As mentioned above, our criteria for selecting a best-fitting CMD is
that the Fe abundance calculated from the Fe I lines is consistent
with [Fe/H] used to produce the isochrone.  This criteria is met where
the solutions cross the dotted black lines in Figures~\ref{fig:Fe
  plots} through \ref{fig:Fe5 plots}.  When the Fe I solution that crosses the dotted line for a
given age (color) lies between two isochrones, we interpolate an
appropriate isochrone according to the
prescription recommended in \cite{2006ApJ...642..797P}.  We then
have 7 possible solutions where isochrones intercept the dotted
black line in Figure~\ref{fig:Fe plots} --- one abundance solution
for each age.

We next isolate the best-fitting CMD out of these 7 using diagnostics
commonly used in standard stellar abundance analyses. These
diagnostics concern the stability of the [Fe/H] solutions, which
should not depend on the parameters of the individual line (excitation
potentials, wavelengths, or reduced EWs\footnote{Reduced EW $\equiv$
  log(EW / wavelength) }).  In principle, as in
individual RGB stars, these diagnostics reflect the accuracies in the
physical properties of the atmospheres used in the synthesis.  The
abundance versus excitation potential (EP) diagnostic is sensitive to
the temperature of stars, while the abundance versus reduced EW
diagnostic is sensitive to the microturbulent velocities of stars.
The abundance versus wavelength diagnostic is potentially sensitive to
the age of the CMD, because stars of different temperatures dominate
the IL flux at different wavelengths.  Unlike in RGB stars, in an IL spectrum,
correlations with EP and wavelength can be caused by an inaccurate
temperature distribution of stars in the CMD, which, for example,
could be the result inaccurate modeling of HB morphology in the
isochrones. Likewise, correlations with reduced EW can be the result
of inaccurate proportions of stars of different gravities as well as a
symptom of an inaccurate microturbulent velocity law. These effects
are difficult to unravel without additional constraints on the CMD,
and identifying these is not the primary goal of this work. We
therefore take the existence of any correlations merely as an
indication that a given isochrone is  less representative of
the true CMD then one with weaker correlations.  We attempt to 
identify a best-fitting CMD solution by selecting an isochrone that
minimizes these correlations.  We use linear least squares fits
between [Fe/H] and these parameters to identify and quantify the
strength of any existing correlations.  Plots illustrating the
behavior of the Fe abundances with EP, wavelength, and reduced EW are
shown in Figures~\ref{fig:g108 diagnostics} through \ref{fig:g351
  diagnostics}.  From these plots, we obtain 5 diagnostics: the slope
of [Fe/H] with EP, wavelength and reduced EW, and the standard
deviation of the [Fe/H] solution for Fe I lines and Fe II lines.

For any one of these diagnostics, there is not a statistically
significant difference between the quality of the solution from CMDs
within a range of $\pm$5 Gyrs.  For example, the slope of the [Fe/H] vs.
EP relationship in Figure~\ref{fig:g315 diagnostics} for G315 looks essentially the same  for CMDS between
ages 5 and 15 Gyrs. However, while the difference in these diagnostics
may be small over a wide range in CMD age, we do find that they change
monotonically, and are strongly correlated with each other.  This suggests
that there is clearly a preferred age and [Fe/H] range of CMD for each GC.

\begin{figure}
\centering
\includegraphics[scale=0.4]{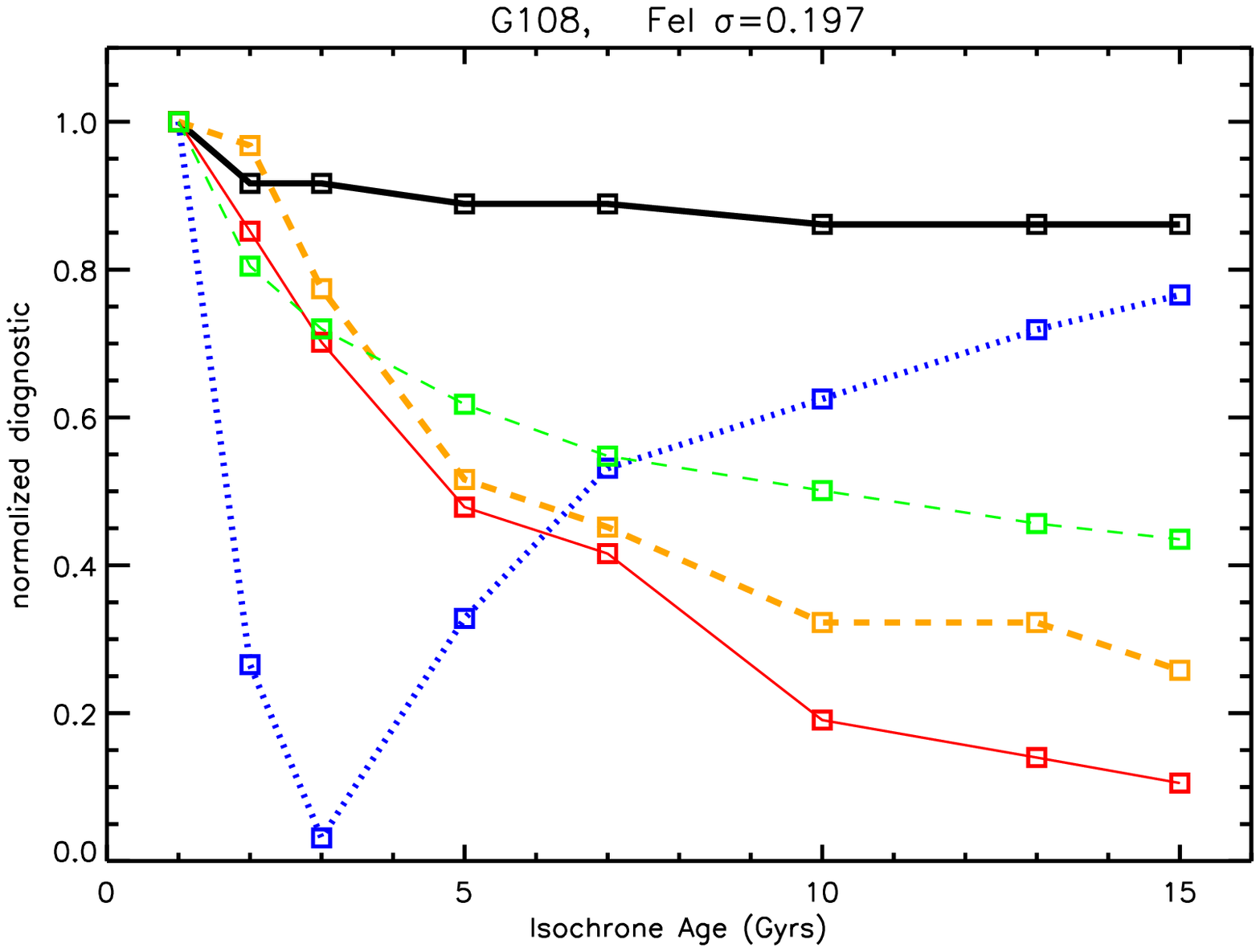}

\caption{Fe line abundance diagnostics for G108.  All diagnostics are normalized to their maximum values so that they can be shown on the same scale.  The thick black line is Fe I $\sigma$, thick dashed orange line is Fe II $\sigma$, dotted blue line is the slope in [Fe/H] vs. $\lambda$, thin dashed green line is the slope in [Fe/H] vs. reduced EW, and the thin red solid line is the slope in [Fe/H] vs. EP.   }
\label{fig:alldiag} 
\end{figure}

\begin{figure}
\centering
\includegraphics[scale=0.4]{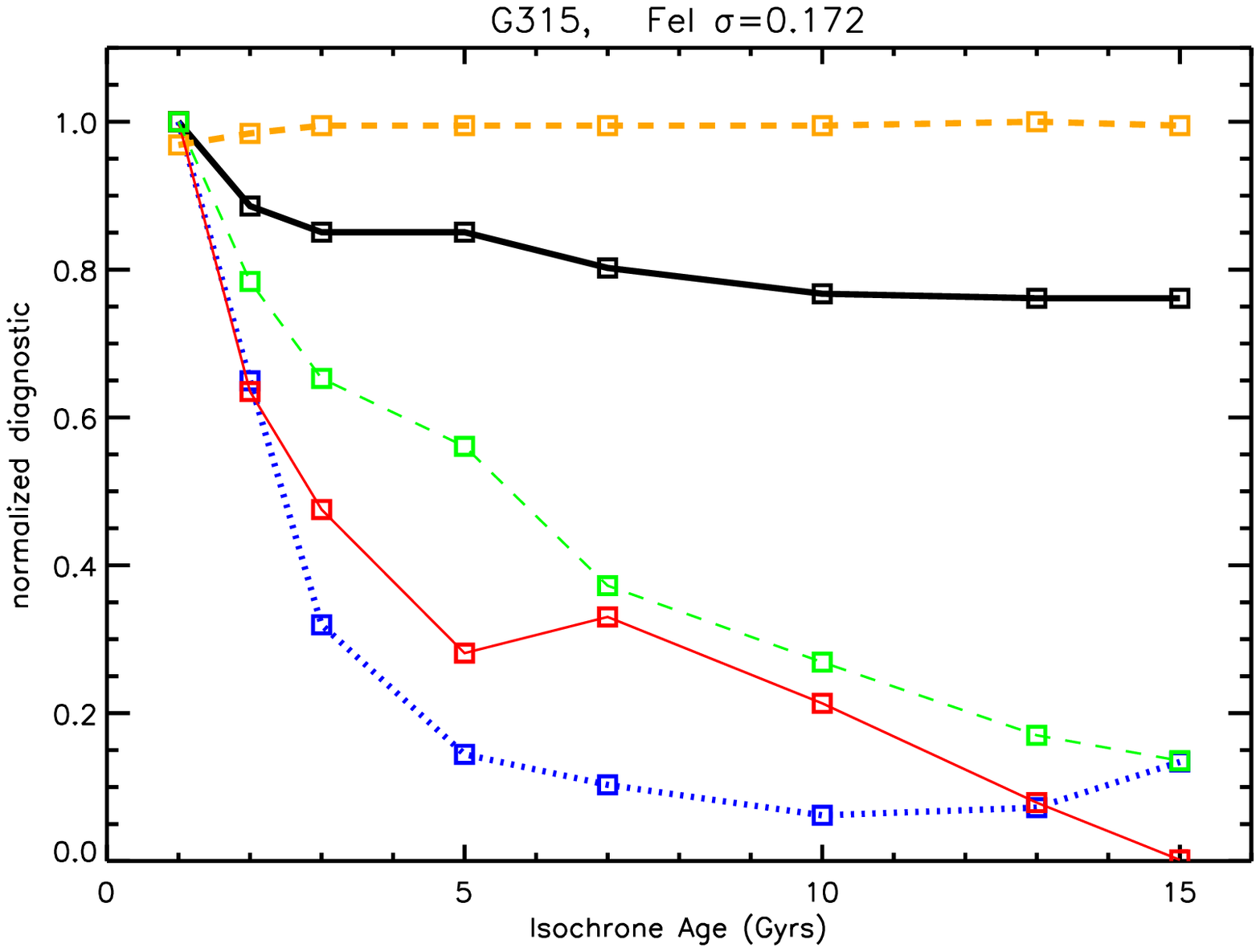}
\caption{Same as Figure~\ref{fig:alldiag} for G315. }
\label{fig:alldiag2} 
\end{figure}

\begin{figure}
\centering
\includegraphics[scale=0.4]{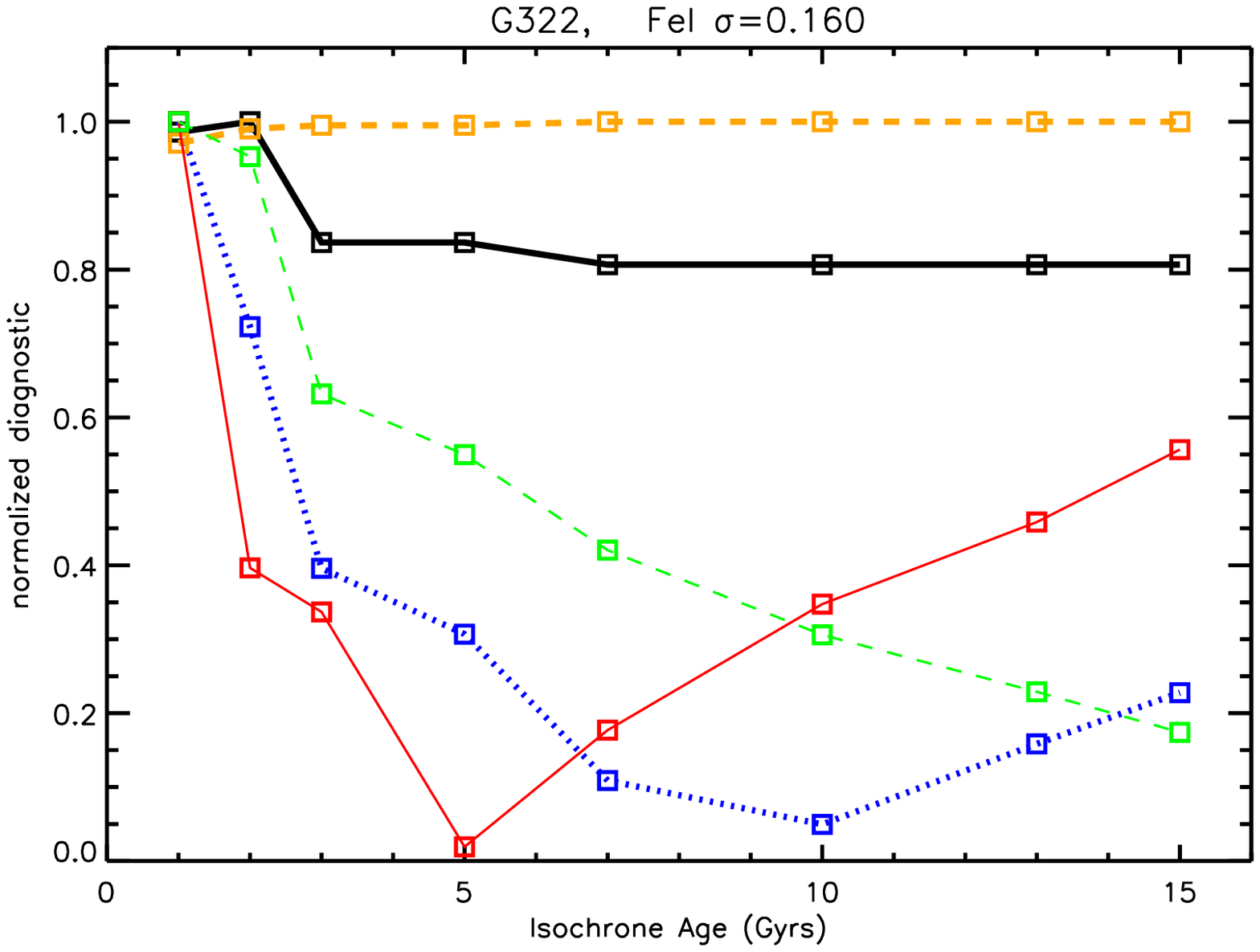}
\caption{Same as Figure~\ref{fig:alldiag} for G322. }
\label{fig:alldiag3} 
\end{figure}

\begin{figure}
\centering
\includegraphics[scale=0.4]{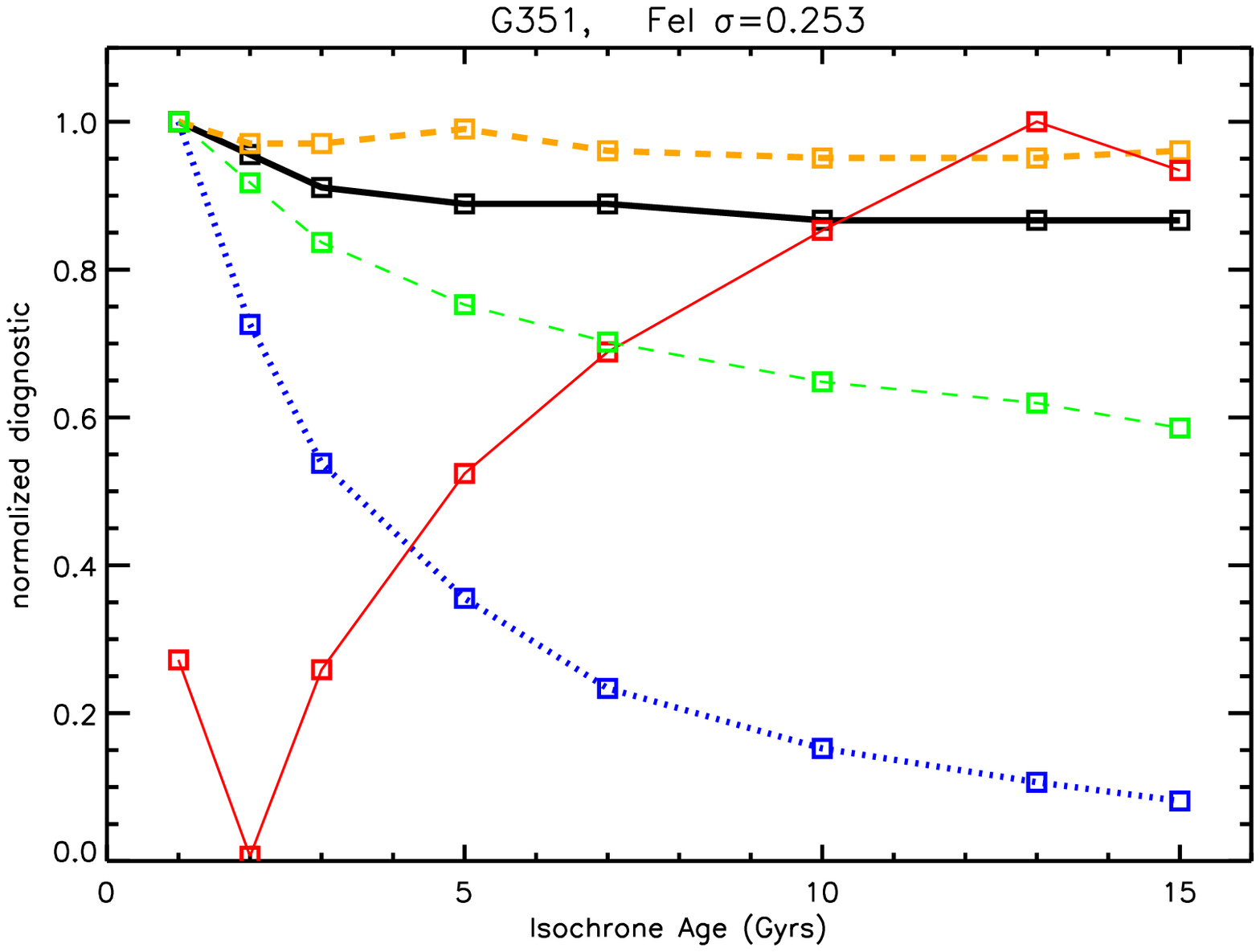}
\caption{Same as Figure~\ref{fig:alldiag} for  G351. }
\label{fig:alldiag4} 
\end{figure}

\begin{figure}
\centering
\includegraphics[scale=0.4]{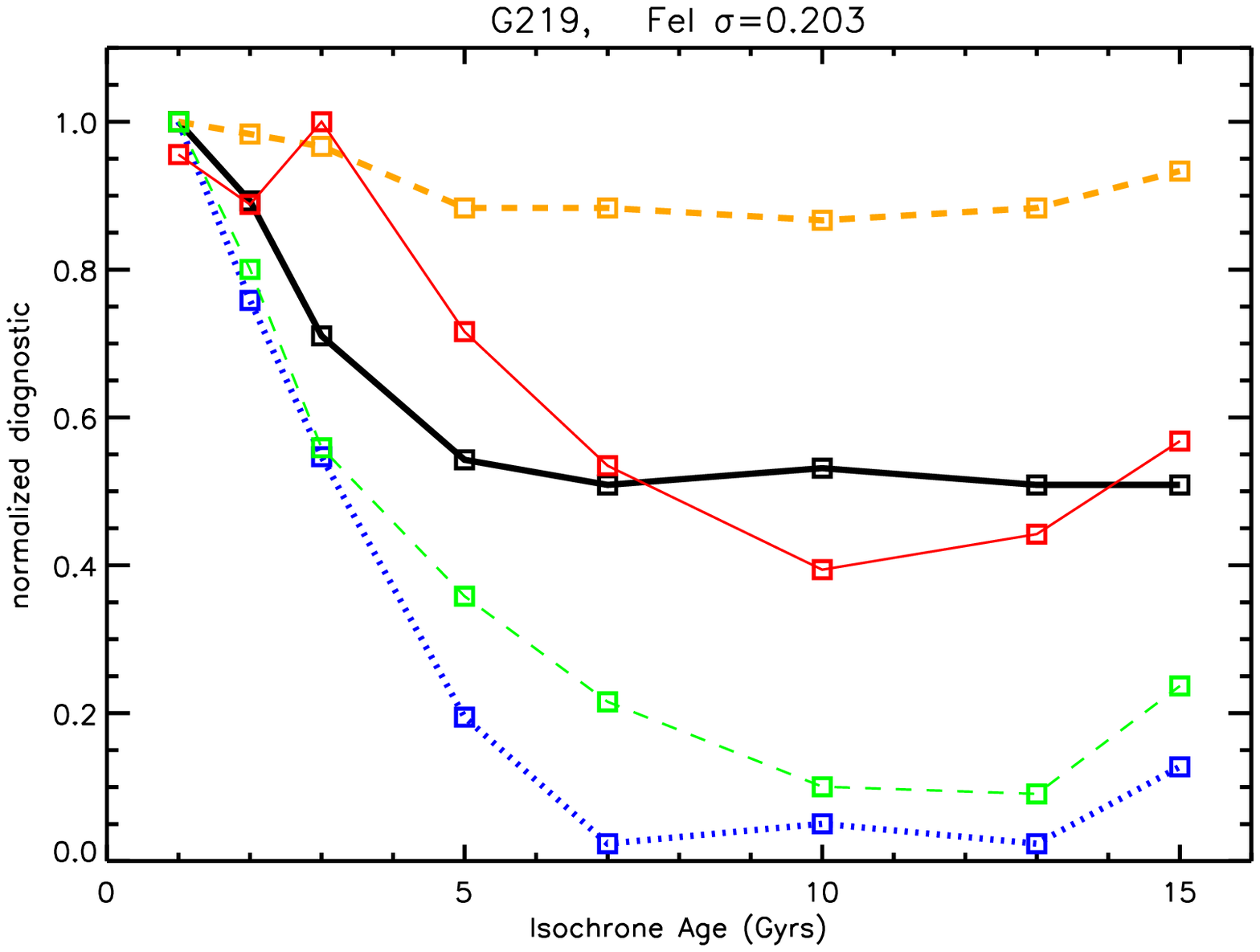}
\caption{Same as Figure~\ref{fig:alldiag} for G219. }
\label{fig:alldiag5} 
\end{figure}

To see this more clearly, we plot these diagnostics for the M31 sample in
Figures~\ref{fig:alldiag} through ~\ref{fig:alldiag5}.  These plots show
all five diagnostics as a function CMD age for the 7 CMDs that satisfy
the original selection criteria (see Figures~\ref{fig:Fe plots} through \ref{fig:Fe5 plots}).
From Figures~\ref{fig:alldiag} through ~\ref{fig:alldiag5}, it is clear that for
all of the GCs in our sample, all five diagnostics simultaneously imply
that better solutions are obtained for CMDs with ages $>$7 Gyr.  For these GCs, as for old Milky Way GCs analyzed as part of our
training set (S. Cameron et al.  2009), we find the acceptable CMD ages
typically cover a range of 5 Gyrs (e.g.  10$-$15 Gyrs or 7$-$13 Gyrs).
This is not surprising because the CMDs themselves change very little over
those ranges in age.  We discuss these age constraints in detail in the
next section.

\begin{figure}
\centering
\includegraphics[angle=90,scale=0.3]{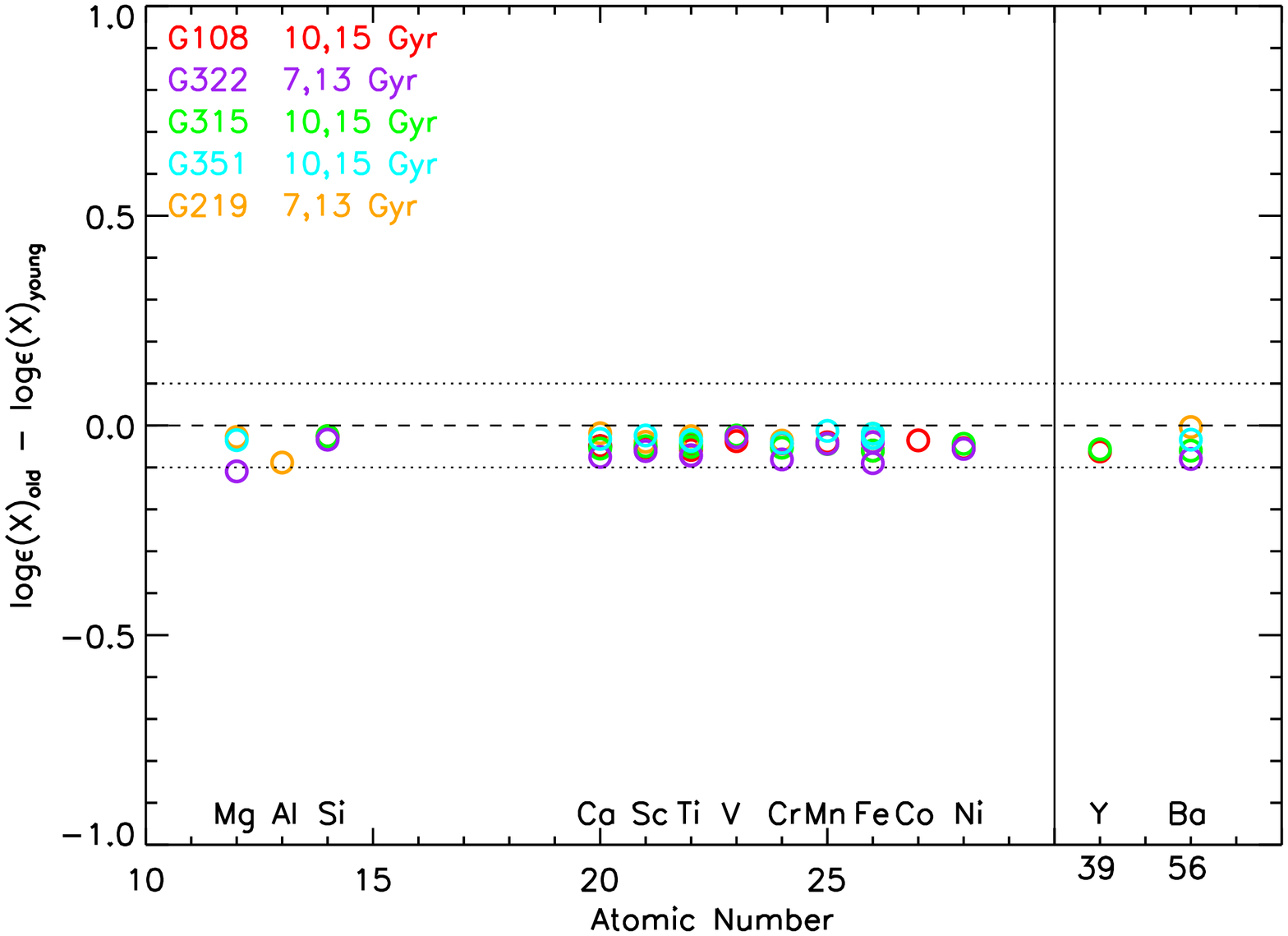}
\includegraphics[angle=90,scale=0.3]{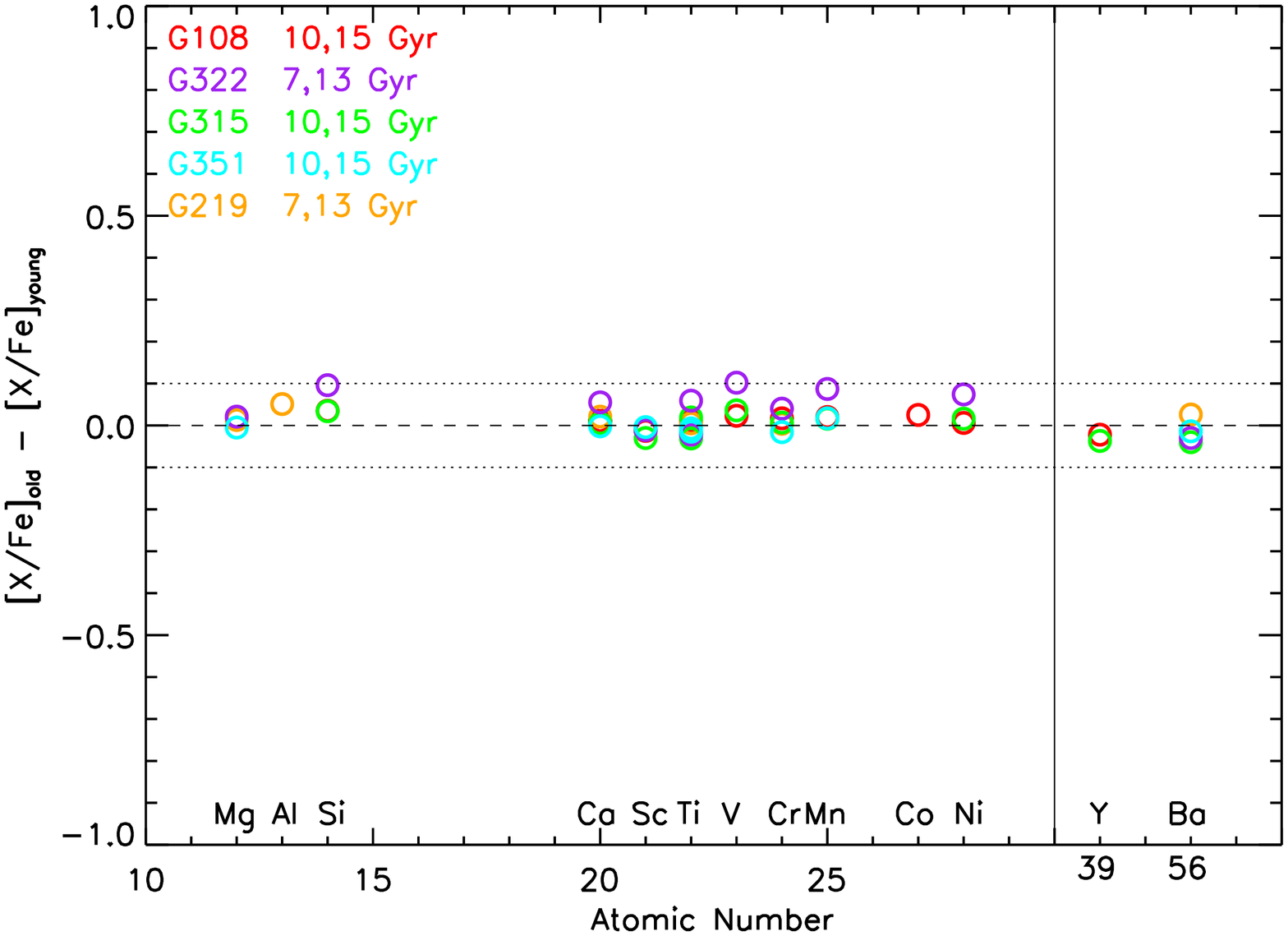}
\caption{Robustness of the abundances over the $\sim$5 Gyr range of acceptable isochrone ages. Top plot shows the difference in the log$\epsilon$(X) of older and younger solutions.  Bottom plot shows the corresponding difference in abundance ratios.  Dotted lines around zero mark a change of $\pm$ 0.1 dex. }%
\label{fig:youngold} 
\end{figure}

From this range of acceptable ages, we select one CMD to use for a
final analysis run of all elements for which we measure lines.  For
old GCs, such as the present sample, our abundance results are quite
insensitive to which CMD in this age range is used.  Again, this is
not surprising as the CMDs in this age range are very similar.  This
weak dependence is quantified in Figure~\ref{fig:youngold}, in which the upper
plot shows the small difference in abundance  between the oldest and youngest
CMDs in the acceptable age range for each GC (discussed in
\textsection~\ref{sec:Fe}).  Nearly all elements
change by $\leq$0.1 dex, and older CMD ages always give smaller
abundances.  The lower plot of Figure~\ref{fig:youngold}
demonstrates that the derived abundance {\em ratios} are even more
robust. Since the change in the abundances of most elements tracks the
change in Fe, the net difference in the abundance ratios is $<$0.05
dex in almost all cases.

\

\section{Results: Chemical Abundances}
\label{sec:abundances}

\begin{figure*}
\centering
\includegraphics[scale=0.45]{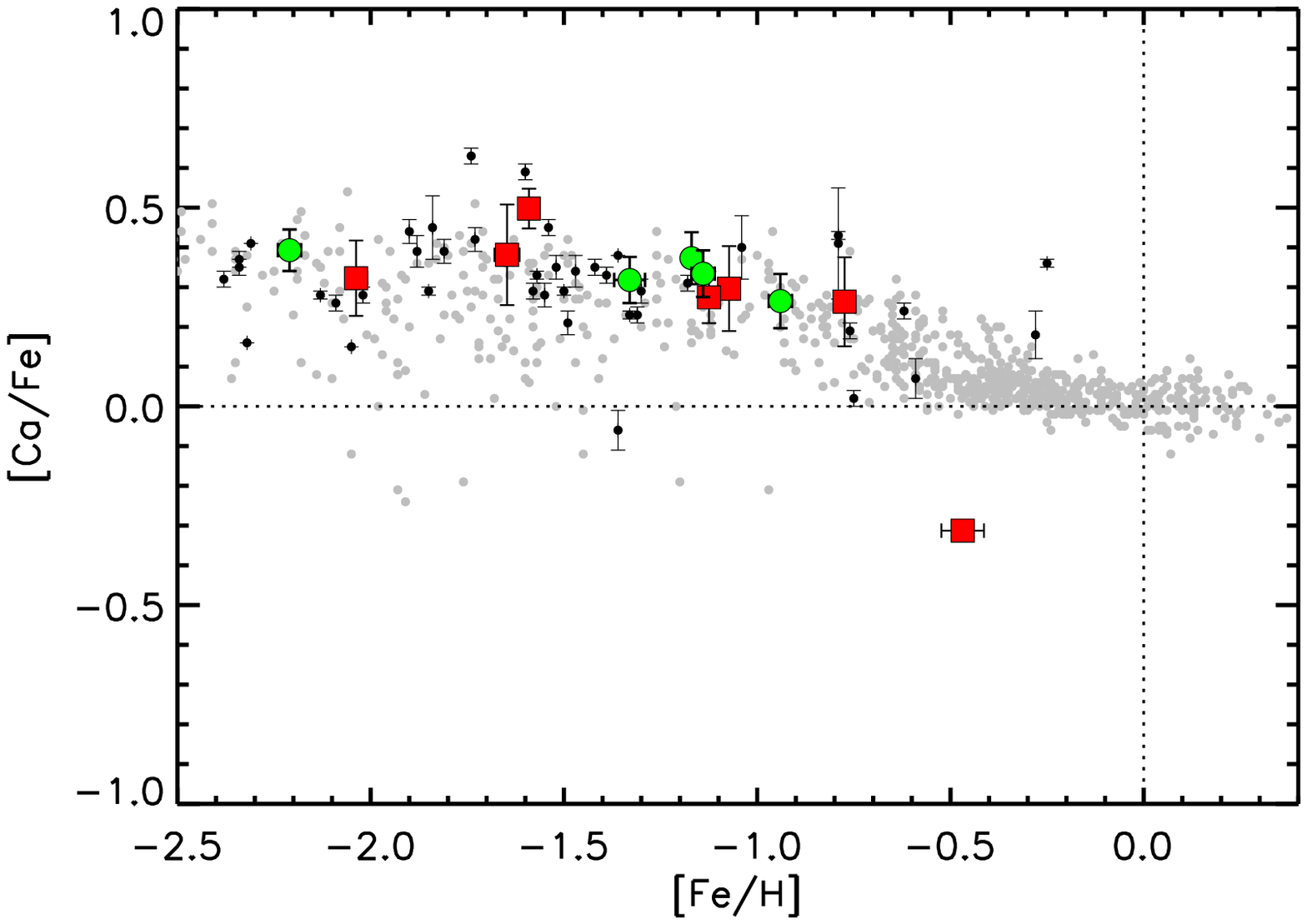}
\includegraphics[scale=0.45]{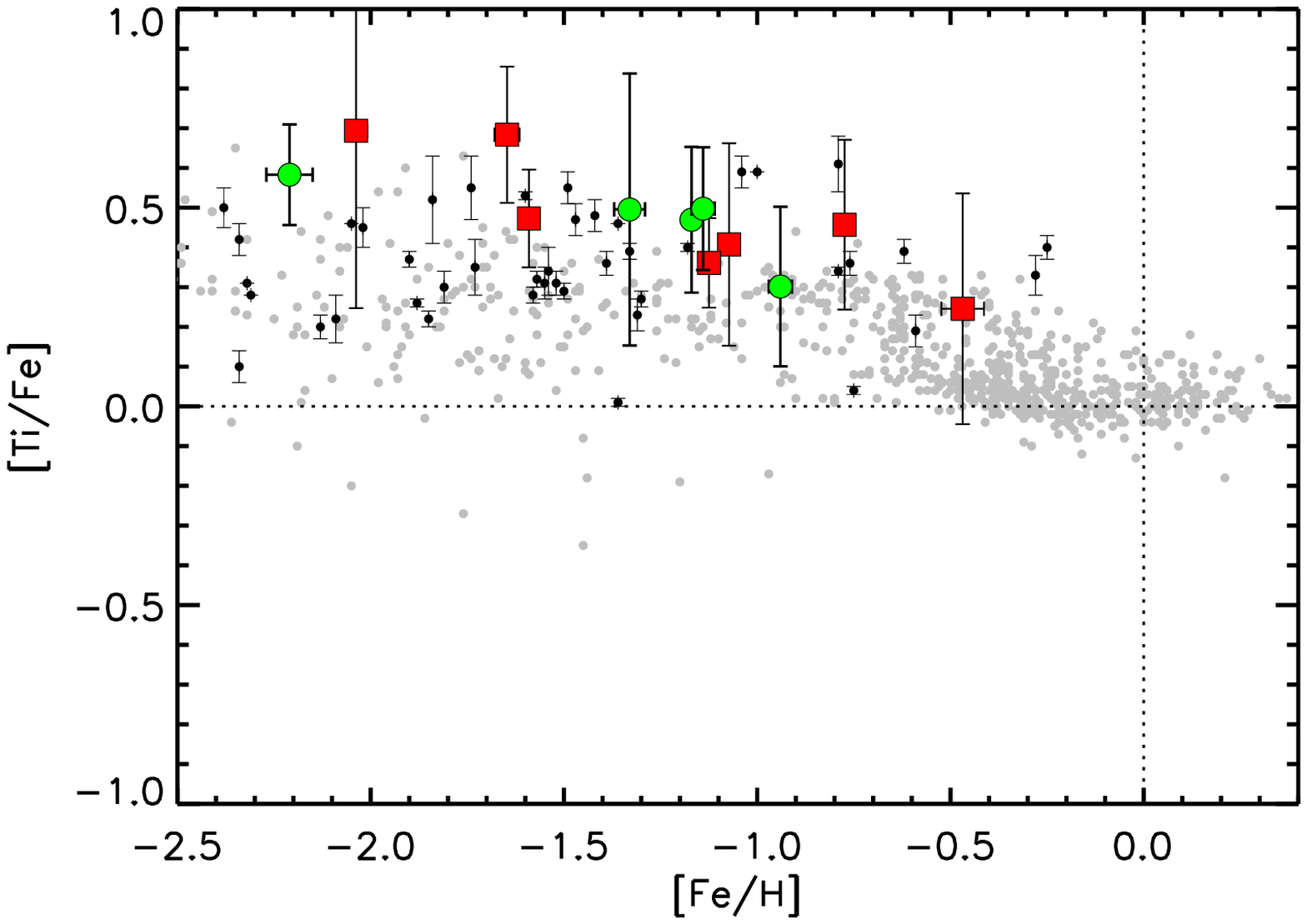}
\includegraphics[scale=0.45]{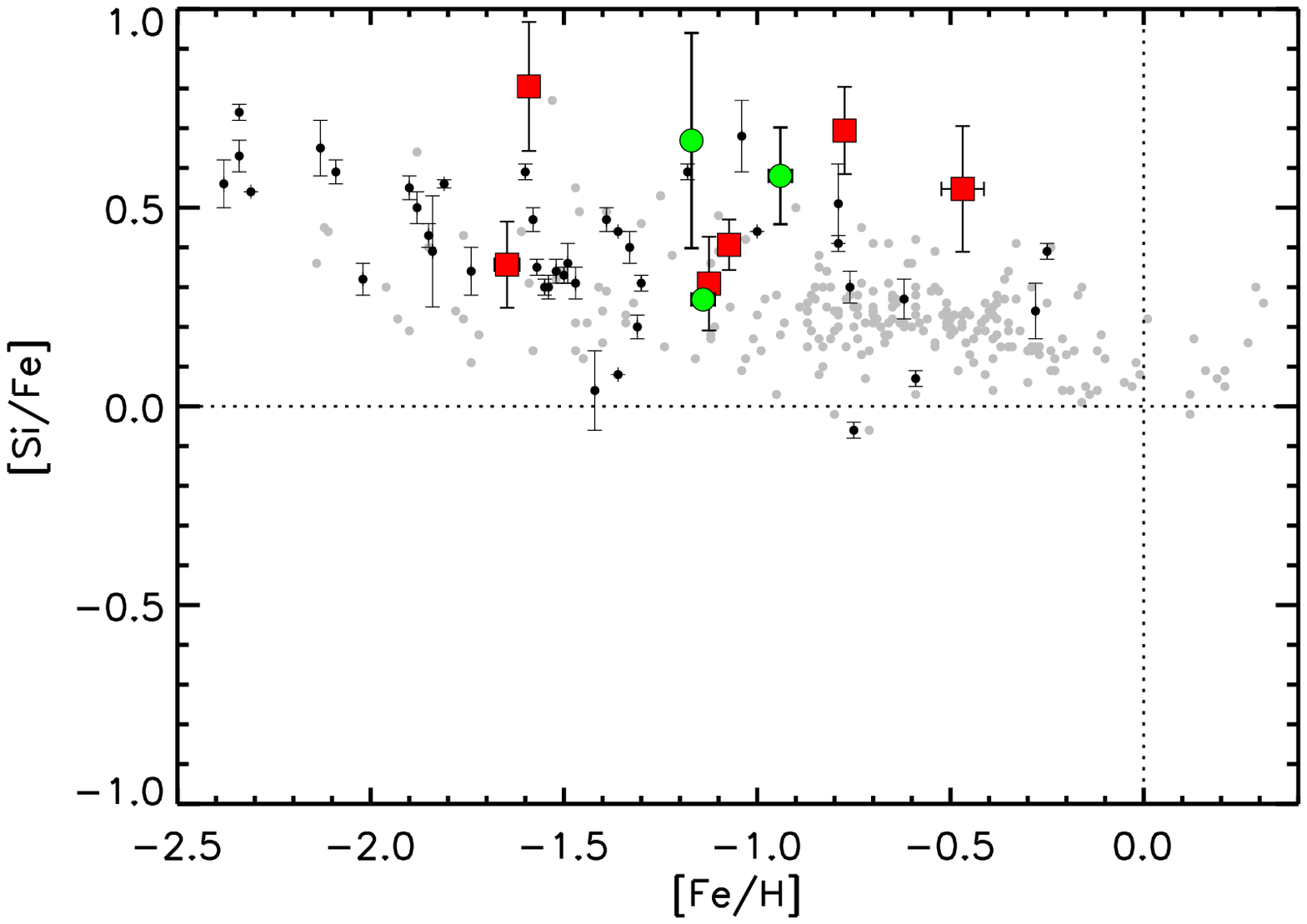}
\includegraphics[scale=0.45]{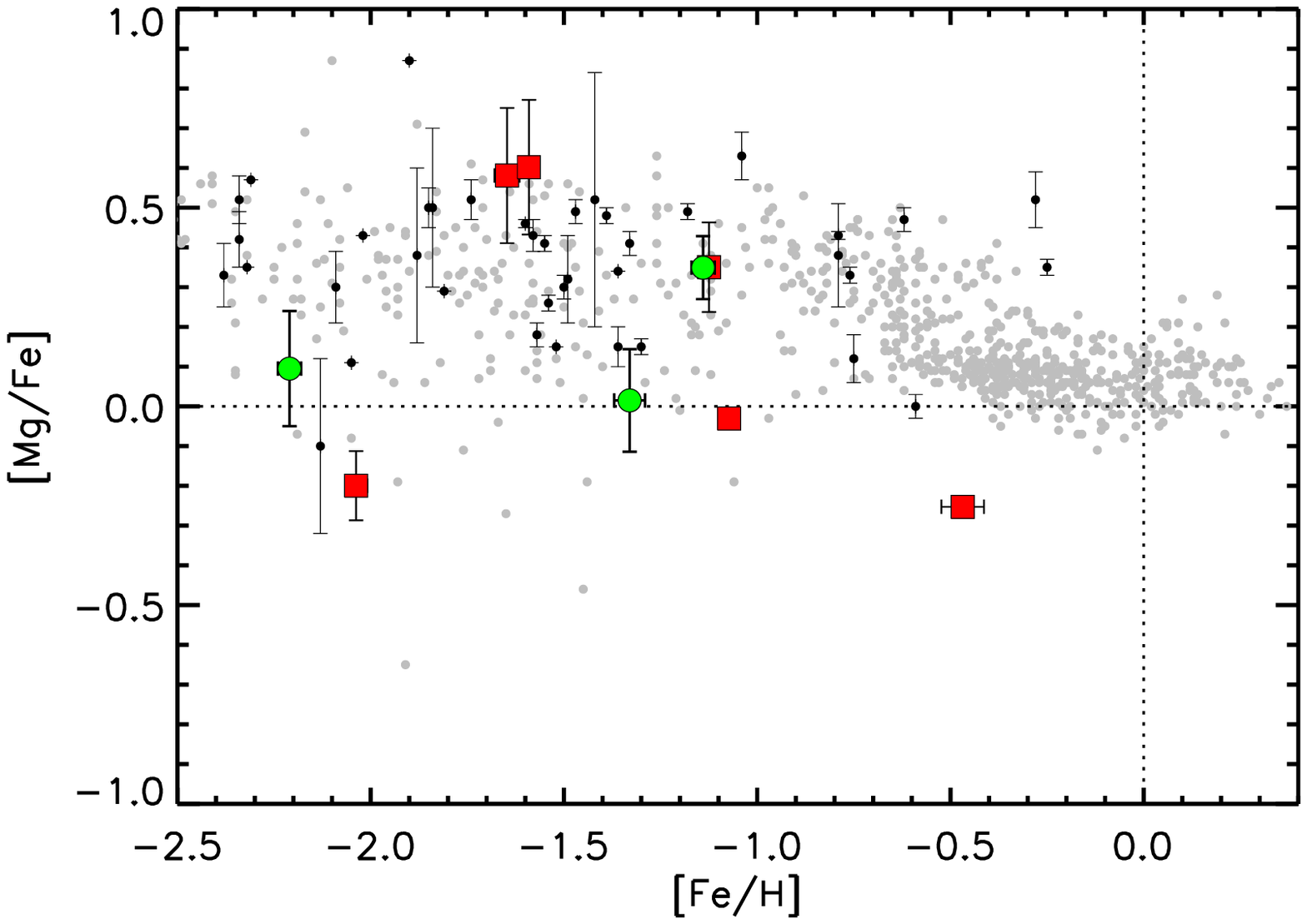}
\caption{$\alpha-$element ratios. Gray points show Milky Way Stars and black points show mean Milky Way GCs abundances from single stars. Data for single stars is from  \cite{2004AJ....128.1177V}, \cite{2005AJ....130.2140P} and references therein. When possible, these abundance ratios have been adjusted to be consistent with the solar abundance distribution of \cite{2005ASPC..336...25A} that was used in our analysis.  Red squares show the abundances from our  Milky Way training set ILS analysis and green circles show  M31 ILS abundances. The average of [Ti/Fe]$_{\rm{I}}$ and [Ti/Fe]$_{\rm{II}}$ are plotted vs. [Fe/H]$_{\rm{I}}$, where we have measured them. }
\label{fig:alphas} 
\end{figure*}

\begin{figure}
\centering
\epsscale{0.4}
\includegraphics[scale=0.46]{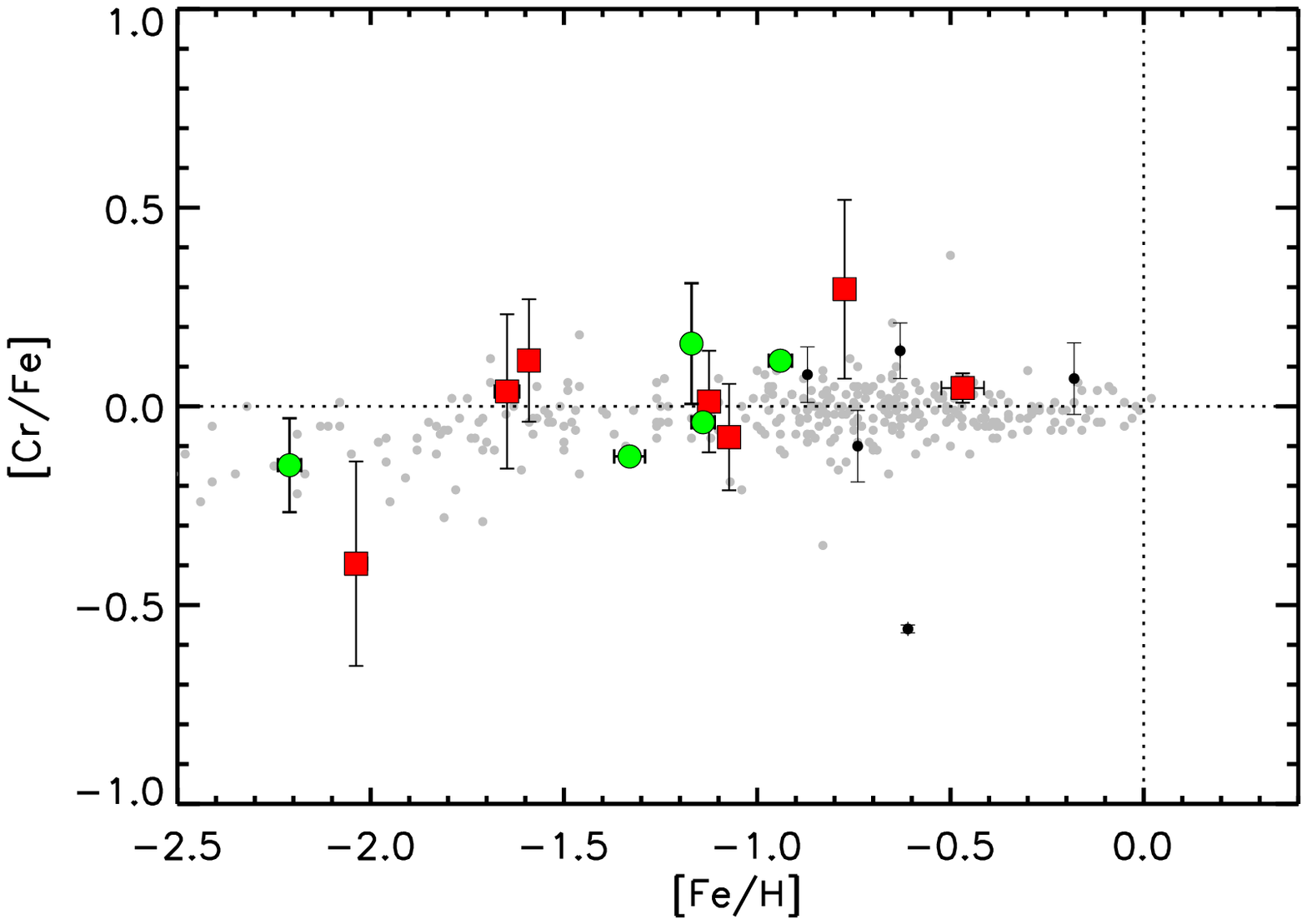}
\includegraphics[scale=0.46]{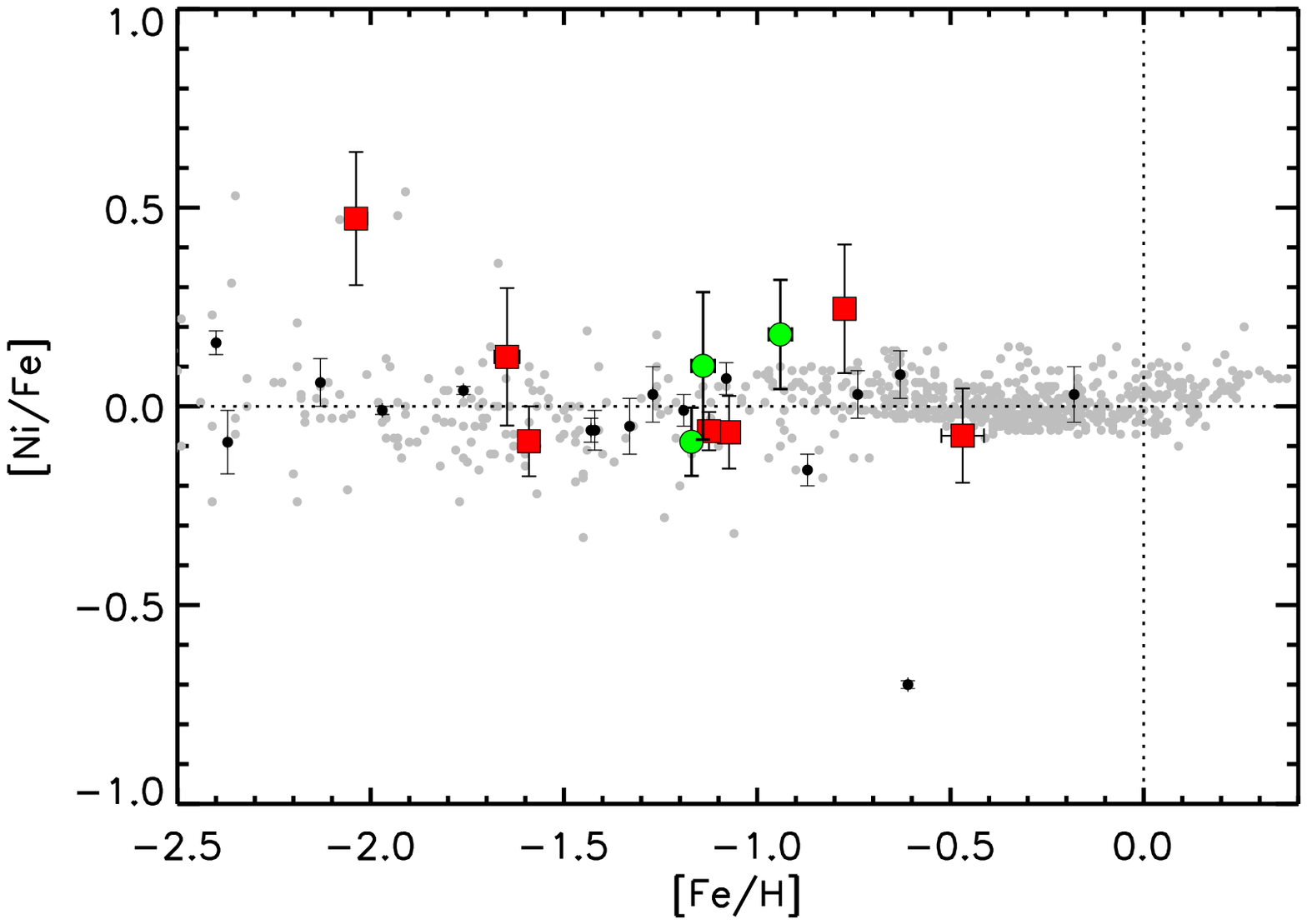}
\includegraphics[scale=0.46]{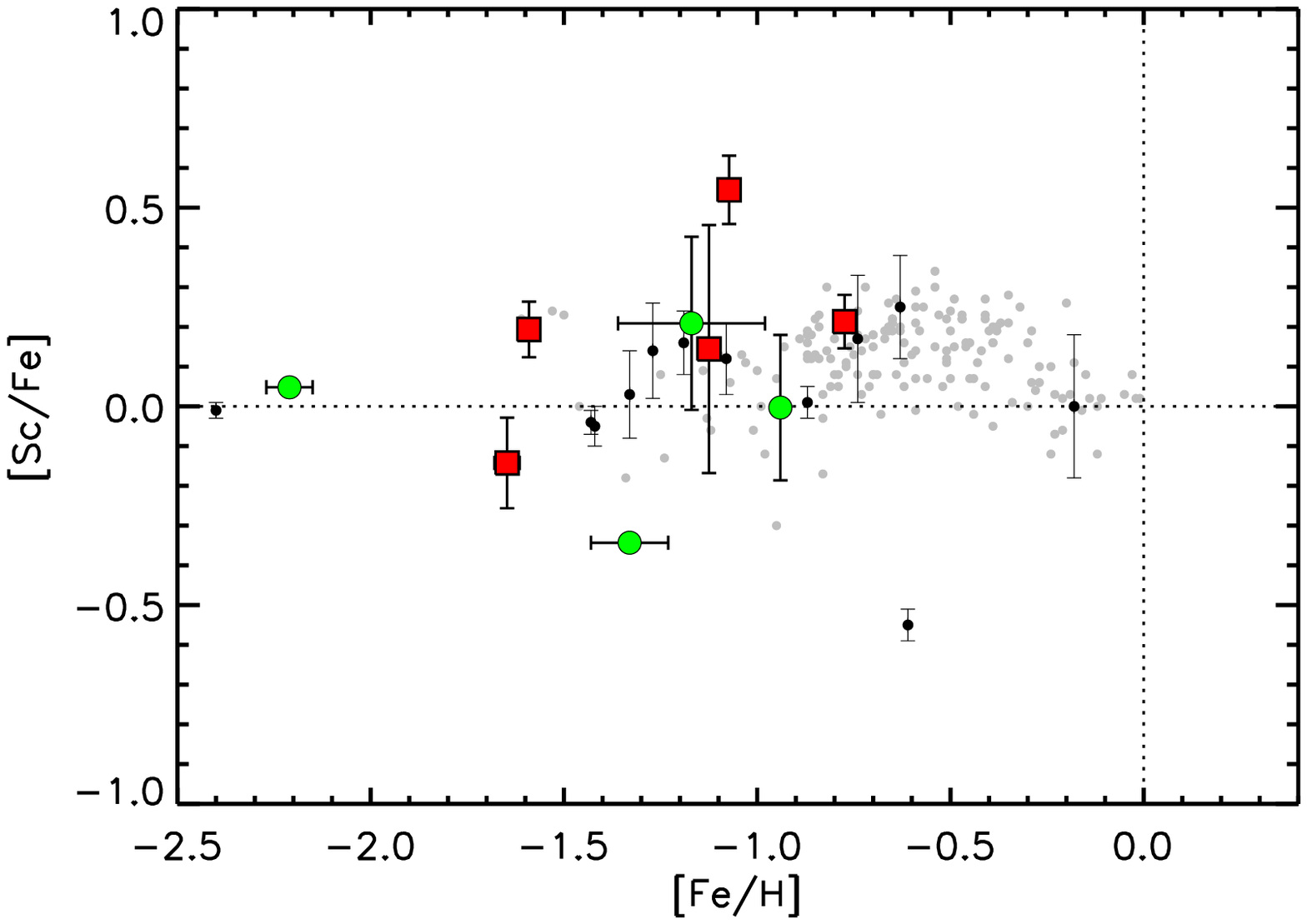}
\caption{Abundances for Fe peak elements Cr, Ni and Sc. Symbols are the same as Figure~\ref{fig:alphas}.  GC data from \cite{1995AJ....109.2586K},  \cite{1997AJ....114.1964S}, \cite{1999ApJ...523..739C}, \cite{1999AJ....118.1273I}, \cite{2000AJ....119..840S}, \cite{2001AJ....122.1438I}, \cite{2002AJ....123.3277R}, \cite{2002AJ....124.1511L}, \cite{2004AJ....127.2162S}, \cite{2004A&A...416..925C},  \cite{2004AJ....127.1545C}, \cite{2004AJ....127..373T}, and \cite{2005AJ....129..251L}}
\label{fig:Fe peak} 
\end{figure}

\begin{figure}
\centering
\includegraphics[scale=0.46]{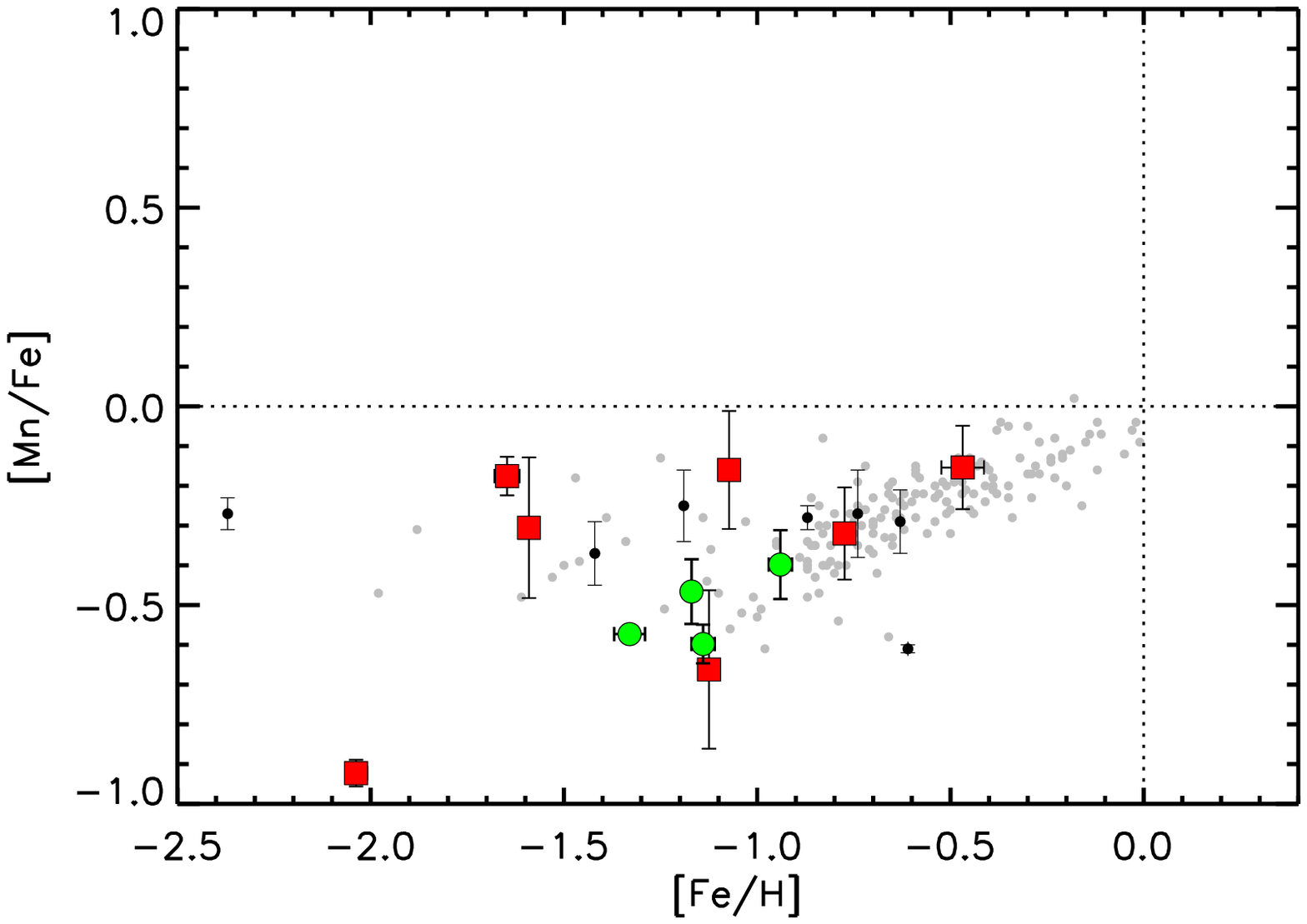}
\includegraphics[scale=0.46]{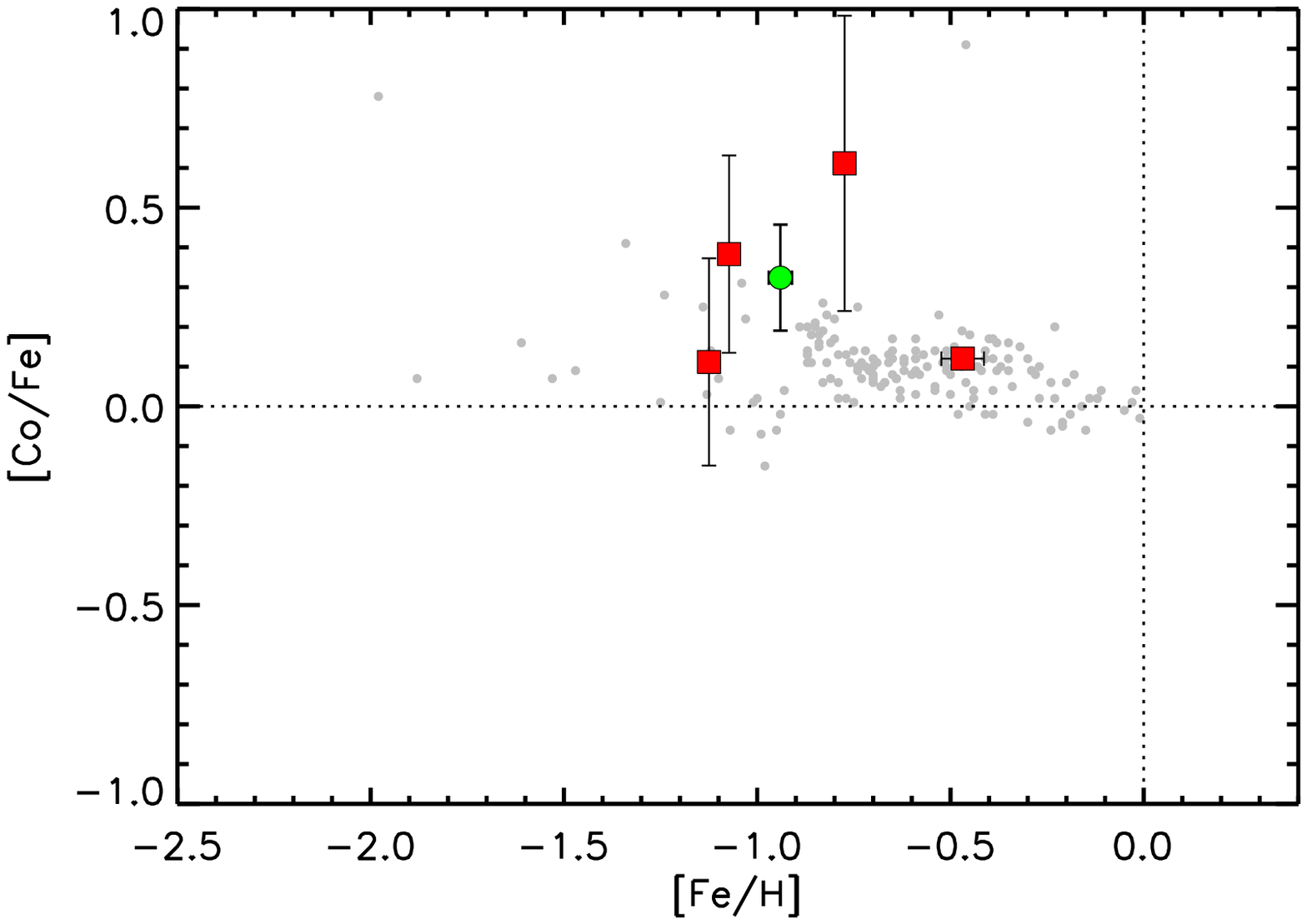}
\includegraphics[scale=0.46]{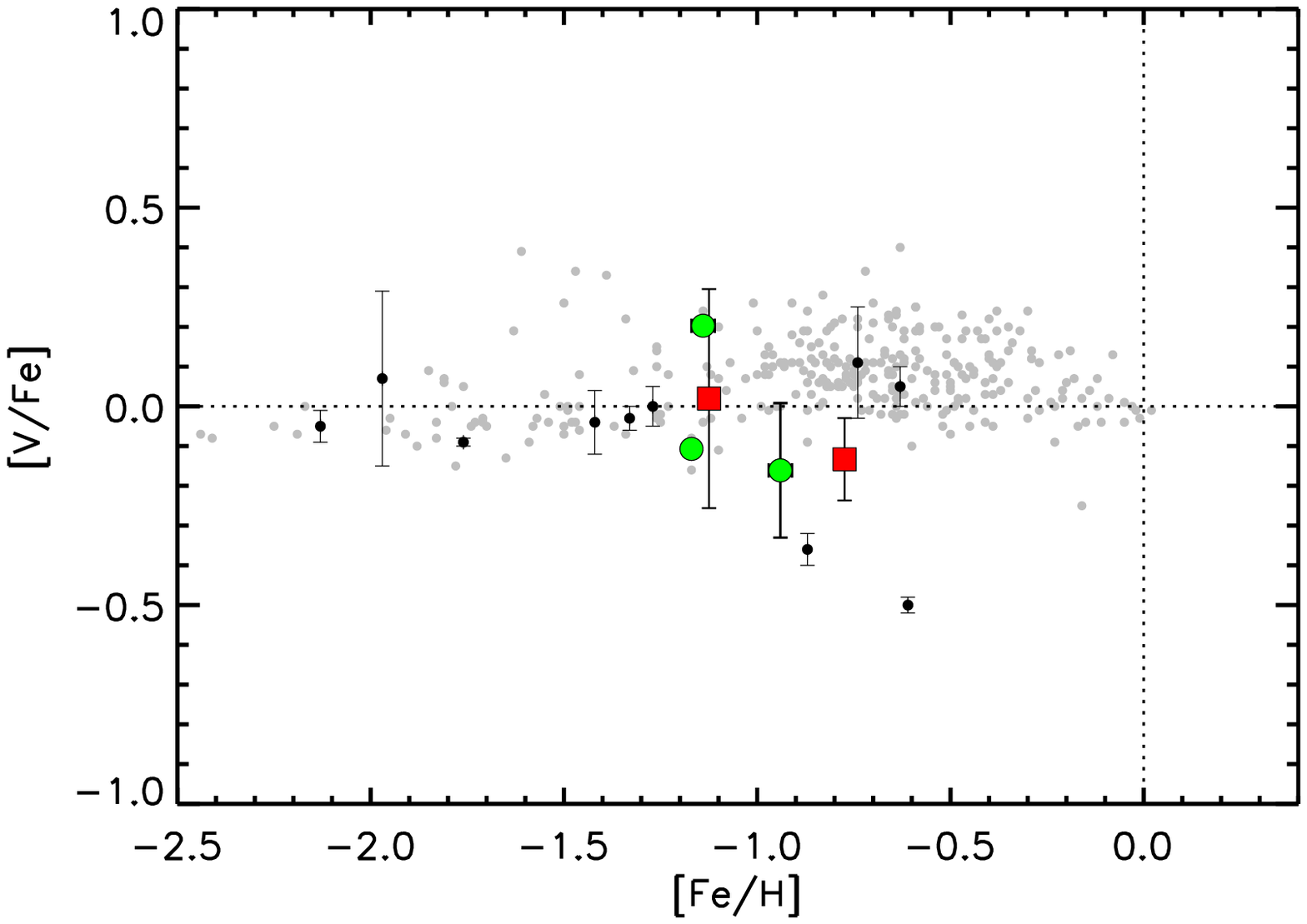}
\caption{Abundances of Mn, Co and V. Symbols and references are the same as Figure~\ref{fig:Fe peak}. }
\label{fig:Fe peak2} 
\end{figure}

\begin{figure}
\centering
\epsscale{0.4}
\includegraphics[scale=0.46]{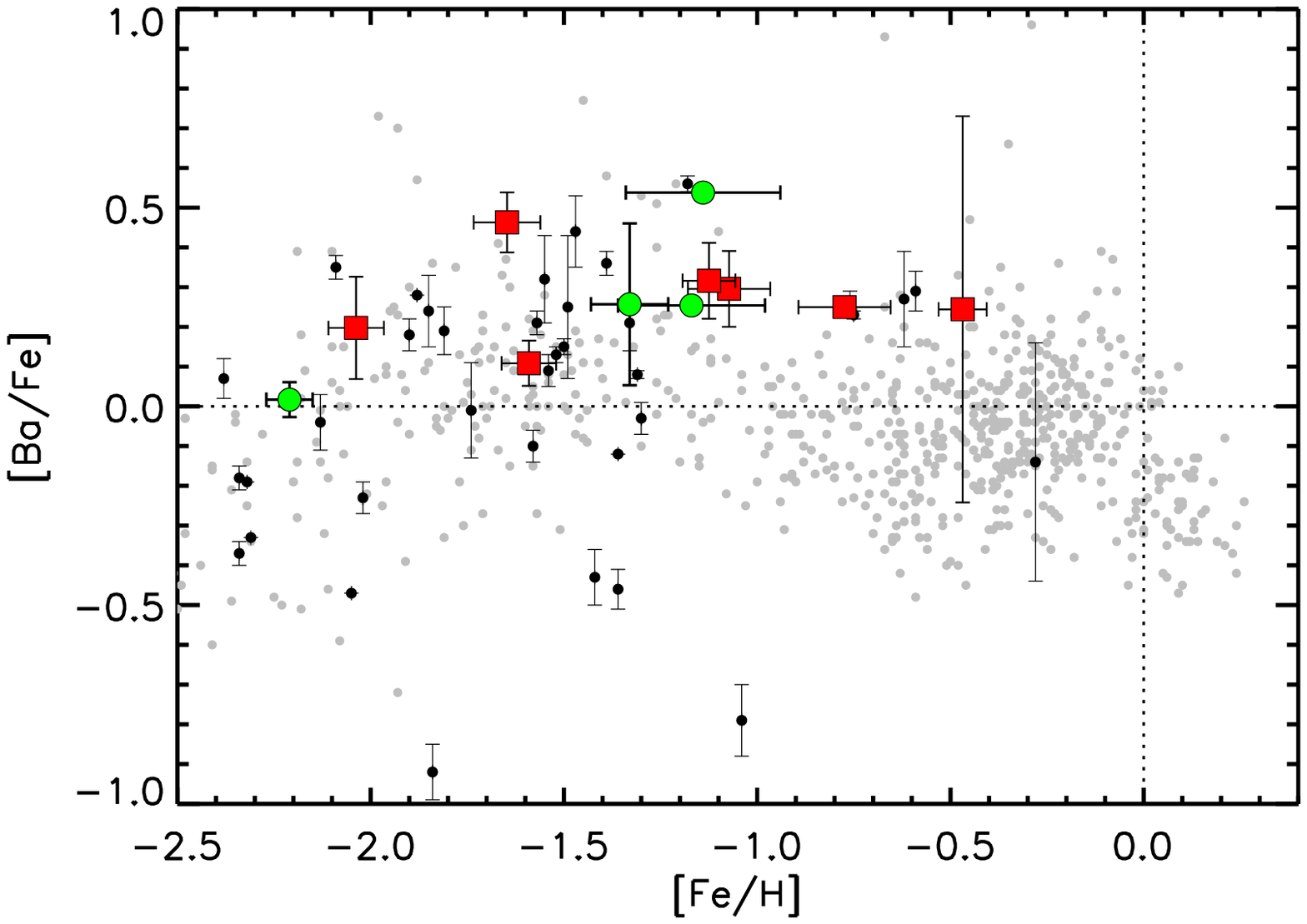}
\includegraphics[scale=0.46]{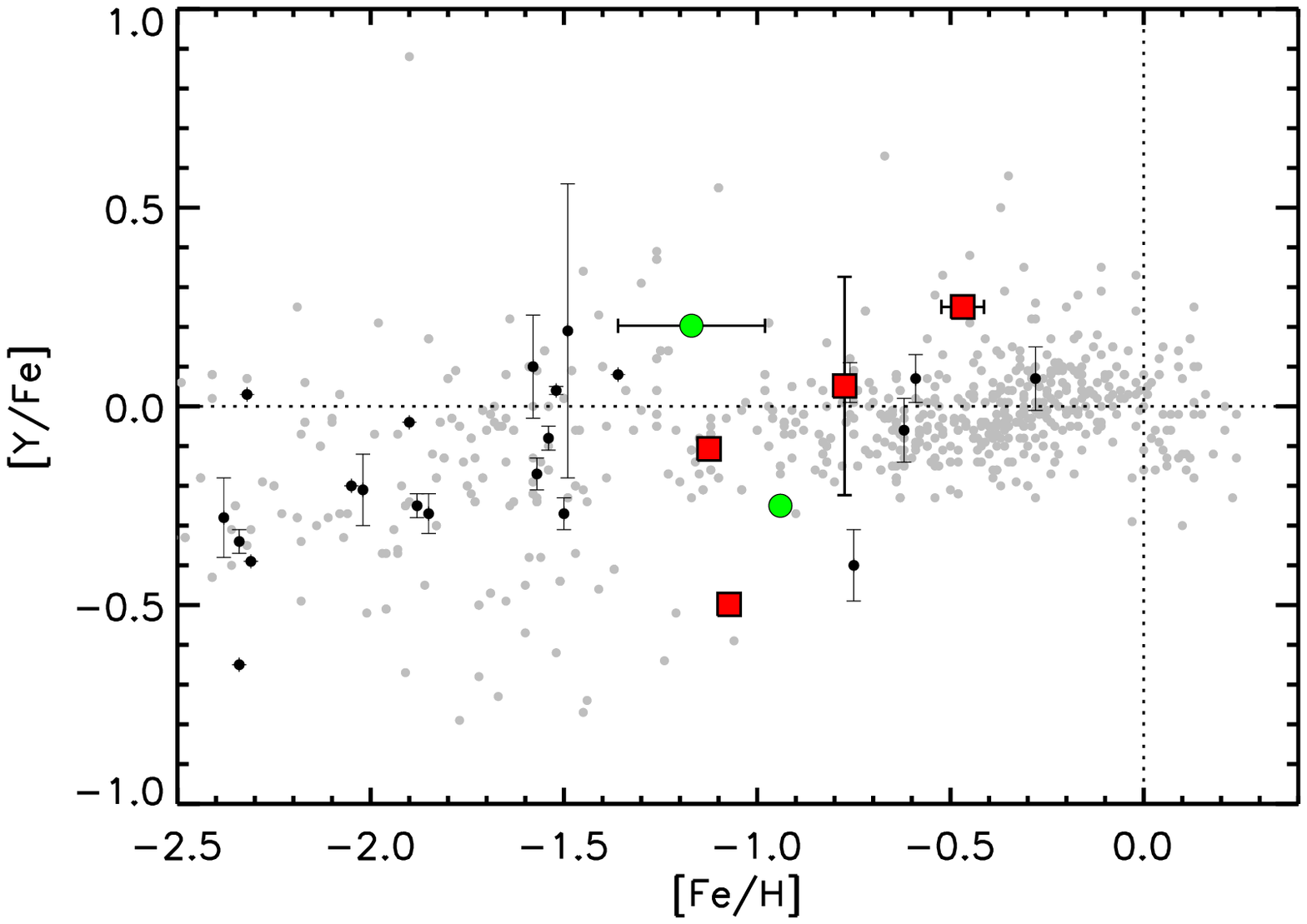}
\caption{ Abundances of neutron capture elements.  Symbols and references are the same as in Figure \ref{fig:alphas}. }
\label{fig:neutron} 
\end{figure}

We have measured abundances from the available clean lines of $\alpha-$elements, Fe and Fe$-$peak elements, and neutron capture elements for each GC. Final abundances, the number of analyzed spectral lines, line-to-line scatter, and the age of the best-fitting isochrone are reported in Tables~\ref{tab:108abund} through \ref{tab:351abund}. All abundance ratios relative to Fe use the solar abundance distribution of \cite{2005ASPC..336...25A}, with a solar log$\epsilon$(Fe)$=7.50$.  Abundance ratios of Sc II, Ti II, Y II, and Ba II are reported with respect to [Fe/H]$_{\rm{II}}$.   Figures~\ref{fig:alphas} through \ref{fig:neutron} show the M31 abundance ratios (green circles) compared to our Milky Way training set IL abundances (red squares).   Error bars for IL abundances in Figures~\ref{fig:alphas} through \ref{fig:neutron} correspond to the statistical error of the deviation in abundances from the N$_{lines}$ available for each species, as reported in  Tables~\ref{tab:108abund} through \ref{tab:351abund}. Note that these errors are often larger for the Milky Way training set abundances, which is a result of the smaller luminosity sampling of these GCs (5-30 \% of the total flux) and, in some cases, lower SNR spectra.

\subsection{Iron and Ages}
\label{sec:Fe}

As in abundance analyses of individual stars, the first element we
analyze is Fe because the large number of available transitions
provide a wide variety of very useful diagnostics and consistency
checks.  As outlined in \textsection~\ref{sec:best cmd}, we first use
the Fe lines to constrain our best CMDs.  We find this sample of M31
GCs is old, with preferred ages $>$7 Gyrs, which is consistent with
age estimates from previous photometric and low resolution
spectroscopic work \citep{2005AJ....129.2670R,1991ApJ...370..495H}.
We also find this sample of GCs spans a metallicity range of
[Fe/H]$\sim-0.9$ to $-2.2$, which is within the range of [Fe/H] of
Milky Way GCs system.  Below, we briefly discuss the best age and
[Fe/H] for the individual GCs in this sample.

We find the preferred range of CMD ages for G108 to be 10-15 Gyrs.  We
pick a best CMD age of 15 Gyrs for our final abundance determinations.
For the best solutions we find negligible trends of Fe I abundance
with EP and a slight correlation remaining with wavelength and reduced
EW, as shown in Figure~\ref{fig:g108 diagnostics}.  We note that the
remaining correlation between Fe abundance and reduced EW suggests
that the microturbulent velocity law we have applied is not perfect
for every star in the synthetic CMD. However, while the correlation
may still be present in the best solution, the overall trend in
correlations over all CMDs is still very clear, and it is easy to
select the most appropriate CMD.  It is important to note that the
scatter in microturbulence values around the relation we have adopted
is at least $\pm0.2$ \rkms for studies of both dwarf and RGB stars in
the Milky Way \citep[e.g.][]{2005A&A...433..185B,2006ApJ...636..821F}.
We have experimented with adjusting the microturbulence law within
this range, which could potentially reduce the Fe abundance standard
deviation. However, as long as the dependence of Fe abundance on
reduced EW is small, this will not significantly alter the mean Fe
abundance.  We have not been able to find a microturbulence law that
improves the overall abundance solution (i.e. improves all five
diagnostics discussed in~\textsection~\ref{sec:best cmd}). A more
detailed investigation of this issue is beyond the goals of this
study. To preserve the self-consistency of our solutions between GCs,
we do not alter the original microturbulent velocities of the
best-fitting CMD.  A detailed discussion of the small ($<$0.1 dex)
systematic error between the IL spectra abundance analysis and that
for individual stars is included in S. Cameron et al. (2009).
For the purposes of this work, we avoid systematic error issues by
concentrating on relative comparisons between IL abundances determined
in the same way for M31 and Milky Way GCs in our training set.  To
summarize our results, the final abundance solution for G108 is
[Fe/H]$=-0.94 \pm 0.03$, where the uncertainty is the standard error
in the abundances from all Fe lines ($\sigma/\sqrt{N_{lines}-1}$).

The preferred CMD age range for G315 is also 10-15 Gyrs.  Four of the
five diagnostics show best solutions at the oldest ages, therefore we
use the 15 Gyr solution as our best CMD.  The 15 Gyr solution for G315
has negligible trends of Fe I abundance with EP, wavelength, and
reduced EW, which can be seen in Figure~\ref{fig:g315 diagnostics}.
The final abundance for G315 is [Fe/H]$=-1.17 \pm 0.02$.

As can be seen in Figures~\ref{fig:g322 diagnostics}
and~\ref{fig:alldiag3}, for G322 we find the EP correlation to be
slightly stronger at the oldest ages, resulting in a slightly younger
preferred age range of 7-13 Gyrs.  We use a best CMD age of 13 Gyrs,
and find a negligible correlation with wavelength and reduced EW, but
a slight correlation between Fe abundance and EP for this solution.
The final abundance is [Fe/H]$=-1.14 \pm 0.03$.

G219 is the only GC in this sample which appears to be slightly
younger than the others.
Figure~\ref{fig:alldiag5} shows that four out of five Fe line
diagnostics are best for ages of 7-13 Gyrs.  We pick a best CMD age of
10 Gyrs, which is in the middle of this preferred age range.
Figure~\ref{fig:g219 diagnostics} shows that this best solution ---the
10 Gyr CMD--- still shows slight correlations of Fe abundance with EP
and reduced EW.  We find an [Fe/H]=$-2.21 \pm 0.03$ for G219, which
makes it one of the most metal-poor GCs in the Local Group confirmed
by high resolution spectra to date.

The preferred CMD age for G351 is 10-15 Gyrs. We note that G351 has a larger Fe I standard deviation than any other GC in the sample for any solution, which can be seen in Figure~\ref{fig:g351 diagnostics}.     We choose a best CMD age of 15 Gyrs, and note that this solution has small correlations of Fe abundance with EP and wavelength, as well as a fairly significant correlation with reduced EW.  We find [Fe/H]$=-1.33 \pm 0.04$.

\begin{deluxetable}{rrrrrr}
\tablecolumns{6}
\tablewidth{0pc}
\tablecaption{G108-B045 Abundances \label{tab:108abund}}
\tablehead{
\colhead{Species}  &\colhead{log$_{10}\epsilon($X$)$} & \colhead{$\sigma$}& \colhead{Error}  & \colhead{[X/Fe]\tablenotemark{1}}& \colhead{$N_{lines}$}\\ \colhead{} & \colhead{15 Gyrs} }
\startdata
 Si  I &   7.15&   0.17&   0.12&    $+$0.58&      3  \\
 Ca  I &   5.64&   0.17&   0.07&    $+$0.26&      7  \\
 Sc II &   2.54&   0.26&   0.18&   $-$0.00&      3  \\
 Ti  I &   4.10&   0.21&   0.10&    $+$0.14&      5  \\
 Ti II &   4.86&   0.34&   0.17&    $+$0.47&      5  \\
  V  I &   2.90&   0.24&   0.17&   $-$0.16&      3  \\
 Cr  I &   4.82&  \nodata&  \nodata&    $+$0.12&      1  \\
 Mn  I &   4.05&   0.15&   0.09&   $-$0.40&      4  \\
 Fe  I &   6.56&   0.22&   0.03&   $-$0.94&     49  \\
 Fe II &   6.99&   0.01&   0.01&   $-$0.51&      2  \\
 Co  I &   4.30&   0.19&   0.13&    $+$0.32&      3  \\
 Ni  I &   5.47&   0.27&   0.14&    $+$0.18&      5  \\
  Y II &   1.45&  \nodata&  \nodata&   $-$0.25&      1  \\
\enddata
\tablenotetext{1}{For Fe this quantity is [Fe/H].}

\end{deluxetable}

\begin{deluxetable}{rrrrrr}
\tablecolumns{6}
\tablewidth{0pc}
\tablecaption{G315-B381 Abundances \label{tab:315abund}}
\tablehead{
\colhead{Species}  &\colhead{log$_{10}\epsilon($X$)$} & \colhead{$\sigma$}& \colhead{Error}  & \colhead{[X/Fe]\tablenotemark{1}}& \colhead{$N_{lines}$}\\ \colhead{} & \colhead{15 Gyrs} }
\startdata
 Si  I &   7.01&   0.27&   0.27&    $+$0.67&      2  \\
 Ca  I &   5.51&   0.18&   0.07&    $+$0.37&      9  \\
 Sc II &   2.29&   0.38&   0.22&    $+$0.21&      4  \\
 Ti  I &   4.07&   0.15&   0.11&    $+$0.34&      3  \\
 Ti II &   4.53&   0.30&   0.15&    $+$0.60&      5  \\
  V  I &   2.72&  \nodata&  \nodata&   $-$0.11&      1  \\
 Cr  I &   4.63&   0.15&   0.15&    $+$0.16&      2  \\
 Mn  I &   3.75&   0.14&   0.08&   $-$0.47&      4  \\
 Fe  I &   6.33&   0.17&   0.02&   $-$1.17&     61  \\
 Fe II &   6.53&   0.50&   0.19&   $-$0.97&      6  \\
 Ni  I &   4.97&   0.19&   0.09&   $-$0.09&      6  \\
  Y II &   1.44&  \nodata&  \nodata&    $+$0.20&      1  \\
 Ba II &   1.45&  \nodata&  \nodata&    $+$0.25&      1  \\
\enddata
\tablenotetext{1}{For Fe this quantity is [Fe/H].}

\end{deluxetable}

\begin{deluxetable}{rrrrrr}
\tablecolumns{6}
\tablewidth{0pc}
\tablecaption{G322-B386 Abundances \label{tab:abundance_table_322}}
\tablehead{
\colhead{Species}  &\colhead{log$_{10}\epsilon($X$)$} & \colhead{$\sigma$}& \colhead{Error}  & \colhead{[X/Fe]\tablenotemark{1}}& \colhead{$N_{lines}$}\\ \colhead{} & \colhead{13 Gyrs} }
\startdata
 Mg  I &   6.70&   0.08&   0.08&    $+$0.35&      2  \\
 Si  I &   6.70&  \nodata&  \nodata&    $+$0.27\rlap{$^2$}&      1  \\
 Ca  I &   5.46&   0.16&   0.06&    $+$0.33&      8  \\
 Ti  I &   4.21&   0.14&   0.10&    $+$0.49&      3  \\
 Ti II &   4.33&   0.21&   0.12&    $+$0.51&      4  \\
  V  I &   3.02&  \nodata&  \nodata&    $+$0.20&      1  \\
 Cr  I &   4.68&  \nodata&  \nodata&   $-$0.04\rlap{$^3$}&      1  \\
 Mn  I &   3.61&   0.07&   0.05&   $-$0.60&      3  \\
 Fe  I &   6.36&   0.16&   0.03&   $-$1.14&     35  \\
 Fe II &   6.43&   0.41&   0.21&   $-$1.07&      3  \\
 Ni  I &   5.15&   0.26&   0.19&    $+$0.10&      3  \\
 Ba II &   1.63&  \nodata&  \nodata&    $+$0.54&      1  \\
\enddata
\tablenotetext{1}{For Fe this quantity is [Fe/H].}
\tablenotetext{2}{A correction of $-0.1$ dex has been applied (see \textsection ~\ref{sec:alphas})}
\tablenotetext{3}{A correction of $-0.25$ dex has been applied (see \textsection ~\ref{sec:fepeak})}

\end{deluxetable}

\begin{deluxetable}{rrrrrr}
\tablecolumns{6}
\tablewidth{0pc}
\tablecaption{G219-B358 Abundances \label{tab:358abund}}
\tablehead{
\colhead{Species}  &\colhead{log$_{10}\epsilon($X$)$} & \colhead{$\sigma$}& \colhead{Error}  & \colhead{[X/Fe]\tablenotemark{1}}& \colhead{$N_{lines}$}\\ \colhead{} & \colhead{10 Gyrs} }
\startdata
 Mg  I &   5.42&   0.14&   0.14&   $+$0.09&      2  \\
 Al  I &   4.24&   0.05&   0.05&    $+$0.69\rlap{$^2$}&      2  \\
 Ca  I &   4.49&   0.14&   0.05&    $+$0.39&      8  \\
 Sc II &   0.95&  \nodata&  \nodata&    $+$0.05&      1  \\
 Ti II &   3.33&   0.22&   0.13&    $+$0.58&      4  \\
 Cr  I &   3.28&   0.29&   0.12&   $-$0.15&      7  \\
 Fe  I &   5.29&   0.21&   0.03&   $-$2.21&     47  \\
 Fe II &   5.35&   0.12&   0.05&   $-$2.15&      4  \\
 Ba II &   0.04&   0.08&   0.04&    $+$0.02&      4  \\
\enddata
\tablenotetext{1}{For Fe this quantity is [Fe/H].}
\tablenotetext{2}{A non-LTE correction of $+0.6$ dex has been applied.}

\end{deluxetable}

\subsection{Alpha Elements}
\label{sec:alphas}

As described in \textsection~\ref{sec:introduction}, [$\alpha$/Fe] abundance ratios are a valuable tool for studying the star formation history of a galaxy.
$\alpha-$elements are produced primarily in SNII that occur on
timescales of 1-20 million years, which corresponds to the lifetimes
of massive stars.  While Fe-peak elements are produced in both type Ia (SNIa)
and SNII, the SNIa
contribution dominates on timescales of $\gtrsim$10$^{9}$ years
\citep[e.g.][]{1992AJ....103.1621S}.   Thus, at early times many
$\alpha-$elements are produced while total Fe abundances are low,
resulting in an enhanced [$\alpha$/Fe] abundance ratio at low [Fe/H].
After the onset of SNIa, the total Fe abundance increases at a faster
rate than that of $\alpha-$elements, decreasing the [$\alpha$/Fe]
ratio \citep{1979ApJ...229.1046T}. 
Most GCs in the Milky Way have [$\alpha$/Fe] ratios that are enhanced with respect to solar abundance ratios,  similar to Milky Way halo stars at comparable [Fe/H].   
This implies that Milky Way GCs formed when the ISM was dominated by enrichment by SNII.
Like Milky Way GCs, we find all the GCs in our M31 sample to be enhanced in Ca, Ti and Si.

Abundances for Ca I come from 7$-$9 clean lines per GC, with rms scatter about the mean of 0.1$-$0.2 dex, which is similar to the scatter in our Fe abundances.  We measure 4$-$10 Ti I and Ti II lines per GC, with slightly higher line-to-line scatter that may be due to weak blends.   We are able to confidently measure 1$-$3 Si I  lines in three of the GCs; lines in G219 and G351 were too noisy or badly blended to use.  
The one Si I line we measure in G322 is partially blended.  To estimate the effect of weak blends on the Si abundance from the EW, we have used the SYNTH routine from MOOG,  which we have modified to synthesize IL spectra.   We estimate that at most our EW abundance measurement is $\sim$0.1 dex too high after visual inspection of synthesized IL spectra of different abundances.

In Milky Way field stars, the  [Mg/Fe] ratios behave similarly to Ca,
Ti and Si.  However, in Milky Way GCs,  [Mg/Fe]  shows
inter- and intra-cluster abundance variations
\citep[see][]{2004ARA&A..42..385G} with respect to other
$\alpha-$elements and  star-to-star differences within individual
GCs.  Analysis of our full training set of Milky Way GCs has
revealed lower [Mg/Fe] than [Ca/Fe], [Ti/Fe], or [Si/Fe] in three out
of the six GCs where it was measured (see
\citetalias{2008ApJ...684..326M} and Cameron et al. 2009).

We have been able to measure [Mg/Fe] in three of the GCs in M31.
Similar to the Milky Way training set GCs, in M31 we measure
[Mg/Fe] to be lower ( [Mg/Fe]$< +$0.1 dex) than other
$\alpha-$elements within individual GCs in two out of the three GCs
where it was measured.  In the other two clusters the Mg I lines had
strengths $>$ 150 m\AA, and were therefore not analyzed in this work.

\begin{figure}
\centering
\includegraphics[angle=90,scale=0.35]{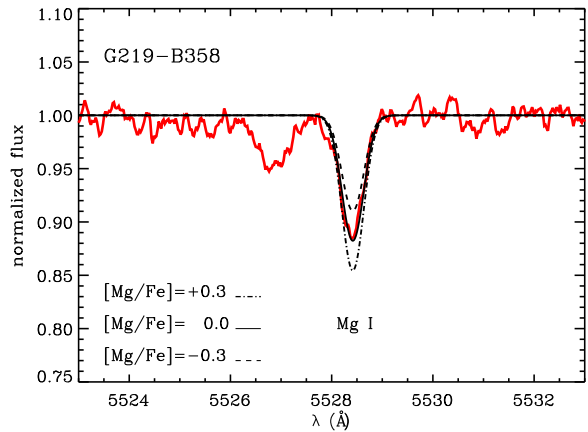}
\caption{Spectrum synthesis test for Mg I 5528 \rAA line in G219.  Smoothed data is shown in red, and overplotted synthesized spectra from top to bottom correspond to [Mg/Fe]$=-0.3,0.0,+0.3$. The closest matching profile is for [Mg/Fe]$=0.0$.}
\label{fig:219mg} 
\end{figure}

We have performed spectral synthesis tests of the Mg I lines to see
if the [Mg/Fe] depletion can be explained by line-to-line measurement
error.  An example of this test is shown in Figure~\ref{fig:219mg},
for the unblended 5528 \rAA Mg I line in the most metal-poor GC G219.
From the EW of this line, we measure an abundance of [Mg/Fe]$=+0.01$.
Overplotted are synthesized spectra with abundances of [Mg/Fe]$=+0.3,
0.0, -0.3$. G219 is extremely metal poor, and no other elements
contribute to the line EW. It is clear from Figure~\ref{fig:219mg}
that the closest matching abundance is [Mg/Fe]$=0.0$, and that the
line is inconsistent with the [$\alpha$/Fe]$=+0.3$ measured for Ca and
Ti, clearly showing that measurement error cannot explain the Mg
deviation from the other $\alpha$-element abundances in this cluster.
In \textsection~\ref{sec:correlations}, we further discuss light
element variations in the IL of GCs, and in \textsection~\ref{sec:lick
  alpha} we discuss further implications for low resolution IL
abundances.

\begin{deluxetable}{rrrrrr}
\tablecolumns{6}
\tablewidth{0pc}
\tablecaption{G351-B405 Abundances \label{tab:351abund}}
\tablehead{
\colhead{Species}  &\colhead{log$_{10}\epsilon($X$)$} & \colhead{$\sigma$}& \colhead{Error}  & \colhead{[X/Fe]\tablenotemark{1}}& \colhead{$N_{lines}$}\\ \colhead{} & \colhead{15 Gyrs} }
\startdata
 Mg  I &   6.22&   0.22&   0.13&    $+$0.01&      4  \\
 Ca  I &   5.30&   0.16&   0.06&    $+$0.32&      9  \\
 Sc II &   1.67&  \nodata&  \nodata&   $-$0.34&      1  \\
 Ti  I &   3.94&   0.40&   0.23&    $+$0.37&      4  \\
 Ti II &   4.49&   0.25&   0.25&    $+$0.62&      2  \\
 Cr  I &   4.44&  \nodata&  \nodata&   $-$0.13\rlap{$^2$}&      1  \\
 Mn  I &   3.49&  \nodata&  \nodata&   $-$0.57&      1  \\
 Fe  I &   6.17&   0.26&   0.04&   $-$1.33&     42  \\
 Fe II &   6.46&   0.17&   0.10&   $-$1.04&      2  \\
 Ni I &  \nodata&   \nodata&   \nodata&   $-$0.2\rlap{$^3$}&      1  \\

 Ba II &   1.39&   0.29&   0.20&   $+$0.26&      3  \\
\enddata
\tablenotetext{1}{For Fe this quantity is [Fe/H].}
\tablenotetext{2}{A correction of $-0.25$ dex has been applied (see \textsection ~\ref{sec:fepeak})}
\tablenotetext{3}{Estimate from spectral synthesis of 6767 \rAA  (see Figure ~\ref{fig:351ni})}

\end{deluxetable}

\begin{deluxetable}{r|r|r}
\tablecolumns{5}
\tablewidth{0pc}
\tablecaption{Mean IL $\alpha$ Abundances for Milky Way Training Set and Current Sample of M31 GCs  \label{tab:alpha_table}}
\tablehead{
\colhead{}&\colhead{Milky Way}&\colhead{M31}}
\startdata

$[$Ca/Fe$]$	 & $+$0.35 $\pm$	0.08 	&$+$0.34 $\pm$ 0.05\\
$[$Ti/Fe$]$	& $+$0.46 $\pm$	0.15 	&$+$0.47 $\pm$ 0.10\\
$[$Si/Fe$]$	& $+$0.52 $\pm$	0.20	&$+$0.51 $\pm$ 0.21 \\
$[$Mg/Fe$]$ 	& $+$0.18 $\pm$	0.39	&$+$0.15 $\pm$ 0.17 \\

\hline
\\
$[$$\alpha$$_{\rm{CaTiSiMg}}$/Fe$]$ & $+$0.38 $\pm$	0.15	&$+$0.37 $\pm$ 0.16\\
$[$$\alpha$$_{\rm{CaTiSi}}$/Fe$]$	& $+$0.44 $\pm$	0.09	&$+$0.44 $\pm$ 0.09\\

\enddata

\end{deluxetable}

We can address the star formation history of M31 by calculating an
average [$\alpha$/Fe] ratio for each GC similar to that in
\cite{2005AJ....130.2140P}.  While these authors use Ca, Ti, Si, and
Mg in their average [$\alpha$/Fe], we note that [Mg/Fe] is probably
not a good [$\alpha$/Fe] indicator in the IL of GCs, for the reasons
discussed above. Therefore, for the mean [$\alpha$/Fe] for each GC
discussed here, we include only Ca I, Ti I, Ti II, and Si I. The
mean [$\alpha$/Fe]$=+0.36 \pm 0.20$, $+0.50 \pm 0.16$, $+0.40 \pm
0.12$, $+0.44 \pm 0.16$, and $+0.49 \pm 0.13$ for G108, G315, G322,
G351 and G219, respectively, which is significantly and consistently
enhanced relative to solar in all five M31 GCs.  We can also calculate
the mean ratio for the individual $\alpha-$elements across the sample
of GCs, and compare these values to similar means in our sample of
Milky Way training set GCs.  This comparison of abundances derived
only with our IL spectra method avoids any potential sources of
systematic error.  The mean values for Ca, Ti, Si, and Mg are
presented in Table~\ref{tab:alpha_table}.  In addition, we present the
mean and deviation of all the [$\alpha$/Fe] ratios including and
excluding [Mg/Fe]. Table~\ref{tab:alpha_table} shows that GCs in both
the Milky Way and M31 have extremely consistent [$\alpha$/Fe] ratios.
GCs in both galaxies are also consistent with the Milky Way halo
average $\alpha$-enhancement of $+0.35$
\citep[e.g.][]{1997ARA&A..35..503M}.  The obvious implication of the
[$\alpha$/Fe] abundances in this small sample of M31 GCs is that M31
was (or its now-merged components were) dominated by enrichment by
SNII when these GCs formed.

\subsection{Aluminum}
\label{sec:al}

Al abundances have been used to put constraints on chemical evolution
models of the Milky Way because Al enrichment is particularly
sensitive to the details of SNII explosions
\citep[see][]{2006A&A...451.1065G}.  Al abundances for Milky Way disk
stars with [Fe/H]$<-2$ have some contribution from explosive burning
in SNII \citep{2008A&A...481..481A}, so that like [$\alpha$/Fe], Al
abundances reach a plateau value of [Al/Fe]~$\sim+0.3$
\citep[e.g.][]{2005A&A...433..185B}.  However, some stars in Milky Way
GCs are found to be even more enriched, reaching levels as high as
[Al/Fe]$\sim+1$ \citep{2001A&A...369...87G}.  Also, like Mg, Al
exhibits inter- and intra-cluster variations, which we discuss further
in \textsection~\ref{sec:correlations}, suggesting that the
influences on Al abundance are more complicated in GCs than in the field.

We have only been able to make one measurement of Al I in this first
sample of M31 GCs.  This measurement was made from the 3944 \rAA Al I
line in the metal-poor GC G219. We were able to make two independent
measurements of this 3944 \rAA line because it was present at the ends
of two adjacent orders.  Both measurements give a consistent result of
[Al/Fe]$=+0.09 \pm 0.05$.  However, we note that Al I abundances
derived from the 3944 \rAA resonance line can be problematic;
\cite{1995AJ....109.2757M} find that the line is significantly blended
with CH lines in some giant stars, which causes derived abundances to
be too high, and according to \cite{1997A&A...325.1088B} Al I
abundances from this line will be underestimated by $\sim$0.6 dex due
to non-LTE effects.  \cite{2008A&A...481..481A} find that this
correction may be even larger in metal-poor hot stars.  We do not see
evidence in the IL spectrum for contamination of the 3944 \rAA line by
significant CH blends.  An appropriate non-LTE correction of $+0.6$
dex raises the abundance to [Al/Fe]$=+0.69$.  This is higher than the
[Al/Fe]$\sim$0 halo average at an [Fe/H]=$-2.2$, but consistent with
the significantly enhanced Al I found in some individual GC stars.
This enhanced [Al/Fe] is also consistent with Al abundances we derive
for three Milky Way GCs in our training set.

\begin{figure}
\centering
\includegraphics[scale=0.46]{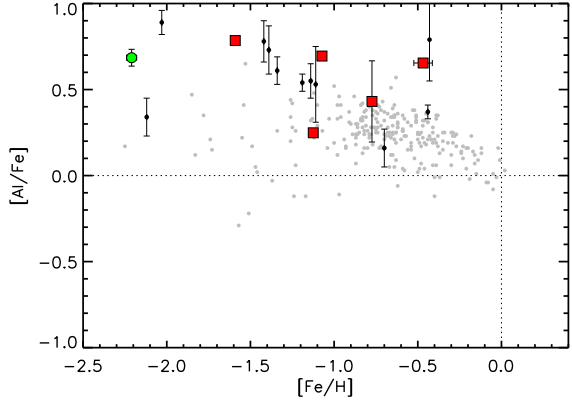}
\caption{ Aluminum abundances.  Symbols are the same as in Figure~\ref{fig:alphas}.  GC data is from references compiled in \cite{2006AJ....131.1766C}. A non-LTE correction of 0.6 dex has been added to G219 abundance.}
\label{fig:AL} 
\end{figure}

The training set [Al/Fe] measurements provide much stronger evidence that the enhanced [Al/Fe] we measure in GC IL spectra is real because they are not measured from the potentially problematic 3944 \rAA Al I feature. 
The training set  [Al/Fe] (see Figure~\ref{fig:AL}) were measured from the 6696/6698 \rAA Al I doublet, which should not be contaminated by blends. In addition,  \cite{1997A&A...325.1088B} find that non-LTE effects are much smaller for these lines.  The 6696 \rAA feature is not detectable in G219, even though we measure an abundance as high as [Al/Fe]$=+0.7$ with the non-LTE correction.  We performed spectral synthesis tests to check that the lower limit of [Al/Fe]$=+0.7$ is consistent with a 6696 \rAA line that would be too weak to detect in the IL spectra.  We indeed find that an [Al/Fe]$=+0.7$ is not high enough for the feature to be seen after convolution with the velocity broadening of G219 because it results in a line depth of $<0.01\%$ of the continuum level.

The 6696/6698 \rAA features were also too weak in the other M31 GC spectra to get reliable abundance measurements from EWs, with the exception of G108, in which it unfortunately falls too near the end of an order for a good measurement to be made. The 3944 \rAA line was saturated in all of the more metal-rich GCs.

\subsection{Fe-peak Elements}
\label{sec:fepeak}

Ni, Cr, Mn Co, Sc, and V abundances are interesting because their production generally tracks that of Fe,  \citep[e.g.][]{1999ApJS..125..439I}, resulting in [X/Fe]$\sim0$.  Therefore, deviations from [X/Fe]$\sim0$ are particularly interesting because they imply special conditions may have been present, such as variations in the mass function of supernovae, variations of mixing of supernovae ejecta into the local ISM, metallicity dependent or explosion dependent supernova yields, or contributions from special types of supernovae \citep[e.g.][]{1997ARA&A..35..503M}.

Ni abundances in Milky Way field and GC stars generally follow the expected Fe-peak element trend of [Ni/Fe]$=0$ at all metallicities. We are able to measure [Ni/Fe] in three M31 GCs, and find it to be consistent with the Milky Way abundance trend.
 Ni was measured from 3-6 Ni I lines in G108, G315, and G322.  The three GCs have a mean $<$[Ni/Fe]$>=+0.06 \pm 0.14$, which is essentially identical to the mean of the Milky Way training set GCs, which have $<$[Ni/Fe]$>=+0.05 \pm 0.21$.

\begin{figure}
\centering
\includegraphics[angle=90,scale=0.35]{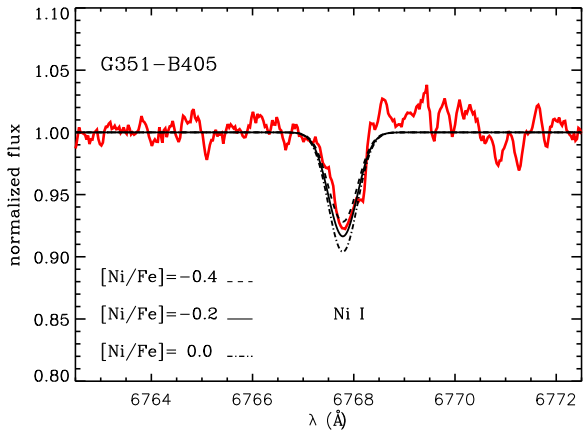}
\caption{Spectrum synthesis test for Ni I 6767 \rAA line in G351.  Smoothed data is shown in red, and overplotted synthesized spectra from top to bottom correspond to [Ni/Fe]$=-0.4,-0.2,0.0$.  The spectrum is noisy and continuum placement uncertain, but [Ni/Fe] is approximately $\sim-0.2$ with an uncertainty of $\sim$0.2 dex.}
\label{fig:351ni} 
\end{figure}

In the most metal-poor cluster G219, all Ni I lines are too weak for EWs to be measured reliably. In G315 most Ni I lines are too weak or noisy for EW measurements, and the 7393 \rAA line is blended with telluric absorption lines. We estimate from spectral synthesis tests of the noisy 6767 \rAA line in G351, that [Ni/Fe]$\sim-0.2$.  The spectral synthesis and spectrum are shown in Figure~\ref{fig:351ni}, where the uncertainty in abundance due to the noisy continuum can be fully appreciated.

[Cr/Fe]$=0$ in Milky Way stars for [Fe/H]$>-2$, but [Cr/Fe]$<0$ for
[Fe/H]$<-2$.  The deviation from [Cr/Fe]$=0$ at low metallicity may be
due to different chemical enrichment for the lowest metallicity stars
in the Milky Way halo \citep{1997ARA&A..35..503M}.  We measure Cr I
abundances for the four most metal-rich M31 GCs that are consistent
with the solar [Cr/Fe] average in both our Milky Way training set
GCs and in Milky Way stars at these metallicities.  Few [Cr/Fe]
measurements exist for GCs with [Fe/H]$<-2$.  When observed, [Cr/Fe]
in GCs also follows the decreasing [Cr/Fe] abundance trend observed in
Milky Way halo stars (see \cite{2001ApJ...548..592S} for M92 and NGC
2419, and \cite{2006A&A...453..547L} for M15 and Fornax GCs).  We are
able to measure Cr I in the low-metallicity M31 cluster G219, and
find [Cr/Fe]$=-0.15$, which is also consistent with the decreasing
halo abundance trend below [Fe/H]$\sim-2$.  Cr abundances were
calculated from one Cr I line in G108, G322 and G351, two Cr I lines
in G315, and seven Cr I lines in G219.  Most other Cr I lines in the
more metal-rich GCs have EWs over 150 m\rAA and were not analyzed.
The Cr feature at 5409 \AA, which is the only Cr line we measure in
G322 and G351, is partially blended with weak Ti I and Fe I lines,
so that the EW abundance may be slightly high.  We used spectral
synthesis tests of the Ti I and Fe I blends around the 5409 \rAA Cr
I feature to estimate the effect of the blends on the derived Cr
abundance.  We find that the Cr I abundance derived with the original
EW measurement may be approximately $\sim$0.25 dex too high.  A
correction of $-0.25$ dex to our EW abundances of [Cr/Fe]$=+0.21$ and
$+0.12$, would result in [Cr/Fe]$=-0.04$, and $-0.13$ for G322 and
G351, respectively.  These [Cr/Fe] are consistent with the
[Cr/Fe]$\sim0$ of Milky Way GCs.

Mn is a particularly interesting Fe$-$peak element because unlike most
Fe$-$peak elements the [Mn/Fe] ratio is not solar at most
metallicities.  From [Fe/H]$=-1$ to $-2.5$ dex, [Mn/Fe]$\sim-0.4$, and
increases to [Mn/Fe]$=0$ at [Fe/H]$=0$.  This trend is similar but
opposite to that of $\alpha-$elements
\citep[e.g.][]{1997ARA&A..35..503M}.  Possible explanations for this
trend are: Mn is underproduced in SNII and overproduced in SNIa, or
that SNIa Mn production is dependent on the metallicity of the
progenitor star \citep{1989A&A...208..171G}.  Recent observations of
[Mn/Fe] in stars in the Sagittarius Dwarf Galaxy have shown that only
the latter explanation can simultaneously reproduce the data in both
the Milky Way and Sagittarius
\citep{2003ApJ...592L..21M,2008A&A...491..401C}.  This demonstrates
the importance of obtaining abundance ratios in a variety of
environments for a full understanding of chemical evolution.
Because \cite{2008PhST..133a4013B} find that Mn abundances may be
underestimated due to non-LTE effects over most of this metallicity
range, we focus on a relative comparison of our abundances with others
calculated under similar assumption of LTE.

 We find that the Mn I abundances in the M31 GCs are consistent with our training set abundances and the Milky Way [Mn/Fe] abundance trend.  Our abundances for Mn are measured from 3$-$4 Mn I lines in each the three most metal-rich GCs, G108, G315 and G322.  [Mn/Fe] is measured from the Mn I 6021 \rAA line only in G351. We do not see evidence for blends that would change the result by more than $\sim$0.1 dex.

The mean [Sc/Fe] in Milky Way stars and GCs is approximately solar.     We measure a mean [Sc/Fe]$\sim0$ for GCs in M31, with a larger scatter between lines for individual GCs and between the five GCs ($\sigma\sim0.2$ dex)  than we see for other, easier-to-measure Fe$-$peak elements ($\sigma\sim0.1$). We find a similar mean and scatter in our training set GC abundances. We measure abundances for 1$-$4 Sc II lines in each of the five GCs.

Co abundances in stars and GCs in the Milky Way track that of Fe for [Fe/H]$>-2$, so that over this range in metallicity the [Co/Fe]$\sim0$. We measure Co I from three lines for the most metal-rich GC G108.  Co I features in the other four GCs are too weak to measure reliably, with the exception of the feature at 4121 \AA, which we find to be significantly blended.  The G108 [Co/Fe]$=+0.32 \pm 0.13$ is in agreement with Milky Way training set abundances at this metallicity.

Milky Way stellar and GC V abundances typically track the abundance of Fe, resulting in [V/Fe]$\sim0$. 
We measure a mean V abundance from G108, G315, and G322 of [V/Fe]$=-0.02$, which is consistent with our measurements of [V/Fe] in the Milky Way training set GCs.  Abundances for G108 come from 3 V I lines, and abundances for G315 and G322 each come from the V I 6081 \rAA line.  These V I lines are all weak  (EWs$\sim$30-40 m\AA), but it appears  unlikely that there are any blends that would cause misleading abundance measurements.

In summary,  the Fe-peak element abundances in this sample of M31 GCs are similar to those measured in Milky Way GCs, and consistent with enrichment dominated by SNII.

\subsection{Neutron Capture Elements}
\label{sec:ncapture}

The relative abundances of neutron capture elements are important because they are particularly sensitive to the details of star formation, but also because the nucleosynthetic sources and yields of the rapid ($r-$process) and slow ($s-$process) neutron capture reactions remain uncertain \citep[see][]{2004AJ....128.1177V, 1997ARA&A..35..503M}.  
In particular, the difference between the Ba and Y abundances for stars in the Milky Way and stars in nearby dwarf spheroidal galaxies provides strong evidence for differences between the star formation histories of large galaxies and their satellites \citep{2001ApJ...548..592S, 2003AJ....125..684S,2005AJ....129.1428G}. 
 Perhaps more interesting is that these abundance differences are at times inconsistent with simple nucleosynthetic explanations, and thus can provide new constraints on uncertain reaction sites \citep[e.g.][]{2004AJ....128.1177V}.

We are able to measure abundances for the strong lines of Ba II and Y II in some of the M31 GCs. Unfortunately, we are unable to measure abundances for weaker Eu II and La II lines due to the high velocity dispersions of this sample of GCs (see \textsection~\ref{sec:vdisp}).
Typical Ba abundances in Milky Way GC stars are between [Ba/Fe]$\sim0-0.5$ for [Fe/H]$>-1$ and [Ba/Fe]$<0$ for [Fe/H]$<-1$. Our Ba abundances for the M31 GCs are consistent with these trends and with what we find for our training set GCs; we also measure the lowest [Ba/Fe] for the lowest metallicity cluster G219.  Ba II abundances come from 1-4 lines in each of the GCs except for the most metal-rich cluster G108, for which all line strengths were over 150 m\AA.

For [Fe/H]$>-1.5$, Milky Way GCs typically have [Y/Fe]$\sim0$. We measure a mean value of [Y/Fe]$\sim0$ for G108 and G315 in M31, as well, which is also consistent with what we find for the Milky Way training set. These Y abundances come from the 4883 \rAA Y II feature.  Because the Y abundances are derived from a single line, we have performed spectral synthesis tests to confirm that the 4883 \rAA  line is relatively unaffected by blends.

\section{Discussion}
\label{sec:discussion}

\begin{deluxetable*}{rr|rrrr}
\centering
\tablecolumns{6}
\tablewidth{0pc}
\tablecaption{Metallicity Comparisons  \label{tab:fetable}}
\tablehead{
\colhead{}&\colhead{ILS}&\colhead{1}&\colhead{2}&\colhead{3}&\colhead{4}\\\colhead{Cluster} &  \colhead{[Fe/H]} &  \colhead{[Fe/H]}&  \colhead{[Fe/H]}&  \colhead{[Fe/H]} &  \colhead{[Fe/H]}   }
\startdata

G108-B045 & $-$0.94 $\pm$ 0.03	&$-$0.85	&$-$0.94 $\pm$ 0.27  & $-$1.05 $\pm$ 0.25  & $-$0.71 $\pm$ 0.11	 \\
G219-B358 & $-$2.21 $\pm$ 0.03	&$-$1.92	&$-$1.83 $\pm$ 0.22	 & \nodata  &$-$2.00 $\pm$ 0.11  \\
G315-B381 & $-$1.17 $\pm$ 0.02	&\nodata	&$-$1.22 $\pm$ 0.43  	&\nodata &\nodata  \\
G322-B386 & $-$1.14 $\pm$ 0.02	 &$-$1.09 	 &$-$1.21 $\pm$ 0.38	&$-$1.62 $\pm$ 0.14 & $-$1.03 $\pm$ 0.22	 \\
G351-B405 & $-$1.33 $\pm$ 0.04&	$-$1.77	 &$-$1.80 $\pm$ 0.31 	& \nodata &\nodata	\\

\enddata
\tablerefs{From CMD photometry: (1.)  \cite{2005AJ....129.2670R}.  From Lick index spectroscopy: (2.)  \cite{1991ApJ...370..495H}  (3.)  \cite{2002AJ....123.2490P} (4.) \cite{2005A&A...434..909P}, calculated from the relation in \cite{2003MNRAS.339..897T} : [Fe/H]$=$[Z/H]$-0.94$[$\alpha$/Fe].}

\end{deluxetable*}

Although this is only the first set in a larger sample of GCs from our M31 study, we already have several interesting results.
First, in \textsection~\ref{sec:history} we discuss the chemical enrichment history of the present sample of M31 GCs and compare it to the Milky Way and dwarf galaxy GC systems. 
In \textsection~\ref{sec:correlations}  and \textsection~\ref{sec:lick} we discuss the implications of our measurements for both GC formation and evolution and IL abundance work at low resolution.  In the final two sections we comment on constraints that can be put on horizontal branch morphology and reddening of unresolved GCs using high resolution IL spectra.

\subsection{Chemical History of M31 GCs}
\label{sec:history}

Overall, these five M31 GCs are old and have chemical properties similar to those of most Milky Way GCs.  All five GCs are enhanced in the $\alpha-$elements Ca, Ti, and Si to the same extent as Milky Way GCs.  Fe$-$peak element ratios are consistent with Milky Way abundance trends. These results are consistent with existing simple Galactic chemical enrichment scenarios \citep[e.g.][]{1979ApJ...229.1046T}, in so far as it suggests that the gas in both the Milky Way and M31 halos was dominated by enrichment by SNII when these GCs formed. The similar levels of $\alpha-$enhancement in this small sample so far suggests that M31 and the Milky Way are likely to have had similar IMFs and star formation rates at early times.  Since we do not see evidence for the low [$\alpha$/Fe] observed in GCs in the LMC or the disrupting Sagittarius Dwarf Galaxy, it does not seem likely that any of these five GCs are associated with recent satellite accretion events.

\subsection{Variation of Light Elements}
\label{sec:correlations}

Variations in the abundances of light elements between individual
stars within GCs have been observed in all GCs studied in detail since
the phenomenon was discovered in the Milky Way GCs M3 and M13 by
\cite{1978ApJ...223..487C}.  Of the light elements, Mg and O are
observed to be depleted in some GC stars, while Na and Al are
overabundant.  These abundance variations are related and have since
been called the Na$-$O and Mg$-$Al anticorrelations
\citep[see][]{2004ARA&A..42..385G}. Abundances varying in this way are
predicted from high temperature (T $>10^{7}$) C$-$N$-$O cycle
H$-$burning  \citep{1990SvAL...16..275D,1998A&A...333..926D}.  While
these reaction products can in principle be brought up to the stellar
surfaces during ``deep mixing'' on the RGB, the observation of the
abundance variations even in GC main sequence stars has suggested that
they are instead the result of pollution by GC intermediate$-$mass AGB
stars, for example, as discussed in \cite{2008A&A...479..805V},
although this solution is not universally accepted (see modeling of 
NGC 6752 in  \cite{2004MNRAS.353..789F}).

Star-to-star Mg variations have been more difficult to measure than
those of O, Al, and Na in many Milky Way GCs
\citep{2004A&A...416..925C}.  \cite{2004AJ....127.2162S} successfully
measured [Mg/Fe] variations in a large sample of stars in the Milky
Way GCs M3 and M13, with values that range from [Mg/Fe]$=-0.2$ to
$+0.4$.  This variation is detectable in the IL as the mean [Mg/Fe]
will be lower than the mean [$\alpha$/Fe], as shown in
\textsection~\ref{sec:alphas}.  We can also expect that our IL
measurements correspond to a kind of average [Mg/Fe] for each GC,
which is complicated by the extent of the Mg depletion present within
the GC, and the flux weighting of stars of different types.  As a
simple consistency check, we note that our IL [Mg/Fe] measurements
fall within the range [Mg/Fe]$=-0.2$ to $+0.4$ that is expected for
individual stars (\cite{2004AJ....127.2162S}).  The scatter in IL
[Mg/Fe] between GCs tends to be high in both the Milky Way and M31,
and does not appear to correlate with any property of the GCs.  These
factors suggest that it is difficult to predict an expected value of
[Mg/Fe] in the IL spectra of a GC at this time.

AGB star pollution in GCs can also affect the abundance of [Al/Fe].
 Abundances in some GC stars are observed to be as high [Al/Fe]$>+1$, significantly higher than the abundances seen in halo field stars. 
We determine an abundance for G219 of [Al/Fe]$=+0.69$ (see \textsection~\ref{sec:al}), which is close to the high value of [Al/Fe]$\sim1$ measured in some GC stars. 
Three of our training set Milky Way GCs have [Al/Fe]$>+$0.6, which was measured from the more reliable Al 6696 \rAA line, so we believe that we are seeing evidence of the Mg$-$Al anticorrelation in IL measurements of [Al/Fe] in both the Milky Way and M31.

Unfortunately, we were unable to measure oxygen abundances in this sample of GCs.  This was either due to blends with telluric absorption lines or because the lines were too weak to reliably measure from EWs with the reasonably high velocity dispersions of these GCs. Na lines were either too weak or very saturated. Future IL light abundances for GCs with smaller velocity dispersions should be more useful for investigating the Na$-$O anticorrelation.

\subsection{Comparisons to Lick Indexes}
\label{sec:lick}

\begin{figure}
\centering
\includegraphics[scale=0.43]{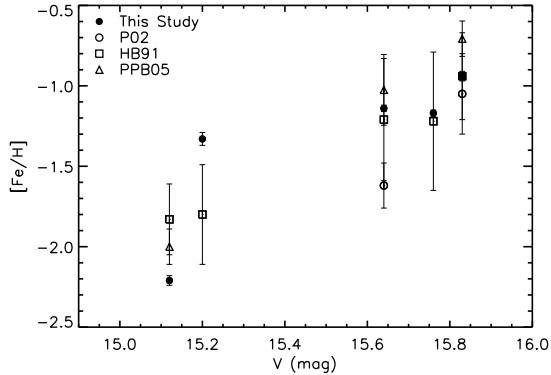}
\caption{Comparison of high resolution ILS [Fe/H] with line index measurements by \cite{2002AJ....123.2490P} (P02, open circles), \cite{1991ApJ...370..495H} (HB91, open squares) and \cite{2005A&A...434..909P} (PPB05, open triangles) plotted as a function of  V magnitude.}
\label{fig:mag} 
\end{figure}

Much progress has been made using low resolution Lick index systems to
measure global properties of GC systems \citep[see review
by][]{2006ARA&A..44..193B}.  Low resolution metallicities have helped
establish the general trends in GC populations in other galaxies, and
their importance for tracing galaxy formation and evolution.  In this
section, we compare previous measurements of Lick index metallicities
for our sample of GCs in M31 with the goal of understanding where and
why abundances from low and high resolution spectra may differ.  The
Lick system or similar methods are critical to studying extragalactic
GCs because low resolution spectra analyses will always need to be
applied to systems that are so distant that high resolution analyses
are impossible.

In making these comparisons we caution that the definition of [Fe/H]
is slightly ambiguous.  Line index metallicities are typically
calibrated to the \cite{1984ApJS...55...45Z} [Fe/H] scale as
established for Milky Way GCs\footnote{The \cite{1984ApJS...55...45Z}
  scale is based on the integrated photometric parameter $Q_{39}$ and
  measurements of absorption from Ca H and K, CN, Fe, and the Mg$b$
  region from low resolution integrated spectra and was calibrated to
  early high resolution abundance results.}.  This metallicity scale
is advantageous because it covers the metallicity range spanned by
Milky Way GCs and can be easily applied to distant objects, providing
a consistent IL metallicity scale for both Milky Way and extragalactic
GCs.  However, by definition this scale is based on blends of lines
from multiple elements.  This limits the information that can be
reliably determined for individual element abundances, including
Fe. Another difficulty in these comparisons is that the calibration is
based on the Milky Way GC system, which has a very consistent
abundance pattern and HBR $-$ [Fe/H] relation that is not necessarily
the same in GCs in other galaxies.  Since the calibration is based on
spectral regions with blends of several elements, it may be less
accurate if targets don't have Milky Way-like abundance ratios.

\subsubsection{``[Fe/H]''}
\label{sec:lick fe}

Because of the ambiguities described above, the comparison of Lick
index metallicities to the high resolution [Fe/H] determined in our
analysis is interesting both when the abundances agree and when they
disagree.  Our IL [Fe/H] results for all five GCs are summarized in
Table 11, along with metallicity estimates in the literature.

A comparison of the high resolution [Fe/H] derived from our analysis
and low resolution Lick index [Fe/H] estimates is shown in
Figure~\ref{fig:mag}, plotted as a function of V magnitude for
convenience.  High resolution [Fe/H] are plotted as solid symbols and
the line index measurements of \cite{1991ApJ...370..495H},
\cite{2002AJ....123.2490P} and \cite{2005A&A...434..909P} correspond
to open squares, circles, and triangles, respectively.
Figure~\ref{fig:mag} shows that the Lick index [Fe/H] estimates agree
within the errors for the higher metallicity clusters G108, G315, and
G322.  However, bigger differences appear at the lowest metallicities;
our measurement of [Fe/H]$=-2.21 \pm 0.03$ for G219 is $\sim0.2-0.4$
dex lower than previous estimates, and our measurement of
[Fe/H]$=-1.33 \pm 0.04$ for G351 is $\sim$0.5 dex higher than low
resolution results (see Table~\ref{tab:fetable}).

Larger discrepancies at the lowest metallicities are expected for
several reasons.  First, line index strengths change little for
[Fe/H]$<-1.6$, so calibrations to the Lick system at low metallicity
are uncertain \citep[e.g. see][]{2002A&A...395...45P}.  In the case of
G219, a difference between [Fe/H]=$-1.8$ and [Fe/H]=$-2.2$ would be
particularly difficult to detect at low resolution, emphasizing the
importance of high resolution measurements at these lowest metallicity
GCs.  The discrepancy between the high resolution and Lick index
abundance for G351 is a little more difficult to understand,
especially since the CMD metallicity estimate of
\cite{2005AJ....129.2670R} is similar to the Lick index result.  This
discrepancy stands out in our sample, however the difference is still
within $\sim$1.3$\sigma$ of the quoted error of
\cite{1991ApJ...370..495H}, and our own rms error for this cluster is
slightly larger than for others in our sample.  As a simple reality
check, a visual comparison of the G351 spectrum and our other 4 GC
spectra (see Figure~\ref{fig:spectra}) suggest that G219 is
substantially more metal poor than G351, while line index measurements
put both G351 and G219 at [Fe/H]$\sim-1.8$.

It is possible that the Lick index analysis is complicated by a
combination of factors that include both low-metallicity degeneracies
and poor modeling of blue horizontal branches (see discussion of the
effect of HB morphology on our results in \textsection~\ref{sec:hbr}).
We note G351 has been observed to have a bimodal horizontal branch
\citep{2005AJ....129.2670R}.  We also note that G351 is one of the GCs
we find to have depleted [Mg/Fe], suggesting that abundance ratio
calibration degeneracies may be a particular problem for this cluster
in the sample.

\subsubsection{[$\alpha$/Fe]}
\label{sec:lick alpha}

\begin{figure}
\centering
\includegraphics[scale=0.43]{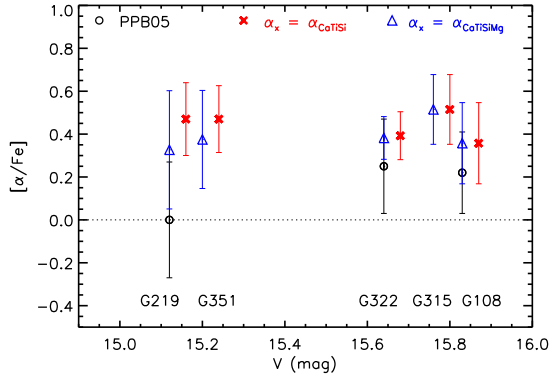}
\caption{Comparison of  [$\alpha$/Fe] from high resolution ILS with line index estimates by \cite{2005A&A...434..909P} (PPB05, open circles) plotted as a function of  V magnitude. Red crosses correspond to mean [$\alpha$/Fe] measured from Ca I, Ti I, Ti II and Si I only. Crosses are plotted with a $+0.05$ offset in V for visibility. Blue triangles correspond to mean [$\alpha$/Fe] from Ca I, Ti I, Ti II, Si I and Mg I.}
\label{fig:alphamag} 
\end{figure}

Recent progress has been made in developing SSP IL spectra models with variable element ratios for comparison with Lick index absorption features \citep{2003MNRAS.339..897T, 2005ApJS..160..176L, 2007ApJS..171..146S}.  In particular, the models of  \cite{2003MNRAS.339..897T}  have made the [$\alpha$/Fe] estimates of a portion of M31 GCs possible, including three of those studied here \citep{2005A&A...434..909P}.  In this case, [$\alpha$/Fe] is determined from a comparison with models of Lick indexes with Fe$-$dominated  and Mg$-$dominated absorption features.  A significant result of the study of \citep{2005A&A...434..909P} is that GCs in M31 with ages $>$8 Gyrs have a mean [$\alpha$/Fe]$=+0.18 \pm 0.05$ with a dispersion of 0.37 dex, which is $0.1\sim0.2$ dex lower than what  \cite{2005A&A...434..909P} find for Milky Way GCs.  However, \cite{2005AJ....129.1412B} find that the discrepancy between [$\alpha$/Fe] in M31 and Milky Way GCs may be an SSP model$-$dependent result. 

Since we have found that [Mg/Fe] is likely to be depleted compared to other
$\alpha-$elements within GCs due to  AGB star self$-$pollution, we expect
that [$\alpha$/Fe] ratios in GCs determined from indexes with Mg$-$dominated absorption features could be lower than the [$\alpha$/Fe] abundances that we determine
from Ca, Si, or Ti. 

To test this, we use the mean [$\alpha$/Fe] from Ca, Si and Ti lines
for each GC (discussed in \textsection~\ref{sec:alphas}) and compare
to the [$\alpha$/Fe] estimated by \cite{2005A&A...434..909P} from Lick
indexes in Figure~\ref{fig:alphamag}. Measurements by
\cite{2005A&A...434..909P} are plotted as open circles, and our mean
[$\alpha$/Fe] excluding Mg are plotted as red crosses.  We measure a
systematically higher value for [$\alpha$/Fe] than
\cite{2005A&A...434..909P} obtain for [$\alpha$/Fe] in all three GCs.
The largest discrepancy is for G219, for which we find
[$\alpha_{\rm{CaSiTi}}$/Fe]$=+0.47$ and \cite{2005A&A...434..909P}
estimate [$\alpha$/Fe]$=0.0$.  We find the discrepancy for G219 is
reduced, but not resolved, if we include our [Mg/Fe]$=+0.04$
measurement in the mean [$\alpha$/Fe], which is shown by the blue
triangle in Figure~\ref{fig:alphamag}.

We note that, in addition to the degeneracy in the Lick indexes at low
abundances, the effect of different [$\alpha$/Fe] ratios in SSP
modeling at low abundances is also very weak, further obscuring the
resolution of the Lick index measurements at low abundances, as
already discussed by \cite{2003A&A...400..823M} and
\cite{2005A&A...434..909P}.  It is likely that this is the cause of
the remaining discrepancy in the [$\alpha$/Fe] estimate from low
resolution for G219.

While our present sample size of high resolution IL abundances in
extragalactic GCs is still small, it already removes some of the
discrepancies between M31 and Milky Way GCs from Lick indexes that
have been discussed in the literature
\citep{2005A&A...434..909P,2005AJ....129.1412B,
  2006ARA&A..44..193B}. We find that for this sample of GCs the true
[$\alpha$/Fe] from Ca, Si, and Ti appear similar for the two galaxies,
and that an accurate estimate of this value in GCs for interpretation
of the chemical enrichment history of a galaxy must come from elements
other than Mg due to the peculiarities of Mg abundances in GCs.  Our
results indicate that the unexpected, low [$\alpha$/Fe] ratios in
metal-poor GCs may be an artifact of the uncertainties in line index
systems at low metallicities.

\subsection{Horizontal Branch Morphology}
\label{sec:hbr}

In general, the position of a star on the HB is a function of
metallicity; it is expected that metal-poor GCs will have bluer HB
morphologies than GCs with higher abundances.  The fact that a number
of Milky Way GCs at the same metallicity are observed to have very
different HB morphologies has led to the conclusion that at least one
important ``second$-$parameter'' plays a role in HB morphology
\citep[see review by][]{2005ARA&A..43..387G}.  SSP models that
reproduce the HBR$-$[Fe/H] relationship of the Milky Way are difficult
to establish because HB morphology is sensitive to a variety of
factors (e.g. mass loss, age, helium abundance, and others)
\citep[e.g.][]{1994ApJ...423..248L,2006A&A...452..875R}. Universal SSP
models will be even more difficult to establish if the HBR$-$[Fe/H]
relationship is different in other galaxies. There is evidence that
this might indeed be the case; observations of GCs in both M31 and
Fornax suggest the HBR$-$[Fe/H] relationship is offset to lower
metallicities than in Milky Way GCs
\citep{2005AJ....129.2670R,1998AJ....115.2369S}.

Low resolution studies have found it important to consider the 
effect of HB morphology on IL spectra of unresolved GCs because the presence
of old, hot stars on the blue HB can mimic light from young main
sequence stars, resulting in young or intermediate age determinations
for GCs that are actually old \citep[e.g.][]{1995A&A...302..718D,
  2002MNRAS.336..168B, 2004ApJ...608L..33S}. In the next section we
discuss the effect of HB morphology on abundance and age
determinations with our high resolution IL method.

\subsubsection{ Effect on Age and [Fe/H] Determinations}
\label{sec:hb abundances}

We have used our training set Milky Way GCs to perform tests to
assess the effect of inaccurate proportions of red and blue HB stars
on our results.
The Teramo group has produced isochrones with two different values for
mass loss during stellar post-main sequence evolution, a parameter which
influences HB morphology.  Of the two possible values, we have used
isochrones with the less extreme mass loss parameter.
 This mass loss parameter produces very blue ($B-V\lesssim$ 0.1) HB
 stars for our 13 and 15 Gyr CMDs at [Fe/H] $\lesssim -1.8$, but does
not produce bimodal HBs, as is common in many intermediate metallicity GCs.
Because we are using the less extreme mass loss parameter,
the fear is that the HB may not be as blue as appropriate in some
cases.

\begin{figure*}
\centering
\includegraphics[scale=0.72]{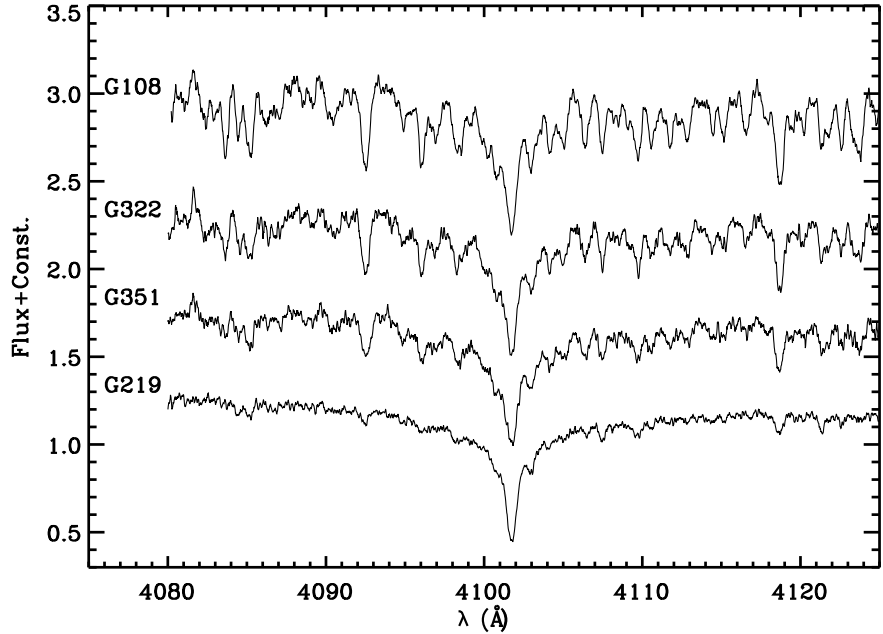}
\caption{M31 GC spectra near H$\delta$ region, shown in order of increasing HBR from top to bottom.  }
\label{fig:Hdelta} 
\end{figure*}

To test the effect of this potential error, we have
added blue HB stars by hand into our synthetic CMDs to ascertain the consequences of underpredicting the number of blue HB stars with our choice of isochrones.  In these tests we conserve the total number of stars and total flux of the HB.
We find that even though blue HB stars can contribute $10-15\%$ of the
total flux of the population at wavelengths below 5000 \AA, most of the Fe I
lines are found over the range
4500-7000 \AA, which is not as strongly influenced by the blue HB stars.
 Empirically, the addition of
these stars into the CMD changes the derived Fe I abundances by
$<+$0.05 dex.
 This is not unexpected, because \citetalias{2008ApJ...684..326M}
 showed that the Fe I EW strengths are dominated ($\sim80\%$) by the luminous, cooler
 stars on the RGB, AGB and red HB, with little effect from hot turn
 off or hot blue straggler stars,
particularly at redder wavelengths.

As an example of an extreme test case, we replaced all of the red HB
stars with the same number of blue HB stars in the 10 Gyr best-fitting
CMD for G219.  We know from the HST CMDs of \cite{2005AJ....129.2670R}
that G219 has both red HB stars and a large number of blue HB stars
that are not entirely represented in the best-fitting CMD.  Using this
extreme CMD, we find that the Fe I abundance changes by $<$0.05 dex,
the scatter in the Fe abundances increases, and the correlations of Fe
abundance with EP, wavelength, and reduced EW get larger. From this
extreme example, it is clear that not only do blue HB stars have
little impact on the Fe I solution, but that accurate modeling of the
red HB results in a much more self-consistent solution from all of the
Fe I lines.
Therefore, we can expect that having roughly the correct number of red HB
stars in the synthetic CMDs will be more important for our abundance
determinations than the correct number of  blue HB stars.
Given the small effect of blue HB stars on Fe I abundances, and given
that our analysis only constrains the CMD age for old GCs
to a 5 Gyr age range, we find that the accuracy of the
blue HB morphology has a no significant (or even detectable) impact on our
results.
We conclude that the ages and metallicities that we derive for old GCs
with our abundance analysis method will not be significantly affected
by synthetic CMDs that have inaccurate blue HB morphologies.

As a further consistency check, we can look for qualitative
information about the temperature distribution of the preferred CMD
solutions in the Fe abundance vs. EP plots discussed in \textsection
~\ref{sec:best cmd}.  A symptom of too few hot, blue HB stars in the
isochrone compared to the real GC would be increasing Fe abundances 
for lines with larger EP.  For G108, G315, and G322 we find no significant
trend in Fe abundance with EP, which implies that the distribution of
stellar temperatures in the CMD solutions for these GCs are very
accurate.

For G219 we find decreasing Fe abundances with increasing EP, which
suggests in this case that we actually do have too many hot stars in the
isochrone for this metal-poor GC.  However, we also find higher Fe
abundances at larger reduced EWs, which implies that we may have too
many dwarf  (low microturbulent velocity) stars in the isochrone being
used in the analysis.  Since the dwarf stars have higher temperatures
than the giant stars, for G219 the excess of hot dwarf stars
may cause the observed trends. Note that the
dependence of Fe abundance on reduced EW may be associated with a 
small error in the microturbulence for some stars, as discussed in
\textsection~\ref{sec:Fe}.  However, the mean Fe solution
is not strongly affected by these weak trends, and so we do not 
pursue this issue further.

The only GC in the sample that shows the symptom of not enough hot
stars in the Fe abundance vs. EP plot is G351, although it also shows
the largest positive dependence of Fe abundance on reduced EW.  The
diagnostics suggest that there are both too many dwarf stars and not
enough hot stars in the best CMD for G351.  Also, the standard
deviation for [Fe/H] is larger than for the other four GCs.  While
the solution for this cluster shows somewhat less self-consistency
than for the other 4 clusters from the diagnostics discussed above, we
emphasize that the overall scatter in the solution is still quite
small, suggesting that our solution has not been dramatically affected
by unavoidable problems (e.g. interloping stars or internal age 
or abundance variations) and that the statistical uncertainty is
a meaningful estimate of the overall accuracy of the analysis.
Moreover, any difficulties are not likely to be due to the HB
morphology.

\subsubsection{Consistency with Photometry}
\label{sec:hb spectra}

We are fortunate that the details of GC HB morphology do not have a
large effect on the abundances or ages derived from high resolution IL
spectra using this method.  However, we can see the effect of HB stars
on the temperature sensitive Balmer lines. We are still testing a
method to constrain the HB morphology of unresolved GCs using the
Balmer lines in high resolution IL spectra (J. Colucci et al. 2009, in
preparation).
For the purposes of this work, we simply check for qualitative
consistency of the Balmer line profiles in the IL spectra with the HST
CMDs of \cite{2005AJ....129.2670R}.  HBR ratios from \cite{2005AJ....129.2670R} for G108, G322, G351,
and G219 are listed in Table~\ref{tab:clusterdata}, where a value of
zero would correspond to a GC with a strongly red HB, and a value of
unity would correspond to a GC with a strongly blue HB.  From the IL
spectra in the region of H$\delta$, shown in Figure~\ref{fig:Hdelta},
it is clear that G219 and G351 have more prominent Balmer line wings,
and thus a larger contribution of flux from hot stars than G108,
consistent with expectations.

\cite{2005AJ....129.2670R} measure an HBR$-$[Fe/H] relation from HST
CMDs for 18 M31 GCs that is offset to lower metallicities than the
HBR$-$[Fe/H] relation in Milky Way GCs. One explanation proposed is
that the most metal-poor M31 GCs are $\sim1-2$ Gyr younger than
similar GCs in the Milky Way.  We note that our lower preferred CMD
age range for G219 is consistent with this, although we are unable to put strong constraints on absolute age.

\subsection{Reddening Constraints from Broadband $B-V$ Colors}
\label{sec:colors}

\begin{deluxetable}{rr|rr}

\tablecolumns{6}
\tablewidth{0pc}
\tablecaption{$E(B-V) $ Comparisons  \label{tab:red}}
\tablehead{
\colhead{}&\multicolumn{3}{c}{$E(B-V)$}\\\colhead{Cluster} &  \colhead{ILS Limit} &  \colhead{1}&  \colhead{2}   }
\startdata

G108-B045 &	0.13& 0.18& 0.10 \\
G219-B358 &   0.06 & 0.06&0.06\\
G315-B381 &  0.08 &0.17&\nodata\\
G322-B386 & 	 0.15&0.21&0.13\\
G351-B405 & 	0.06&0.14&0.08\\

\enddata
\tablerefs{(1.)  \cite{2008MNRAS.385.1973F}  (2.) \cite{2005AJ....129.2670R}}

\end{deluxetable}

\begin{figure}
\centering
\includegraphics[angle=90,scale=0.35]{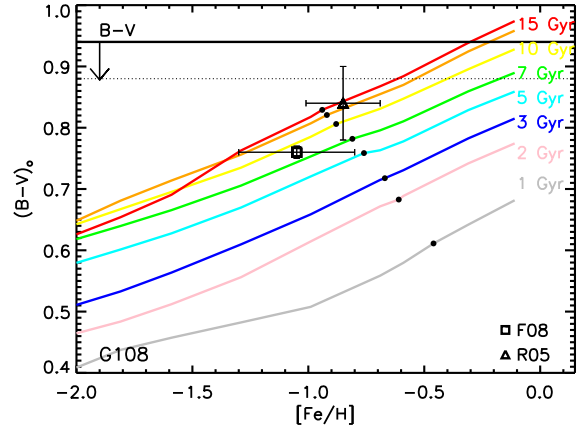}
\caption{Integrated $(B-V)_{0}$ colors calculated from synthetic CMDs are shown as a function of [Fe/H] for each age.  Black points show the [Fe/H] and age solutions determined from this analysis for G108.  The horizontal solid line corresponds to the observed $B-V$ color for the GC from \cite{2004A&A...416..917G}, with an arrow and dotted line to show the $E(B-V)$ correction due to Galactic reddening of \cite{1998ApJ...500..525S}.  $(B-V)_{0}$ calculated with reddening and metallicity of  \cite{2008MNRAS.385.1973F} (F08) and  \cite{2005AJ....129.2670R} (R05) are plotted as open squares and triangles, respectively. }
\label{fig:color1} 
\end{figure}

\begin{figure}
\centering
\includegraphics[angle=90,scale=0.35]{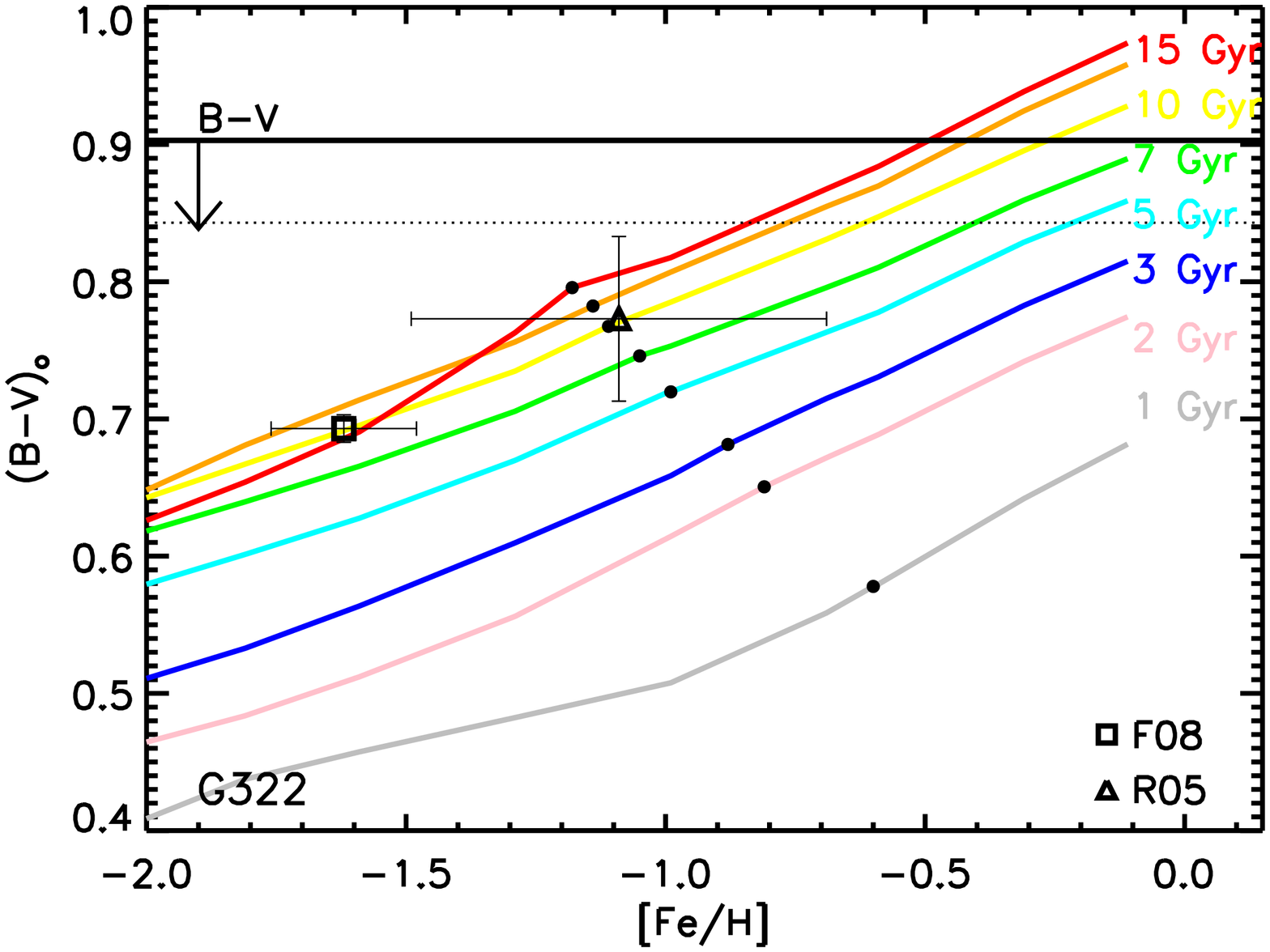}
\caption{ Same as Figure~\ref{fig:color1} for G322.}
\label{fig:color2} 
\end{figure}

\begin{figure}
\centering
\includegraphics[angle=90,scale=0.35]{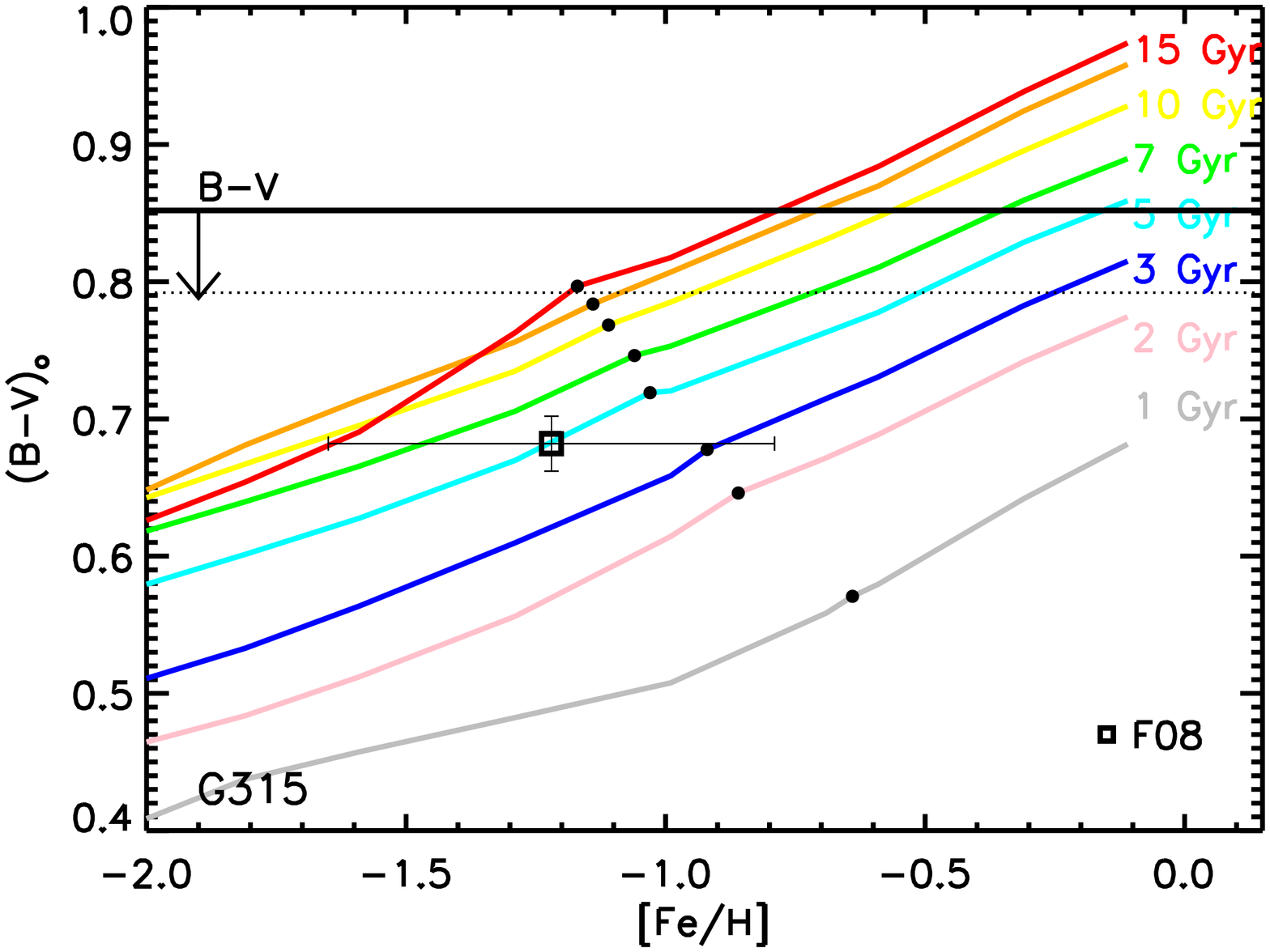}
\caption{ Same as Figure~\ref{fig:color1} for G315.}
\label{fig:color3} 
\end{figure}

\begin{figure}
\centering
\includegraphics[angle=90,scale=0.35]{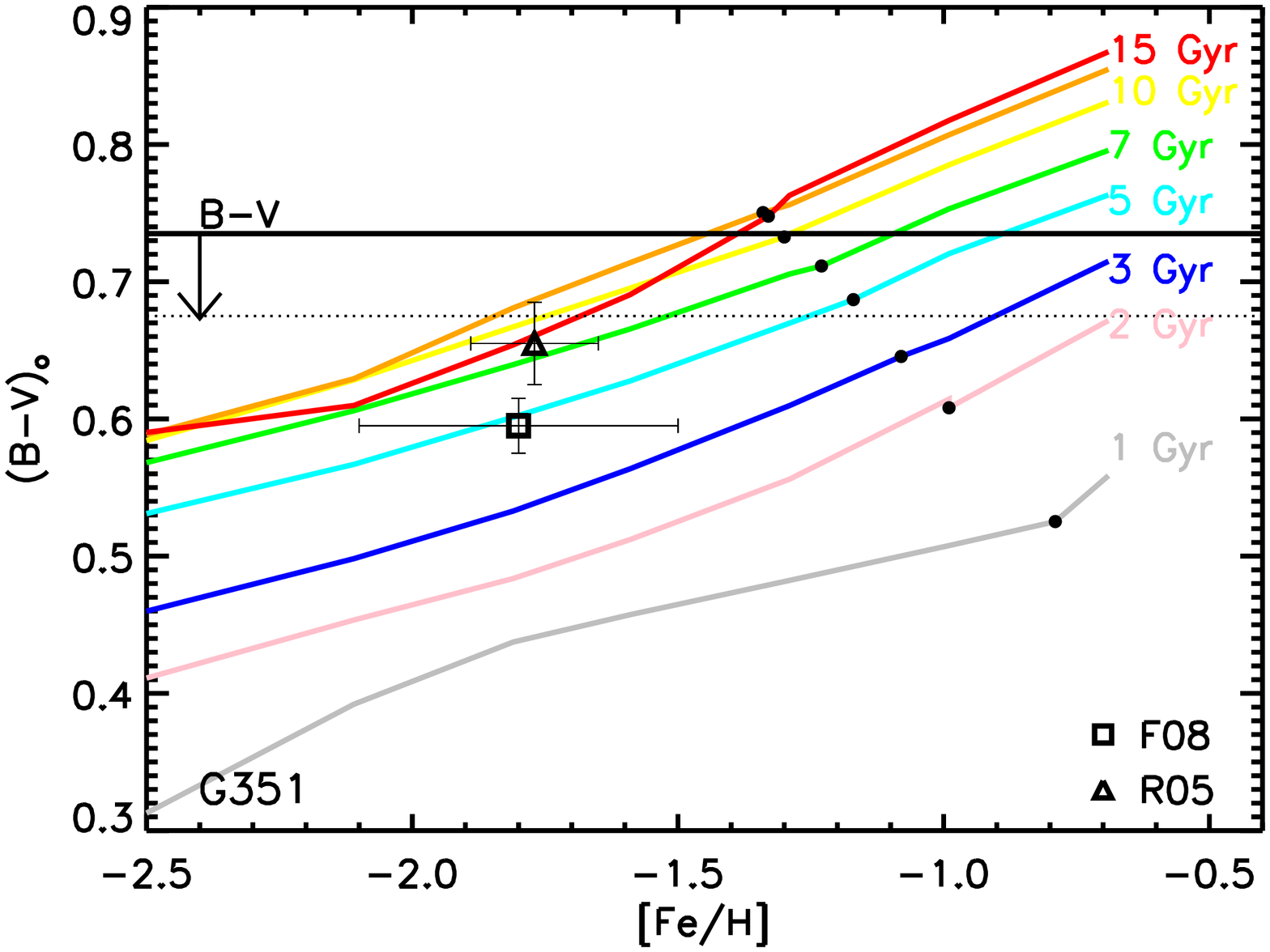}
\caption{ Same as Figure~\ref{fig:color1} for G351.}
\label{fig:color4} 
\end{figure}

\begin{figure}
\centering
\includegraphics[angle=90,scale=0.35]{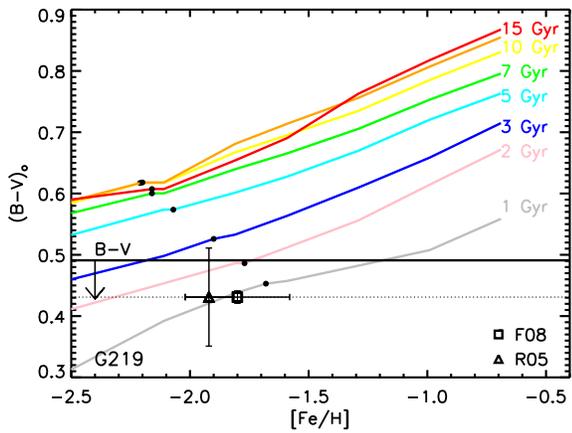}
\caption{ Same as Figure~\ref{fig:color1} for G219.}
\label{fig:color5} 
\end{figure}

In principle, an interesting additional constraint on our potential
CMD solutions can be derived by comparing total  broadband
$(B-V)_{0}$ colors with existing photometry. This comparison could
help us eliminate potential CMD solutions where the integrated colors
of the CMDs are inconsistent with the photometry (see
\citetalias{2008ApJ...684..326M}). In practice, however, we find that reddening
estimates for GCs in M31 are too uncertain to provide constraints on
CMDs of the GCs discussed here.  Rather, our abundance analysis
constrains the CMD with enough fidelity that we can actually put some
limits on the reddening for these individual M31 GCs. We note that
GCs have previously been
used in this way to probe reddening in galaxies
\citep[e.g.][]{2000AJ....119..727B}.

The colors for our synthetic CMDs using Teramo isochrones are shown in
Figures~\ref{fig:color1} through~\ref{fig:color5}.  The broadband
$(B-V)_{0}$ colors calculated from our synthetic CMDs are plotted
against the [Fe/H] adopted in the Teramo isochrones for each GC.  The
trend of $(B-V)_{0}$ with [Fe/H] for constant age is emphasized by the
solid colored lines.  Black points correspond to the CMD and [Fe/H]
solutions for each age described in \textsection~\ref{sec:best cmd}
for each GC.  The black solid line in each figure corresponds to the
observed $B-V$ color for the GC from the Revised Bologna Catalog
\citep{2004A&A...416..917G}. Black arrows show the minimum reddening
for the M31 line-of-sight of $E(B-V)=0.06$, which is due to Galactic
extinction alone \citep{1998ApJ...500..525S}.

For each individual GC we show the recent reddening determination by
\cite{2008MNRAS.385.1973F}, who have derived these from correlations
between optical and infrared colors and metallicity using a large set
of M31 GC low resolution spectroscopic abundances.  The $(B-V)_{0}$
calculated with the reddenings of \cite{2008MNRAS.385.1973F} are
plotted as open squares, using the low resolution
spectroscopic [Fe/H] solutions for the x-axis location.
For G108, G322, G351 and G219
we also show the reddening adopted by the deep HST study of  \cite{2005AJ....129.2670R},
which is an average of several previous measurements. Reddenings of
\cite{2005AJ....129.2670R} are plotted as open triangles at the [Fe/H]
determined from their HST CMDs.

For each GC we can place an upper limit on the reddening that is
consistent with the preferred CMDs for each GC discussed in
\textsection~\ref{sec:Fe}.  This limit is calculated using the
$(B-V)_{0}$ color of the youngest CMD in the preferred age range and
the $B-V$ color from \cite{2004A&A...416..917G}.  The limits are
listed in Table ~\ref{tab:red} with the values of
\cite{2008MNRAS.385.1973F} and \cite{2005AJ....129.2670R} for
comparison.  We note that synthetic CMDs created with the Teramo
isochrones that employ a higher mass loss rate than those used here,
and thus have bluer HBs overall, would result in slightly lower
predicted $(B-V)_{0}$.  For the ages relevant here, the effect HB
morphology is at most $\Delta(B-V)_{0}\sim-0.03$, and an indication of
the uncertainty in our reddening limit.

In general, we find that our preferred CMDs and the reddenings they
imply agree best with the [Fe/H] and reddenings of
\cite{2005AJ....129.2670R}, rather than with the reddenings inferred
from low resolution spectroscopic metallicities. For G108, G322 and
G351 the $E(B-V)$ derived by \cite{2008MNRAS.385.1973F} are higher
than those adopted by \cite{2005AJ....129.2670R}, but it is clear from
Figures~\ref{fig:color1} through~\ref{fig:color5} that they are still
consistent with an old ($>$10 Gyr) population at the metallicities
applied by \cite{2008MNRAS.385.1973F}.

Although G219 has been verified to be $\sim$10 Gyr or older by HST
photometry, by low resolution spectral line indexes, and by high
resolution spectra abundances in this work, it has $B-V$ colors
substantially too blue for this age.  This inconsistency has been
noted in the Bologna catalog.  In Table~\ref{tab:red} we list the
Galactic $E(B-V)=0.06$ as the only reliable constraint.  We also note
that the difference between the observed and predicted $B-V$ for G219
is too large to be explained by a deficit of blue HB stars in the
synthetic CMD alone, but may be reduced with some contribution from
relatively rare, but luminous, post-AGB or UV-bright stars
\citep[see][]{2001PASP..113.1162M}.

For G351, the $B-V$ of the 10 Gyr CMD is consistent with $E(B-V)=0.0$,
lower than the Galactic $E(B-V)=0.06$, which we again list as the most
reliable constraint in Table~\ref{tab:red}.  This difference in color
may be at least partially explained by missing blue HB stars in the
synthetic CMD as compared to the observed HST CMD in
\cite{2005AJ....129.2670R}.

G315 is the only GC discussed here for which there are no HST
constraints available. We find an $E(B-V)\leq0.08$ is consistent with
our analysis.  The larger value determined by
\cite{2008MNRAS.385.1973F} is hard to understand because their derived
reddening implies a $(B-V)_{0}$ that is inconsistent with an old
population, even though the high and low resolution [Fe/H] are very
similar.  It is possible that the discrepancy can arise from
inaccurate photometry used by \cite{2008MNRAS.385.1973F} to determine
the $E(B-V)$.

\section{Summary}
\label{sec:summary}

We have applied a new method for detailed abundance analysis of high
SNR, high resolution IL spectra of unresolved GCs to a sample of five
GCs in M31.  From over 60-100  resolved spectral lines in each
cluster we have derived abundances from EWs for Mg, Al, Si, Ca, Sc,
Ti, V, Cr, Mn, Fe, Co, Ni, Y and Ba.  We have used our abundance
analysis to put independent constraints on the ages and reddening of
these M31 GCs, and used the high resolution IL spectra to measure
velocity dispersions to high precision.

We find these 5 M31 GCs to be similar to the Milky Way GCs system in
several respects.  First, they are $>$10 Gyrs old, and span a range in
[Fe/H] of $-2.2$ to $-0.9$. Second, their Ca, Si, and Ti abundances
are enhanced to similar levels as Milky Way GCs, and suggesting that
the gas reservoirs from which they formed were dominated by products
of SNII at the time of their formation.  Finally, the Fe$-$peak and
the neutron capture abundance ratios studied here also follow Milky
Way abundance trends.

We have confirmed that light element abundance variations between
stars within GCs can effect abundances derived from high resolution IL
spectra for Mg and possibly Al.  We suggest that part of the large
scatter in [$\alpha$/Fe] measurements of extragalactic GCs using low
resolution line indexes may be due to the effects of Mg$-$dominated
absorption features on line indexes.

We have demonstrated that a significant number of quantitative
constraints on galaxy and GC formation and evolution can be made for
unresolved GCs using this new abundance analysis technique.  We have
shown for the first time that abundance ratios fundamental to
understanding galaxy formation can be obtained for other nearby,
massive galaxies.  While we have intentionally targeted ``typical''
GCs for this first sample in M31, a larger selection of GCs is crucial
for a complete picture of the GC system and formation history of M31.
Future work is needed to investigate other possible differences in the
M31 GC system compared to the Milky Way, i.e.  young GCs, high
metallicity disk GCs, and enhanced nitrogen abundances.

\acknowledgements
This research was supported by NSF grant AST-0507350. J.E.C. thanks E. Kirby for help with the IL spectral synthesis code.

\clearpage
\LongTables
\begin{deluxetable}{rrrrrrrrr}
\tablecolumns{9}
\tablewidth{0pc}
\tablecaption{Line Parameters and Integrated Light Equivalent Widths for M31 GCs \label{tab:linetable}}
\tablehead{\colhead{Species} & \colhead{$\lambda$}   & \colhead{E.P.} & \colhead{log gf}   & \colhead{EW(m\AA)} & \colhead{EW(m\AA)} & \colhead{EW(m\AA)} & \colhead{EW(m\AA)} & \colhead{EW(m\AA)}\\ 
\colhead{} & \colhead{(\AA)} & \colhead{(eV)} &\colhead{} & \colhead{G108}&\colhead{G322}&\colhead{G315}&\colhead{G351}&\colhead{G219}}

\startdata

 Mg  I &4571.102 &  0.000 & -5.691 &  \nodata &120.0 &  \nodata & 86.7 & 34.5\\
 Mg  I &4703.003 &  4.346 & -0.666 &  \nodata &139.2 &  \nodata & 99.8 &  \nodata\\
 Mg  I &5528.418 &  4.346 & -0.341 &  \nodata &  \nodata &  \nodata &147.3 & 58.2\\
 Mg  I &5528.418 &  4.346 & -0.341 &  \nodata &  \nodata &  \nodata &143.6 &  \nodata\\
\\
 Al  I &3944.016 &  0.000 & -0.638 &  \nodata &  \nodata &  \nodata &  \nodata &134.9\\
 Al  I &3944.016 &  0.000 & -0.638 &  \nodata &  \nodata &  \nodata &  \nodata &140.1\\
\\
 Si  I &7405.790 &  5.610 & -0.660 & 58.4 &  \nodata &  \nodata &  \nodata &  \nodata\\
 Si  I &7415.958 &  5.610 & -0.730 & 66.4 &  \nodata & 50.1 &  \nodata &  \nodata\\
 Si  I &7423.509 &  5.620 & -0.580 & 84.3 & 66.5 & 82.4 &  \nodata &  \nodata\\
\\
 Ca  I &4318.659 &  1.899 & -0.295 &  \nodata &109.6 &  \nodata &  \nodata &  \nodata\\
 Ca  I &4425.444 &  1.879 & -0.358 &127.9 &  \nodata &  \nodata &  \nodata & 46.6\\
 Ca  I &5581.979 &  2.523 & -0.555 & 99.0 & 77.6 & 86.3 & 64.2 &  \nodata\\
 Ca  I &5588.764 &  2.526 &  0.358 &  \nodata &138.0 &137.6 &122.8 & 59.8\\
 Ca  I &5590.126 &  2.521 & -0.571 &101.4 & 96.6 & 89.6 & 74.2 &  \nodata\\
 Ca  I &5601.286 &  2.526 & -0.690 &112.3 & 91.9 & 89.7 & 53.3 & 21.3\\
 Ca  I &5857.459 &  2.933 &  0.240 &  \nodata &124.0 &145.0 &111.9 & 43.1\\
 Ca  I &6122.226 &  1.886 & -0.320 &  \nodata &  \nodata &  \nodata &149.8 & 69.0\\
 Ca  I &6162.180 &  1.899 & -0.090 &  \nodata &  \nodata &  \nodata &  \nodata & 89.0\\
 Ca  I &6166.440 &  2.520 & -1.142 & 84.4 & 64.6 & 71.0 &  \nodata &  \nodata\\
 Ca  I &6439.083 &  2.526 &  0.390 &  \nodata &  \nodata &  \nodata &147.8 & 57.9\\
 Ca  I &6471.662 &  2.526 & -0.686 & 98.9 &  \nodata & 78.8 & 70.0 &  \nodata\\
 Ca  I &6493.781 &  2.521 & -0.109 &  \nodata &  \nodata &124.8 & 94.3 & 42.7\\
 Ca  I &6572.795 &  0.000 & -4.310 & 82.5 &  \nodata &  \nodata &  \nodata &  \nodata\\
 Ca  I &7148.150 &  2.709 &  0.137 &  \nodata &133.5 &146.4 &131.3 &  \nodata\\
\\
 Sc II &4246.837 &  0.315 &  0.240 &  \nodata &  \nodata &141.7 &  \nodata & 78.3\\
 Sc II &4670.413 &  1.357 & -0.580 & 84.3 & 86.4 & 86.9 &  \nodata &  \nodata\\
 Sc II &5526.821 &  1.768 &  0.020 & 79.4 & 62.2 & 76.5 & 52.2 &  \nodata\\
 Sc II &6604.600 &  1.357 & -1.480 & 44.6 &  \nodata &  \nodata &  \nodata &  \nodata\\
\\
 Ti  I &4991.072 &  0.836 &  0.380 &  \nodata &  \nodata &  \nodata &134.2 &  \nodata\\
 Ti  I &4999.510 &  0.826 &  0.250 &143.5 &122.0 &134.8 &129.7 &  \nodata\\
 Ti  I &5039.964 &  0.021 & -1.130 &115.9 &  \nodata &  \nodata &  \nodata &  \nodata\\
 Ti  I &5210.392 &  0.048 & -0.884 &  \nodata &140.1 &  \nodata &  \nodata &  \nodata\\
 Ti  I &5401.379 &  0.818 & -2.890 &  \nodata &  \nodata &  9.4 &  \nodata &  \nodata\\
 Ti  I &5648.565 &  2.495 & -0.260 & 17.3 &  \nodata &  \nodata &  \nodata &  \nodata\\
 Ti  I &5866.451 &  1.067 & -0.840 & 88.4 & 71.7 &  \nodata & 46.3 &  \nodata\\
 Ti  I &6743.127 &  0.900 & -1.630 & 53.4 &  \nodata & 55.3 & 30.5 &  \nodata\\
\\
 Ti II &4395.040 &  1.084 & -0.660 &  \nodata &  \nodata &  \nodata &  \nodata &106.5\\
 Ti II &4395.848 &  1.243 & -2.170 & 98.7 &  \nodata &  \nodata &  \nodata &  \nodata\\
 Ti II &4399.778 &  1.237 & -1.270 &137.7 &  \nodata &136.6 &  \nodata &  \nodata\\
 Ti II &4418.342 &  1.237 & -2.460 & 85.2 &  \nodata & 73.6 &  \nodata &  \nodata\\
 Ti II &4501.278 &  1.116 & -0.760 &  \nodata &150.1 &  \nodata &  \nodata &  \nodata\\
 Ti II &4563.766 &  1.221 & -0.960 &  \nodata &155.6 &136.4 &126.3 & 68.9\\
 Ti II &4571.982 &  1.572 & -0.530 &  \nodata &  \nodata &  \nodata &156.5 & 76.7\\
 Ti II &4589.953 &  1.237 & -1.790 & 82.6 & 91.7 & 77.2 &  \nodata & 44.5\\
 Ti II &5381.010 &  1.566 & -2.080 & 92.0 & 60.6 & 75.3 &  \nodata &  \nodata\\
\\
  V  I &6039.730 &  1.060 & -0.650&  39.5 & \nodata &  \nodata &  \nodata &  \nodata\\
  V  I &6081.430 &  1.050 & -0.580&  50.6 & 41.8 & 32.2  &  \nodata &  \nodata\\
  V  I &6274.658 &  0.270 & -1.670 & 37.6 &  \nodata &  \nodata &  \nodata &  \nodata\\
\\
 Cr  I &4254.346 &  0.000 & -0.114 &  \nodata &  \nodata &  \nodata &  \nodata & 83.5\\
 Cr  I &4274.806 &  0.000 & -0.231 &  \nodata &  \nodata &  \nodata &  \nodata & 67.3\\
 Cr  I &4274.806 &  0.000 & -0.231 &  \nodata &  \nodata &  \nodata &  \nodata & 67.1\\
 Cr  I &5206.044 &  0.941 &  0.019 &  \nodata &  \nodata &  \nodata &  \nodata & 84.5\\
 Cr  I &5208.432 &  0.941 &  0.158 &  \nodata &  \nodata &  \nodata &  \nodata & 87.2\\
 Cr  I &5208.432 &  0.941 &  0.158 &  \nodata &  \nodata &  \nodata &  \nodata & 93.3\\
 Cr  I &5409.799 &  1.030 & -0.720 &  \nodata &144.7 &134.0 &118.7 & 36.8\\
 Cr  I &7400.188 &  2.900 & -0.111 & 85.9 &  \nodata & 76.9 &  \nodata &  \nodata\\
\\
 Mn  I &4754.039 &  2.282 & -0.086 &103.9 & 73.1 & 80.4 &  \nodata &  \nodata\\
 Mn  I &6013.520 &  3.070 & -0.250 & 63.7 & 38.1 & 38.7 &  \nodata &  \nodata\\
 Mn  I &6016.620 &  3.070 & -0.216   &73.4 &\nodata & 44.7 &  \nodata &  \nodata\\
 Mn  I &6021.820 &  3.070 & 0.034 &65.1 &44.5 &62.5 &36.6&  \nodata\\
\\ 
 Fe  I &3878.027 &  0.958 & -0.896 &  \nodata &  \nodata &  \nodata &  \nodata & 85.3\\
 Fe  I &3899.719 &  0.087 & -1.515 &  \nodata &  \nodata &  \nodata &  \nodata &143.7\\
 Fe  I &4063.605 &  1.557 &  0.062 &  \nodata &  \nodata &  \nodata &  \nodata &133.0\\
 Fe  I &4071.749 &  1.608 & -0.008 &  \nodata &  \nodata &  \nodata &  \nodata &120.5\\
 Fe  I &4114.451 &  2.831 & -1.303 & 72.4 &  \nodata & 50.7 &  \nodata &  \nodata\\
 Fe  I &4132.067 &  1.608 & -0.675 &  \nodata &  \nodata &  \nodata &  \nodata & 93.4\\
 Fe  I &4132.908 &  2.845 & -1.005 &  \nodata &  \nodata &  \nodata &  \nodata & 45.3\\
 Fe  I &4147.675 &  1.485 & -2.071 &  \nodata &109.8 &  \nodata &  \nodata &  \nodata\\
 Fe  I &4154.505 &  2.831 & -0.688 &  \nodata &  \nodata &  \nodata &  \nodata & 33.6\\
 Fe  I &4156.806 &  2.831 & -0.808 &  \nodata &  \nodata &131.7 &  \nodata &  \nodata\\
 Fe  I &4157.788 &  3.417 & -0.403 & 82.9 &  \nodata & 71.8 &  \nodata &  \nodata\\
 Fe  I &4174.917 &  0.915 & -2.938 &  \nodata &100.8 & 75.8 &  \nodata & 45.1\\
 Fe  I &4174.917 &  0.915 & -2.938 &  \nodata &  \nodata &113.5 &  \nodata &  \nodata\\
 Fe  I &4175.643 &  2.845 & -0.827 &105.6 &102.9 & 90.2 &  \nodata & 20.0\\
 Fe  I &4181.764 &  2.831 & -0.371 &  \nodata &  \nodata &  \nodata &  \nodata & 50.3\\
 Fe  I &4182.387 &  3.017 & -1.180 & 82.1 &  \nodata & 74.0 &  \nodata &  \nodata\\
 Fe  I &4187.047 &  2.449 & -0.514 &  \nodata &136.1 &  \nodata &  \nodata & 56.1\\
 Fe  I &4191.437 &  2.469 & -0.666 &  \nodata &  \nodata &  \nodata &  \nodata & 59.5\\
 Fe  I &4195.340 &  3.332 & -0.492 &  \nodata &  \nodata &  \nodata &123.5 &  \nodata\\
 Fe  I &4199.105 &  3.047 &  0.156 &  \nodata &  \nodata &135.9 &  \nodata &  \nodata\\
 Fe  I &4202.040 &  1.485 & -0.689 &  \nodata &  \nodata &  \nodata &  \nodata & 96.3\\
 Fe  I &4206.702 &  0.052 & -3.960 &140.1 &  \nodata &115.6 & 87.1 &  \nodata\\
 Fe  I &4216.191 &  0.000 & -3.357 &  \nodata &  \nodata &  \nodata & 93.3 & 47.6\\
 Fe  I &4222.221 &  2.449 & -0.914 &120.7 &110.4 &119.6 &  \nodata & 50.2\\
 Fe  I &4227.440 &  3.332 &  0.266 &  \nodata &  \nodata &  \nodata &  \nodata & 73.7\\
 Fe  I &4233.612 &  2.482 & -0.579 &  \nodata &  \nodata &142.9 &  \nodata &  \nodata\\
 Fe  I &4250.130 &  2.469 & -0.380 &  \nodata &  \nodata &  \nodata &  \nodata & 69.4\\
 Fe  I &4250.797 &  1.557 & -0.713 &  \nodata &  \nodata &  \nodata &  \nodata & 86.4\\
 Fe  I &4260.486 &  2.399 &  0.077 &  \nodata &  \nodata &  \nodata &  \nodata & 94.9\\
 Fe  I &4271.164 &  2.449 & -0.337 &  \nodata &  \nodata &  \nodata &  \nodata & 78.3\\
 Fe  I &4271.774 &  1.485 & -0.173 &  \nodata &  \nodata &  \nodata &  \nodata &138.1\\
 Fe  I &4282.412 &  2.176 & -0.779 &  \nodata &  \nodata &  \nodata &131.8 & 86.8\\
 Fe  I &4325.775 &  1.608 &  0.006 &  \nodata &  \nodata &  \nodata &  \nodata &139.1\\
 Fe  I &4337.055 &  1.557 & -1.704 &  \nodata &127.5 &  \nodata &  \nodata &  \nodata\\
 Fe  I &4369.779 &  3.047 & -0.803 &  \nodata & 88.9 &  \nodata &  \nodata &  \nodata\\
 Fe  I &4404.761 &  1.557 & -0.147 &  \nodata &  \nodata &  \nodata &  \nodata &142.2\\
 Fe  I &4415.135 &  1.608 & -0.621 &  \nodata &  \nodata &  \nodata &  \nodata &121.6\\
 Fe  I &4427.317 &  0.052 & -2.924 &  \nodata &  \nodata &  \nodata &  \nodata & 96.0\\
 Fe  I &4430.622 &  2.223 & -1.728 &  \nodata &  \nodata &138.1 &  \nodata &  \nodata\\
 Fe  I &4442.349 &  2.198 & -1.228 &  \nodata &135.2 &144.2 &  \nodata &  \nodata\\
 Fe  I &4443.201 &  3.071 & -1.043 &  \nodata &  \nodata &  \nodata &  \nodata & 21.7\\
 Fe  I &4447.728 &  2.223 & -1.339 &  \nodata &114.4 &119.3 &  \nodata &  \nodata\\
 Fe  I &4466.562 &  0.110 & -0.600 &  \nodata &128.0 &139.1 &  \nodata &  \nodata\\
 Fe  I &4494.573 &  2.198 & -1.143 &  \nodata &141.3 &  \nodata &  \nodata &  \nodata\\
 Fe  I &4602.949 &  1.485 & -2.208 &142.7 &105.5 &113.7 & 97.8 & 45.4\\
 Fe  I &4632.918 &  1.608 & -2.901 &107.6 &  \nodata & 98.3 &  \nodata &  \nodata\\
 Fe  I &4691.420 &  2.990 & -1.523 &  \nodata &146.1 & 76.1 & 78.6 &  \nodata\\
 Fe  I &4736.783 &  3.211 & -0.752 &106.3 &107.2 &113.8 &  \nodata &  \nodata\\
 Fe  I &4871.325 &  2.865 & -0.362 &  \nodata &157.9 &  \nodata &  \nodata &  \nodata\\
 Fe  I &4872.144 &  2.882 & -0.567 &  \nodata &158.2 &  \nodata &128.7 &  \nodata\\
 Fe  I &4890.763 &  2.875 & -0.394 &  \nodata &  \nodata &  \nodata &154.0 &  \nodata\\
 Fe  I &4891.502 &  2.851 & -0.111 &  \nodata &  \nodata &  \nodata &141.9 &  \nodata\\
 Fe  I &4903.316 &  2.882 & -0.926 &  \nodata &130.6 &137.8 &123.8 &  \nodata\\
 Fe  I &4918.998 &  2.865 & -0.342 &  \nodata &  \nodata &  \nodata &141.7 & 60.9\\
 Fe  I &4920.514 &  2.832 &  0.068 &  \nodata &  \nodata &  \nodata &  \nodata & 90.6\\
 Fe  I &4966.095 &  3.332 & -0.871 &125.4 &  \nodata &113.8 & 84.8 & 78.8\\
 Fe  I &4994.138 &  0.915 & -2.969 &137.1 &125.1 &112.6 &128.7 &  \nodata\\
 Fe  I &5001.870 &  3.881 &  0.050 &102.5 &  \nodata & 92.2 &109.0 &  \nodata\\
 Fe  I &5014.951 &  3.943 & -0.303 &102.9 &  \nodata & 81.2 &  \nodata &  \nodata\\
 Fe  I &5049.827 &  2.279 & -1.355 &  \nodata &157.2 &134.8 &111.7 &  \nodata\\
 Fe  I &5051.640 &  0.915 & -2.764 &  \nodata &158.7 &144.0 &124.4 &  \nodata\\
 Fe  I &5068.771 &  2.940 & -1.041 &132.4 &111.5 &106.9 & 85.8 &  \nodata\\
 Fe  I &5074.753 &  4.220 & -0.160 & 85.2 &  \nodata & 87.1 &  \nodata &  \nodata\\
 Fe  I &5083.345 &  0.958 & -2.842 &  \nodata &131.4 &122.3 &109.3 &  \nodata\\
 Fe  I &5110.435 &  4.260 & -3.758 &  \nodata &  \nodata &  \nodata &136.0 &  \nodata\\
 Fe  I &5123.730 &  1.011 & -3.058 &  \nodata &  \nodata &  \nodata &148.4 &  \nodata\\
 Fe  I &5127.368 &  0.915 & -3.249 &  \nodata &113.9 &  \nodata &  \nodata &  \nodata\\
 Fe  I &5216.283 &  1.608 & -2.082 &142.5 &122.4 &140.3 & 95.9 &  \nodata\\
 Fe  I &5225.534 &  0.110 & -4.755 &114.2 &  \nodata &  \nodata &  \nodata &  \nodata\\
 Fe  I &5232.952 &  2.940 & -0.057 &  \nodata &  \nodata &  \nodata &141.9 &  \nodata\\
 Fe  I &5269.550 &  0.859 & -1.333 &  \nodata &  \nodata &  \nodata &  \nodata &156.0\\
 Fe  I &5281.798 &  3.038 & -0.833 &142.6 &  \nodata &101.9 & 75.1 &  \nodata\\
 Fe  I &5283.629 &  3.241 & -0.524 &  \nodata &148.2 &  \nodata &  \nodata &  \nodata\\
 Fe  I &5383.380 &  4.312 &  0.645 &126.4 & 97.9 &103.7 & 94.4 & 34.0\\
 Fe  I &5393.176 &  3.241 & -0.715 &117.1 &103.3 & 91.0 & 74.3 &  \nodata\\
 Fe  I &5397.141 &  0.915 & -1.982 &  \nodata &  \nodata &  \nodata &  \nodata & 99.8\\
 Fe  I &5405.785 &  0.990 & -1.852 &  \nodata &  \nodata &  \nodata &  \nodata &108.9\\
 Fe  I &5424.080 &  4.320 &  0.520 &149.5 &108.2 &111.9 &115.4 & 39.2\\
 Fe  I &5429.706 &  0.958 & -1.881 &  \nodata &  \nodata &  \nodata &  \nodata &122.5\\
 Fe  I &5434.534 &  1.011 & -2.126 &  \nodata &  \nodata &  \nodata &  \nodata & 98.7\\
 Fe  I &5446.924 &  0.990 & -3.109 &  \nodata &  \nodata &  \nodata &  \nodata &103.7\\
 Fe  I &5497.526 &  1.011 & -2.825 &  \nodata &143.7 &  \nodata &135.7 & 66.6\\
 Fe  I &5501.477 &  0.958 & -3.046 &147.7 &134.1 &122.2 & 97.5 &  \nodata\\
 Fe  I &5506.791 &  0.990 & -2.789 &  \nodata &155.3 &  \nodata &136.0 & 65.6\\
 Fe  I &5569.631 &  3.417 & -0.500 &128.1 &104.9 &101.2 &  \nodata & 30.3\\
 Fe  I &5572.851 &  3.396 & -0.275 &  \nodata &141.8 &141.9 &  \nodata & 38.9\\
 Fe  I &5576.099 &  3.430 & -0.900 &103.6 & 85.3 & 81.3 &  \nodata &  \nodata\\
 Fe  I &5586.771 &  4.260 & -0.096 &  \nodata &  \nodata &145.2 &  \nodata & 57.5\\
 Fe  I &5763.002 &  4.209 & -0.450 & 74.3 & 73.5 & 85.5 &  \nodata &  \nodata\\
 Fe  I &6136.624 &  2.453 & -1.410 &  \nodata &  \nodata &  \nodata &  \nodata & 67.0\\
 Fe  I &6137.702 &  2.588 & -1.346 &  \nodata &150.5 &  \nodata &105.5 & 54.6\\
 Fe  I &6151.623 &  2.180 & -3.330 & 51.6 &  \nodata & 47.2 &  \nodata &  \nodata\\
 Fe  I &6173.341 &  2.220 & -2.863 & 71.7 & 61.7 & 60.9 & 54.6 &  \nodata\\
 Fe  I &6180.209 &  2.730 & -2.628 &  \nodata &  \nodata & 50.3 & 26.0 &  \nodata\\
 Fe  I &6187.995 &  3.940 & -1.673 & 40.7 &  \nodata & 26.7 &  \nodata &  \nodata\\
 Fe  I &6200.321 &  2.610 & -2.386 & 73.8 & 54.5 & 62.1 &  \nodata &  \nodata\\
 Fe  I &6219.287 &  2.200 & -2.428 & 99.1 &  \nodata & 81.9 & 69.4 &  \nodata\\
 Fe  I &6229.232 &  2.830 & -2.821 & 40.9 &  \nodata & 29.9 &  \nodata &  \nodata\\
 Fe  I &6230.736 &  2.559 & -1.276 &  \nodata &  \nodata &  \nodata &125.2 & 44.7\\
 Fe  I &6246.327 &  3.600 & -0.796 &101.0 &  \nodata & 91.9 & 91.7 &  \nodata\\
 Fe  I &6252.565 &  2.404 & -1.767 &120.9 &  \nodata &104.3 &107.5 & 44.5\\
 Fe  I &6254.253 &  2.280 & -2.435 &114.4 &  \nodata &105.7 & 82.6 &  \nodata\\
 Fe  I &6265.141 &  2.180 & -2.532 & 96.8 &  \nodata & 94.8 & 72.4 &  \nodata\\
 Fe  I &6270.231 &  2.860 & -2.543 & 46.0 &  \nodata & 43.7 & 22.1 &  \nodata\\
 Fe  I &6297.799 &  2.220 & -2.669 &101.4 &  \nodata &  \nodata &  \nodata &  \nodata\\
 Fe  I &6335.337 &  2.200 & -2.175 &  \nodata &  \nodata &108.9 & 81.7 & 31.0\\
 Fe  I &6336.830 &  3.690 & -0.667 &  \nodata &  \nodata & 89.9 & 69.1 &  \nodata\\
 Fe  I &6355.035 &  2.840 & -2.328 &  \nodata &  \nodata & 75.2 & 51.8 &  \nodata\\
 Fe  I &6393.612 &  2.430 & -1.505 &  \nodata &  \nodata &129.3 &134.7 & 47.3\\
 Fe  I &6411.658 &  3.650 & -0.646 &  \nodata &  \nodata & 94.4 & 73.5 &  \nodata\\
 Fe  I &6421.360 &  2.280 & -1.979 &  \nodata &  \nodata &  \nodata & 83.1 &  \nodata\\
 Fe  I &6430.856 &  2.180 & -1.954 &  \nodata &  \nodata &117.6 & 86.3 &  \nodata\\
 Fe  I &6481.878 &  2.280 & -2.985 & 74.3 &  \nodata &  \nodata &  \nodata &  \nodata\\
 Fe  I &6494.994 &  2.400 & -1.246 &  \nodata &  \nodata &  \nodata &151.1 & 73.8\\
 Fe  I &6498.945 &  0.960 & -4.675 & 94.2 &  \nodata &  \nodata &  \nodata &  \nodata\\
 Fe  I &6546.252 &  2.750 & -1.536 &128.4 &  \nodata &  \nodata &  \nodata &  \nodata\\
 Fe  I &6569.224 &  4.730 & -0.380 & 61.6 & 52.7 &  \nodata &  \nodata &  \nodata\\
 Fe  I &6593.874 &  2.430 & -2.377 &  \nodata & 85.1 & 74.7 & 64.5 &  \nodata\\
 Fe  I &6677.997 &  2.690 & -1.395 &  \nodata &  \nodata &123.1 &  \nodata & 30.5\\
 Fe  I &6703.576 &  2.760 & -3.059 & 42.3 &  \nodata & 39.9 &  \nodata &  \nodata\\
 Fe  I &6710.323 &  1.480 & -4.807 & 44.3 &  \nodata &  \nodata &  \nodata &  \nodata\\
 Fe  I &6750.164 &  2.420 & -2.592 & 89.7 & 71.8 & 71.1 & 86.3 &  \nodata\\
 Fe  I &6806.856 &  2.730 & -2.633 & 32.6 &  \nodata &  \nodata &  \nodata &  \nodata\\
 Fe  I &6839.835 &  2.560 & -3.378 & 28.8 &  \nodata &  \nodata &  \nodata &  \nodata\\
 Fe  I &6841.341 &  4.610 & -0.733 &  \nodata &  \nodata & 41.1 &  \nodata &  \nodata\\
 Fe  I &7130.925 &  4.300 & -0.708 & 86.9 &  \nodata &  \nodata &  \nodata &  \nodata\\
 Fe  I &7411.162 &  4.280 & -0.287 &  \nodata & 83.7 & 72.6 &  \nodata &  \nodata\\
 Fe  I &7445.758 &  4.260 &  0.053 &104.9 & 92.2 & 98.8 &  \nodata &  \nodata\\
 Fe  I &7461.527 &  2.560 & -3.507 & 55.7 &  \nodata & 45.3 &  \nodata &  \nodata\\
 Fe  I &7491.652 &  4.280 & -1.067 & 70.1 & 55.8 &  \nodata &  \nodata &  \nodata\\
 Fe  I &7531.153 &  4.370 & -0.557 &  \nodata &  \nodata & 64.1 &  \nodata &  \nodata\\
\\
 Fe II &4178.859 &  2.583 & -2.489 &  \nodata & 61.6 & 51.1 &  \nodata &  \nodata\\
 Fe II &4178.859 &  2.583 & -2.489 &  \nodata &  \nodata & 78.5 &  \nodata &  \nodata\\
 Fe II &4233.169 &  2.583 & -1.900 &  \nodata &  \nodata &131.2 &  \nodata &  \nodata\\
 Fe II &4508.289 &  2.856 & -2.318 & 93.1 & 73.2 & 65.4 &  \nodata &  \nodata\\
 Fe II &4515.343 &  2.844 & -2.422 &102.1 & 84.3 & 97.1 &  \nodata &  \nodata\\
 Fe II &4541.523 &  2.856 & -3.030 &  \nodata &  \nodata & 60.3 &  \nodata &  \nodata\\
 Fe II &4583.839 &  2.807 & -1.890 &130.1 &131.4 &140.2 &131.0 & 61.2\\
 Fe II &4923.930 &  2.891 & -1.307 &135.5 &118.8 &125.8 &121.8 & 74.7\\
 Fe II &5018.450 &  2.891 & -1.292 &  \nodata &147.0 &  \nodata &136.7 & 84.1\\
 Fe II &5534.848 &  3.245 & -2.790 & 59.6 &  \nodata &  \nodata &  \nodata &  \nodata\\
 Fe II &6456.391 &  3.903 & -2.075 &  \nodata &  \nodata & 56.0 &  \nodata &  \nodata\\
\\
 Co  I &6770.970 &  1.880 & -1.970 & 49.5 &  \nodata &  \nodata &  \nodata &  \nodata\\
 Co  I &6814.961 &  1.956 & -1.900 & 45.8 &  \nodata &  \nodata &  \nodata &  \nodata\\
 Co  I &6872.440 &  2.010 & -1.850 & 51.9 &  \nodata &  \nodata &  \nodata &  \nodata\\

\\
 Ni  I &6586.319 &  1.951 & -2.810 & 56.2 &  \nodata &  \nodata &  \nodata &  \nodata\\
 Ni  I &6643.638 &  1.676 & -2.300 & 97.8 & 97.7 & 83.3 &  \nodata &  \nodata\\
 Ni  I &6767.784 &  1.826 & -2.170 & 91.4 & 79.3 & 67.5 &  \nodata &  \nodata\\
 Ni  I &7122.206 &  3.542 &  0.040 & 98.6 &  \nodata &  \nodata &  \nodata &  \nodata\\
 Ni  I &7393.609 &  3.606 & -0.270 &  \nodata &  \nodata & 66.8 &  \nodata &  \nodata\\
 Ni  I &7414.514 &  1.986 & -2.570 &101.8 &  \nodata & 63.3 &  \nodata &  \nodata\\
 Ni  I &7422.286 &  3.635 & -0.140 &  \nodata & 63.3 & 62.8 &  \nodata &  \nodata\\
 Ni  I &7525.118 &  3.635 & -0.520 &  \nodata &  \nodata & 37.1 &  \nodata &  \nodata\\
\\
  Y II &4883.690 &  1.084 &  0.070 & 72.0 &  \nodata & 73.5 &  \nodata &  \nodata\\
\\
 Ba II &4554.036 &  0.000 &  0.163 &  \nodata &  \nodata &  \nodata &  \nodata & 85.0\\
 Ba II &4934.095 &  0.000 & -0.157 &  \nodata &  \nodata &  \nodata &  \nodata & 90.5\\
 Ba II &5853.688 &  0.604 & -1.010 &  \nodata &  \nodata & 87.8 & 86.5 &  \nodata\\
 Ba II &6141.727 &  0.704 & -0.076 &  \nodata &142.4 &  \nodata &112.9 & 67.5\\
 Ba II &6496.908 &  0.604 & -0.377 &  \nodata &  \nodata &  \nodata &141.8 & 59.4\\
\enddata

\tablecomments{~ Lines listed twice correspond to those measured in adjacent orders with overlapping wavelength coverage.}

\end{deluxetable}

\end{document}